\documentclass[iop,apj]{emulateapj}

 \usepackage{apjfonts}
\usepackage{graphicx}
 \usepackage{natbib}

\shorttitle{Turbulent mixing chemistry in disks}
\shortauthors{Semenov \& Wiebe}

\begin{document}

\title{Chemical evolution of turbulent protoplanetary disks and the Solar nebula}

\author{D. Semenov\altaffilmark{1}}
\affil{Max Planck Institute for Astronomy, K\"onigstuhl 17, D--69117 Heidelberg, Germany}
\altaffiltext{1}{Visiting Astronomer, Astronomy Department, University of California, Berkeley, CA 94720-3411, USA}   
\email{semenov@mpia.de}

\and

\author{D. Wiebe}
\affil{Institute of Astronomy of the RAS, Pyatnitskaya St. 48, 119017 Moscow, Russia}
\email{dwiebe@inasan.ru}

\begin{abstract}
This is the second paper in a series where we study the influence
of transport processes on the chemical evolution of protoplanetary disks.
Our analysis is based on a 1+1D flared $\alpha$-model of a $\sim5$~Myr
\object{DM Tau}-like system, coupled to a large gas-grain chemical network. To account for
production of complex molecules, the chemical network is supplied with
an extended set of surface reactions and photo-processes in ice mantles.
Our chemo-dynamical disk model covers a wide
range of radii, 10--800~AU (from a Jovian planet-forming zone to the outer disk edge). 
Turbulent transport of gases and ices is implicitly modeled in full 2D along with the time-dependent chemistry, 
using the mixing-length approximation. Two regimes are
considered, with high and low efficiency of turbulent mixing. 
The results of the chemical model with suppressed turbulent
diffusion are close to those from the laminar model, but not completely.
A simple analysis for the laminar chemical model to highlight potential sensitivity of a
molecule to transport processes is performed.
It is shown that the higher the ratio of the characteristic chemical timescale
to the turbulent transport timescale for a given molecule, the higher the
probability that its column density will be affected by diffusion.
We find that turbulent transport enhances abundances and column
densities of many gas-phase species and ices, particularly, complex ones. 
For such species a chemical steady-state is not reached due to long timescales
associated with evaporation and surface photoprocessing and recombination ($t\ga10^5$years).
When a grain with an icy mantle is transported from a cold disk
midplane into a warm intermediate/inner region, heavy radicals become
mobile on the surface, enriching the mantle with complex ices, which
are eventually released into the gas phase. The influence of turbulent mixing on disk chemistry is more pronounced
in the inner, planet-forming disk region where gradients of temperature and high-energy radiation intensities are steeper 
than in the outer region.
In contrast, simple radicals and molecular ions, which chemical
evolution is fast and proceeds solely in the gas phase, are not much affected by
dynamics. All molecules are divided into three groups according to the
sensitivity of their column densities to the turbulent diffusion. The molecules
that are unresponsive to transport include, e.g., C$_2$H, C$^+$, CH$_4$,
CN, CO, HCN, HNC, H$_2$CO, OH, as well as water and ammonia ice. Their column densities
computed with the laminar and 2D-mixing model differ by a
factor of $\la 2-5$ (``steadfast'' species). Molecules which vertical
column densities in the laminar and dynamical models differ by up to 2
order of magnitude include, e.g., C$_2$H$_2$, some carbon chains, CS, H$_2$CS,
H$_2$O, HCO$^+$, HCOOH, HNCO, N$_2$H$^+$, NH$_3$, CO ice, H$_2$CO ice, CH$_3$OH
ice, and electrons (``sensitive'' species). Molecules which column densities are 
altered by diffusion by more than 2 orders of magnitude include, e.g., C$_2$S,
C$_3$S, C$_6$H$_6$, CO$_2$, O$_2$, SiO, SO, SO$_2$, long carbon chain ices,
CH$_3$CHO ice, HCOOH ice, O$_2$ ice, and OCN ice (``hypersensitive'' species).
The chemical evolution of assorted molecules in the laminar and turbulent models is 
thouroughly analyzed and compared with previous studies. 
We find that column densities of observed gas-phase molecules in the DM Tau disk are well
reproduced by both the laminar and the chemo-dynamical disk models. The observed abundances of 
many reduced and oxidized cometary ices are also successfully reproduced by the both models.
We indicate several observable or potentially detectable tracers of
transport processes in protoplanetary disks and the Solar nebula, 
e.g., elevated concentrations of heavy hydrocarbon ices, complex organics, CO$_2$,
O$_2$, SO, SO$_2$, C$_2$S, C$_3$S compared to CO and the water ice.
A combination of UV photodesorption, grain growth, and turbulent mixing leads to 
non-negligible amount of molecular gases in the cold disk midplane. 
\end{abstract}

\keywords{accretion, accretion disks --- astrochemistry --- molecular processes --- 
protoplanetary disks --- stars: \object{DM Tau} --- turbulence}

\section{Introduction}
One of the most exciting questions in astrophysics is 
the genesis of prebiotic molecules served as life-building blocks in the Solar system and 
preserved in meteorites and comets. Nowadays protoplanetary disks 
are believed to be birth places of planetary systems, so we may expect prebiotic molecules to be the eventual outcome of the disk evolution. However, our understanding 
of the chemical composition and evolution of protoplanetary disks is far from being complete. Apart from CO 
and its isotopologues, and occasionally HCO$^+$, DCO$^+$, CN, HCN, DCN, CCH, H$_2$CO, and 
CS, the molecular content of protoplanetary disks 
remains largely unknown \citep[e.g.,][]{DGG97,Kastner_ea97,Aikawa_ea03,Thi_ea04,Pietu_ea07,Qi_ea08,Henning_ea10}.
Molecular line data are limited in their sensitivity and resolution. This 
means that spatial distribution of molecular abundances in disks is still poorly determined
\citep[e.g.,][]{Pietu_ea05,Dutrey_ea07,Panic_ea09}, hampering a detailed comparison with existing chemical models, which is often
based on global data (e.g. integrated line profiles). 

Multi-molecule, multi-transition interferometric observations, coupled to line
radiative transfer and chemical modeling, allowed to constrain disk sizes,
kinematics, distribution of temperature, surface density, and molecular column
densities (see reviews by \citet{Bergin_ea07} and \citet{DGH07}).
The measured line intensities are indicative of vertical temperature
gradients in disks \citep[e.g.,][]{DDG03,Qi_ea06}, though several disks with large
inner cavities do not show evidence for such a gradient \citep[e.g.,
GM~Aur and \object{LkCa 15};][]{Dutrey_ea08,Hughes_ea09}. A significant reservoir of very cold CO,
HCO$^+$, CN and HCN gases has been found in the disk of \object{DM Tau} at
temperatures $\la$ 6-17~K, which cannot be explained by conventional chemical
models without invoking a non-thermal desorption or transport mechanism
\citep[e.g.,][]{Semenov_ea06,Aikawa_07,Hersant_ea09}.
Non-thermal broadening of emission lines of $\sim 0.1$~km\,s$^{-1}$ has been
reported \citep[e.g.,][]{Bergin_ea07,DGH07,Hughes_ea11a}, which is likely due to subsonic turbulence driven by the
magnetorotational instability \citep{MRI}.

Recently, with space-borne ({\it Spitzer}) and ground-based (Keck, VLT, Subaru) infrared telescopes,
molecules have been detected in very inner zones of 
planet-forming systems, at $r \la 1-10$~AU.
Rotational-vibrational emission lines from CO, CO$_2$, C$_2$H$_2$,
HCN, OH, H$_2$O imply a rich chemistry driven
by endothermic reactions or reactions with activation barriers and photoprocesses
\citep{Lahuis_ea06,Carr_Najita08,Salyk_ea08,Pontoppidan_ea08,Pascucci_ea09,
vdPlas_ea09,Salyk_ea11a}.
Through {\it ISO} and {\it Spitzer} infrared spectroscopy abundant
ices in cold disk regions consisting of water ice and
substantial amounts ($\sim 1-30$\%) of volatile materials like CO, CO$_2$,
NH$_3$, CH$_4$, H$_2$CO, and HCOOH have been detected
\citep[e.g.,][]{Pontoppidan_ea05,Terada_ea07a,Zasowski_ea08}.

The conditions of planets formation in the early Solar system have been
revealed by a detailed analysis of chemical and mineralogical composition of
meteoritic samples and cometary dust particles
\citep[e.g.,][]{Bradley_05}. The recent {\it Stardust} and {\it Genesis}
space missions have returned first samples of pristine materials, likely of
cometary origin, showing a complex structure of high-temperature crystalline
silicates embedded in low-temperature condensates
\citep{Brownlee_ea04,Flynn_ea06,Brownlee_ea08}.
Comets have been assembled around or beyond Neptune and expelled
gravitationally outward, however the presence of
Mg-rich crystalline silicates in cometary dust indicates annealing of
amorphous presolar grains at temperatures above 800~K
\citep{Wooden_ea99,Wooden_ea07}.
The presence of crystalline silicates
in outer regions of protoplanetary disks has also been revealed
\citep[e.g.,][]{vanBoekel_ea04,Juhasz_ea10a}. 

Recently, an anti-correlation between the
age of a disk and the X-ray hardness/luminosity of a central star, and the
observed crystallinity fraction has been inferred, making the overall picture
even more complicated \citep[e.g.,][]{Glauser_ea09}.

An isotopic analysis of refractory condensates in unaltered chondritic
meteorites shows strong evidence that the inner part of the Solar Nebula has
been almost completely mixed during the first several Myr of evolution
\citep[e.g.,][]{Boss2004,Ciesla_09}. This mixing (either advective or
turbulent), along with high-energy irradiation, could have also been important
for fractionation ratios in both gas-phase and solid compounds in the nebula
\citep[e.g.,][]{Clayton_93,Clayton_Mayeda96,LeeT_ea98,Hersant_ea01,Lyons_Young05}.
The rich variety of organic compounds in meteorites, including amino acids, suggest that
these complex species have formed just prior or during the formation of planets
in heavily irradiated, warm regions of the Solar Nebula
\citep[e.g.,][]{Ehrenfreund_Charnley00,Busemann_ea06}. Combustion and pyrolysis
of hydrocarbons at high temperatures, coupled with outward transport, has been
inferred to explain the omni-presence of kerogene-like (mainly aromatic)
carbonaceous material in meteoritic and cometary samples \citep{Morgan_ea91}.

These intriguing findings are partly understood in modern astrochemical models
of protoplanetary disks
\citep{wl00,Aea02,Mea02,vZea03,IHMM04,Kamp_Dullemond94,Semenov_ea05,Aikawa_ea06,
TG07,Agundez_ea08,Woods_Willacy08,Visser_ea09,Walsh_ea10}. The major result of the
chemical modeling is that disks have a layered chemical structure
due to heavy freeze-out of gas-phase molecules in the cold midplane and their
photodissociation in the atmosphere. Vertical column densities
of CO, HCO$^+$, N$_2$H$^+$, CN, HCN,
HNC, CS, etc. are reproduced with modern chemical models
\citep[e.g.,][]{Aea02,Semenov_ea05,Willacy_ea06,Dutrey_ea07,Schreyer_ea08,Henning_ea10}.

While most of the chemical studies are still based on laminar disk models, evidences for mixing, mentioned above, call for a more sophisticated treatment. A few such
dynamical studies have been presented. Models of the early Solar nebula with radial
transport by advective flows have been developed 
\citep[e.g.,][]{Morfill_Voelk84,G01,G02,Wehrstedt_Gail02,Boss2004,Keller_Gail04}.
\citet{IHMM04} for the first time simultaneously modeled the influence of turbulent diffusion in the vertical direction and
advection flows in the radial direction on the chemical composition of the
inner disk region. They found that dynamical processes significantly
affect the chemical evolution of sulfur-bearing species. \citet{Willacy_ea06}
have shown that 1D vertical mixing modifies chemical composition of the outer
disk region and that the mixing results better agree to observations.
\citet[][Paper~I]{Semenov_ea06} and \citet{Aikawa_07} have found that turbulent transport
allows explaining the presence of a large amount of cold ($\la15$~K) CO gas in the
disk of DM Tau.
\citet{TG07} have used a 2D disk chemo-hydrodynamical model and showed that in
the disk midplane matter moves outward, carrying out the angular momentum, while
the accretion flows toward the star are located at elevated
altitudes. Consequently, gas-phase species produced by warm chemistry in the inner
nebula can be steadily transported into the cold outer region and freeze out.
\citet{TG07} have claimed that global radial advective flows dominate over diffusive
mixing for the disk chemical evolution.
A radial advection model has also been utilized by \citet{Nomura_ea09}, who have
demonstrated that inward radial transport enhances abundances of organic
molecules (produced mainly on dust surfaces in cold outer regions).
\citet{Woods_Willacy08} have elaborated a disk chemical model with
improved heating and cooling balance and accurate modeling of the UV radiation
field and found that the $^{12}$C/$^{13}$C fractionation in the Solar system
comets can be explained by the reprocessing of presolar materials in a warm
nebular region. \citet{Hersant_ea09} have studied various mechanisms to retain gas-phase
CO in very cold disk regions. They concluded that efficient photodesorption
in moderately obscured disk regions ($A_{\rm V}<5^{\rm m}$) greatly enhances gas-phase CO
concentrations, while the role of vertical mixing is less important.
Finally, \citet{Heinzeller_ea11} have investigated the disk chemical evolution with
radial advection, vertical mixing, and vertical wind transport processes.  They have found that the disk wind has
a negligible effect on disk chemistry, whereas the radial accretion alters the molecular abundances
in the cold midplane, and the vertical turbulent mixing affects the chemistry in the warm molecular layer.
The abundances of NH$_3$, CH$_3$OH, C$_2$H$_2$ and sulfur-containing species are the most enhanced by the transport.

In this paper we continue our detailed study of chemo-dynamical interactions
in protoplanetary disk started in Paper~I with an intent to find out if chemistry can be used as a diagnostic of dynamical processes in a protoplanetary accretion disk. For the first time we utilize a large-scale disk
physical model along with an extended gas-grain chemical network coupled
to 2D turbulent transport. 
A wide range of temperatures, densities, and X-ray/
UV radiation intensities encountered in this dynamical model allows us
to follow formation and destruction of various molecules, possibly
detectable with ALMA and {\it Herschel}. The primary aim of the present study is to 
characterize the importance of turbulent diffusion for the chemical evolution in various disk domains and
for various chemical families.
We argue that even though the overall efficiency of the diffusive
transport in the outer disk is dominated by vertical
mixing, one has to consider vertical and radial mixing simultaneously in the
planet-forming region. Many complex ices and their gas-phase counterparts
are enhanced by turbulent diffusion, in particular sulfur-bearing and other heavy 
(complex) species.

The organization of the paper is the following. In Section~\ref{model}
we describe the adopted disk physical model and the chemical network. In
Sect.~\ref{timescales} basic chemical and dynamical timescales in
protoplanetary disks are outlined and discussed. A general scheme to estimate
possible sensitivity of a given molecule to transport processes is presented in Sect.~\ref{mim}.
Influence of the 2D-turbulent diffusion on the chemical evolution of dominant
ions as well as C-, O-, N-, S-bearing species and complex organic
molecules is studied in detail in Sects.~\ref{ions}-\ref{complex}. 
Detailed comparison with the previous studies and future improvements of chemo-dynamical models 
are discussed in Sect.~\ref{sec:diss:studies}.
We discuss importance of turbulent diffusion for the presence of cold gases in disk
midplanes in Sect.~\ref{sec:diss:cold_gases}. We verify feasibility of our chemical
and physical disk models by comparing the calculated and observed column
densities in the DM Tau disk and abundances of cometary ices in the Solar system (Sect.~\ref{sec:diss:obs}). 
Finally, detected or potentially detectable molecular tracers of transport processes in
protoplanetary disks are summarized in Sect.~\ref{sec:diss:tracers}. Summary and
conclusions follow.

\section{Model}
\label{model}

\subsection{Disk structure}
\label{disk_model}
\begin{figure*}
\includegraphics[angle=90,width=0.75\textwidth]{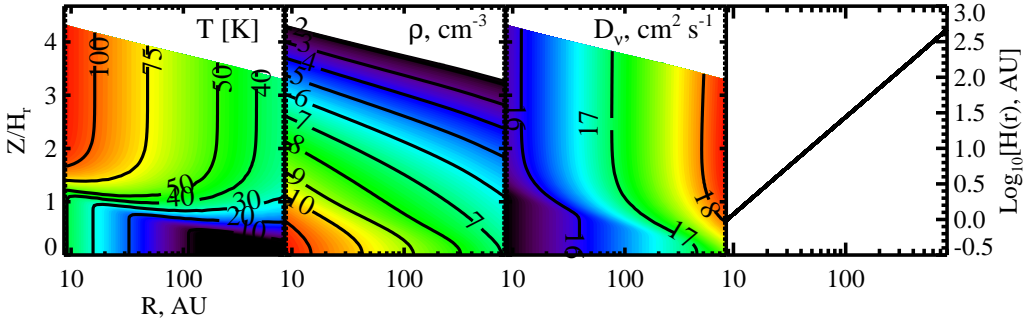}
\caption{(Left to right) Distributions of the temperature, particle
density ($\log_{10}$ scale), diffusion coefficient in cm$^2$\,s$^{-1}$
($\log_{10}$ scale), and pressure scale height in AU ($\log_{10}$ scale)
in the DM~Tau disk model. In the first 3 panels the Y-axis is given in units of
the pressure scale height.}
\label{fig:disk_struc}
\end{figure*}

We focus on the DM Tau system because it is one of the most observationally studied 
and molecularly-rich protoplanetary disk around a T Tauri star.
The adopted flaring disk structure is based on a 1+1D steady-state $\alpha$-model similar
to \citet{DAea99} model.  
The DM Tau is a single M0.5 dwarf ($T_{\rm eff}=3720$~K), 
with a mass of $0.65M_\odot$, and a radius of $1.2R_\odot$
\citep[][]{Mazzitelli89,Simon_ea00}. The non-thermal FUV radiation field from DM Tau is
represented by the scaled ISRF of \citet{G}, 
with the un-attenuated intensity at 100~AU of $\chi_*(100)=410$ \citep[e.g.,][]{Bergin_ea04}. 
For the X-ray luminosity of the star we adopt a value of $10^{30}$~erg\,s$^{-1}$, which is constrained by 
recent measurements with Chandra and XMM in the range of 0.3-10~keV (M.~Guedel, priv. comm.).

The disk has an inner radius $r_0=0.03$~AU (dust sublimation front, $T\approx 1\,500$~K),
an outer radius $r_1=800$~AU, an accretion rate
$\dot{M}= 4\,10^{-9}\,M_\odot$\,yr$^{-1}$, a viscosity parameter
$\alpha = 0.01$, and a mass of $M=0.066\,M_\odot$ \citep{Dutrey_ea07,Henning_ea10}.
The DM Tau disk age is about 5--7~Myr \citep[][]{Simon_ea00}, so we adopted 5~Myr as a limiting time in our chemical simulations.
According to the {\it Spitzer} IRS observations \citep{Calvet_ea05},
the inner DM Tau disk is cleared of small dust ($\la 3$~AU) and is in a pre-transitional phase already.
Therefore, in the chemical simulations a disk region beyond 10~AU is considered, where dust grain evolution
seems to be slow and grain growth is moderate \citep{Birnstiel_ea10a}.
In chemical modeling, the dust grains are assumed to  be uniform $0.1\,\mu$m amorphous olivine particles (with density of
$\rho_d=3$~g\,cm$^{-3}$). This is the size representing a mean radius in the dust size distribution.
Equal gas and dust temperatures are assumed. 
Gas becomes hotter than dust only in an upper, tenuous and heavily irradiated disk layer, which often has 
negligible contribution to molecular column densities.

The turbulence in disks is likely driven by the magnetorotational instability (MRI), which is operative even in a 
weakly ionized medium, and is essentially a 3-D phenomenon \citep[e.g.,][]{MRI}. This turbulence causes anomalous viscosity that 
enables efficient redistribution of the angular momentum and accretion of disk matter onto the star. However, inner disk midplane 
can be almost neutral and decoupled from magnetic fields, forming a region with reduced, inherited turbulence (``dead zone''), 
\citet[see, e.g.,][]{gammie,sano,Red2,Wunsch_ea06}. With modern computational facilities coupled chemo-MHD 3-D models are only 
manageable for extremely limited chemical networks and a restricted disk domain \citep[see, for example,][]{Turner_ea07}. We have
followed the parametrization of \citet{ShakuraSunyaev73}, where turbulent viscosity $\nu$ is related to local disk
properties such as the characteristic (vertical) spatial scale $H(r)$, the sound speed $c_{\rm s}(r,z)$, 
and the dimensionless parameter $\alpha$:
\begin{equation}
 \nu(r,z) = \alpha\,c_{\rm s}(r,z)\,H(r).
\end{equation}
From observational constraints $\alpha$ is $\sim 0.001-0.1$ \citep[][]{Andrews_Williams07,Guilloteau_ea11a}, similar to 
values obtained from MHD modeling of the MRI \citep[e.g.,][]{Dzyurkevich_ea10}. 
However, the magnitude of MHD viscous stresses changes throughout the disk, and thus in this simplistic parametrization the 
$\alpha$-parameter should also be variable. Unfortunately, without detailed MHD studies it is hard to characterize $\alpha$, 
so we adopt the constant value of $0.01$. Since only a disk region beyond 10~AU is studied,
a ``dead zone'' where effective $\alpha$ is very low, $\la 10^{-4}$, is avoided. 

Consequently, the diffusion coefficient is calculated as
\begin{equation}
D_{\rm turb}(r,z) = \nu(r,z)/Sc,
\end{equation}
where $Sc$ is the Schmidt number describing efficiency of turbulent diffusivity \citep[see
e.g.][]{ShakuraSunyaev73,SchraeplerHenning04}.
In our simulations we assume that gas-phase species
and dust grains are well mixed, and transported with the same diffusion coefficient. We treat diffusion of mantle materials
similarly to gas-phase molecules, without relating it to individual grain dynamics. 
Hence two chemo-dynamical models are considered: (1) the high-efficiency mixing model with $Sc=1$ and (2) the 
low-efficiency mixing model with $Sc=100$. In the $Sc=1$ model the diffusion coefficient in outer disk regions is 
$\sim 10^{18}$\,cm$^2$\,s$^{-1}$, similar to \citet{Willacy_ea06}. 
The second model represents a hypothetical case when
mixing of molecules occurs much slower than turbulent eddy turn-over speeds yet faster than in the pure laminar gas.
The temperature and density structure, diffusion coefficient $D_{\rm turb}$, and vertical 
pressure scale height $H(r)$ of the DM Tau-like disk model are shown in Fig.~\ref{fig:disk_struc}.

\subsection{Chemical network}
\label{chem_model}
The adopted gas-grain chemical model is described in our recent papers on benchmarking of disk chemical
models \citep{Semenov_ea10} and observations of CCH in \object{DM Tau}, \object{LkCa 15}, and \object{MWC 480} 
\citep{Henning_ea10}. A brief summary is provided below. 
The chemical network is based on the osu.2007 ratefile with recent updates to
reaction rates\footnote{See: \url{http://www.physics.ohio-state.edu/$\sim$eric/research.html}}. 
(Note that in Paper~I we used UMIST\,95 ratefile.)
A new class of X-ray-driven reactions leading to production of O$^{++}$, C$^{++}$, N$^{++}$,
S$^{++}$, Fe$^{++}$, Si$^{++}$ is added. Their neutralization reactions by electrons and charge transfer reactions 
with molecules are adopted from \citet{Staeuber_ea05}. The photoionization cross sections are
taken from \citet{Verner_ea93}, as described in \citet{Maloney_ea96}. Secondary electron impact ionization 
cross sections are taken from \citet{MS_05}. 

To calculate UV ionization and dissociation rates, the mean FUV intensity at
a given disk location is obtained by adding the stellar $\chi_*(r)=410\,(r, {\rm AU})/(100~{\rm AU})^{2}$ 
and interstellar $\chi_0$ components that are scaled down by the visual extinction in the vertical
direction and in the direction to the central star (1D plane-parallel
approximation). Several tens of photoreaction rates are updated using the new calculations of \citet{vDea_06},
which are publicly available\footnote{\url{http://www.strw.leidenuniv.nl/~ewine/photo/}}. 
The self-shielding of H$_2$ from photodissociation is calculated by  Eq.~(37) from \citet{DB96}.
The shielding of CO by dust grains, H$_2$, and its self-shielding is calculated using precomputed table of
\citet[][Table~11]{Lea96}.

We model the attenuation of cosmic rays (CRP) by
Eq.~(3) from \citet{Red2}, using the standard CRP ionization rate $\zeta_{\rm CR}=1.3\,10^{-17}$~s$^{-1}$.  
Ionization due to
the decay of short-living radionuclides is taken into account, $\zeta_{\rm RN}=6.5\cdot10^{-19}$~s$^{-1}$
\citep{FG97}. The stellar
X-ray radiation is modeled using observational results of
\citet{Glassgold_ea05} and the approximate expressions~(7--9) from the 2D Monte Carlo simulations of
\citet{zetaxa,zetaxb}. The typical X-ray photon energy is 3~keV, and the X-ray emitting source is located at 12
stellar radii above the midplane. The X-ray ionization rates exceed that of the CRPs in the disk regions above
the midplane.

The gas-grain interactions include sticking of neutral species and electrons to uniformly-sized dust grains with 100\%
probability, 
release of frozen molecules by thermal, CRP-, and UV-induced desorption, dissociative recombination and
radiative neutralization of ions on charged grains, and grain re-charging. We do not allow H$_2$ to stick to grains 
because the binding energy of H$_2$ to pure H$_2$ mantle is low, $\sim 100$~K \citep{Lee_72}, and it freezes out
in substantial quantities only at temperatures below $\approx 4$~K.
Chemisorption of surface molecules is not considered. We considered various UV photodesorption yields between $10^{-5}$
and $10^{-3}$ \citep[e.g.,][]{Greenberg73,Oeberg_ea09a,Oeberg_ea09b} and found that in this range the exact yield value has a 
negligible impact on the modeling results. 
To allow synthesis of complex (organic) molecules, an 
extended list of surface reactions together with desorption energies and a list of photodissociation reactions of surface species 
is adopted from \citet{Garrod_Herbst06}. 
Desorption energies for assorted molecules are listed in Table~\ref{tab:des_E}. 
We assume that each $0.1\mu$m spherical olivine grain provides $\approx 2\,10^6$ surface sites, and that surface recombination
proceeds solely through the Langmuir-Hinshelwood formation mechanism. Upon a surface recombination, there is a 5\% chance for
the products to leave the grain.
Following interpretations of experimental results on
the formation of molecular hydrogen on dust grains \citep{Katz_ea99}, we employ the standard rate equation approach to the 
surface chemistry without H and H$_2$ tunneling either through the potential walls of the surface sites or through reaction
barriers.
As it has been shown by \citet{Vasyunin_ea09}, when surface recombination rates are slow, the stochastic effects are of no
importance 
for the surface chemistry.

Overall, the disk chemical network consists of 657 species made of 13 elements, and 7306 reactions.
In contrast to the Paper~I, we do not apply our ``Automatic Reduction Technique'' (ART) to reduce the size of this network
(thanks to increased performance of our chemo-dynamical code).
The ``low metals'' initial abundances of \citet{Lea98} are utilized (Table~\ref{tab:inabun}).
The choice of initial abundances does not affect the
resulting molecular abundances and column densities due to relatively high
densities in and long evolutionary timescales of protoplanetary disks, which essentially reset a chemical ``clock''
\citep[e.g.,][]{Willacy_ea06}.

\subsection{Modeling chemistry with transport}
\label{chem_code}
\begin{deluxetable}{ll}
\tablecaption{Desorption energies\label{tab:des_E}}
\tablehead{
\colhead{Species} & \colhead{Energy, K}
}
\startdata
 C              &  800 \\
 C$_2$             &  1600 \\
 C$_2$H            &  2140 \\
 C$_2$H$_2$           &  2590 \\
 C$_2$S            &  2700 \\
 C$_3$             &  2400 \\
 C$_3$H$_2$           &  3390 \\
 C$_6$H$_6$           &  7590 \\
 C$_8$H$_2$           &  7390 \\
 CH$_2$OH          &  5080 \\
 CH$_2$CO         &  2200 \\
 CH$_3$CHO         &  2870 \\
 CH$_3$OH          &  5530 \\
 CH$_4$            &  1300 \\
 CN             &  1600 \\
 CO             &  1150 \\
 CO$_2$            & 2580 \\
 CS             &  1900 \\
 H              &  624 \\
 H$_2$             &  552 \\
 H$_2$S             & 2740 \\
 H$_2$CO           &  2050 \\
 H$_2$CS           &  2700 \\
 H$_2$O            &  5700 \\
 HCN            &  2050 \\
 HCOOH          &  5570 \\
 HNC            &  2050 \\
 HNCO           &  2850 \\
 HNO            &  2050 \\
 N              &  800  \\
 N$_2$             &  1000 \\
 NH             &  2380 \\
 NH$_2$            &  3960 \\
 NH$_3$            &  5530 \\
 NO             &  1600 \\
 O              &  800 \\
 O$_2$             &  1000 \\
 OCN            &  2400  \\
 OH             &  2850  \\
 S              &  1100  \\
 SO             &  2600  \\
 SO$_2$            &  3400 \\
\enddata
\end{deluxetable}

\begin{deluxetable}{ll}
\tablecaption{Initial abundances\label{tab:inabun}}
\tablehead{
\colhead{Species} & \colhead{Relative abundance}
}
\startdata
H$_2$&   $0.499$     \\
H    &   $2.00 (-3)$  \\
He   &   $9.75 (-2)$  \\
C    &   $7.86 (-5)$  \\
N    &   $2.47 (-5)$  \\
O    &   $1.80 (-4)$  \\
S    &   $9.14 (-8)$  \\
Si   &   $9.74 (-9)$  \\
Na   &   $2.25 (-9)$  \\
Mg   &   $1.09 (-8)$  \\
Fe   &   $2.74 (-9)$  \\
P    &   $2.16 (-10)$ \\
Cl   &   $1.00 (-9)$  \\
\enddata
\end{deluxetable}

\begin{deluxetable}{llll}
\tablecaption{Characteristic Timescales: Inner Disk (10 AU)\label{tab:tau_inner}}
\tablehead {\colhead{Processes} & \colhead{Midplane} & \colhead{Warm layer$^*$} & \colhead{Atmosphere$^*$}\\
\colhead{} & \colhead{[yr]} & \colhead{[yr]} & \colhead{[yr]}}
\startdata
Mixing &       3.4 (3)&       3.4 (3)&       1.3 (3)\\
Gas-phase &   1.4 (-5)&  1.3 (-4)&    1.0 (-2)\\
UV &       >1.0 (7) &       3.3 (4)&       5.9 (2)\\
Accretion &     1.2 (-2)&      1.1 (-1)&       5.4 (0)\\
Desorption &   5.8 (-7)&   6.0 (-7)&   <1.0 (-7)\\
Surface &       >1.0 (7) &       >1.0 (7) &   <1.0 (-7)\\
\enddata
\tablenotemark{*}\tablenotetext{*}{The warm layer and atmosphere are located at the $z/H_{\rm r}=0.8$ and $1.75$, respectively.}
\end{deluxetable}

\begin{deluxetable}{llll}
\tablecaption{Characteristic Timescales: Outer Disk (250 AU)\label{tab:tau_outer}}
\tablehead {\colhead{Processes} & \colhead{Midplane} & \colhead{Warm layer$^*$} & \colhead{Atmosphere$^*$}\\
\colhead{} & \colhead{[yr]} & \colhead{[yr]} & \colhead{[yr]}}
\startdata
Mixing &       1.0 (6) &       2.5 (5)&       1.4 (5)\\
Gas-phase &     2.0 (-2)&      1.8 (-1)&       2.9 (0)\\
UV &       >1.0 (7)&       1.2 (6)&       3.1 (1)\\
Accretion &       2.7 (1)&       1.8 (2)&       2.3 (3)\\
Desorption &       1.0 (6)&       4.3 (0)&   <1.0 (-7)\\
Surface &       >1.0 (7)&       >1.0 (7)&       1.4 (5)\\
\enddata
\tablenotemark{*}\tablenotetext{*}{The warm layer and atmosphere are located at the $z/H_{\rm r}=0.8$ and $1.75$, respectively.}
\end{deluxetable}

\begin{figure*}
\includegraphics[width=0.3\textwidth]{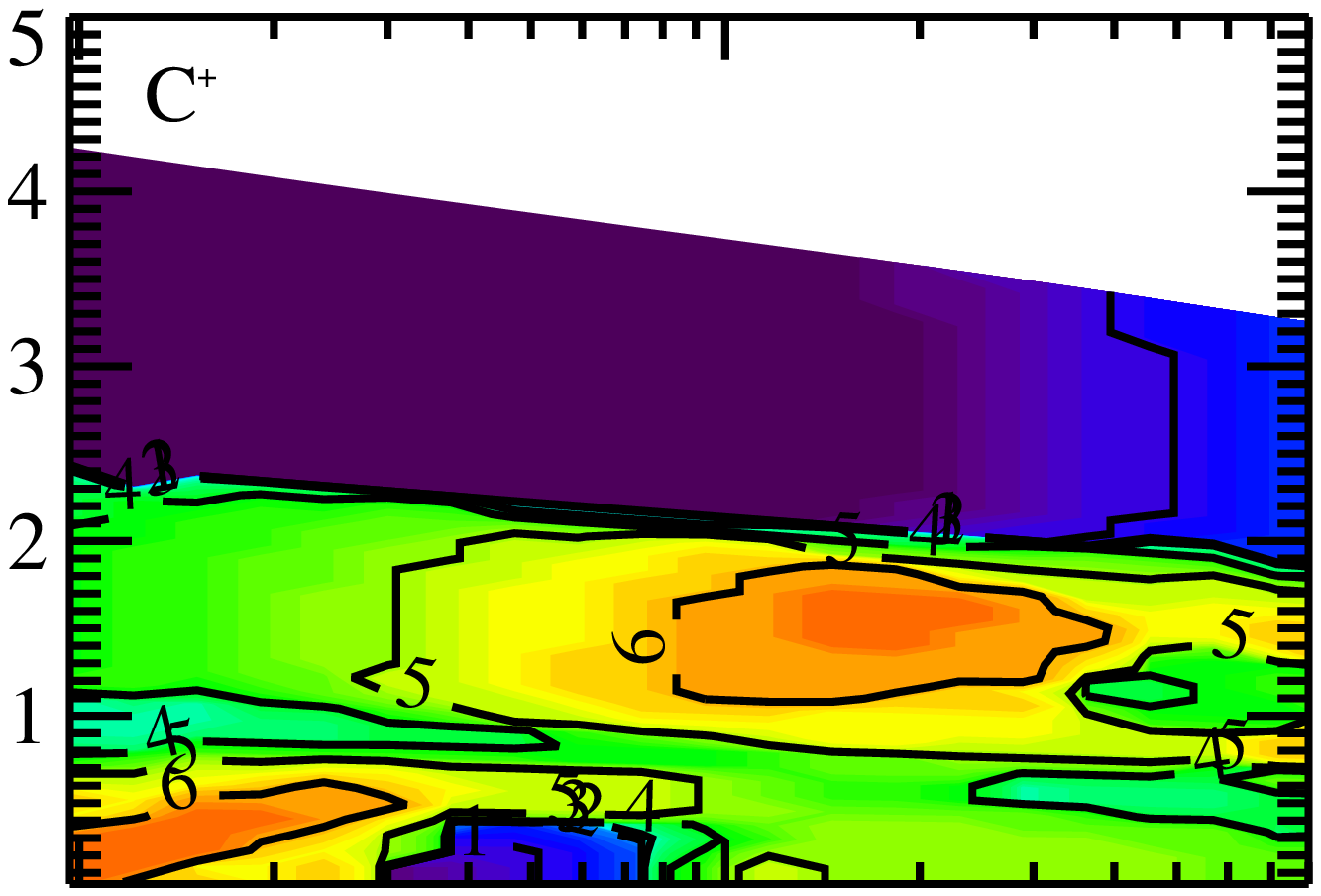}
\includegraphics[width=0.3\textwidth]{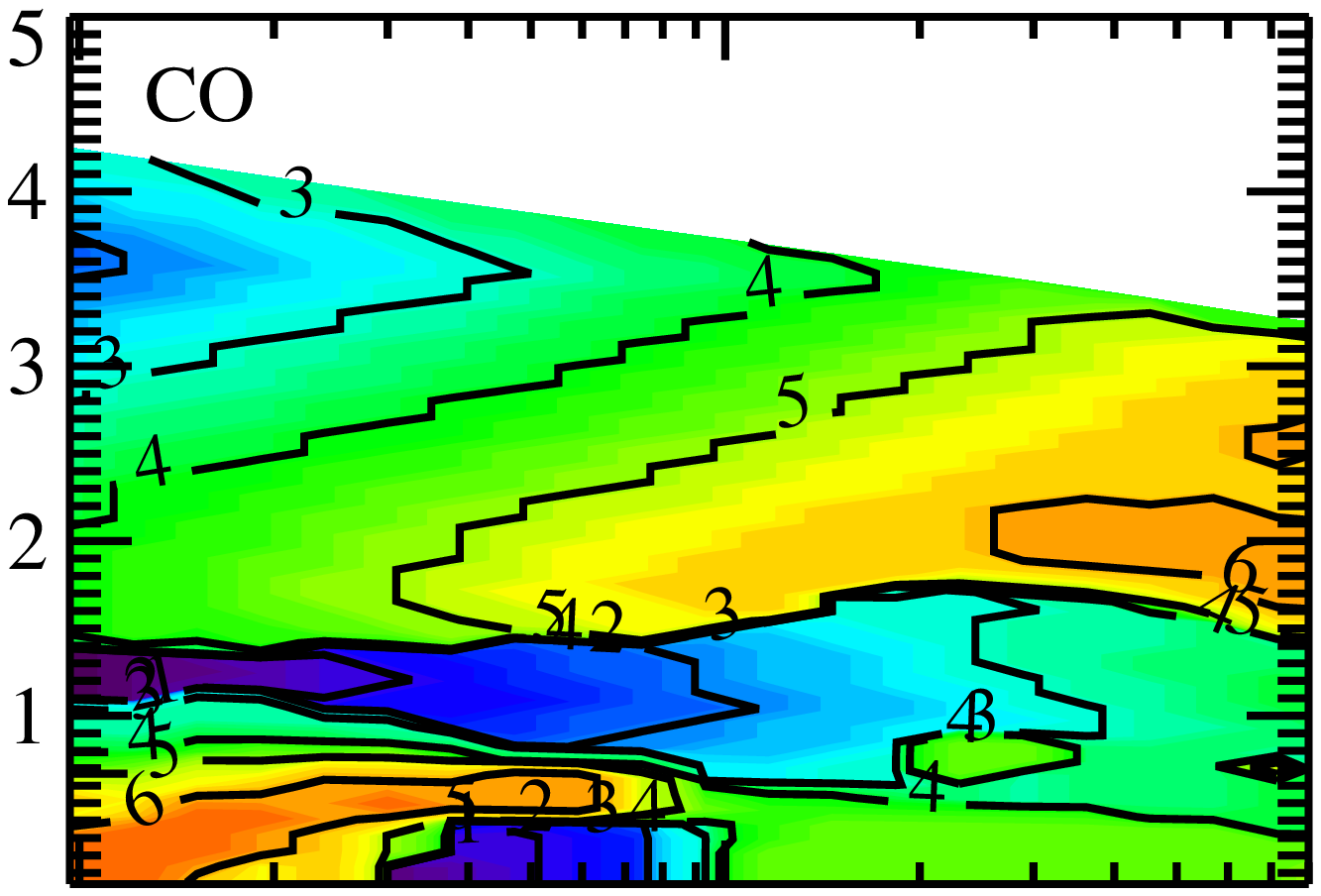}\\
\includegraphics[width=0.3\textwidth]{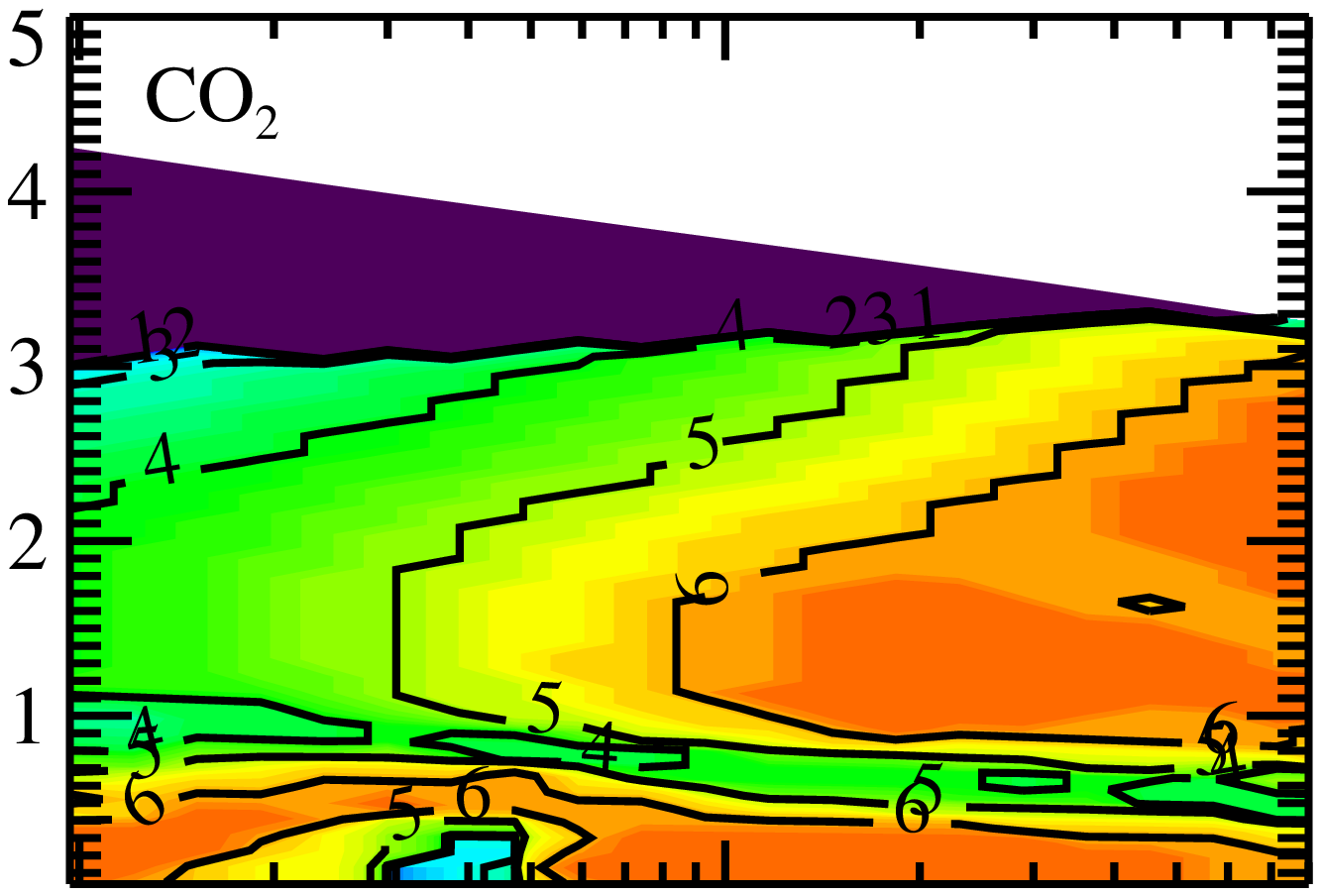}
\includegraphics[width=0.3\textwidth]{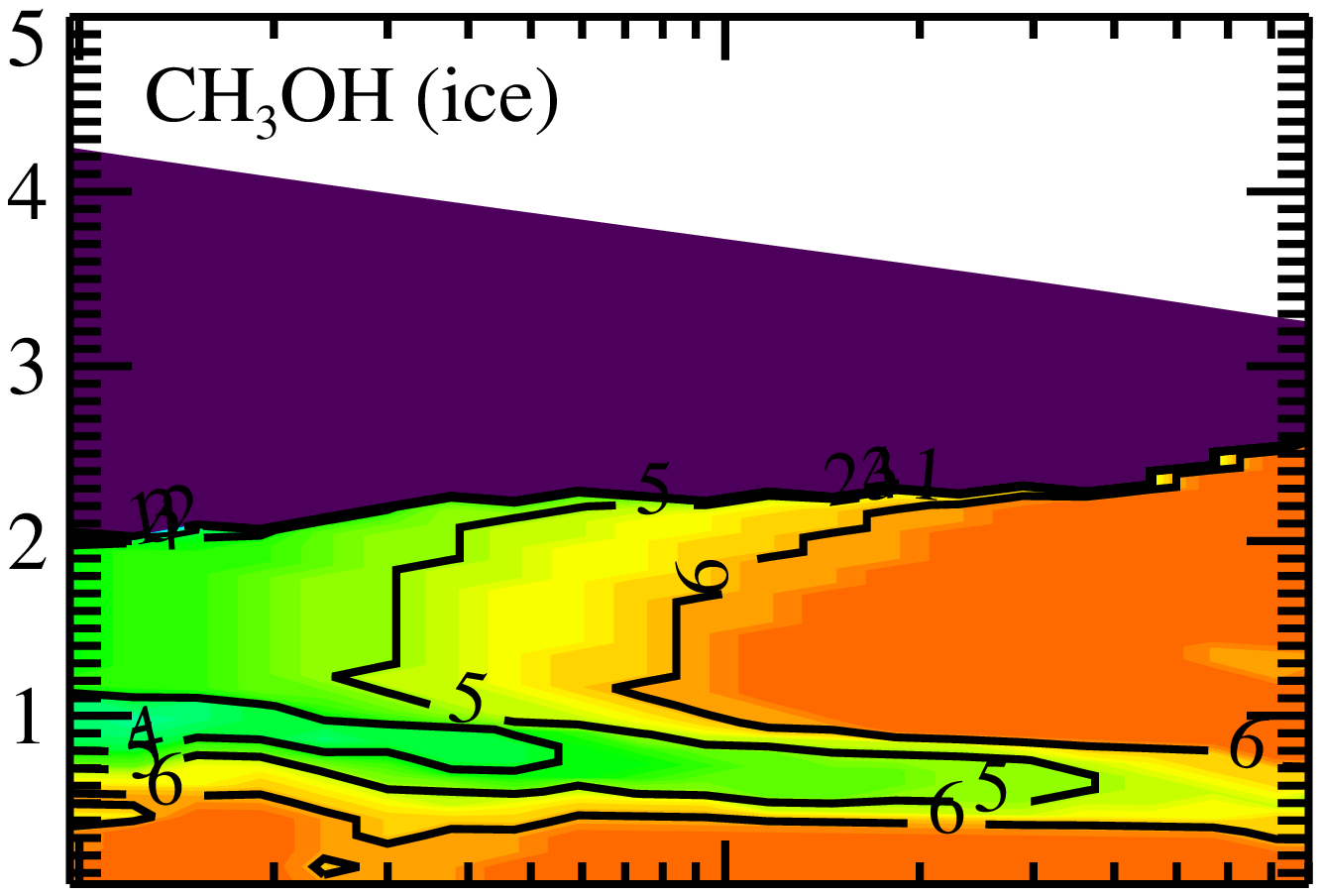}\\
\includegraphics[width=0.3\textwidth]{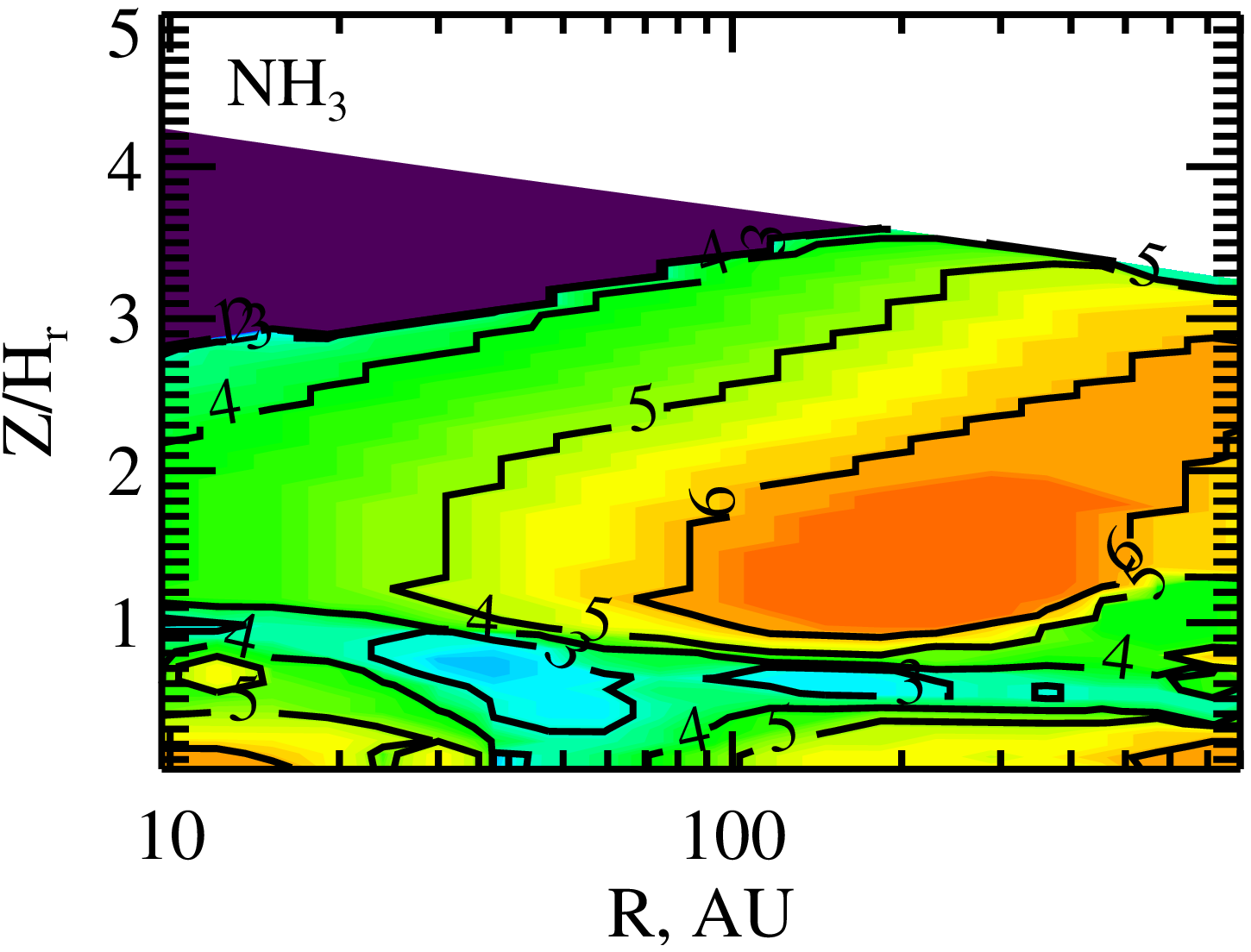}
\includegraphics[width=0.3\textwidth]{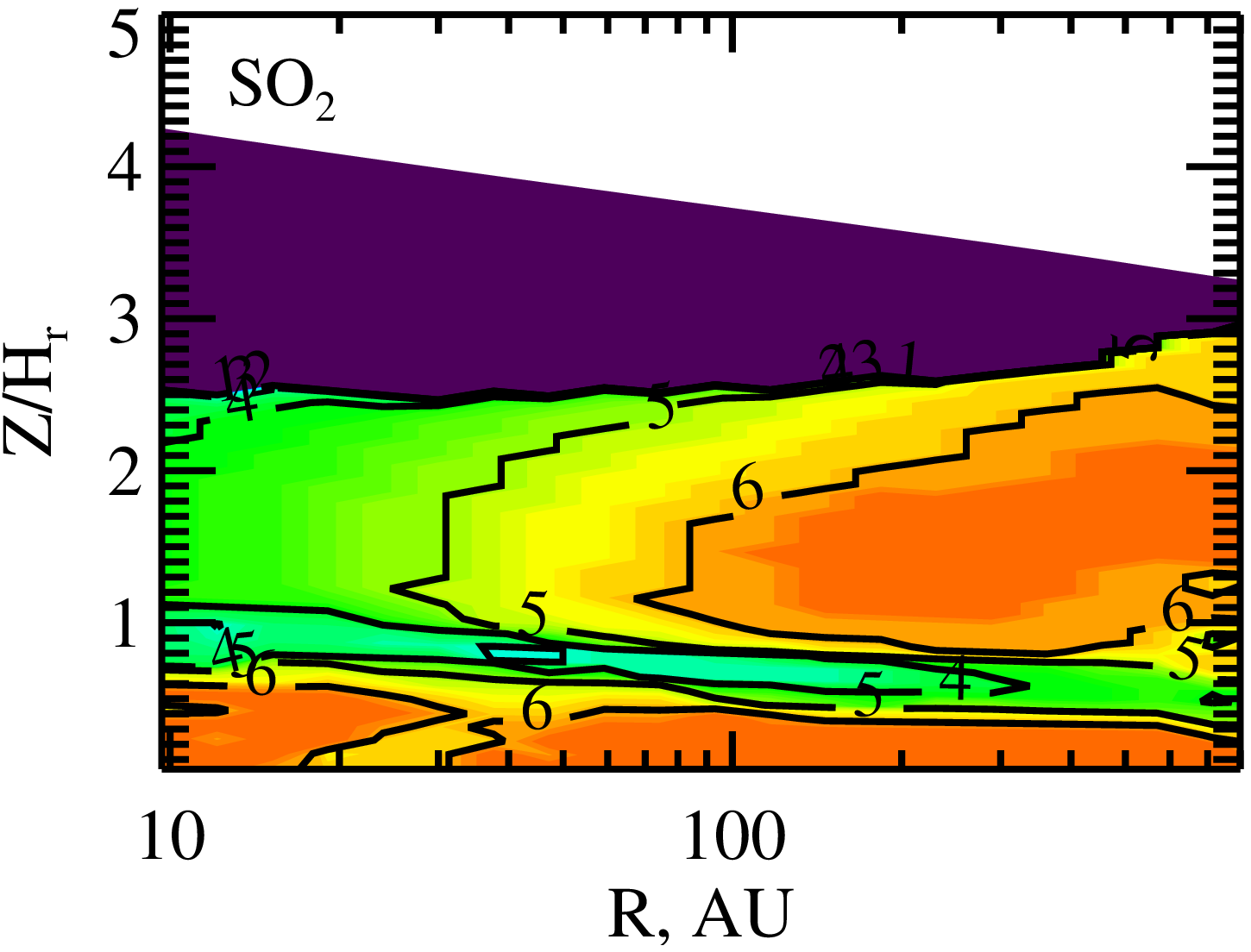}
\caption{(Top to bottom) Distributions of timescales to reach a chemical
steady-state during the 5~Myr of evolution for C$^+$, CO, CO$_2$,
CH$_3$OH ice, NH$_3$, and SO$_2$. For relative abundances below $10^{-25}$
this timescale is assumed to be 1~year.}
\label{fig:chem_ss}
\end{figure*}

The disk physical structure and the chemical model described above are used to solve
chemical kinetics equations together with turbulent transport terms. Since a while such kind of models have been employed 
in atmospheric chemistry, planetary atmosphere chemistry, and chemistry of molecular clouds 
\citep[e.g.,][]{Garsia_Solomon83,Xie_ea95,Yate_Millar03}.  
We utilize the mixing approach of \citet{Xie_ea95}, which is based on the Fickian's diffusion laws:
\begin{equation}
 \label{chem_kin_transport}
\frac{dn_i}{dt}(r,z) = F_i(r,z)-L_i(r,z)-\bigtriangledown \cdot \left(D_{\rm turb}(r,z)n_{\rm H}(r,z)\bigtriangledown 
        \frac{n_i(r,z)}{n_{\rm H}(r,z)}\right). 
\end{equation}
Here $n_i$ is concentrations of the $i$-th species (cm$^{-3}$), $F_i$ and $L_i$ are formation and destruction (loss)
terms. In the following we will use relative abundances for considered species $X_i= n_i/n_{\rm H}$ where $ n_{\rm H}$ is the
total hydrogen nucleus number density.
In this formalism the turbulence mixing rate for a certain species depends on its chemical gradient.

The formation and destruction of molecules are governed by the chemical kinetics:
\begin{equation}
\frac{dn_i}{dt} = \sum_{l,m}k_{lm}n_ln_m - n_i\sum_{i\neq l}k_ln_l +
k_i^{\rm des}n_i^{s} - k_i^{\rm acc}n_i
 \end{equation}
 \begin{equation}
\frac{dn_i^s}{dt} = \sum_{l,m}k^{s}_{lm}n_l^{s}n_m^{s} - n_i^{s}\sum_{i\neq
l}k_l^{s}n_l^{s} - k_i^{\rm des}n_i^{s}
+ k_i^{\rm acc}n_i
\end{equation}
where $n_i^{s}$ is the surface concentration of the
$i$-th species (cm$^{-3}$),
$k_{lm}$ and $k_{l}$ are the gas-phase
reaction rates (in units of s$^{-1}$ for the first-order kinetics and
cm$^3$\,s$^{-1}$ for the second-order
kinetics), $k_i^{\rm acc}$ and $k_i^{\rm des}$ denote the accretion and
desorption rates
(s$^{-1}$), and $k_{lm}^s$ and $k_{l}^s$ are surface reaction rates (cm$^3$\,s$^{-1}$).

The stiff equations of chemical kinetics are integrated simultaneously with the diffusion terms in the Eulerian
description, using a fully implicit 2D integration scheme. 
As boundary conditions for mixing, we assume that there is no inward and outward diffusion across 
boundaries of the disk domain, and that there is no flux through the midplane.
We do not employ approximate, operator-splitting integration schemes, in
which transport and chemical processes are treated separately, and instead integrate PDE system directly. 
Our ``ALCHEMIC'' FORTRAN77 code \citep{Semenov_ea10} 
is based on the Double-precision Variable-coefficient Ordinary Differential equation solver
with the Preconditioned Krylov (DVODPK) method GMRES for the solution of linear
systems\footnote{\url{http://www.netlib.org/ode/vodpk.f}}. The approximate Jacobi matrix is generated automatically 
from the supplied chemical network (without transport terms) and serves as a lefthand preconditioner. For astrochemical models
dominated by hydrogen reactions the Jacobi matrix is sparse, with $\la
1\%$ of non-zero elements. The corresponding linearized system of algebraic equations is solved using a high-performance sparse 
unsymmetric MA48 solver from the Harwell Mathematical Software Library\footnote{\url{http://www.hsl.rl.ac.uk/}}.
All the equations are solved on a non-uniform staggered grid consisting of 41 radial points (from 10 to 800~AU) and 91 
vertical points. This resolution is found to be optimal for mixing problems in a protoplanetary accretion disk, keeping the
computation tractable and still providing enough accuracy for the analysis. 
With a typical $10^{-6}$ relative and $10^{-15}$ absolute errors, the 2D-mixing model with high mixing efficiency
has about 15 million non-zero Jacobi matrix elements and takes about 48~hours of CPU time (Xeon 3.0~GHz, 4~Gb RAM,
gfortran~4.4-x64) 
to calculate the disk chemical structure within 5~Myr.

Our main set of chemical simulations consists of three runs: (1) the laminar disk chemical model
(no transport processes are taken into account), (2) the fast 2D-mixing
model ($Sc=1$), and (3) the slow 2D-mixing transport model ($Sc=100$). The chemical evolutionary time span is 5~Myr.

\section{When turbulence affects disk chemistry}
\label{analysis}
Before drilling into complex numerics, it is reasonable to analyze general conditions at which turbulent diffusion may affect the
chemical evolution in a protoplanetary disk.

\subsection{Chemical and dynamical timescales}
\label{timescales}
\begin{figure*}
\includegraphics[angle=90,width=0.95\textwidth]{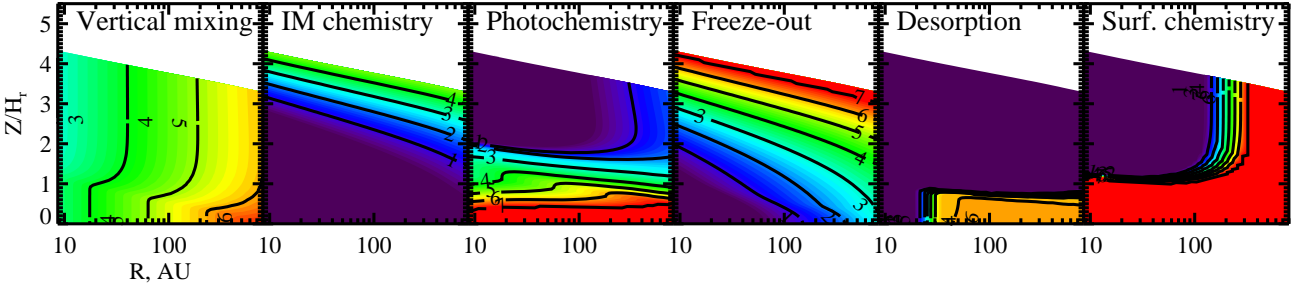}
\caption{(From left to right) Distributions of the characteristic timescales: turbulent diffusion,
ion-molecule chemistry, photochemistry, freeze-out, desorption, and surface chemistry ($\log_{10}$ scale).}
\label{fig:timescales}
\end{figure*}

In computational fluid dynamics with
reacting flows Damk\"{o}hler number $Da$ is often used as a measure of the influence of dynamical processes on the chemical
evolution. This number is simply the ratio
of a physical timescale to a chemical timescale:
\begin{equation}
\label{eq:da}
 Da = \tau_{\rm phys}/\tau_{\rm chem}.
\end{equation}
When $Da \la 1$, chemical evolution of a molecule is slow and therefore may be sensitive to changes in
physical conditions due to the medium flow. In contrast, when $Da \gg 1$, chemical evolution is fast and is not affected
by transport processes. 

The physical (or dynamical) timescale is often written as $\tau_{\rm phys}=L/V$, where $V$ is a characteristic
value of velocity fluctuations set by turbulence and $L$ is their correlation length \citep[e.g.,][]{Cant_Mastorakis08}.
For our $\alpha$-disk model, characteristic physical timescale is a turbulent mixing timescale: 
\begin{equation}
\label{eq:tau_phys}
\tau_{\rm phys}(r,z)=H(r)^2/D_{\rm turb}(r,z).   
\end{equation}
The distribution of $\tau_{\rm phys}(r,z)$ in the adopted disk model is shown in Fig.~\ref{fig:timescales} (1st panel).
In the outer disk region, $r\sim 100-800$~AU $\tau_{\rm phys}$ exceeds $10^5$~years, while in the Jovian planet-forming zone
the mixing timescale is faster, $\la 10^4$~years. Due to the vertical temperature gradient this timescale is slightly
shorter in the superheated disk upper region (see Fig.~\ref{fig:disk_struc}, 1st panel). Further we compare the mixing
timescale with timescales of key chemical processes.

The typical timescale of a first-order kinetics reaction (e.g., photodissociation) with a rate $k$ 
is $\tau_{\rm chem} \sim k^{-1}$. For a second-order reaction involving species A and B with a rate $k_{\rm AB}$ the 
reaction timescale for the species A is $\tau_{\rm chem} \sim 1/k_{\rm AB}n_{\rm B}$. Lets assess the characteristic timescale
of ion-molecule chemistry in disks, using HCO$^+$ as an example. The ion-molecule reactions are rapid even at very cold
temperatures 
and usually have no barriers 
\citep[e.g.,][]{OSU03,Woodall_ea07,Wakelam_09b}.
We assume that the HCO$^+$ evolution is governed by the following formation and destruction reactions,
CO + H$_3^+$ $\rightarrow$ HCO$^+$ + H$_2$ ($k_1=1.61\,10^{-9}$~cm$^3$\,s$^{-1}$) and 
dissociative recombination HCO$^+$ + e$^-$ $\rightarrow$ CO + H ($k_2=2.4\,10^{-7}(T/300)^{-0.69}$~cm$^3$\,s$^{-1}$).  
Then the corresponding HCO$^+$ ion-molecule (IM) chemistry timescale can be estimated as 
\begin{equation}
\tau_{\rm IM} \sim (k_1n_{\rm CO}n_{{\rm H}_3^+}/n_{{\rm HCO}^+}-k_2n_{{\rm HCO}^+})^{-1}.
\end{equation}
The IM timescale distribution in the disk is shown in Fig.~\ref{fig:timescales} (2nd panel), where we take $n_{\rm
CO}=6\,10^{-5}n_{\rm H}$, $n_{{\rm H}_3^+}=10^{-10}n_{\rm H}$, and $n_{{\rm HCO}^+}=10^{-9}n_{\rm H}$.
As can be clearly seen,
ion-molecule chemistry is very rapid, with a typical timescale of $\la 10-10^3$~years even in low density disk regions.
This is also true for neutral-neutral reactions without barriers or with small barriers, involving radicals and open-shell
species. They have comparable 
timescales even in the outer cold disk region \citep{OSU03}. In the warm ($T>50-100$~K) inner disk region other neutral-neutral 
reactions with considerable barriers of $\ga 1\,000$ become competitive. Overall, ion-molecule and neutral-neutral
reactions without large barriers proceed faster than the turbulent transport.

Chemical evolution in upper disk layers is determined by photochemistry, which is driven by intense high-energy
stellar (UV, X-rays) and interstellar (UV, CRP) radiation. The corresponding timescale is primarily set by
rates of the UV dissociation and X-rays ionization processes, 
\begin{equation}
\tau_{\rm h\nu} \sim 1/(k_{\rm pd}^{\rm UV}+k_{\rm pi}^{\rm X}). 
\end{equation}
For a CO-like molecule and without shielding, the UV and X-ray photorates 
are $k_{\rm pd}^{\rm UV}=2\,10^{-10}\left(\exp^{-1.7A_{\rm V}^*}\chi_*+\exp^{-1.7A_{\rm V}^{\rm IS}}\right)$ and
$k_{\rm pi}^{\rm X} = 3(\zeta_{\rm X}+\zeta_{\rm CRP})$, respectively. The calculated photochemistry timescale is 
presented in Fig.~\ref{fig:timescales} (3rd panel). Photochemistry is fast ($\la 1$~year) in the disk atmosphere 
(faster than the turbulent transport) and becomes slow ($\ga 10^6$~years) in dense dark disk regions close to the midplane.

The next important process in disk chemistry is freeze-out of neutral species onto dust grain surfaces in cold and dense
region ($T\la 20-120$~K). The inferred substantial depletions of observed gas-phase molecules in disks compared to the ISM
are generally interpreted as a combined action of photodissociation and freeze-out processes 
\citep[e.g.,][]{Bergin_ea07,DGH07,Semenov_ea10a}. The adsorption (AD) timescale is 
\begin{equation}
 \tau_{\rm AD} \sim k_{\rm AD}^{-1}=1/(\pi a_{\rm gr}^2V_{\rm th}n_{\rm d}),
\end{equation}
where $a_{\rm gr}$ is the grain radius (cm), $V_{\rm th}$ is the kinetic velocity of molecules (cm\,s$^{-1}$), 
and $n_{\rm d}$ is the grain concentration (cm$^{-3}$). The freeze-out timescale for CO in our disk model is depicted in 
Fig.~\ref{fig:timescales} (4th panel). This value is mostly determined by density (assuming homogeneous dust and gas mixture)
and only slightly by temperature and mass of a molecule.
It varies between $\la 1$ and $10^3$~years around midplane. At higher, less dense disk regions $\tau_{\rm AD}$ is longer,
up to $\sim 10^5-10^6$~years, though at such conditions evaporation rate will be much shorter. The adsorption timescale 
is in general shorter than the mixing timescale in the disk regions favorable for freeze-out.

A process, competitive to adsorption, is evaporation of icy mantles in warm and/or irradiated disk regions.
The evaporation timescale in disks is a combination of thermal desorption, CRP-induced desorption, 
photodesorption and possibly other non-thermal desorption mechanisms 
\citep[e.g., explosive desorption;][]{Shalabiea_Greenberg94}. 
Thermal desorption will prevail in warm viscously-heated midplane region ($r\sim1-5$~AU)
and across intermediate molecular layer, whereas CRP-desorption operates in the coldest, dark outer disk midplane, and
photodesorption 
becomes competitive in upper disk layer. Thus, evaporation timescale can be written as:
\begin{equation}
 \tau_{\rm des} \sim 1/(k_{\rm des}^{\rm UV}+k_{\rm des}^{\rm CRP}+k_{\rm des}^{\rm th}),
\end{equation}
where 
\begin{equation}
 k_{\rm des}^{\rm UV}=10^{-3}\pi a_{\rm gr}^2\left(\exp^{-2A_{\rm V}^*}\chi_*+\exp^{-2A_{\rm V}^{\rm IS}}\right),
\end{equation}
\begin{equation}
 k_{\rm des}^{\rm CRP}=2.4\,10^{-2}\zeta_{\rm CRP}\nu_0\exp^{-\gamma/T_{\rm CRP}},
\end{equation}
and 
\begin{equation}
 k_{\rm des}^{\rm th}=\nu_0\exp^{-\gamma/T}.
\end{equation}
Here, $T_{\rm CRP}=(4.36\,10^5+T^3)^{1/3}$ is a peak grain temperature (K) reached upon a hit by relativistic iron nucleus
\citep[see Eq.~6 in][]{Red2}, $\nu_0$ is characteristic vibrational frequency (s$^{-1}$) of a molecule, and $\gamma$ its
desorption 
energy (K). In Fig.~\ref{fig:timescales} (5th panel) we show distribution of the evaporation timescale for CO 
in the \object{DM Tau} disk model. In cold and dark outer midplane, where temperatures are below $\approx 20$~K, desorption of CO
is too slow,
and $\tau_{\rm des}$ exceeds about 1~Myr, whereas in upper disk layer and in the warm planet-forming zone desorption is a rapid
process.

Finally, surface chemistry timescale in the absence of tunneling is controlled by thermal hopping rates of the reactants, 
and the reaction barrier \citep[see Eqs.~12, 14 in][]{Semenov_ea10}.
As an example, we consider a slow first step in surface hydrogenation sequence of CO into CH$_3$OH, namely, H (ice) + CO (ice) 
$\rightarrow$ HCO (ice), which has a barrier of about 2\,500~K caused by bond restructuring. In Fig.~\ref{fig:timescales} (last
panel) 
the timescale of the CO surface chemistry is presented. As can be clearly seen, surface chemistry has a long timescale of 
$\sim10^6$~years around midplane and in cold outer disk region, which is similar to the turbulent mixing timescale. 
Therefore, chemical species produced mainly via surface processes (e.g., ices and complex organics) 
shall be sensitive to turbulent diffusion mixing in disks. The aforementioned characteristic timescales in the inner (10~AU)
and outer (250~AU) disk regions are also compared in Tables~\ref{tab:tau_inner}-\ref{tab:tau_outer}. 

In reality chemical evolution of a large multi-component mixture is controlled by both fast and slow processes, so it is hard to
obtain a single characteristic evolutionary timescale as discussed above. Usually detailed analysis of eigenvalues and
eigenvectors of 
Jacobi matrix or its diagonal and off-diagonal terms is used to isolate fast and slow evolving subsystems in the chemical network
\citep[e.g.,][]{Lovrics_ea06}. However, for a tightly coupled, extended set of chemical reactions, like a typical 
hydrogen- and carbon-dominated astrochemical network, such an analysis is hard to perform. We elaborate a different approach and
use 
the results of our non-mixing disk model to derive individual chemical timescales as the time needed to reach a chemical quasi
steady-state 
for a given species. The steady-state time is the evolutionary moment when molecular abundances in the subsequent time-steps
change by a factor of $\la 3$ or assumed 5~Myr otherwise.
In Fig.~\ref{fig:chem_ss} the chemical steady-state times for a few assorted species in the \object{DM Tau} disk model 
are presented. 

The steady-state timescale distributions exhibit a complex pattern that is not easily related to characteristic chemical
timescales
shown in Fig.~\ref{fig:timescales}. Noteworthy, this pattern is similar for many species in the network, with slow surface
processes, 
$\tau_{\rm chem} \sim 10^5-10^6$~years, determining the chemical timescales in the midplane, 
$z/H_{\rm r} \la 0.8$ (here $ H_{\rm r}$ is the pressure scale height). 
In the warm inner midplane, $r\la 20$~AU, the characteristic chemical timescales are large because not only H and H$_2$ but also
heavy 
radicals become mobile at $T \sim 30-40$~K, leading to active surface chemistry \citep[see, e.g.,][]{Garrod_ea08b}.
Within $\sim 30-50$~AU temperatures in the midplane are between 15--25~K, and hydrogen evaporates too rapidly, 
while other radicals are too heavy to activate surface chemistry. This results in a dearth of rapid surface 
reactions. Therefore, chemical timescales are usually shorter, $\tau_{\rm chem} \sim 10^4-10^5$~years, in this region, and are
governed
by gas-grain interaction rates. Above the cold and dark midplane, at $\approx 1$ pressure scale height, where a warm molecular
layer is 
located, the characteristic chemical timescales are shorter, $\la 10^5$~years, since thermal desorption and photodesorption 
start to prevail there over surface chemistry and accretion. Within the inner $\sim 200$~AU of the molecular layer lukewarm
temperatures
and elevated X-ray ionization enable slow photodestruction of well-bound ices.
In the upper disk region, $\ga 1.5$ pressure scale height, where molecules are effectively destroyed by high-energy UV and X-ray
photons, photochemistry coupled to slow neutral-neutral reactions set chemical timescales that increase outward from 
$\sim 10^4$~years till $\ga 1$~Myr. The outward decrease of the gas density in this region makes recombination rates slower
(as these scale down as density squared), whereas the stellar UV and X-ray fluxes also decrease with the radial distance from the
star. 
Note that chemical timescales are shorter in the CO molecular layer and, in particular, for C$^+$, which is abundant
in the disk atmosphere at $z/H_{\rm r} \la 1.8-2$, while $\tau_{\rm chem}$ is long and comparable to the dynamical timescales in
regions
where CO$_2$, CH$_3$OH ice, and SO$_2$ are produced (see Fig.~\ref{fig:chem_ss}). As we shall see later, indeed chemical evolution
of
these (and many others) species is affected by the turbulent diffusion.

\subsection{Mixing Importance Measure}
\label{mim}
Considerations presented in the previous subsection provide some clues on the mixing sensitivity for various species. However, it
would be instructive to find some general quantitative measure of this sensitivity. We suggest a new Mixing Importance Measure
(MIM) to find necessary conditions for sensitivity of column densities of a given
molecule to the turbulent mixing, based on results of the laminar model. This quantity comes in three varieties, namely, a local
value

\begin{figure*}
\includegraphics[height=0.3\textheight,angle=90]{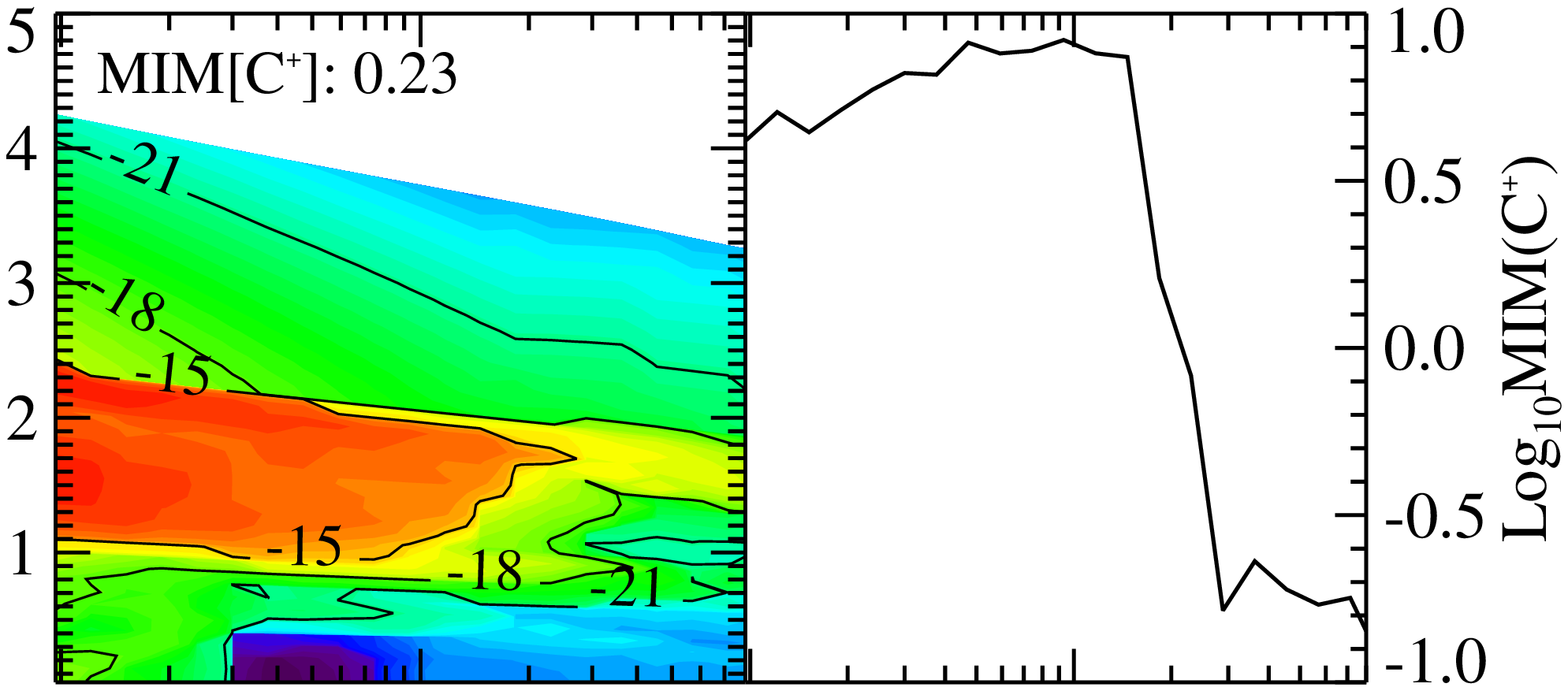}
\includegraphics[height=0.3\textheight,angle=90]{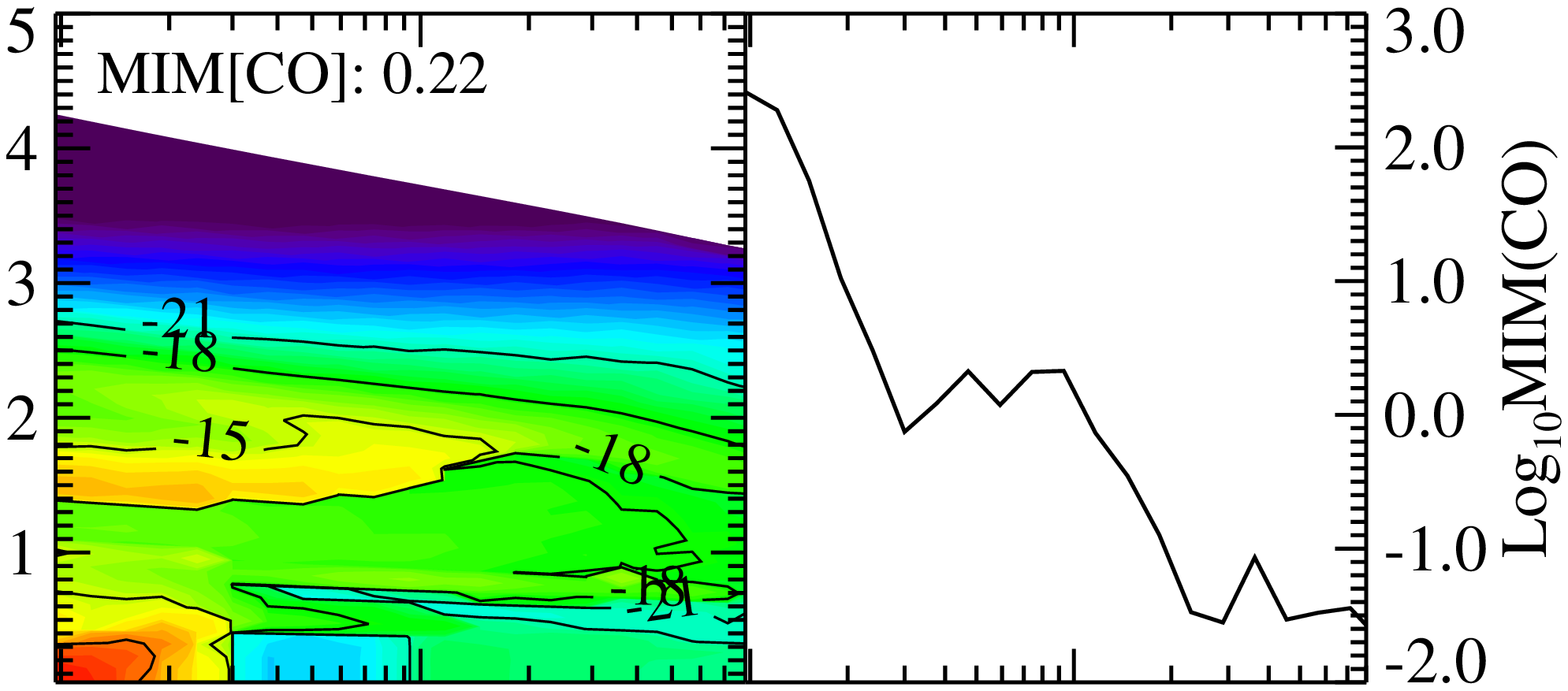}\\
\includegraphics[height=0.3\textheight,angle=90]{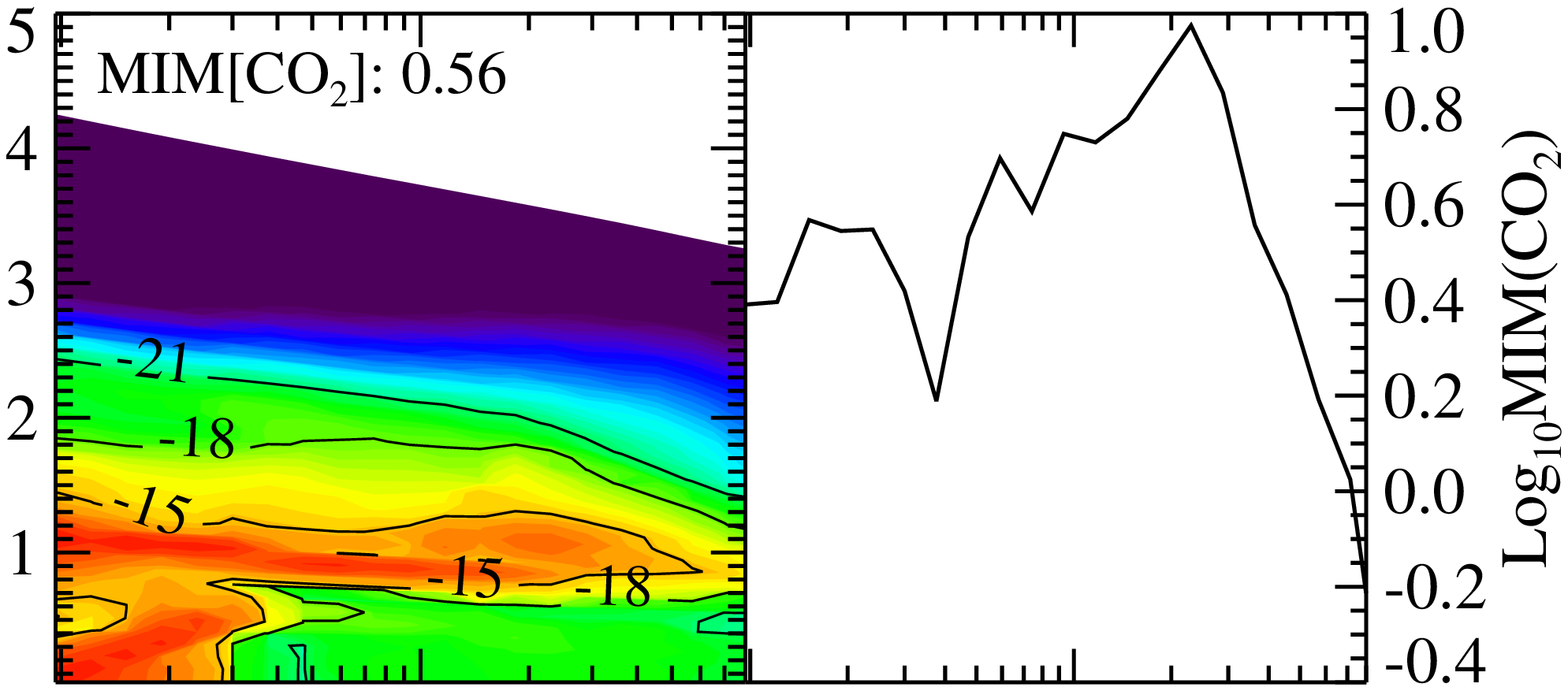}
\includegraphics[height=0.3\textheight,angle=90]{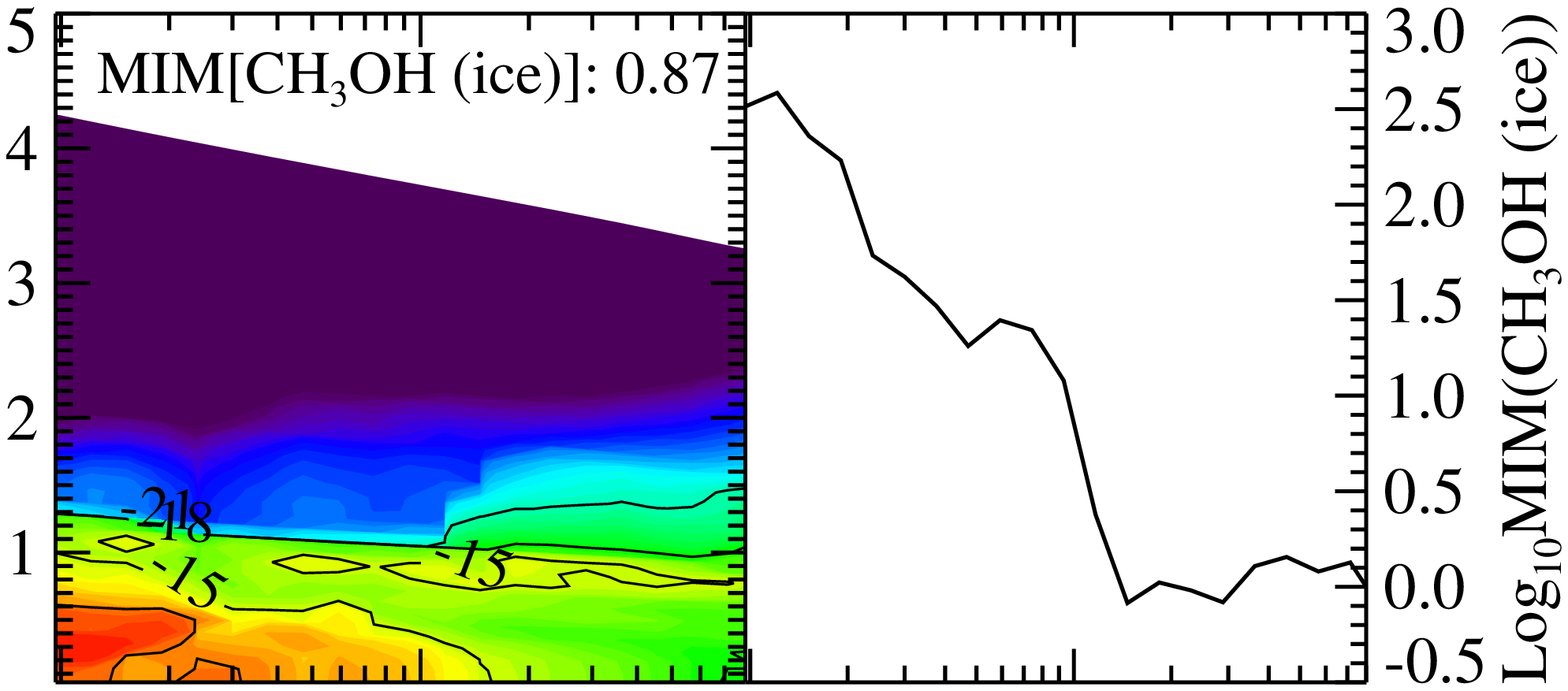}\\
\includegraphics[height=0.3\textheight,angle=90]{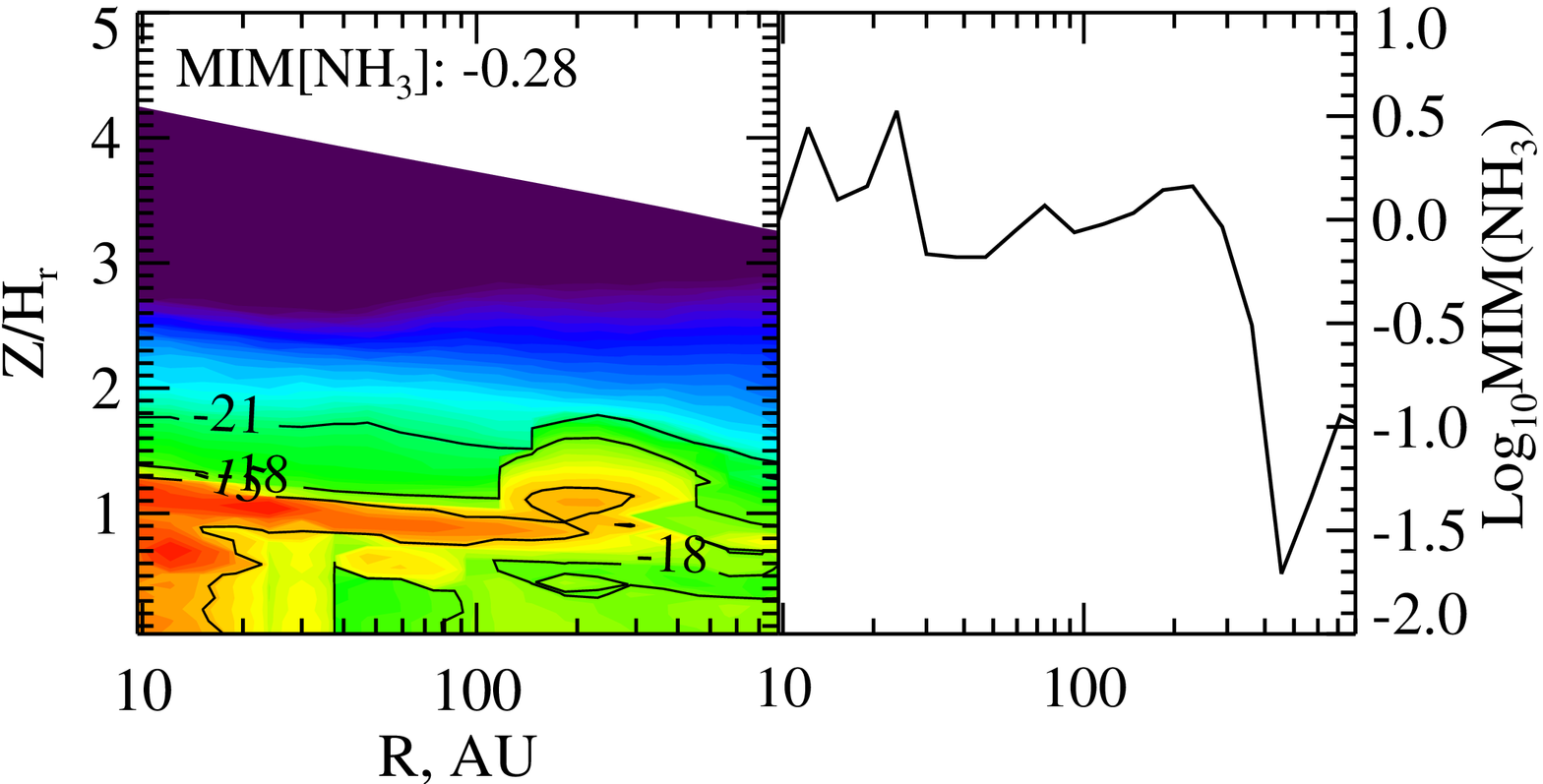}
\includegraphics[height=0.3\textheight,angle=90]{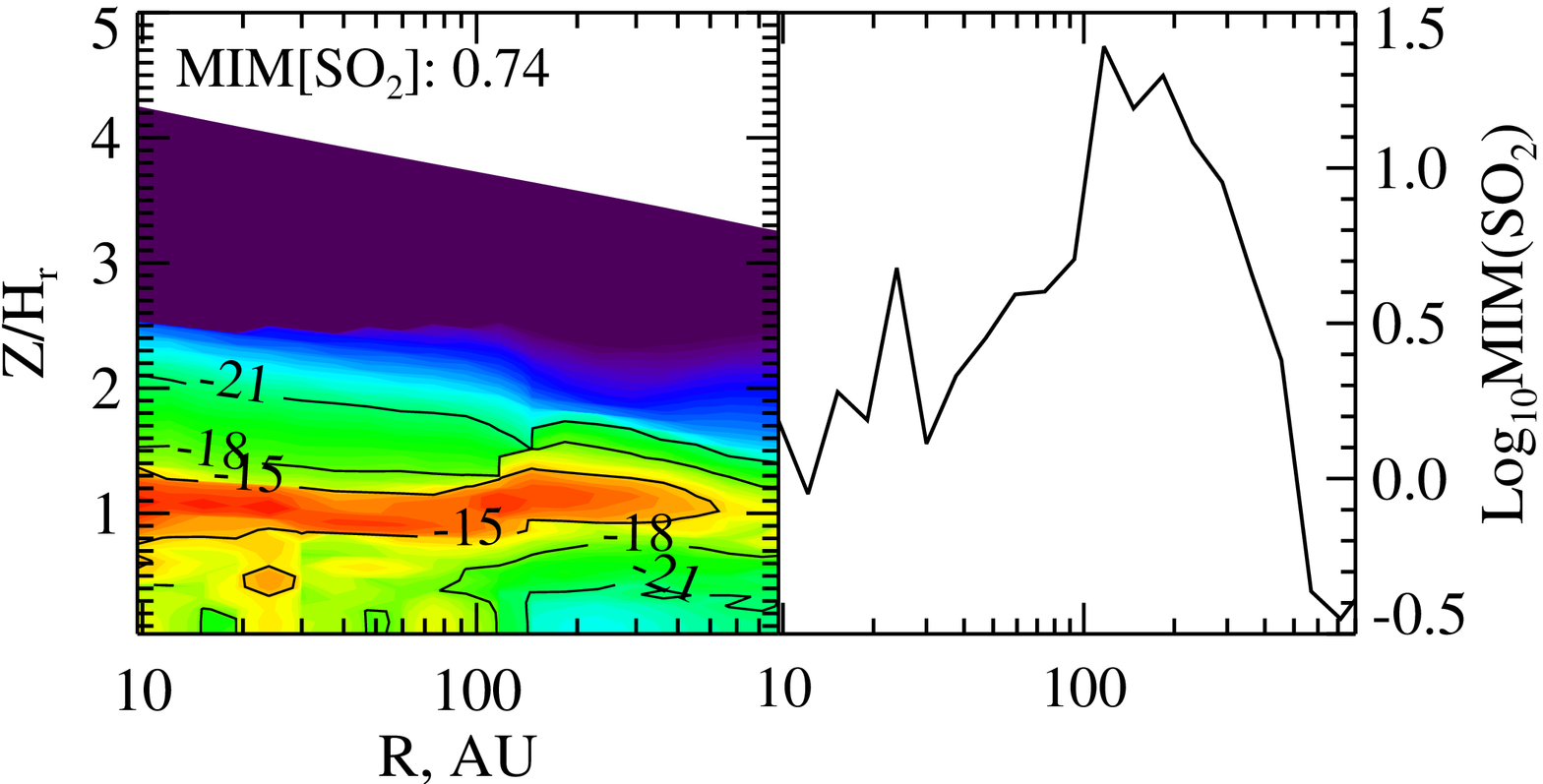}
\caption{(Top to bottom) Distributions of the Mixing Importance Measure
(MIM; left panel) and vertically-integrated MIM (right panel) for C$^+$, CO, CO$_2$,
CH$_3$OH ice, NH$_3$, and SO$_2$. The higher the MIM, the more
sensitive can be the chemical evolution of the given molecule to dynamical
processes. Also given are globally integrated MIM values.}
\label{fig:mim_distr}
\end{figure*}

\begin{equation}
\label{eq:MIMrz}
{\rm MIM}_{r,z} = \frac{1}{Da(r,z)}\frac{n(r,z)}{N(r)},
\end{equation}
a vertically integrated value
\begin{equation}
\label{eq:MIMr}
{\rm MIM}_r = \int_{0}^{z_{\rm max}(r)}{\rm MIM}_{r,z} dz,
\end{equation}
and a global value
\begin{equation}
\label{eq:MIM}
{\rm MIM} = \frac{1}{(r_1-r_0)}\int_{r_0}^{r_1}{\rm MIM}_r dr,
\end{equation}
where $Da(r,z) $ is the Damk\"{o}hler number (Eq.~\ref{eq:da}), $N(r)$ is the total vertical column density 
(cm$^{-2}$) of a given species at the radius $r$, $n(r,z)$ (cm$^{-3}$) is its number density at the disk location $(r,z)$, 
$z_{\rm max}(r)$ is the disk height at the radius $r$, and $r_0$, $r_1$ are the disk inner and outer radii, respectively. 
The characteristic physical timescale is the diffusion timescale given by Eq.~\ref{eq:tau_phys}, 
and the characteristic chemical timescale of a molecule is its quasi steady-state timescale defined above (see
Fig.~\ref{fig:chem_ss}).
By definition, the MIM for a molecule allows localizing those disk regions where the chemical evolution is slow 
and which contribute most to the vertical column density (a typical observationally inferred quantity). Thus,
the larger the MIM, the stronger possible changes in vertical column densities of a considered species due to turbulent transport.
However, the straightforward expression~(\ref{eq:MIMrz}) is not a sufficient criterion for making reliable estimates on a
magnitude
of such changes. For example, in the absence of relative abundance gradients even very slow chemistry (resulting in large MIM)
leads to the same column 
densities as the non-mixing (laminar) disk model. On the other hand, if MIM is small everywhere in the disk, 
the mixing will not alter resulting column densities. The local (Eq.~\ref{eq:MIMrz}), vertically-integrated 
(Eq.~\ref{eq:MIMr}), and disk-averaged (Eq.~\ref{eq:MIM}) MIMs for 
several species are shown in Fig.~\ref{fig:mim_distr} ($\log_{10}$ scale). 

The MIM distribution for C$^+$ peaks at elevated disk heights, $\sim 1-2\,H_{\rm r}$, where the C$^+$ concentration is high, and
photochemistry is relatively slow yet efficient, especially on ices (cf. Fig.~\ref{fig:timescales}). 
The importance of turbulent mixing to its chemical 
evolution is low ($\lg(MIM)=0.23$) compared to the other species, particularly, in outer disk region at $r\ga 300$~AU. Two other
species 
with similarly low global MIM values of $\lg(MIM)=0.22$ and $\lg(MIM)=-0.28$ are CO and NH$_3$, respectively. The MIM distribution
for
CO has a distinct peak around inner disk midplane at $r \la 30$~AU, where temperatures are appropriate to its surface conversion
to
CO$_2$, and where the gas-phase CO concentration is high. Another region where the CO chemistry may be affected by the turbulent 
processes is located at the upper CO molecular layer, at $\sim 2\,H_{\rm r}$, in the inner disk zone. There the UV
photodissociation of
CO is slow due to self-shielding and mutual-shielding by H$_2$ and dust, and X-ray ionization of He leads to slow 
destruction of CO by He$^+$ at late times, $t\sim 10^5-10^6$~years. 

In contrast to CO and C$^+$, the MIM for the CO$_2$, NH$_3$, and SO$_2$ chemistry exhibit a somewhat similar pattern 
(Fig.~\ref{fig:mim_distr}). The most likely dynamically-sensitive regions are the warm molecular layer, where photoprocessing
is activated, and inner part of the disk midplane, where surface heavy C-, O- and N-bearing radicals become mobile. This is also 
probably true for S-bearing radicals as well, but our surface network includes only a very limited sulfur chemistry, therefore 
the MIM for SO$_2$ does not reach its 
maximum in the inner midplane, and peaks within the SO$_2$ molecular layer. Finally, the MIM for CH$_3$OH ice is the largest 
in the disk midplane where solid methanol is produced by surface hydrogenation of CO and by CH$_3$ reacting with OH. 
The disk-averaged MIM values are relatively high for CO$_2$ ($\lg(MIM)=0.56$) and especially
for SO$_2$ ($\lg(MIM)=0.74$) and solid CH$_3$OH ($\lg(MIM)=0.87$). Below we will reveal that their column densities are
enhanced by the turbulent diffusion by more than an order of magnitude, unlike those of C$^+$, CO, and NH$_3$.

\section{Results of numerical modeling}
\label{results}
\begin{figure}
\includegraphics[angle=90,width=0.45\textwidth]{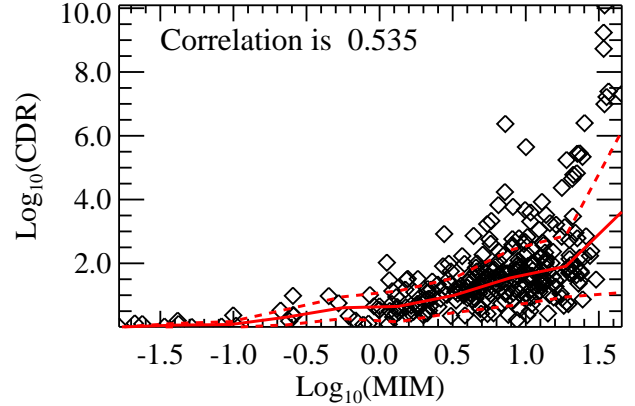}
\caption{Dependence of the ratio of the molecular column densities integrated
over the disk radius and computed with the 2D-mixing and laminar models
($t=5$~Myr) versus MIM. Shown are the molecules with the maximum column densities in the laminar case that exceed
$10^{11}$~cm$^{-2}$.
The corresponding linear correlation coefficient is also shown (top left corner).}
\label{fig:mim_vs_cdr}
\end{figure}

\begin{figure}
\includegraphics[angle=90,width=0.45\textwidth]{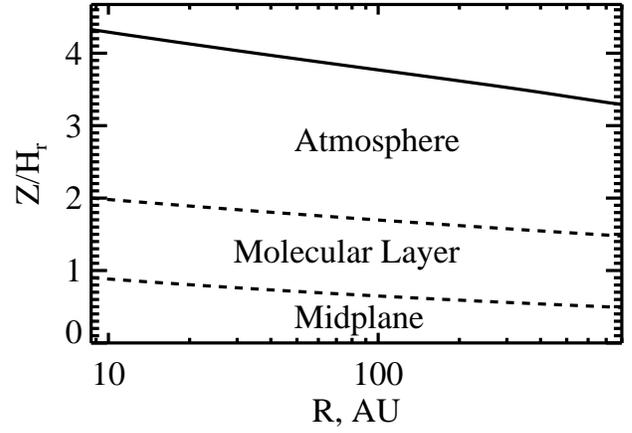}
\caption{Major disk regions are designated by analyzing the distribution patterns
of the C$^+$ and CO abundances. The atmosphere is the disk region dominated by C$^+$ in contrast
to CO, whereas in the molecular layer carbon is mainly locked in the CO gas-phase molecules. The
midplane is the disk region where depletion of gas-phase molecules occurs.}
\label{fig:zones}
\end{figure}

\begin{deluxetable}{lccc}
\tablecaption{Species steadfast to turbulent mixing\label{tab:steadfast}}
\tablehead{
\colhead{Molecule$^*$} & \colhead{$\lg(Se)$}  & \colhead{$\lg(CDR)$} &
\colhead{$\lg(N_{\rm stat}^{\rm max})$}
}
\startdata
         C$^+$    &     0.23 &     0.48 &    16.79\\
         C$_2$    &     0.25 &     0.68 &    14.41\\
        C$_2$H    &     0.40 &     0.64 &    14.29\\
C$_2$H$_2$$^+$    &     0.67 &     0.68 &    12.36\\
            CH    &     0.30 &     0.37 &    12.91\\
        CH$_2$    &     0.33 &     0.45 &    12.73\\
    CH$_3$$^+$    &     0.69 &     0.57 &    12.22\\
    CH$_5$$^+$    &     0.34 &     0.66 &    13.48\\
            CL    &    -0.17 &     0.21 &    13.95\\
            CN    &     0.12 &     0.56 &    14.41\\
            CO    &     0.22 &     0.43 &    18.67\\
            Fe    &     0.31 &     0.45 &    11.31\\
        Fe$^+$    &    -0.62 &     0.19 &    13.69\\
           FeH    &     0.22 &     0.68 &    12.16\\
      grain(0)    &    -0.21 &     0.35 &    12.29\\
         H$_2$    &    -1.08 &     0.02 &    24.30\\
   H$_2$CN$^+$    &     0.13 &     0.49 &    13.06\\
       H$_2$CO    &     0.32 &     0.44 &    13.91\\
   H$_2$CO$^+$    &     0.78 &     0.44 &    11.23\\
        H$_2$S    &     0.11 &     0.44 &    13.23\\
     H$_3$$^+$    &    -0.68 &     0.52 &    14.81\\
   H$_3$CO$^+$    &     0.58 &     0.37 &    11.62\\
    H$_3$O$^+$    &     0.12 &     0.64 &    13.43\\
           HCN    &    -0.00 &     0.65 &    14.50\\
           HCS    &     0.27 &     0.61 &    11.28\\
        He$^+$    &     0.07 &     0.51 &    15.07\\
           HNC    &    -0.04 &     0.66 &    14.24\\
            Mg    &    -1.67 &     0.12 &    12.06\\
        Mg$^+$    &    -1.38 &     0.10 &    13.91\\
       MgH$_2$    &    -1.38 &     0.20 &    13.27\\
         N$_2$    &     0.19 &     0.43 &    17.61\\
            Na    &    -1.62 &     0.10 &    11.84\\
        NH$_2$    &    -0.59 &     0.35 &    14.18\\
    NH$_4$$^+$    &    -0.00 &     0.53 &    12.17\\
            OH    &     0.09 &     0.32 &    14.92\\
            Si    &     0.69 &     0.42 &    13.06\\
           SiH    &     0.49 &     0.35 &    11.21\\
       SiH$_3$    &    -0.59 &     0.38 &    11.05\\
       C ice    &     1.03 &     0.50 &    16.00\\
C$_2$H$_2$ ice  &     0.18 &     0.62 &    13.91\\
 CH$_5$N ice    &     1.26 &     0.67 &    15.54\\
     FeH ice    &    -1.47 &     0.02 &    16.04\\
       H ice    &    -1.00 &     0.37 &    18.55\\
   H$_2$ ice    &     1.15 &     0.43 &    17.08\\
 H$_2$CS ice    &     0.34 &     0.52 &    16.29\\
  H$_2$O ice    &     0.06 &     0.09 &    20.84\\
  H$_2$S ice    &    -0.98 &     0.05 &    17.54\\
     HCN ice    &     0.27 &     0.65 &    19.43\\
    HCSi ice    &    -0.00 &     0.54 &    12.50\\
      HS ice    &     1.01 &     0.38 &    15.54\\
 MgH$_2$ ice    &    -1.29 &     0.01 &    16.64\\
     NaH ice    &    -1.78 &     0.01 &    15.96\\
  NH$_3$ ice    &    -0.08 &     0.11 &    19.84\\
       S ice    &     0.96 &     0.57 &    15.50\\
   S$_2$ ice    &     0.74 &     0.42 &    11.45\\
     SiC ice    &     0.88 &     0.39 &    13.85\\
SiCH$_2$ ice    &     0.94 &     0.23 &    12.95\\
 SiH$_4$ ice    &    -0.11 &     0.08 &    16.59\\
\enddata
\tablenotemark{*}\tablenotetext{*}{Listed are the molecules which vertical
column densities in the disk exceed $10^{11}$~cm$^{-2}$.}
\tablecomments{Col.2 The Mixing Importance Measure (MIM) integrated over the
disk (in log scale). Col.3 The ratio of the vertical column density at $t=5$~Myr
computed with
the 2D-mixing and the laminar chemical models, averaged over the radius (in
log scale). Col.4 The maximal vertical column density for a given molecule
at 5~Myr in the laminar model.}
\end{deluxetable}

\LongTables
\begin{deluxetable}{lccc}
\tablecaption{Species sensitive to turbulent mixing\label{tab:sens}}
\tablehead{
\colhead{Molecule$^*$} & \colhead{$\lg(Se)$}  & \colhead{$\lg(CDR)$} &
\colhead{$\lg(N_{\rm stat}^{\rm max})$}
}
\startdata
             C    &     0.79 &     1.80 &    17.89\\
    C$_2$H$_2$    &    -0.07 &     1.01 &    14.35\\
C$_2$H$_3$$^+$    &     0.35 &     1.18 &    11.43\\
    C$_2$H$_4$    &     0.93 &     1.75 &    11.82\\
        C$_2$N    &     0.76 &     1.83 &    12.53\\
         C$_3$    &     0.52 &     1.12 &    14.93\\
        C$_3$H    &     0.64 &     1.18 &    13.42\\
    C$_3$H$^+$    &     0.74 &     0.91 &    12.38\\
    C$_3$H$_2$    &     0.60 &     1.24 &    13.95\\
C$_3$H$_2$$^+$    &     0.59 &     1.11 &    12.05\\
C$_3$H$_3$$^+$    &     0.86 &     1.04 &    11.83\\
        C$_3$N    &     0.86 &     1.40 &    12.01\\
        C$_3$O    &     0.96 &     1.25 &    11.17\\
         C$_4$    &     0.73 &     1.17 &    13.13\\
        C$_4$H    &     0.56 &     1.11 &    13.74\\
    C$_4$H$_2$    &     0.32 &     1.09 &    13.89\\
    C$_4$H$_3$    &     0.56 &     1.30 &    13.81\\
C$_4$H$_3$$^+$    &     0.59 &     1.13 &    11.81\\
    C$_4$H$_4$    &     1.04 &     1.61 &    12.15\\
C$_4$H$_4$$^+$    &     0.86 &     0.84 &    11.59\\
        C$_4$N    &     0.84 &     1.41 &    12.71\\
         C$_5$    &     0.55 &     1.21 &    14.18\\
        C$_5$H    &     0.72 &     1.40 &    12.98\\
    C$_5$H$^+$    &     0.96 &     1.06 &    11.15\\
    C$_5$H$_2$    &     0.69 &     1.23 &    13.29\\
C$_5$H$_2$$^+$    &     1.10 &     1.62 &    11.07\\
C$_5$H$_2$N$^+$   &     1.14 &     1.48 &    11.19\\
        C$_5$N    &     0.81 &     1.63 &    12.39\\
         C$_6$    &     0.62 &     1.51 &    13.85\\
        C$_6$H    &     0.95 &     1.59 &    12.71\\
    C$_6$H$_2$    &     0.91 &     1.07 &    13.25\\
C$_6$H$_2$$^+$    &     1.14 &     1.84 &    11.10\\
    C$_6$H$_6$    &     1.18 &     1.59 &    11.39\\
         C$_7$    &     0.74 &     1.39 &    13.44\\
        C$_7$H    &     1.12 &     1.37 &    12.22\\
    C$_7$H$_2$    &     1.08 &     1.31 &    12.36\\
         C$_8$    &     1.03 &     1.47 &    13.41\\
        C$_8$H    &     1.11 &     1.30 &    12.31\\
    C$_8$H$_2$    &     1.02 &     0.96 &    12.93\\
         C$_9$    &     1.09 &     1.41 &    13.33\\
        C$_9$H    &     1.09 &     1.11 &    11.99\\
    C$_9$H$_2$    &     1.11 &     1.46 &    12.49\\
        C$_9$N    &     1.13 &     1.98 &    11.05\\
      CH$_2$CN    &     0.75 &     1.12 &    11.80\\
      CH$_2$CO    &     1.05 &     1.78 &    11.29\\
      CH$_2$NH    &     0.53 &     1.05 &    11.56\\
        CH$_3$    &    -0.35 &     0.98 &    15.34\\
      CH$_3$CN    &     0.76 &     1.04 &    11.33\\
        CH$_4$    &     0.60 &     0.91 &    17.69\\
       CH$_5$N    &     0.68 &     1.07 &    11.50\\
       CNC$^+$    &     0.33 &     1.20 &    11.38\\
            CS    &     0.43 &     1.91 &    12.15\\
         e$^-$    &     0.42 &     0.89 &    17.25\\
             H    &     0.93 &     1.81 &    22.69\\
         H$^+$    &     0.88 &     1.66 &    17.07\\
       H$_2$CS    &     0.28 &     0.93 &    12.33\\
   H$_2$NO$^+$    &     0.19 &     1.52 &    11.20\\
        H$_2$O    &     0.37 &     0.95 &    15.52\\
    H$_2$S$_2$    &     0.98 &     1.67 &    11.21\\
      H$_2$SiO    &     0.65 &     1.68 &    11.31\\
       HC$_3$N    &     1.03 &     1.17 &    11.97\\
       HC$_5$N    &     1.12 &     1.56 &    12.15\\
          HCCN    &     0.82 &     1.52 &    11.25\\
           HCL    &     0.83 &     1.62 &    11.16\\
           HCO    &     0.66 &     1.25 &    11.28\\
       HCO$^+$    &     0.48 &     1.35 &    13.98\\
          HNCO    &     0.95 &     1.96 &    12.14\\
           HNO    &     0.18 &     1.48 &    13.67\\
          HNSi    &     0.47 &     1.99 &    11.29\\
            HS    &     0.79 &     1.37 &    11.60\\
        HS$_2$    &     0.97 &     1.67 &    11.21\\
             N    &     0.52 &     0.84 &    17.78\\
    N$_2$H$^+$    &     0.47 &     1.47 &    12.39\\
            NH    &     0.09 &     1.13 &    13.36\\
      NH$_2$CN    &     0.62 &     1.07 &    12.40\\
        NH$_3$    &    -0.28 &     0.75 &    14.74\\
    NH$_3$$^+$    &     0.34 &     0.89 &    11.43\\
            NO    &     0.36 &     1.43 &    14.57\\
        NO$^+$    &     0.66 &     0.89 &    11.23\\
             O    &     0.08 &     0.73 &    18.65\\
         O$^+$    &     0.80 &     1.52 &    12.24\\
         O$_2$    &     0.58 &     1.98 &    16.91\\
     O$_2$$^+$    &     0.86 &     1.10 &    13.41\\
           OCN    &     0.18 &     1.18 &    13.69\\
           OCS    &     0.49 &     1.75 &    11.98\\
             P    &     0.26 &     0.73 &    12.53\\
         P$^+$    &     0.73 &     1.56 &    12.89\\
             S    &     0.60 &     1.23 &    14.73\\
         S$^+$    &     0.68 &     1.13 &    15.52\\
        Si$^+$    &     0.69 &     1.25 &    14.12\\
       SiH$_4$    &     0.32 &     0.87 &    12.07\\
        SO$^+$    &     0.59 &     1.11 &    12.00\\
C$_1$$_0$ ice   &     1.04 &     1.19 &    15.00\\
   C$_2$ ice    &     0.96 &     1.91 &    15.28\\
  C$_2$H ice    &     0.94 &     1.91 &    15.22\\
C$_2$H$_3$ ice  &     1.34 &     1.89 &    12.79\\
   C$_3$ ice    &     0.96 &     1.80 &    15.81\\
  C$_3$H ice    &     0.78 &     1.77 &    16.39\\
C$_3$H$_2$ ice  &     0.70 &     1.29 &    19.66\\
  C$_3$S ice    &     0.92 &     1.51 &    15.12\\
   C$_4$ ice    &     1.30 &     1.70 &    12.84\\
  C$_4$H ice    &     1.30 &     1.86 &    13.18\\
C$_4$H$_2$ ice  &     0.84 &     1.74 &    14.43\\
C$_4$H$_3$ ice  &     1.28 &     1.85 &    12.74\\
C$_4$H$_4$ ice  &     0.79 &     1.89 &    18.06\\
  C$_4$N ice    &     0.67 &     1.75 &    14.78\\
  C$_4$S ice    &     1.16 &     1.01 &    12.84\\
C$_5$H$_2$ ice  &     0.77 &     1.54 &    18.73\\
C$_6$H$_2$ ice  &     0.61 &     1.34 &    18.59\\
C$_6$H$_6$ ice  &     1.06 &     1.99 &    13.81\\
C$_7$H$_2$ ice  &     0.67 &     1.62 &    17.95\\
C$_8$H$_2$ ice  &     0.70 &     1.82 &    18.44\\
C$_9$H$_2$ ice  &     0.58 &     1.04 &    18.06\\
     CCL ice    &     1.06 &     0.98 &    11.76\\
      CH ice    &     0.93 &     1.20 &    16.16\\
  CH$_2$ ice    &     0.79 &     1.37 &    17.73\\
CH$_2$CO ice    &     1.36 &     1.37 &    15.26\\
CH$_2$NH ice    &     1.37 &     1.45 &    13.37\\
CH$_2$NH$_2$ ice  &     1.37 &     1.76 &    12.95\\
  CH$_3$ ice    &     0.40 &     1.34 &    17.77\\
CH$_3$CN ice    &     0.71 &     1.78 &    14.65\\
CH$_3$NH ice    &     1.38 &     1.70 &    12.95\\
CH$_3$OH ice    &     0.87 &     1.70 &    16.08\\
  CH$_4$ ice    &     0.26 &     1.51 &    19.95\\
      CL ice    &     0.63 &     0.88 &    15.20\\
      CN ice    &     0.58 &     1.00 &    15.53\\
      CO ice    &     0.92 &     0.85 &    19.51\\
      CS ice    &     0.66 &     1.87 &    14.93\\
 H$_2$CN ice    &     1.33 &     1.71 &    13.25\\
 H$_2$CO ice    &     1.03 &     0.93 &    16.73\\
H$_2$S$_2$ ice  &     0.63 &     1.16 &    12.84\\
H$_2$SiO ice    &     1.20 &     1.46 &    12.60\\
HC$_2$NC ice    &     0.93 &     1.47 &    12.99\\
 HC$_3$N ice    &     1.31 &     1.95 &    15.57\\
 HC$_5$N ice    &     0.71 &     1.47 &    14.82\\
 HC$_7$N ice    &     0.99 &     0.94 &    14.49\\
 HC$_9$N ice    &     1.04 &     0.72 &    14.42\\
     HCS ice    &     1.29 &     0.72 &    11.06\\
     HNC ice    &     0.60 &     1.41 &    18.32\\
    HNCO ice    &     0.24 &     1.10 &    14.80\\
    HNSi ice    &     0.81 &     1.14 &    12.52\\
  HS$_2$ ice    &     0.64 &     1.17 &    12.84\\
       N ice    &     1.20 &     1.27 &    16.13\\
  N$_2$O ice    &     1.30 &     1.70 &    14.13\\
    NaOH ice    &     1.39 &     0.87 &    11.13\\
      NH ice    &     0.46 &     1.01 &    17.17\\
  NH$_2$ ice    &     0.07 &     1.47 &    17.59\\
NH$_2$CHO ice   &     1.12 &     1.44 &    13.23\\
NH$_2$CN ice    &     0.45 &     1.26 &    13.54\\
      NO ice    &     1.04 &     1.56 &    15.55\\
       O ice    &     1.04 &     1.53 &    16.49\\
     OCN ice    &     1.35 &     1.63 &    12.55\\
      OH ice    &    -0.59 &     0.98 &    18.05\\
       P ice    &     0.11 &     0.94 &    14.65\\
      PH ice    &     1.02 &     1.93 &    13.69\\
  PH$_2$ ice    &     0.86 &     1.65 &    13.75\\
      Si ice    &     1.48 &     1.49 &    12.29\\
 SiC$_2$ ice    &     1.25 &     0.91 &    12.76\\
SiC$_2$H ice    &     1.19 &     0.93 &    12.22\\
 SiH$_2$ ice    &     1.24 &     1.57 &    14.06\\
 SiH$_3$ ice    &     1.24 &     1.05 &    14.03\\
     SiN ice    &     0.83 &     0.74 &    13.40\\
     SiO ice    &     1.28 &     0.78 &    14.99\\
      SO ice    &     0.72 &     1.72 &    14.86\\

\enddata
\tablenotemark{*}\tablenotetext{*}{Listed are the molecules which vertical
column densities in the disk exceed $10^{11}$~cm$^{-2}$.}
\tablecomments{Col.2 The Mixing Importance Measure (MIM) integrated over the
disk (in log scale). Col.3 The ratio of the vertical column density at $t=5$~Myr
computed with
the 2D-mixing and the laminar chemical models, averaged over the radius (in
log scale). Col.4 The maximal vertical column density for a given molecule
at 5~Myr in the laminar model.}
\end{deluxetable}

\LongTables
\begin{deluxetable}{lccc}
\tablecaption{Species hypersensitive to turbulent mixing\label{tab:hyps}}
\tablehead{
\colhead{Molecule$^*$} & \colhead{$\lg(Se)$}  & \colhead{$\lg(CDR)$} &
\colhead{$\lg(N_{\rm stat}^{\rm max})$}
}
\startdata
     C$_1$$_0$    &     1.18 &     2.01 &    12.09\\
        C$_2$S    &     0.71 &     2.27 &    11.80\\
    C$_3$H$_3$    &     1.12 &     2.64 &    11.39\\
    C$_3$H$_4$    &     1.03 &     3.31 &    11.89\\
        C$_3$S    &     1.00 &     2.31 &    11.35\\
        C$_7$N    &     1.12 &     2.57 &    11.58\\
  CH$_3$C$_4$H    &     1.10 &     2.18 &    11.21\\
        CO$_2$    &     0.56 &     2.23 &    14.94\\
    H$_2$O$^+$    &     0.95 &     2.09 &    12.53\\
    H$_2$O$_2$    &     0.91 &     2.53 &    13.54\\
       HC$_7$N    &     1.13 &     2.34 &    11.34\\
         HCOOH    &     0.81 &     2.07 &    12.36\\
      HSiO$^+$    &     0.75 &     2.75 &    11.38\\
         N$^+$    &     0.67 &     2.00 &    11.51\\
        N$_2$O    &     0.44 &     2.92 &    12.30\\
        NO$_2$    &     0.85 &     2.26 &    12.06\\
         O$_3$    &     0.05 &     2.02 &    15.06\\
        OH$^+$    &     1.00 &     2.68 &    12.86\\
           SiO    &     0.76 &     3.33 &    13.17\\
            SO    &     0.74 &     3.23 &    13.33\\
        SO$_2$    &     0.81 &     3.83 &    12.45\\
C$_2$H$_4$ ice  &     0.99 &     2.61 &    16.71\\
C$_2$H$_6$ ice  &     0.86 &     6.37 &    18.21\\
  C$_2$S ice    &     0.95 &     2.79 &    12.86\\
C$_3$H$_3$N ice  &     1.14 &     2.04 &    11.92\\
C$_3$H$_4$ ice  &     1.00 &     5.65 &    19.05\\
   C$_5$ ice    &     1.28 &     5.24 &    12.63\\
  C$_5$H ice    &     1.00 &     3.67 &    13.38\\
C$_5$H$_4$ ice  &     1.57 &     7.39 &    14.81\\
   C$_6$ ice    &     1.33 &     4.78 &    12.51\\
  C$_6$H ice    &     0.91 &     3.77 &    13.08\\
C$_6$H$_4$ ice  &     1.10 &     2.21 &    14.51\\
   C$_7$ ice    &     1.39 &     5.34 &    12.22\\
  C$_7$H ice    &     1.11 &     3.93 &    12.65\\
C$_7$H$_4$ ice  &     1.53 &     8.73 &    13.81\\
   C$_8$ ice    &     1.32 &     4.65 &    12.72\\
  C$_8$H ice    &     1.01 &     2.83 &    13.24\\
C$_8$H$_4$ ice  &     1.53 &     9.24 &    14.19\\
   C$_9$ ice    &     1.35 &     4.83 &    12.79\\
  C$_9$H ice    &     1.14 &     3.52 &    13.60\\
C$_9$H$_4$ ice  &     1.54 &    10.09 &    13.83\\
     CCP ice    &     1.30 &     2.74 &    11.47\\
CH$_2$PH ice    &     1.54 &     6.99 &    12.26\\
CH$_3$C$_3$N ice &     1.37 &     2.18 &    13.23\\
CH$_3$C$_4$H ice &     1.04 &     2.01 &    14.79\\
CH$_3$C$_6$H ice &     1.14 &     2.13 &    13.66\\
CH$_3$CHO ice   &     1.17 &     3.26 &    13.15\\
  CO$_2$ ice    &     1.03 &     3.60 &    19.62\\
      CP ice    &     1.12 &     2.66 &    14.68\\
      Fe ice    &     1.27 &     3.14 &    13.35\\
H$_2$C$_3$O ice &     1.34 &     2.96 &    15.94\\
H$_2$O$_2$ ice  &     1.40 &     2.20 &    16.97\\
H$_5$C$_3$N ice &     1.55 &     7.22 &    13.16\\
    HCCP ice    &     1.41 &     2.85 &    11.08\\
     HCL ice    &     0.64 &     2.86 &    15.60\\
   HCOOH ice    &     1.32 &     3.47 &    16.71\\
     HCP ice    &     1.21 &     3.21 &    14.22\\
 HNC$_3$ ice    &     0.91 &     2.76 &    11.09\\
     HNO ice    &     0.81 &     2.38 &    17.90\\
     HPO ice    &     1.37 &     5.45 &    13.28\\
      Mg ice    &     1.66 &     7.32 &    12.74\\
     MgH ice    &     1.25 &     4.37 &    14.39\\
   N$_2$ ice    &     0.86 &     4.24 &    18.37\\
N$_2$H$_2$ ice  &     0.76 &     2.62 &    16.65\\
      Na ice    &     1.35 &     5.44 &    12.99\\
  NO$_2$ ice    &     1.25 &     2.84 &    13.20\\
      NS ice    &     1.19 &     2.37 &    14.27\\
   O$_2$ ice    &     0.99 &     2.25 &    17.31\\
  O$_2$H ice    &     1.35 &     2.79 &    11.28\\
   O$_3$ ice    &     0.91 &     2.58 &    18.07\\
      PN ice    &     1.30 &     2.25 &    13.69\\
      PO ice    &     1.09 &     3.05 &    14.17\\
     SiH ice    &     1.44 &     2.24 &    12.99\\
     SiS ice    &     1.30 &     2.14 &    11.63\\
C$_3$H$_3$ ice  &     1.40 &     6.40 &    11.31\\

\enddata
\tablenotemark{*}\tablenotetext{*}{Listed are the molecules which vertical
column densities in the disk exceed $10^{11}$~cm$^{-2}$.}
\tablecomments{Col.2 The Mixing Importance Measure (MIM) integrated over the
disk (in log scale). Col.3 The ratio of the vertical column density at $t=5$~Myr
computed with
the 2D-mixing and the laminar chemical models, averaged over the radius (in
log scale). Col.4 The maximal vertical column density for a given molecule
at 5~Myr in the laminar model.}
\end{deluxetable}

As we have already mentioned, we present results for three models. In the laminar model no diffusion is taken into account. 
In the fast diffusion model $Sc=1$ is assumed, while in the slow diffusion model we use $Sc=100$. In Fig.~\ref{fig:mim_vs_cdr} 
the disk-averaged MIMs (Eqs.~\ref{eq:MIMrz}--\ref{eq:MIM}) are plotted 
versus the Column Density Ratios (CDRs) for fast mixing and laminar cases and for all species 
potentially observable with ALMA (with maximum column densities above $10^{11}$~cm$^{-2}$). The CDR is defined as:
\begin{equation}
 \label{eq:CDR}
 \lg(CDR)=\max\left(|\lg(N_{\rm 2D}(r)-\lg(N_{\rm laminar}(r)|\right)
\end{equation}
As we have mentioned in Sect.~\ref{mim}, high MIM value is only the necessary condition for molecular concentrations 
to be altered by the turbulence. This is clearly demonstrated in the Figure~\ref{fig:mim_vs_cdr}, where 
for species with the low MIM values of $\la 0.25$ the maximum ratio between the vertical column densities calculated with the
efficient 2D-mixing and the laminar models does not exceed an order of magnitude. For example, C$^+$, CO, and NH$_3$
belong to this group of species. For larger MIMs the scatter in CDR values rapidly increases, reaching 
10 orders of magnitude. The CH$_3$OH ice, SO$_2$, and CO$_2$ falls into this MIM range, with the CDRs in the range of 
1.7--3.8~dex. The most extreme group of species populating the right top corner in Fig.~\ref{fig:mim_vs_cdr} are 
carbon chains sitting in dust grain mantles (e.g., C$_9$H$_4$ ice). The linear Pearson correlation coefficient calculated
for the MIM-CDR distribution is $\approx 0.5$ -- a medium/large confidence correlation.
Thus our straightforwardly-defined MIM, which is easy to calculate as it only requires a non-mixing chemical model, 
can be useful in {\em a priori} analysis of a sensitivity of a given chemical species to 
the effects of transport. 

To facilitate further interpretation of our results 
for the reader, we divide the disk into 3 chemically distinct regions \citep[e.g.,][]{Aea02, Red2}, using the abundance 
distributions of C$^+$ and CO as a proxy (see Fig.~\ref{fig:zones}). The hot and UV-irradiated atmosphere at 
$z\ga 1.5-2\,H_{\rm r}$ is molecularly poor,  with ionized carbon being the primal C-bearing species. Deeper in the disk, 
in the warm molecular layer, C$^+$ is converted into CO and a multitude of other molecules that remain in the gas phase 
(located between $\sim 0.7$ and $\sim 2$ pressure scale heights). The dense and cold disk midplane region is marked
by the onset of molecular depletion from the gas phase ($\la 0.5-0.9$ scale heights). Also, in what follows we 
refer to the regions located at radii $\la 100$~AU as to the ``inner'' disk regions, and the ``outer'' disk regions
otherwise.

The individual families of primal ions, C-, O-, N-, S-bearing and complex (organic) species are depicted in 
Figs.~\ref{fig:ions}--\ref{fig:complex} and discussed in detail below. The major influence of turbulent transport on the
disk chemical evolution can be summarized as follows. 
First, two-dimensional mixing behaves as a combination of vertical and radial mixing processes,
that have to be considered simultaneously.
Vertical mixing is more important as it affects the evolution of gas-phase and surface species of any kind, 
whereas the effect of radial mixing is pronounced mostly for the evolution of ices
(e.g. the CO$_2$ ice from the OH and CO ices). The reason is that radial temperature gradient is weaker, and thus is only
relevant for the evolution of polyatomic ices formed via surface reactions of heavy radicals. On the other hand, steep 
vertical gradients of temperature and high-energy radiation intensity 
cause much sharper transition from the ice-dominated chemistry in the disk midplane to the oasis of the gas-phase chemistry 
in the warm molecular layer.

Second, the inefficient diffusion in the slow mixing model leads to molecular abundances and column densities
that almost coincide with those calculated with the laminar (non-mixing) model, but not always. 
The large Schmidt number $Sc=100$ assumed for this model makes its characteristic dynamical timescale too long,
$\sim 1-100$~Myr, to be competitive with most of the slowest chemical timescales deterimed by surface processes.

Third, diffusion in the fast mixing model enhances abundances and column densities of many species
by several orders of magnitude (Tables~\ref{tab:steadfast}--\ref{tab:hyps}). We divide all the considered species
into three groups.  Species with fast chemistry (faster than the diffusion timescale of $ \sim 10^5-10^6$ years) are rather 
insensitive to turbulent transport. Further we call them ``steadfast'' species (Table~\ref{tab:steadfast}). 
These include simple radicals and ions (e.g., C$^+$, Mg$^+$, CO,  OH, C$_2$H, H$_3$O$^+$, HCN, N$_2$) and few
abundant ices (e.g., water and ammonia ices). Their column densities calculated with the laminar model and the fast 
mixing model differ by no more than a factor of 3-5, which is comparable to intrinsic uncertainties in
molecular concentrations caused by uncertainties in the reaction rates \citep[e.g.,][]{Vasyunin_ea08}.

On the other hand, turbulent diffusion alters abundances of
many polyatomic species, in particular complex (organic) molecules and their ices produced (at least partially) on dust
grain  surfaces (e.g., CO$_2$, HCOOH, CH$_3$CHO ice, carbon chain and cyanopolyyne ices, O$_2$ ice, SO, SO$_2$). These species 
are included in ``sensitive'' and ``hypersensitive'' groups. 
Column densities of ``sensitive'' species (Tables~\ref{tab:sens}) are altered by diffusion by up to 2 orders of magnitude, and 
even stronger for a ``hypersensitive'' group (Table~\ref{tab:hyps}).
Mixing steadily transports ice-coated grains in
warmer regions, allowing more efficient surface processing due to enhanced hopping rates of heavy
radicals. In warm intermediate layer these ``rich'' ices eventually  evaporate, and in the inner disk they can also be
photodissociated by CRP/X-ray-induced UV photons. The importance of
mixing is higher in an inner, planet-forming disk zone, where thermal, density, and high-energy radiation gradients
are stronger than in the outer region. Finally, even efficient 2D transport in the fast mixing model cannot completely erase the
layered chemical structure in the disk, leaving the midplane a gas-phase molecular ``desert''.

\subsection{Major atomic and molecular ions and the ionization degree}
\label{ions}
\begin{figure*}
\includegraphics[width=0.48\textwidth]{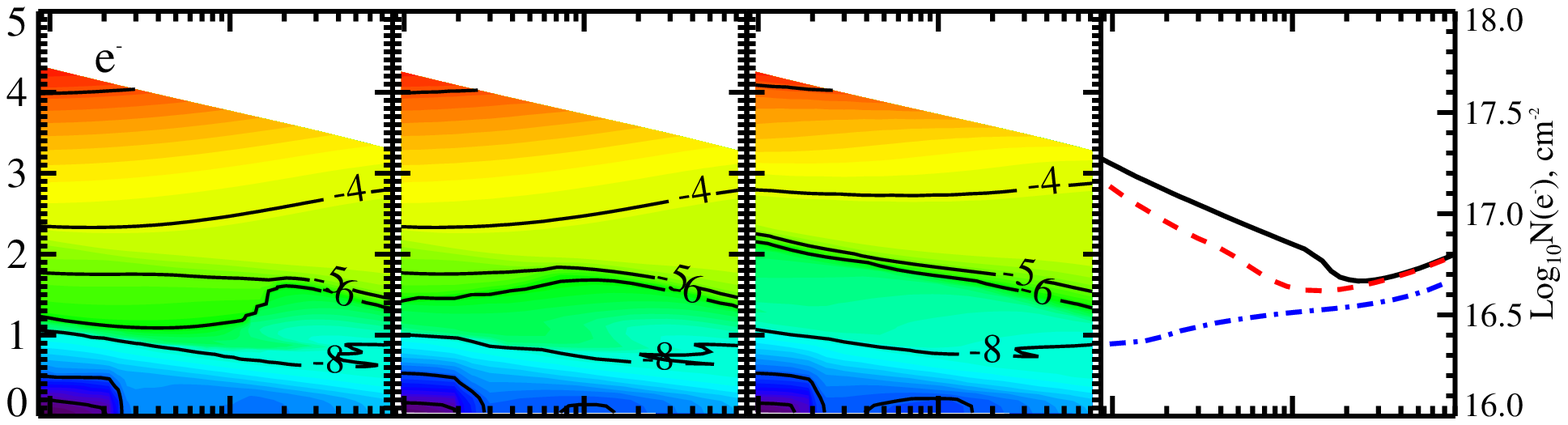}
\includegraphics[width=0.48\textwidth]{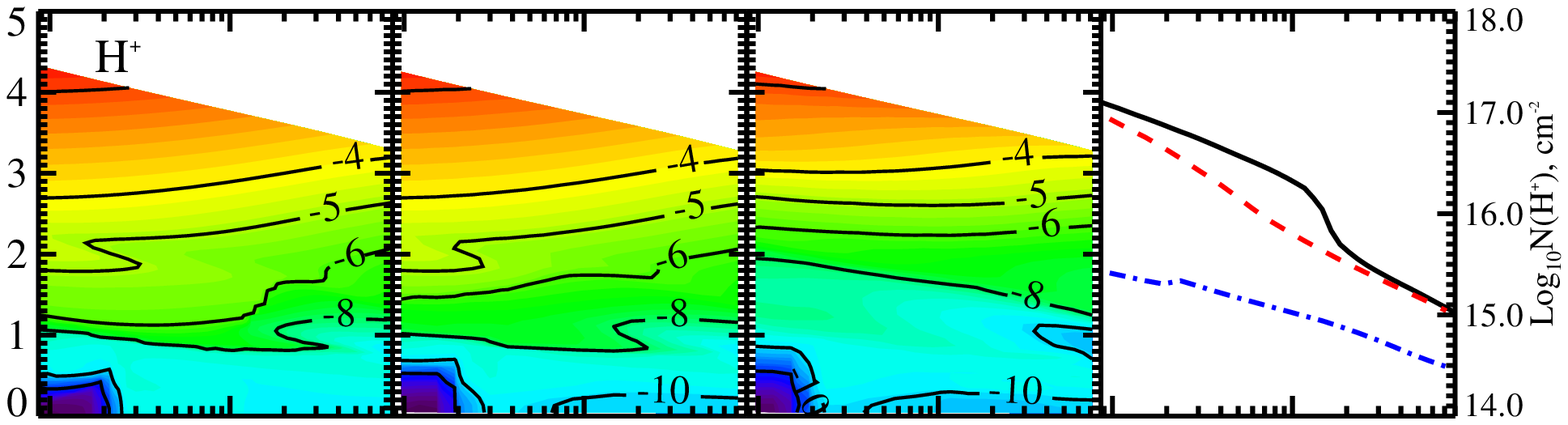}\\
\includegraphics[width=0.48\textwidth]{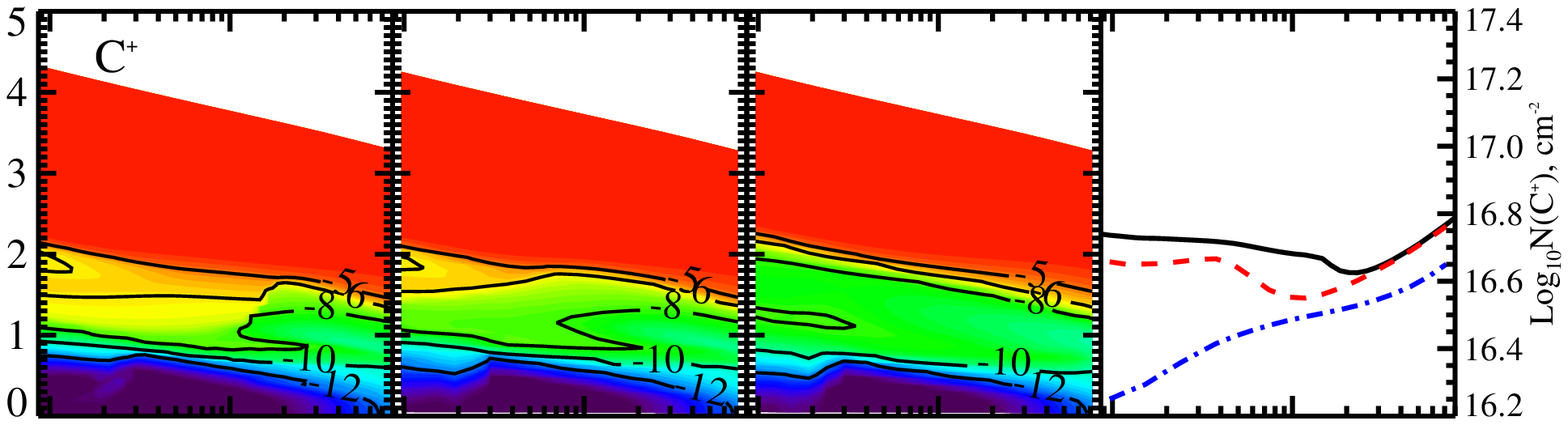}
\includegraphics[width=0.48\textwidth]{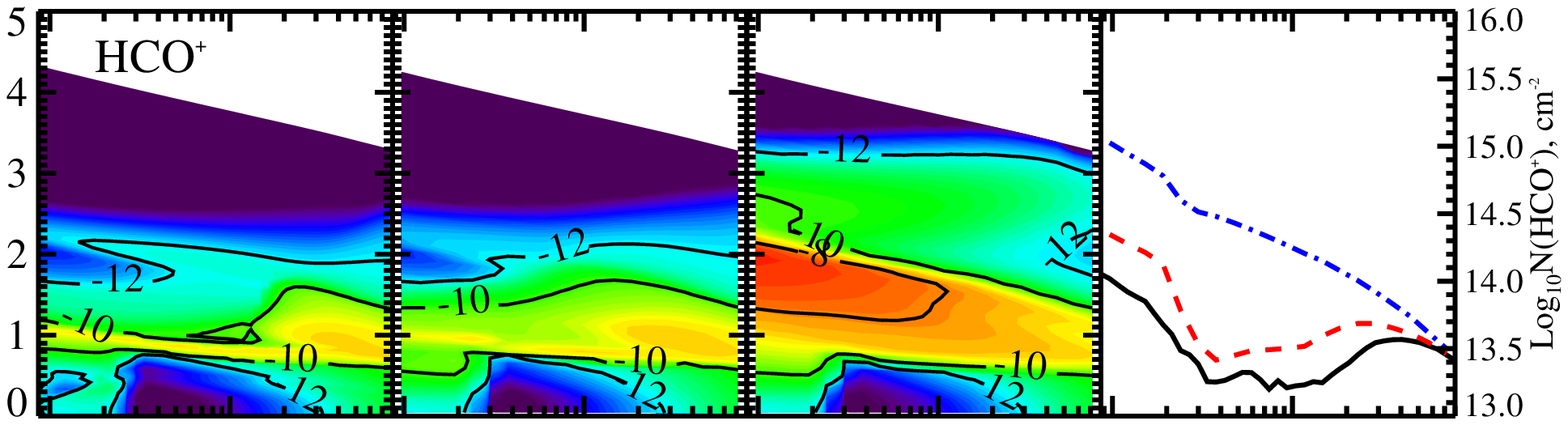}\\
\includegraphics[width=0.48\textwidth]{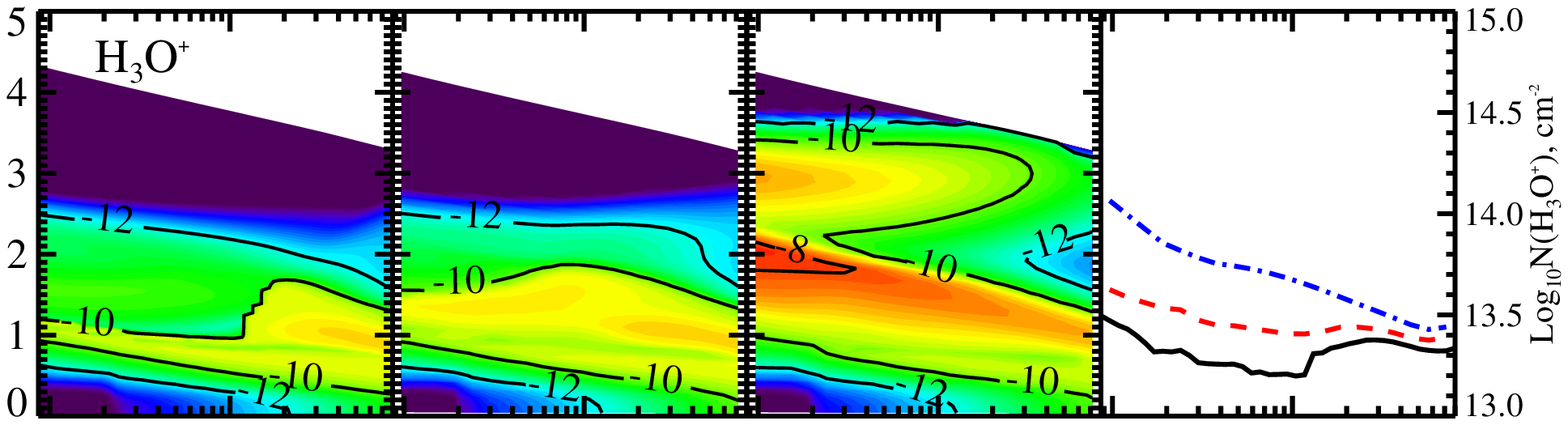}
\includegraphics[width=0.48\textwidth]{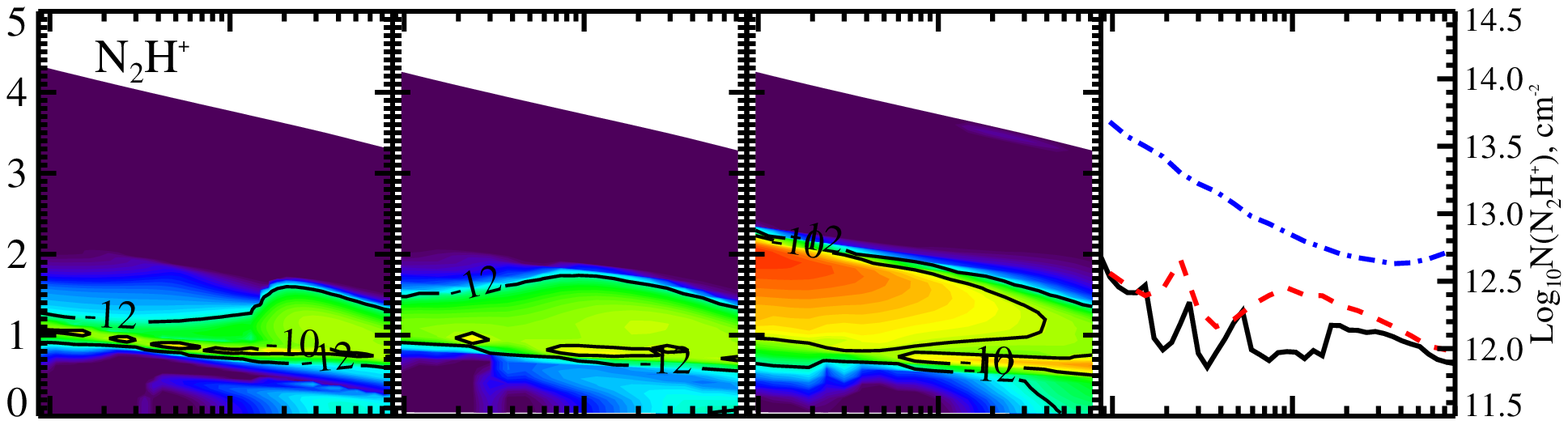}\\
\includegraphics[width=0.48\textwidth]{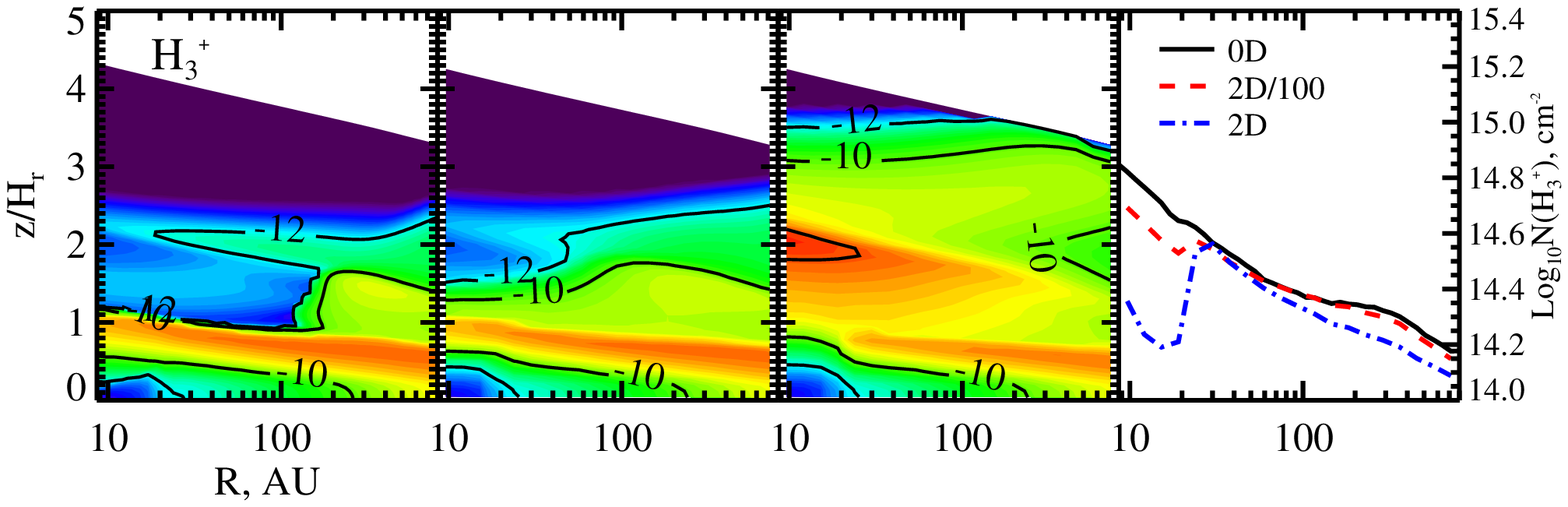}
\includegraphics[width=0.48\textwidth]{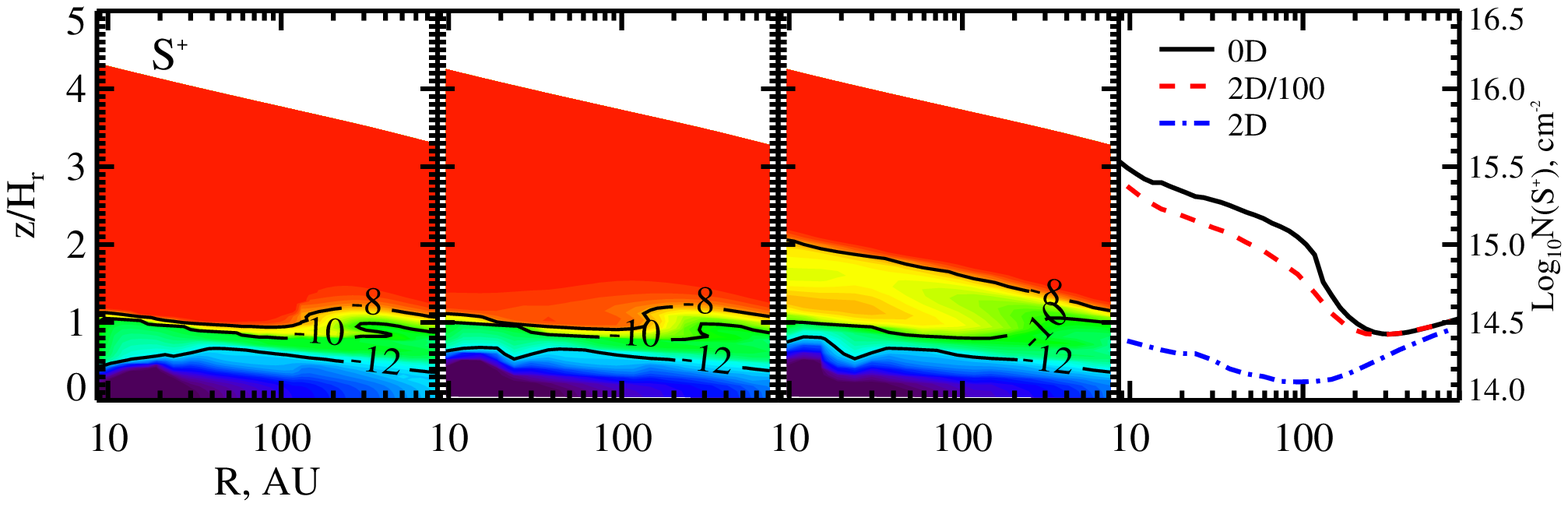}\\
\caption{Abundances and column densities of electrons and selected ions
in the DM Tau disk at 5~Myr. (Top to bottom) The $\log$ of relative abundances
(with respect to the total number of hydrogen nuclei) and vertical column
densities of electrons, H$^+$, C$^+$, HCO$^+$, H$_3$O$^+$,
N$_2$H$^+$, H$_3^+$, and S$^+$. (Left to right) Results for
the three disk models are shown: (1) laminar chemistry, 2) the 2D-mixing
chemistry ($Sc=100$), and (3) the 2D-mixing chemistry with $Sc=1$.}
\label{fig:ions}
\end{figure*}

\LongTables
\begin{deluxetable*}{llrrll}
\tabletypesize{\footnotesize}
\tablecaption{Key chemical processes: ionization degree\label{tab:key_reac_ion}}
\tablehead{
\colhead{Reaction} & \colhead{$\alpha$} & \colhead{$\beta$} & \colhead{$\gamma$}& \colhead{t$_{\rm min}$} & \colhead{t$_{\rm
max}$}\\
\colhead{} & \colhead{[(cm$^3$)\,s$^{-1}$]} & \colhead{} & \colhead{[K]} & \colhead{[yr]} & \colhead{[yr]}
}
\startdata
H$_2$   + X-rays  $\rightarrow$  e$^-$   + H$_2$$^+$   & $0.93\zeta_{\rm X}$   & $0$   & $0$   & $1.00$   & $1.20\,(4)$   \\
H$_2$   +  C.R.P.  $\rightarrow$  e$^-$   + H$_2$$^+$   & $0.93\zeta_{\rm CR}$   & $0$   & $0$   & $1.00$   & $5.00\,(6)$   \\
H   + X-rays  $\rightarrow$  H$^+$   + e$^-$   & $0.46\zeta_{\rm X}$   & $0$   & $0$   & $1.00$   & $5.00\,(6)$   \\
C$^+$   + X-rays  $\rightarrow$  C$^+$$^+$   + e$^-$   & $0.5\zeta_{\rm X}$   & $0$   & $0$   & $1.00$   & $5.00\,(6)$   \\
S$^+$   + X-rays  $\rightarrow$  S$^+$$^+$   + e$^-$   & $\zeta_{\rm X}$   & $0$   & $0$   & $1.00$   & $5.00\,(6)$   \\
H$_2$   + X-rays  $\rightarrow$  e$^-$   + H   + H$^+$   & $0.02\zeta_{\rm X}$   & $0$   & $0$   & $1.00$   & $5.00\,(6)$   \\
C   + UV  $\rightarrow$  e$^-$   + C$^+$   & $0.31\,(-9)$   & $0$   & $3.33$   & $1.00$   & $5.00\,(6)$   \\
H$_2$CO   + UV  $\rightarrow$  HCO$^+$   + H   + e$^-$   & $0.48\,(-9)$   & $0$   & $3.21$   & $4.18\,(2)$   & $1.20\,(4)$   \\
H$_2$$^+$   + UV  $\rightarrow$  H$^+$   + H   & $0.57\,(-9)$   & $0$   & $2.37$   & $1.00$   & $55.90$   \\
H$_3$$^+$   + UV  $\rightarrow$  H$^+$   + H$_2$   & $0.20\,(-7)$   & $0$   & $1.80$   & $1.00$   & $5.00\,(6)$   \\
H$^+$   + O  $\rightarrow$  O$^+$   + H   & $0.70\,(-9)$   & $0$   & $2.32\,(2)$   & $1.00$   & $5.00\,(6)$   \\
H$^+$   + O$_2$  $\rightarrow$  O$_2$$^+$   + H   & $0.12\,(-8)$   & $0$   & $0$   & $1.00$   & $5.00\,(6)$   \\
H$^+$   + S  $\rightarrow$  S$^+$   + H   & $0.13\,(-8)$   & $0$   & $0$   & $1.00$   & $5.00\,(6)$   \\
He$^+$   + CO  $\rightarrow$  C$^+$   + O   + He   & $0.16\,(-8)$   & $0$   & $0$   & $1.00$   & $5.00\,(6)$   \\
C$^+$   + H$_2$  $\rightarrow$  CH$_2$$^+$   & $0.40\,(-15)$   & $-0.20$   & $0$   & $1.00$   & $5.00\,(6)$   \\
S$^+$   + H$_2$  $\rightarrow$  H$_2$S$^+$   & $1.00\,(-17)$   & $-0.20$   & $0$   & $1.00$   & $5.00\,(6)$   \\
H$_2$$^+$   + H$_2$  $\rightarrow$  H$_3$$^+$   + H   & $0.21\,(-8)$   & $0$   & $0$   & $1.00$   & $5.00\,(6)$   \\
H$_2$O$^+$   + H$_2$  $\rightarrow$  H$_3$O$^+$   + H   & $0.61\,(-9)$   & $0$   & $0$   & $1.00$   & $5.00\,(6)$   \\
N$_2$$^+$   + H$_2$  $\rightarrow$  N$_2$H$^+$   + H   & $0.17\,(-8)$   & $0$   & $0$   & $1.00$   & $5.00\,(6)$   \\
NH$^+$   + H$_2$  $\rightarrow$  H$_3$$^+$   + N   & $0.23\,(-9)$   & $0$   & $0$   & $1.96$   & $5.00\,(6)$   \\
C$^+$   + S  $\rightarrow$  S$^+$   + C   & $0.15\,(-8)$   & $0$   & $0$   & $1.00$   & $5.00\,(6)$   \\
H$_3$$^+$   + CO  $\rightarrow$  HCO$^+$   + H$_2$   & $0.16\,(-8)$   & $0$   & $0$   & $1.00$   & $5.00\,(6)$   \\
H$_3$$^+$   + N$_2$  $\rightarrow$  N$_2$H$^+$   + H$_2$   & $0.17\,(-8)$   & $0$   & $0$   & $1.00$   & $5.00\,(6)$   \\
HS$^+$   + H  $\rightarrow$  S$^+$   + H$_2$   & $0.11\,(-9)$   & $0$   & $0$   & $1.00$   & $5.00\,(6)$   \\
N$_2$H$^+$   + CO  $\rightarrow$  HCO$^+$   + N$_2$   & $0.88\,(-9)$   & $0$   & $0$   & $1.00$   & $5.00\,(6)$   \\
NH$_2$$^+$   + N  $\rightarrow$  N$_2$H$^+$   + H   & $0.91\,(-10)$   & $0$   & $0$   & $1.00$   & $5.00\,(6)$   \\
O$^+$   + H  $\rightarrow$  H$^+$   + O   & $0.70\,(-9)$   & $0$   & $0$   & $7.48$   & $5.00\,(6)$   \\
OH$^+$   + N$_2$  $\rightarrow$  N$_2$H$^+$   + O   & $0.36\,(-9)$   & $0$   & $0$   & $2.34\,(4)$   & $5.00\,(6)$   \\
S$^+$   + O$_2$  $\rightarrow$  SO$^+$   + O   & $0.20\,(-10)$   & $0$   & $0$   & $1.00$   & $5.00\,(6)$   \\
S$^+$   + OH  $\rightarrow$  SO$^+$   + H   & $0.46\,(-8)$   & $-0.50$   & $0$   & $1.00$   & $5.00\,(6)$   \\
S$^+$   + NH$_3$  $\rightarrow$  NH$_3$$^+$   + S   & $0.16\,(-8)$   & $-0.50$   & $0$   & $1.00$   & $5.00\,(6)$   \\
S$^+$   + e$^-$  $\rightarrow$  S   & $0.39\,(-11)$   & $-0.63$   & $0$   & $4.18\,(2)$   & $5.00\,(6)$   \\
HCO$^+$   + e$^-$  $\rightarrow$  CO   + H   & $0.24\,(-6)$   & $-0.69$   & $0$   & $1.00$   & $5.00\,(6)$   \\
N$_2$H$^+$   + e$^-$  $\rightarrow$  NH   + N   & $0.64\,(-7)$   & $-0.51$   & $0$   & $1.00$   & $5.00\,(6)$   \\
H$_3$O$^+$   + e$^-$  $\rightarrow$  OH   + H   + H   & $0.26\,(-6)$   & $-0.50$   & $0$   & $1.00$   & $5.00\,(6)$   \\
e$^-$   + grain(0)  $\rightarrow$  grain(-)   & $1.00$   & $0$   & $0$   & $1.00$   & $5.00\,(6)$   \\
HCO$^+$   + grain(-)  $\rightarrow$  CO   + H   + grain(0)   & $1.00$   & $0$   & $0$   & $1.00$   & $5.00\,(6)$   \\
\enddata
\end{deluxetable*}

The magnetorotational instability has been shown to be the most promising mechanism in driving turbulence 
in weakly ionized PPDs \citep[][]{MRI}. However, even high-energy cosmic ray particles
cannot penetrate in cold and dense disk interiors \citep{UN81}, and the ionization degree drops so low that MRI may not be
operational at all
\citep[``dead zone'', e.g.,][]{gammie,sano,Armitageea2003,Ilgner_ea08}. At these conditions slow radiative 
recombination of metallic ions becomes important for the evolution of ionization fraction, if these are still present in the gas 
\citep[e.g.,][]{from2002,Red2,Ilgner_Nelson06a}. On the other hand, the viscous stresses can be transported to the ``dead zone'' 
from above MRI-active accretion layers by diffusive mixing, maintaining a low level of accretion in this region
 \citep[e.g.,][]{fs2003,Wunsch_ea06,Turner_ea07}. In this Section we analyze in detail chemical processes responsible for the
evolution of the ionization state of protoplanetary disks and how these are altered by turbulence.

Turbulent diffusion does not affect column densities of many atomic ions and a few simple molecular ions,
e.g. C$^+$, Mg$^+$, Fe$^+$, He$^+$, H$_3^+$, CH3$^+$, NH$_4^+$ (by less than a factor of 3, see Table~\ref{tab:steadfast}).
The charged species sensitive to mixing include, e.g., hydrocarbons, electrons, H$^+$,  O$^+$, S$^+$, N$_2$H$^+$, and HCO$^+$
(their column densities are altered by factors of $\sim 3$--50; Table~\ref{tab:sens}).
Only a few ions are hypersensitive to mixing, e.g. N$^+$, OH$^+$, H$_2$O$^+$ (CDRs are $\sim 100$; 
Table~\ref{tab:hyps}). This is a manifestation of the fact that the global evolution of the ionization degree is partly
affected by slow surface chemistry (recombination, dissociation) \citep[see also][]{Red2}.
 
In Fig.~\ref{fig:ions} the distributions of molecular abundances and column densities at 5~Myr of several 
major ions and the disk ionization degree calculated with the laminar and the mixing models are shown.  
The global ionization structure of the disk shows a layered structure similar to that of Photon-Dominated Regions
\citep[see also, e.g.,][]{Red2,Bergin_ea07,Roellig_ea07}: (1) heavily irradiated and ionized, hot atmosphere
where the key ions are C$^+$ and H$^+$, (2) partly UV-shielded, warm molecular layer 
where carbon is locked in CO and major charged species are X-ray-produced H$^+$ and polyatomic ions like HCO$^+$ and H$_3^+$,
and (3) dark, dense and cold midplane where most of molecules are frozen out onto dust grains and 
the most abundant charged species are dust grains and H$_3^+$.

Turbulent diffusion lowers abundances of electrons and atomic ions such as C$^+$, S$^+$, and H$^+$ by up to 3 orders of
magnitude, mainly in the inner molecular layer ($r \la 100$~AU, $\sim 1-2H_{\rm r}$), see Fig.~\ref{fig:ions}. 
Consequently, their column densities within $\sim 100$~AU are decreased by factors 3 -- $\ga 100$, and much less in 
the outer disk. The effect is exactly opposite for H$_3$O$^+$, HCO$^+$, and N$_2$H$^+$, which column 
densities and abundances are enhanced by the mixing in this region, albeit only up to an order of magnitude 
(Fig.~\ref{fig:ions}). The mixing remarkably expands their inner molecular layers that are rather confined in the laminar case 
(particularly, for N$_2$H$^+$). The abundance distribution of H$_3^+$ shows similar expansion due to the mixing as
the other polyatomic ions, while its column density is slightly decreased at $r\la 30$~AU. 
Overall, turbulent stirring makes chemical gradients of key ionized species stronger compared to the laminar disk model.
On the other hand, the slow mixing model does not differ significantly from the laminar 
model (compare 1st and 2nd panels in Fig.~\ref{fig:ions}).

To better understand these results, we performed detailed analysis of the chemical evolution of the dominant ions and the
ionization
degree shown in Fig.~\ref{fig:ions} in two disk vertical slices at $r=10$ and 250~AU (the laminar chemical model). 
The most important reactions responsible for
the time-dependent net change of the electron, H$^+$, C$^+$, HCO$^+$, H$_3$O$^+$, N$_2$H$^+$, H$_3^+$, and S$^+$ concentrations
in the midplane, molecular layer, and atmosphere are presented in Table~\ref{tab:key_reac_ion}, 
both for the inner and outer 
disk regions. Time interval when these processes are active is also shown. Note that our list contains only top 15 reactions for
the selected
species per region (midplane, molecular layer, atmosphere) for the entire 5~Myr time span, with all repetitions removed.

The energy deposition into the disk matter leads to ionization of molecular and atomic hydrogen by high-energy cosmic ray 
particles and the stellar X-ray radiation. Typical X-ray
ionization rates in the inner and outer disk atmosphere are $\zeta_{\rm X}\approx 10^{-12}$~s$^{-1}$ and $\approx
2\,10^{-13}$~s$^{-1}$, 
respectively. The H$_2^+$ is rapidly converted to H$_3^+$ by collisions with H$_2$, while in the disk atmosphere both H$_2^+$ 
and H$_3^+$ can be dissociated by UV, producing H$^+$. The line UV-photodissociation of molecular hydrogen is hindered by strong 
self-shielding of this molecule \citep[][]{Solomon_Wickramasinghe69,vD88}. In the atmosphere, the X-rays ionize other neutral
atoms 
(e.g., He), and atomic ions such as  C$^+$ and S$^+$, leading to abundant doubly-charged atomic ions in the very inner region
\citep[see, e.g.][]{Meijerink_ea08,Najita_ea10}. The heavy atomic ions are also produced by charge transfer reactions with H$^+$ 
and C$^+$. The atomic ions further slowly 
radiatively recombine with electrons (e.g., S$^+$ + e$^-$ $\rightarrow$ S) or converted to 
polyatomic ions by radiative recombination reactions (e.g., C$^+$ + H$_2$ $\rightarrow$ CH$_2^+$) 
or via fast ion-molecule pathways involving simple radicals (e.g., S$^+$ + OH $\rightarrow$ SO$^+$ + H
). As a result, the disk fractional ionization in the atmosphere is high, 
$X({\rm e}^-) \approx 10^{-4}$, and dominant charged species are C$^+$ (with the abundance $\la 10^{-4}$) and H$^+$ 
($\sim 10^{-5}-10^{-3}$). The pace of their chemical evolution is restricted to slow radiative recombination and association
processes 
\citep[e.g.,][]{Wakelam_ea10}, $\tau_{\rm chem} \sim 10^3-10^4$~yrs. These timescales are comparable to the vertical mixing
timescales 
($\tau_{\rm phys} \sim 10^3$~yrs) only in the inner disk region (see Figs.~\ref{fig:timescales}, \ref{fig:chem_ss}, and 
Table~\ref{tab:tau_inner}).

In molecular layer located beneath the disk atmosphere the ionization chemistry is mostly regulated by fast ion-molecule reactions
(see Fig.~\ref{fig:timescales}, 2nd panel), e.g. H$_3^+$ + N$_2$ $\rightarrow$ N$_2$H$^+$ + H$_2$. 
Typical X-ray ionization rates in the inner and outer molecular layer are $\zeta_{\rm X}\approx 10^{-12}$~s$^{-1}$ and 
$\approx 3\,10^{-14}$~s$^{-1}$, respectively. 
The partly absorbed X-rays ionize H$_2$, producing H$_2^+$ and H$^+$. In the inner disk molecular layer, the X-ray-driven
chemistry 
destroys $99\%$ of molecular hydrogen  ($r \la 100$~AU). There H$_2$ is converted to atomic hydrogen via fast ion-molecule
reactions
involving OH$^+$ and H$_2$O$^+$ (e.g., H$_2$O$^+$ + H$_2$ $\rightarrow$ H$_3$O$^+$ +H; Table~\ref{tab:key_reac_ion}).
In turn, these ions become abundant due to steady release of elemental oxygen from water, which begins from photodissociation
of the water ice. 
The H$_2^+$ is rapidly converted to H$_3^+$ that protonates abundant radicals (e.g., CO, N$_2$), forming various polyatomic ions 
(e.g., HCO$^+$, N$_2$H$^+$). 
Thus, the abundance distributions of the polyatomic ions are peaked along a rather narrow stripe, $0.5\,H_{\rm r}$, where both
H$_3^+$
and progenitor molecules are abundant. The molecular ion layers are vertically expanded in the outer 
disk at $r\ga200$~AU, where the surface density drops so low that the disk becomes partly transparent to the interstellar UV
radiation.
The N$_2$H$^+$ is destroyed by reactions with abundant CO and, thence, the N$_2$H$^+$ layer is particularly thin 
(see Fig.~\ref{fig:ions}). The N$_2$H$^+$, HCO$^+$, and H$_3$O$^+$ layers are rather sharp-edged from below where their 
parental molecules are locked in dust icy mantles \citep[e.g.,][]{Oeberg_ea07,Oeberg_ea09a,Oeberg_ea09b}, 
whereas the H$_3^+$ layer extends more toward the cold region as H$_2$ does not deplete 
\citep[e.g.,][]{Sandford_ea93,Buch_Devlin94,Kristensen_ea11}. The formation of polyatomic ions is balanced out by dissociative 
recombination with e$^-$ \citep[][]{Florescu_Mitchell06}, and within $\approx 10^4$~years (inner disk) and $\approx 10^5$~years 
(outer disk) a chemical quasi steady-state is reached. This equilibrium timescale is mainly set by slow X-ray-driven destruction
of 
H$_2$ that is needed to produce H$_3^+$, $\tau_{\rm X} \sim 2\,10^3-10^5$~years. Thus the characteristic chemical timescales
exceed 
those due to the mixing, making the ionization state of the inner molecular layer to be vulnerable to the turbulent transport. 
The dominant charged species in the warm molecular layer are H$^+$ as well HCO$^+$, H$_3$O$^+$ and C$^+$, with abundances of 
$10^{-10}-10^{-5}$. The resulting ionization fraction is $X{\rm e}^- \sim 10^{-8}-10^{-5}$, which is sufficient for MRI to be 
operational \citep[e.g.,][]{from2002}. 

Finally, in the cold, dark midplane the X-ray ionization rates further decrease ($\zeta_{\rm X} \sim
10^{-19}-10^{-15}$~s$^{-1}$), 
and the cosmic ray ionization becomes important ($\zeta_{\rm CR} = 1.3\,10^{-17}$~s$^{-1}$). Except of H$_2$,
majority of molecules severely deplete within $\la10^3$~years, leaving only a tiny fraction in the gas 
(see Fig.~\ref{fig:timescales}, 4th panel, and Fig.~\ref{fig:C}). Electron sticking to 
grains is efficient, and negatively charged grains start playing important role as charge carriers 
\citep[e.g.,][]{un1980,Okuzumi_09,PerezBecker_Chiang11a} and sites of dissociative recombination (e.g., 
HCO$^+$   + grain(-)  $\rightarrow$  CO   + H   + grain(0); see Table~\ref{tab:key_reac_ion}). Consequently, the dominant charged
species in the midplane are dust grains and H$_3^+$, with H$^+$ being important in the upper midplane layer (Fig.~\ref{fig:ions}).
 
The fractional ionization drops to as low values as $\la 10^{-12}-10^{-9}$, and a ``dead'' zone for the MRI-driven accretion may 
develop \citep[e.g.,][]{gammie,Turner_ea07,Dzyurkevich_ea10}. Characteristic timescales of the
evolution of the fraction ionization in the disk midplane are $\sim 5\,10^3-10^5$~years, after which a quasi steady-state is
reached.
These values are shorter than the mixing timescale in the midplane (Fig.~\ref{fig:timescales}), and influence of turbulent
transport
on ionization chemistry shall be negligible.

Indeed, as can be clearly seen in Fig.~\ref{fig:ions} (compare 1st and 3rd panels), turbulent diffusion most strongly affects
abundances of ions in the inner warm molecular layer, at $r\la 200$~AU, $H_r \approx 1-2$, where the characteristic diffusive
mixing timescale ($\sim 10^3$~years) is shorther than the chemical timescale ($\sim 10^4$~years) regulated by the slow
X-ray-driven 
destruction of H$_2$ and radiative recombination and association processes involving C$^+$. The diffusion timescale is relatively
short 
in the inner disk because of stronger gradients of density and ionization rates, elevated temperatures, and short distances for
transport 
(see Figs.~\ref{fig:disk_struc} and \ref{fig:timescales}). The vertical uptake of molecular hydrogen from the inner midplane to
the 
inner intermediate layer allows to produce more H$_3^+$, while simultaneously lowering ionization fraction and abundances of 
H$^+$ and C$^+$ due to their enhanced neutralization in via radiative association, charge exchange, and ion-molecule 
reactions (e.g., C$^+$ + H$_2$ $\rightarrow$ CH$_2^+$; Table~\ref{tab:key_reac_ion}). The increased abundances of H$_3^+$ in the
inner 
molecular layer leads to more efficient production of polyatomic ions through protonation reactions, which lead to remarkable
peaks in 
the relative abundances of HCO$^+$, N$_2$H$^+$, H$_3$O$^+$, and other molecular ions. The chemistry of OH$^+$, H$_2$O$^+$, and
N$^+$ 
ions is most sensitive to the X-ray-driven, partly surface, processes \citep[similar results
were obtained by][]{Aresu_ea10a}. Naturally,  OH$^+$, H$_2$O$^+$, and N$^+$  are most sensitive to the mixing in our model.

To summarize, turbulent mixing lowers the disk ionization fraction and the abundances of atomic ions 
in the inner disk region at intermediate heights ($r\la 200$~AU, $\sim 1-2H_r$) by up to 3 orders of magnitude, 
whereas the concentrations of polyatomic ions are enhanced there. 

\subsection{Carbon-containing molecules}
\label{C-species}
\begin{figure*}
\includegraphics[width=0.48\textwidth]{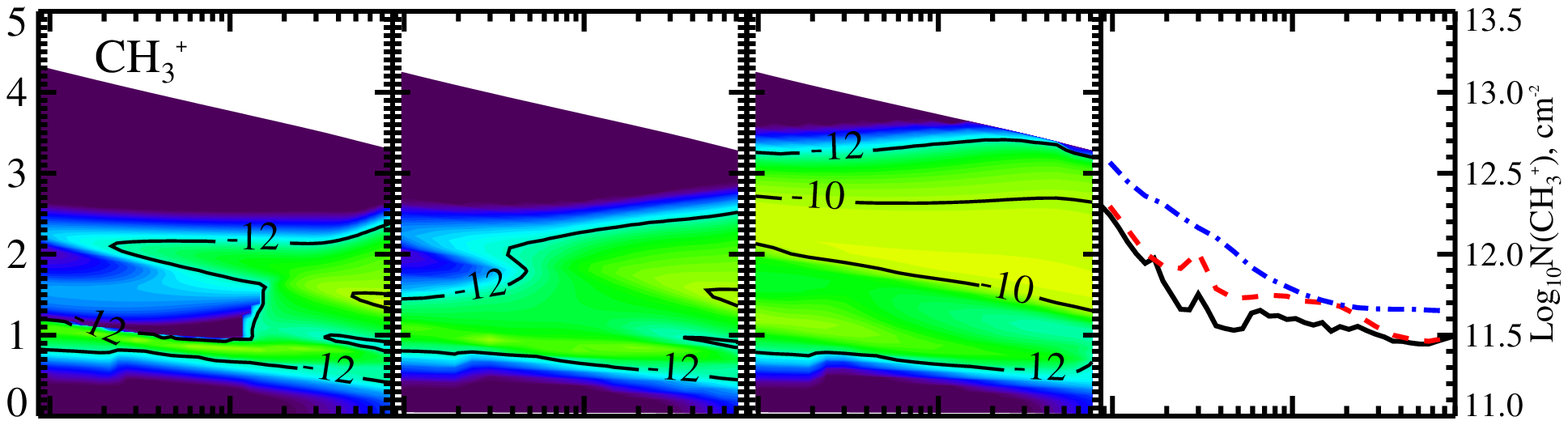}
\includegraphics[width=0.48\textwidth]{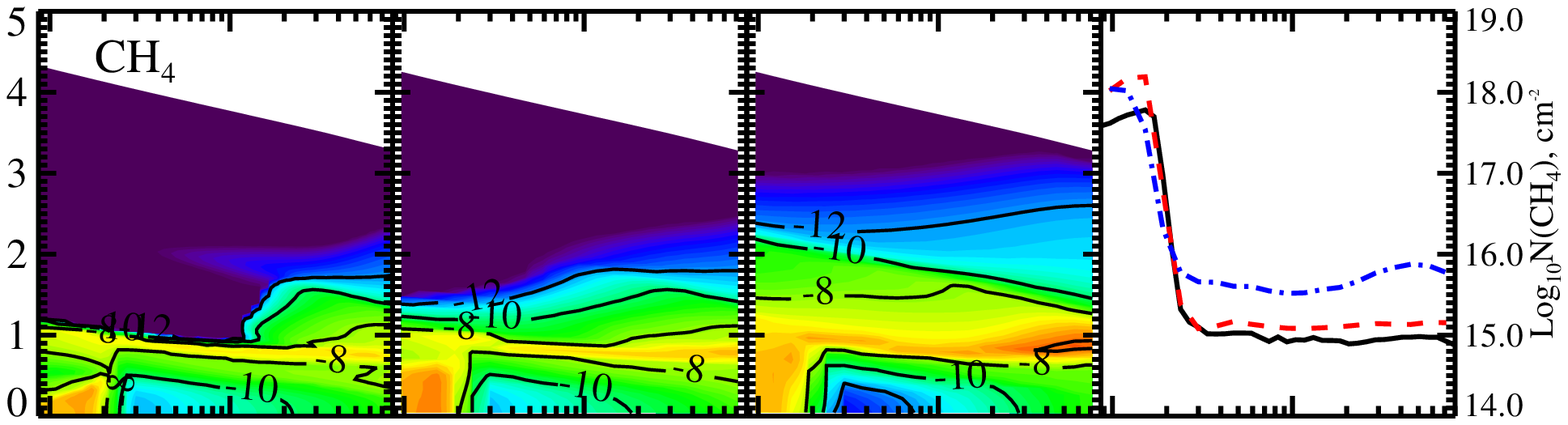}\\
\includegraphics[width=0.48\textwidth]{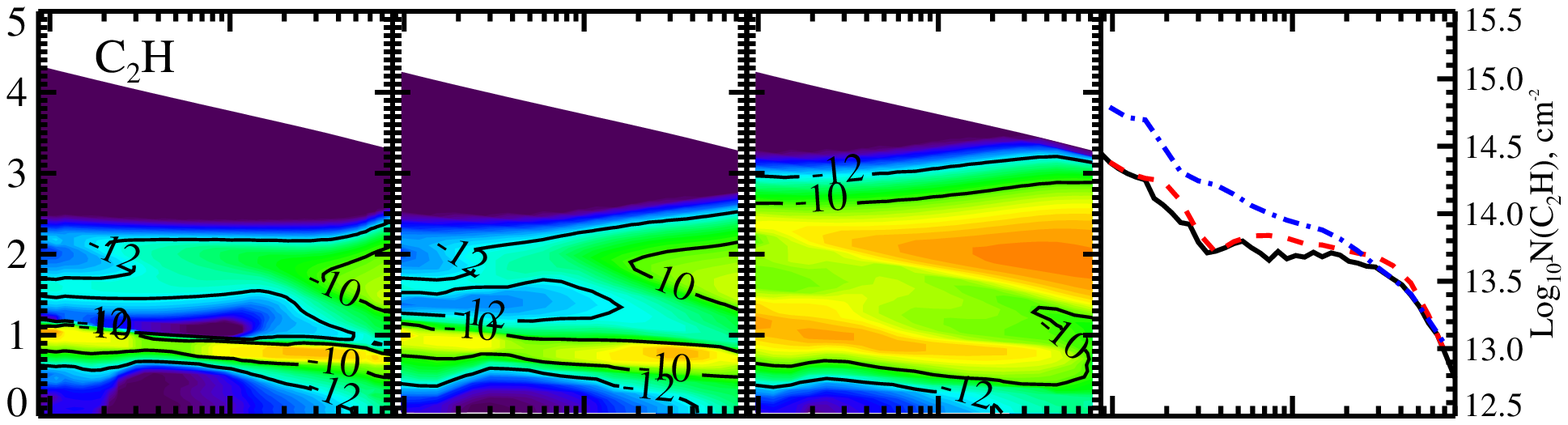}
\includegraphics[width=0.48\textwidth]{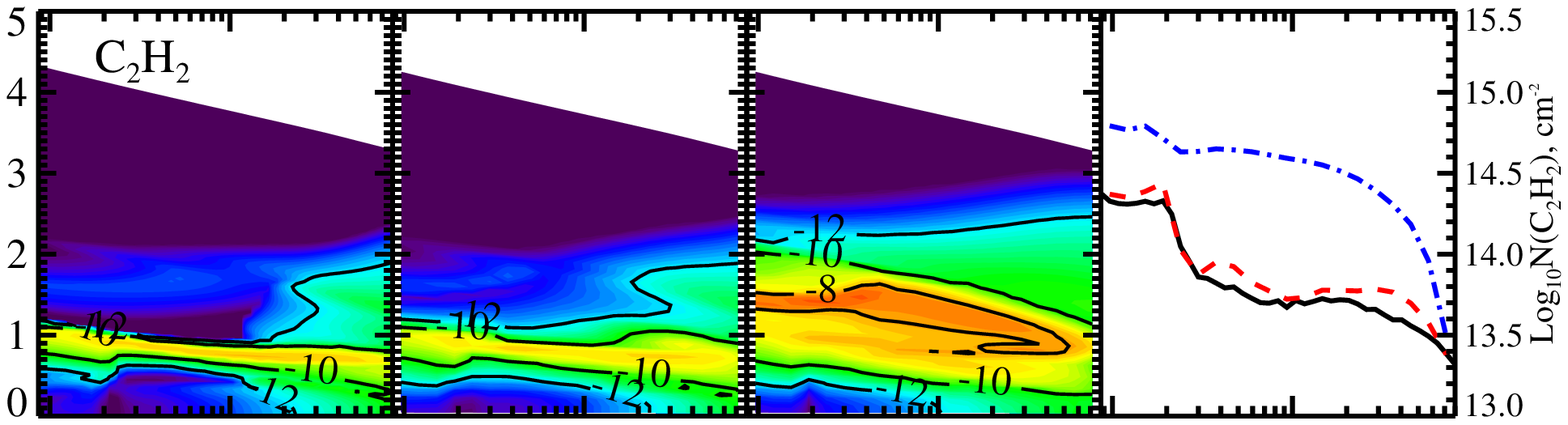}\\
\includegraphics[width=0.48\textwidth]{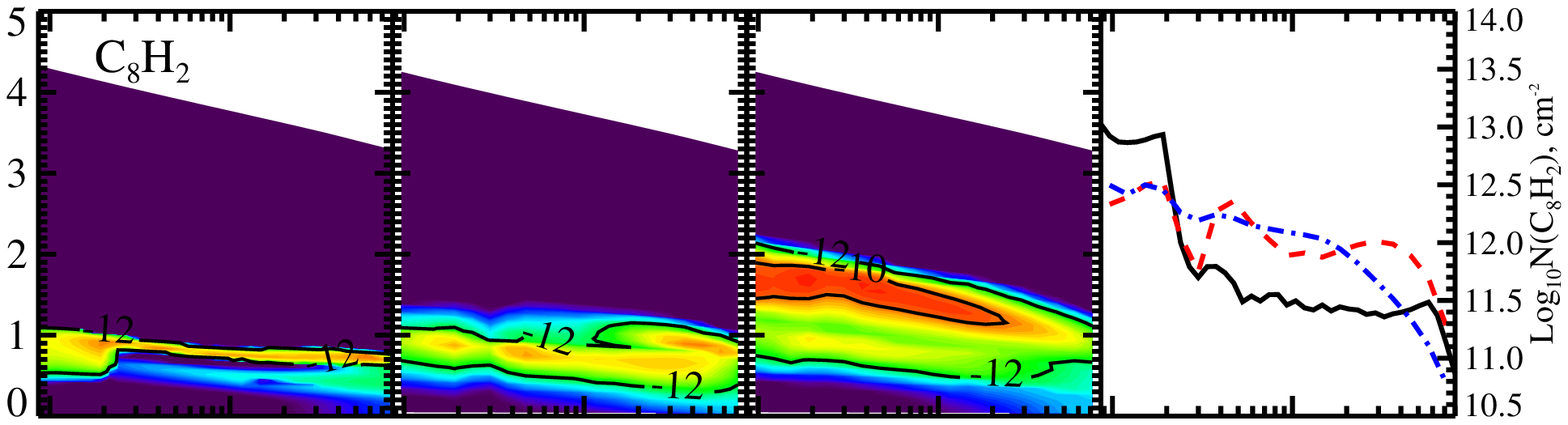}
\includegraphics[width=0.48\textwidth]{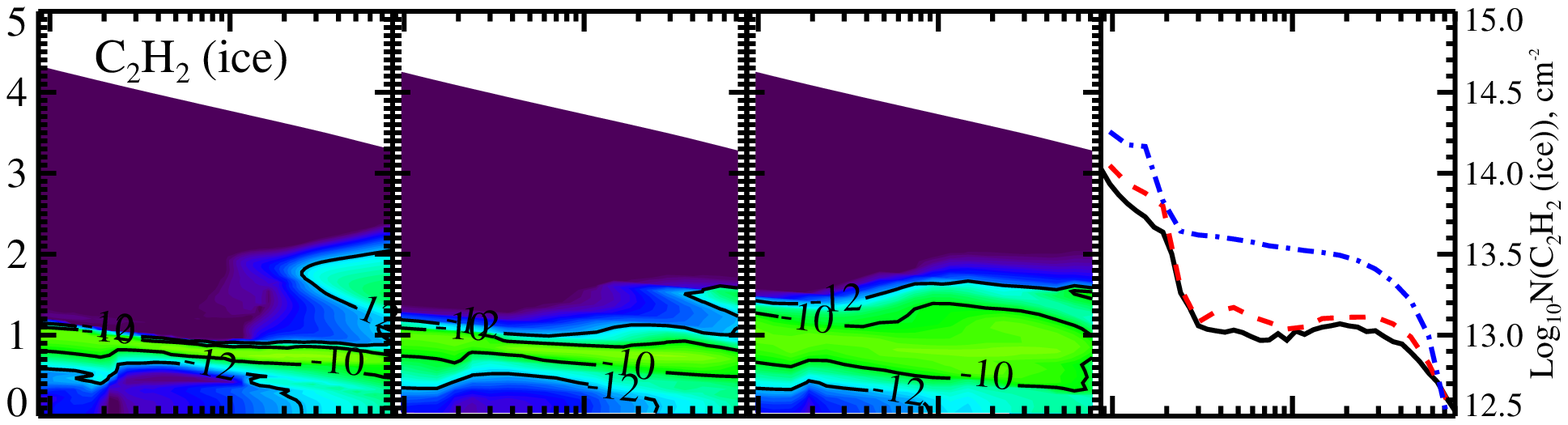}\\
\includegraphics[width=0.48\textwidth]{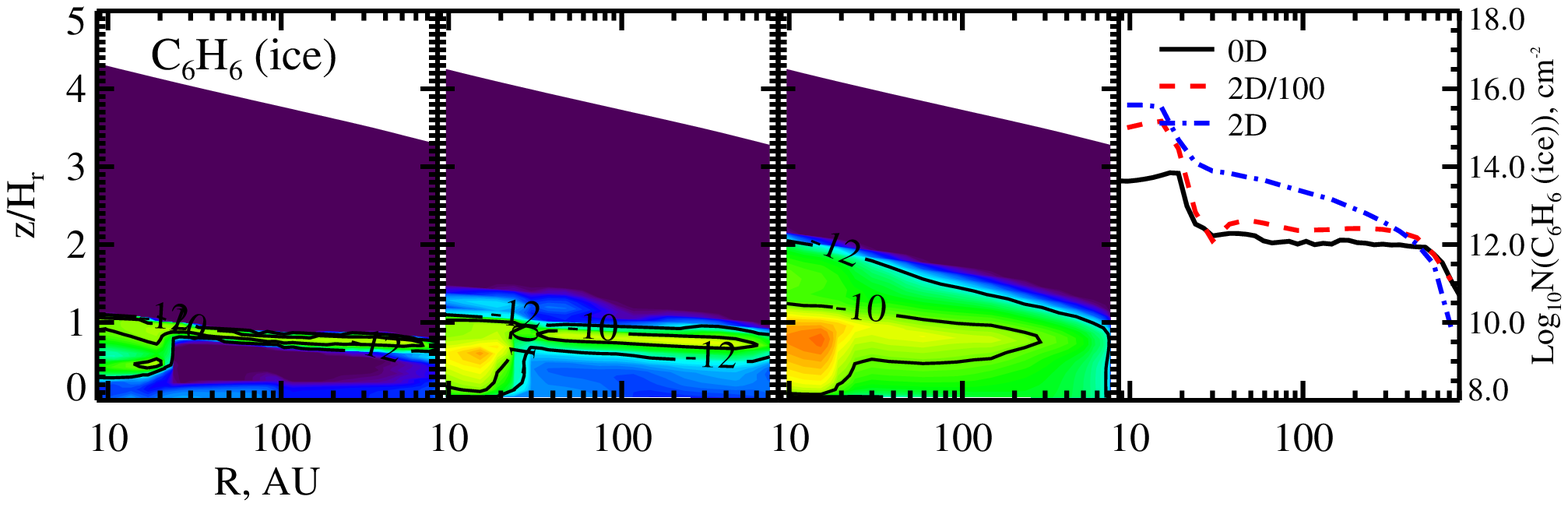}
\includegraphics[width=0.48\textwidth]{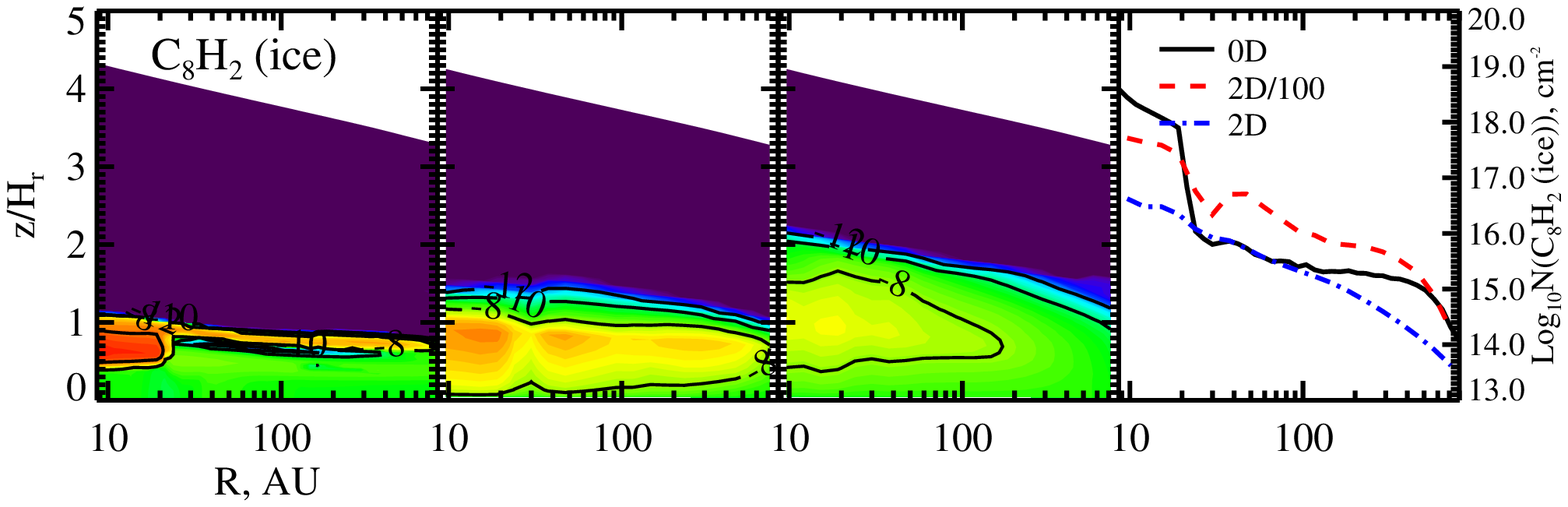}
\caption{The same as in Fig.~\ref{fig:ions} but for the C-containing
species. Results are shown for CH$_3$$^+$, CH$_4$, C$_2$H,
C$_2$H$_2$, C$_8$H$_2$, C$_2$H$_2$ ice, C$_6$H$_6$ ice, and C$_8$H$_2$ ice.
Abundance distribution patterns of many of the carbon chains look similar to
that of C$_8$H$_2$.}
\label{fig:C}
\end{figure*}

\LongTables
\begin{deluxetable*}{llrrll}
\tabletypesize{\footnotesize}
\tablecaption{Key chemical processes: C-bearing species\label{tab:key_reac_C}}
\tablehead{
\colhead{Reaction} & \colhead{$\alpha$} & \colhead{$\beta$} & \colhead{$\gamma$}& \colhead{t$_{\rm min}$} & \colhead{t$_{\rm
max}$}\\
\colhead{} & \colhead{[(cm$^3$)\,s$^{-1}$]} & \colhead{} & \colhead{[K]} & \colhead{[yr]} & \colhead{[yr]}
}
\startdata
C$_6$H$_6$ ice   + h$\nu_{\rm CRP}$  $\rightarrow$  C$_6$H$_4$ ice   + H$_2$ ice   & $3.00\,(3)$   & $0$   & $0$   & $1.00$   &
$5.00\,(6)$   \\
C$_8$H$_2$ ice   + h$\nu_{\rm CRP}$  $\rightarrow$  C$_8$H ice   + H ice   & $1.75\,(3)$   & $0$   & $0$   & $1.00$   &
$5.00\,(6)$   \\
C$_8$H$_4$ ice   + h$\nu_{\rm CRP}$  $\rightarrow$  C$_8$H$_2$ ice   + H$_2$ ice   & $7.50\,(3)$   & $0$   & $0$   & $1.00$   &
$5.00\,(6)$   \\
H$_2$C$_3$O ice   + h$\nu_{\rm CRP}$  $\rightarrow$  C$_2$H$_2$ ice   + CO ice   & $1.80\,(3)$   & $0$   & $0$   & $1.00$   &
$8.18\,(2)$   \\
C$_2$H$_2$ ice   + UV  $\rightarrow$  C$_2$ ice   + H ice   + H ice   & $0.66\,(-10)$   & $0$   & $0$   & $1.00$   & $5.00\,(6)$
\\
C$_6$H$_6$ ice   + UV  $\rightarrow$  C$_6$H$_4$ ice   + H$_2$ ice   & $1.00\,(-9)$   & $0$   & $1.70$   & $1.00$   & $5.00\,(6)$
 \\
C$_8$H$_2$ ice   + UV  $\rightarrow$  C$_8$H ice   + H ice   & $1.00\,(-9)$   & $0$   & $1.70$   & $1.00$   & $5.00\,(6)$   \\
CH$_3$   + UV  $\rightarrow$  CH$_3$$^+$   + e$^-$   & $1.00\,(-10)$   & $0$   & $2.10$   & $1.00$   & $5.00\,(6)$   \\
C$_2$H   + UV  $\rightarrow$  C$_2$   + H   & $0.52\,(-9)$   & $0$   & $2.30$   & $1.00$   & $5.00\,(6)$   \\
C$_2$H$_2$   + UV  $\rightarrow$  C$_2$H   + H   & $0.33\,(-8)$   & $0$   & $2.27$   & $1.00$   & $5.00\,(6)$   \\
C$_2$H$_3$   + UV  $\rightarrow$  C$_2$H$_2$   + H   & $1.00\,(-9)$   & $0$   & $1.70$   & $1.00$   & $1.75\,(5)$   \\
C$_2$H$_4$   + UV  $\rightarrow$  C$_2$H$_2$   + H$_2$   & $0.30\,(-8)$   & $0$   & $2.10$   & $2.14\,(2)$   & $1.75\,(5)$   \\
C$_8$H$_2$   + UV  $\rightarrow$  C$_8$H   + H   & $1.00\,(-9)$   & $0$   & $1.70$   & $1.00$   & $5.00\,(6)$   \\
CH$_4$   + UV  $\rightarrow$  CH$_2$   + H$_2$   & $0.84\,(-10)$   & $0$   & $2.59$   & $1.00$   & $5.00\,(6)$   \\
C$_2$H   + grain  $\rightarrow$  C$_2$H ice   & $1.00$   & $0$   & $0$   & $1.00$   & $5.00\,(6)$   \\
C$_6$H$_6$   + grain  $\rightarrow$  C$_6$H$_6$ ice   & $1.00$   & $0$   & $0$   & $1.00$   & $5.00\,(6)$   \\
C$_2$H$_2$   + grain  $\rightarrow$  C$_2$H$_2$ ice   & $1.00$   & $0$   & $0$   & $1.00$   & $5.00\,(6)$   \\
C$_8$H$_2$   + grain  $\rightarrow$  C$_8$H$_2$ ice   & $1.00$   & $0$   & $0$   & $1.00$   & $5.00\,(6)$   \\
CH$_4$   + grain  $\rightarrow$  CH$_4$ ice   & $1.00$   & $0$   & $0$   & $1.00$   & $5.00\,(6)$   \\
C$_2$H$_2$ ice  $\rightarrow$  C$_2$H$_2$   & $1.00$   & $0$   & $2.59\,(3)$   & $1.00$   & $5.00\,(6)$   \\
C$_8$H$_2$ ice  $\rightarrow$  C$_8$H$_2$   & $1.00$   & $0$   & $7.39\,(3)$   & $1.00$   & $5.00\,(6)$   \\
CH$_4$ ice  $\rightarrow$  CH$_4$   & $1.00$   & $0$   & $1.30\,(3)$   & $1.00$   & $5.00\,(6)$   \\
H ice   + CH$_3$ ice  $\rightarrow$  CH$_4$ ice   & $1.00$   & $0$   & $0$   & $1.00$   & $5.00\,(6)$   \\
H ice   + CH$_3$ ice  $\rightarrow$  CH$_4$   & $1.00$   & $0$   & $0$   & $1.00$   & $5.00\,(6)$   \\
H ice   + C$_2$ ice  $\rightarrow$  C$_2$H   & $1.00$   & $0$   & $0$   & $1.00$   & $5.00\,(6)$   \\
H ice   + C$_2$H ice  $\rightarrow$  C$_2$H$_2$ ice   & $1.00$   & $0$   & $0$   & $1.00$   & $5.00\,(6)$   \\
H ice   + C$_2$H ice  $\rightarrow$  C$_2$H$_2$   & $1.00$   & $0$   & $0$   & $1.00$   & $5.00\,(6)$   \\
H ice   + C$_8$H ice  $\rightarrow$  C$_8$H$_2$ ice   & $1.00$   & $0$   & $0$   & $1.00$   & $5.00\,(6)$   \\
H ice   + C$_8$H ice  $\rightarrow$  C$_8$H$_2$   & $1.00$   & $0$   & $0$   & $1.00$   & $5.00\,(6)$   \\
C$_8$H$_2$ ice   + H ice  $\rightarrow$  C$_8$H$_3$ ice   & $1.00$   & $0$   & $1.21\,(3)$   & $1.00$   & $5.00\,(6)$   \\
CH$_3$$^+$   + H$_2$  $\rightarrow$  CH$_5$$^+$   & $0.13\,(-13)$   & $-1.00$   & $0$   & $1.00$   & $5.00\,(6)$   \\
CH$_2$$^+$   + H$_2$  $\rightarrow$  CH$_3$$^+$   + H   & $0.12\,(-8)$   & $0$   & $0$   & $1.00$   & $5.00\,(6)$   \\
CH$_5$$^+$   + C  $\rightarrow$  CH$_4$   + CH$^+$   & $1.00\,(-9)$   & $0$   & $0$   & $1.00$   & $5.00\,(6)$   \\
C$^+$   + C$_2$H$_6$  $\rightarrow$  CH$_4$   + C$_2$H$_2$$^+$   & $0.17\,(-9)$   & $0$   & $0$   & $4.18\,(2)$   & $5.00\,(6)$
\\
CH$_5$$^+$   + CO  $\rightarrow$  CH$_4$   + HCO$^+$   & $0.32\,(-9)$   & $-0.50$   & $0$   & $1.00$   & $5.00\,(6)$   \\
CH$_4$   + H$^+$  $\rightarrow$  CH$_3$$^+$   + H$_2$   & $0.23\,(-8)$   & $0$   & $0$   & $1.00$   & $5.00\,(6)$   \\
H$^-$   + CH$_3$  $\rightarrow$  CH$_4$   + e$^-$   & $1.00\,(-9)$   & $0$   & $0$   & $1.09\,(2)$   & $5.00\,(6)$   \\
C + C$_8$H$_2$   $\rightarrow$  C$_9$H   + H   & $0.90\,(-9)$   & $0$   & $0$   & $1.00$   & $5.00\,(6)$   \\
C   + CH$_3$  $\rightarrow$  C$_2$H$_2$   + H   & $1.00\,(-10)$   & $0$   & $0$   & $1.00$   & $5.00\,(6)$   \\
C   + CH$_3$C$_6$H  $\rightarrow$  C$_8$H$_2$   + H$_2$   & $0.74\,(-9)$   & $0$   & $0$   & $1.00$   & $1.31\,(6)$   \\
C$_2$H   + O  $\rightarrow$  CO   + CH   & $0.17\,(-10)$   & $0$   & $0$   & $1.00$   & $5.00\,(6)$   \\
C$_3$H  + O $\rightarrow$  C$_2$H   + CO   & $0.17\,(-10)$   & $0$   & $0$   & $1.00$   & $3.82$   \\
C$_3$H$_2$ + O   $\rightarrow$  C$_2$H$_2$   + CO   & $0.50\,(-10)$   & $0.50$   & $0$   & $1.00$   & $5.00\,(6)$   \\
C$_2$H$_2$$^+$   + C  $\rightarrow$  C$_2$H$_2$   + C$^+$   & $0.11\,(-8)$   & $0$   & $0$   & $1.00$   & $5.00\,(6)$   \\
C$_9$H$_3$$^+$   + O  $\rightarrow$  C$_8$H$_2$   + HCO$^+$   & $0.20\,(-9)$   & $0$   & $0$   & $1.00$   & $2.34\,(4)$   \\
C$_2$H$_2$$^+$   + e$^-$  $\rightarrow$  C$_2$H   + H   & $0.29\,(-6)$   & $-0.50$   & $0$   & $1.00$   & $5.00\,(6)$   \\
C$_2$H$_3$$^+$   + e$^-$  $\rightarrow$  C$_2$H$_2$   + H   & $0.14\,(-6)$   & $-0.84$   & $0$   & $1.00$   & $5.00\,(6)$   \\
C$_2$H$_3$$^+$   + e$^-$  $\rightarrow$  C$_2$H   + H   + H   & $0.30\,(-6)$   & $-0.84$   & $0$   & $1.00$   & $5.00\,(6)$   \\
C$_3$H$_2$$^+$   + e$^-$  $\rightarrow$  C$_2$H$_2$   + C   & $0.30\,(-7)$   & $-0.50$   & $0$   & $1.00$   & $1.20\,(4)$   \\
C$_3$H$_2$N$^+$   + e$^-$  $\rightarrow$  C$_2$H$_2$   + CN   & $0.23\,(-6)$   & $-0.50$   & $0$   & $1.00$   & $5.00\,(6)$   \\
C$_3$H$_2$N$^+$   + e$^-$  $\rightarrow$  C$_2$H   + HNC   & $0.75\,(-7)$   & $-0.50$   & $0$   & $55.90$   & $5.00\,(6)$   \\
C$_8$H$_3$$^+$   + e$^-$  $\rightarrow$  C$_8$H$_2$   + H   & $1.00\,(-6)$   & $-0.30$   & $0$   & $1.00$   & $5.00\,(6)$   \\
C$_8$H$_4$$^+$   + e$^-$  $\rightarrow$  C$_8$H$_2$   + H$_2$   & $1.00\,(-6)$   & $-0.30$   & $0$   & $1.00$   & $5.00\,(6)$   \\
CH$_2$CO$^+$   + e$^-$  $\rightarrow$  C$_2$H$_2$   + O   & $0.20\,(-6)$   & $-0.50$   & $0$   & $1.00$   & $5.00\,(6)$   \\
CH$_5$$^+$   + e$^-$  $\rightarrow$  CH$_4$   + H   & $0.14\,(-7)$   & $-0.52$   & $0$   & $1.00$   & $5.00\,(6)$   \\
C$_2$H$_3$$^+$   + grain(-)  $\rightarrow$  C$_2$H   + H   + H   + grain(0)   & $0.59$   & $0$   & $0$   & $7.48$   & $5.00\,(6)$
\\
\enddata
\end{deluxetable*}

In this subsection we analyze in detail chemical and mixing processes responsible for the evolution of the hydrocarbons
in protoplanetary disks. 

There are only several steadfast light hydrocarbons, including 4 neutral 
diatomic and triatomic species (e.g., CH, CH$_2$, C$_2$H), 4 ions (e.g., CH$_3^+$, CH$_5^+$), and the C$_2$H$_2$ ice. 
Their column densities in the laminar model and the fast mixing model differ by a factor of $\la 3$. In contrast, majority of
hydrocarbons
are sensitive to the turbulent transport that alters their column densities by up to 2 orders of magnitude.
The turbulence-sensitive hydrocarbons include 28 neutral molecules (e.g., CH$_3$, CH$_4$, C$_2$H$_2$, C$_3$H$_2$, C$_6$H$_6$,..., 
C$_9$H$_2$), 8 ions (e.g.,, C$_2$H$_3^+$, C$_3$H$_2^+$, C$_6$H$_2^+$), and 22 ices (CH$_4$, C$_2$, C$_2$H, C$_2$H$_3$, 
..., C$_9$H$_2$; 58 species in total). Diffusive mixing modifies their column densities by up to 2 orders of magnitude.
Finally, for 25 hypersensitive hydrocarbons the turbulent diffusion changes the column densities by more than 2 orders of
magnitude
(up to a factor of $10^{10}$ for the C$_9$H$_4$ ice). These species include no ions, 4 neutral hydrocarbons (e.g., C$_{10}$,
C$_3$H$_3$,
CH$_3$C$_4$H), and 21 ices (C$_2$H$_6$, C$_3$H$_4$, C$_5$, C$_5$H$_4$,..., C$_9$H, C$_9$H$_4$). A rough trend is that 
heavier and more saturated hydrocarbons are stronger affected by the turbulent transport than lighter species, 
and that abundances and column densities of hydrocarbon ices are stronger altered compared to their gas-phase counterparts. 
Overall, frozen hydrocarbons are among the most sensitive species to the mixing in our chemical model. 
Unlike the chemistry of the primal ions and the ionization degree discussed in the previous subsection,
the chemical evolution of hydrocarbons should be dominated by slow surface hydrogenation and radical-radical growth processes,
and slow evaporation.

In Fig.~\ref{fig:C} distributions of relative abundances and column densities at 5~Myr of 
CH$_3^+$, CH$_4$, C$_2$H, C$_2$H$_2$, C$_82$H$_2$, C$_2$H$_2$ ice, C$_6$H$_6$ ice and C$_8$H$_2$ ice 
calculated with the laminar and mixing models are presented. The hydrocarbon abundances show a 3-layered structure
similar to that of the polyatomic ions, with rather narrow molecular layers of 0.2-$0.5\,H_r$ at 
$z \approx 1\,H_r$ and typical values of $\sim 10^{-10}-10^{-7}$. Note that relative abundances of hydrocarbon ices are also 
high in the warm molecular layer ($X\sim 10^{-10}-10^{-7}$). The hydrocarbons are fragile to the UV and X-ray
irradiation and thus are absent in the disk atmosphere, apart from the photostable ethynyl radical (C$_2$H) and CH$_3^+$ ion 
that are abundant in the lower disk atmosphere, at $z\sim 2\,H_r$. In the midplane the gas-phase hydrocarbons are depleted,
whereas their ices are moderately abundant. Turbulent mixing expands and enhances molecular layers of gas-phase hydrocarbons at 
almost all radii. For the ices there is no such a clear trend. As a rule, abundances of carbon chains that serve as
intermediate products to surface hydrogenation are lowered by the turbulent transport, e.g. the C$_8$H$_2$ ice, whereas
more saturated hydrocarbons show increased abundances, e.g. the C$_6$H$_6$ ice (Fig.~\ref{fig:C}). Note that the turbulent 
transport affects gas-phase and solid abundances of heavy hydrocarbons (like C$_8$H$_2$) 
in an opposite way, enhancing concentration of a gas-phase species and reducing its solid-state abundances.
We attribute such a behavior to particularly slow evaporation of frozen heavy carbon chains that is
comparable or longer than the dynamical timescale, $\la 1$~Myr, even in the warm molecular layer.

To better understand these results, we performed detailed analysis of the chemical evolution of the hydrocarbons, shown 
in Fig.~\ref{fig:C}, in two disk vertical slices at $r=10$ and 250~AU (the laminar chemical model). 
The most important reactions responsible for
the time-dependent evolution of their abundances in the midplane, the molecular layer, and the atmosphere are presented in 
Table~\ref{tab:key_reac_C}, both for the inner and outer disk regions. 
The final list contains only top 25 reactions for 
CH$_3^+$, CH$_4$, C$_2$H, C$_2$H$_2$, C$_8$H$_2$, C$_2$H$_2$ ice, C$_6$H$_6$ ice and C$_8$H$_2$ ice per region 
(midplane, molecular layer, atmosphere) for the entire 5~Myr time span, with all repetitions removed.

The chemical evolution starts with production of light hydrocarbons in the gas phase by radiative association of C$^+$
with H$_2$, followed by hydrogen addition reactions: C$^+$ $\rightarrow$ CH$_2^+$ $\rightarrow$ CH$_3^+$ $\rightarrow$ CH$_5^+$.  
An alternative route is the radiative association reaction between C and H$_2$, leading to CH$_2$, that can be further
converted to C$_2$H by addition of C.
The protonated methane reacts with electrons, O, CO, C, OH, etc., forming CH$_4$. Methane undergo reactive
collisions with C$^+$, producing C$_2$H$_2^+$ and C$_2$H$_3^+$. All these primal hydrocarbons dissociatively recombine on
electrons or negatively charged grains (in the inner dark midplane), which leads to simple neutral species like CH, CH$_3$,
C$_2$H, and C$_2$H$_2$. The neutral hydrocarbons readily react with ionized and neutral atomic carbon, forming other, heavier
hydrocarbons, e.g. C$_8$H$_2$   + C  $\rightarrow$  C$_9$H   + H   and 
C$_9$H$_3^+$   + O  $\rightarrow$  C$_8$H$_2$   + HCO$^+$.
The growth rate of chemical complexity of carbon chains is regulated by ion-molecule and neutral-neutral exothermic reactions 
with atomic oxygen that lead to formation of CO, e.g. O  + C$_3$H$_2$  $\rightarrow$  C$_2$H$_2$ + CO
 (see also Fig.~7 in \citet{Henning_ea10} and Fig.~4 in \citet{Turner_ea00}). 
The associated chemical timescales are $\la10^3-10^4$~years, see Fig.~\ref{fig:timescales} and Tables~\ref{tab:tau_inner} 
and \ref{tab:tau_outer}, but not everywhere in the disk. In the disk atmosphere, at $z\approx 2\,H_r$, 
the chemical evolution of CH$_2$ and its daughter species, C$_2$H and CH$_3$, is subject to the evolution of H$_2$ and O. 
As we discussed in the previous subsection, at these high altitudes slow destruction of molecular 
hydrogen, water, and CO by He$^+$ locally sets long evolutionary timescales ($\la 10^5$~years) for many species, 
e.g. ethynyl, CH$_2$, and CH$_3$ radicals.

To enhance further production of hydrocarbons in the gas, particular physical 
conditions have to be reached. The heavy carbon-bearing compounds can be produced at high densities and $T\ga 800$~K 
by pyrolysis of precursor hydrocarbons \citep[e.g.,][]{Morgan_ea91} or in a Fischer-Tropsch-like process involving
CO, H$_2$, and catalytic surfaces \citep[][]{TG07}. These conditions are met at sub-AU radii in the very inner disk midplane 
that is not examined in the present study. 
Another possibility for the gas-phase production of hydrocarbons is to bring elemental carbon locked in CO 
back to the gas, while simultaneously maintaining low oxygen abundances. 
Due to self-shielding and mutual-shielding by H$_2$, the UV photodissociation of CO is only important in dilute
disk atmosphere \citep[e.g.,][]{vD88,Visser_ea09b}. However, in the lower part of the inner warm molecular layer 
($r\la200$~AU, $z\approx 1\,H_r$) the He$^+$ ions produced by the X-ray ionization slowly destroy CO, restoring gas-phase 
concentrations of C$^+$ and O ($\tau_{\rm X} \sim 5\,10^3$~years). Relatively high densities 
($n(\rm {H}) \sim 10^8-10^9$~cm$^{-3}$) and lukewarm temperatures ($T \sim 30-75$~K) in this region (Fig.~\ref{fig:disk_struc}) 
allow rapid conversion of releazed oxygen into water. 
Thus, locally in the inner regions of the X-ray-irradiated disks the gas-phase C/O ratio may revert from the Solar value of 
$\approx 0.43$ to a $\ga 1$ value typical of carbon-rich AGB shells (D.~Hollenbach, priv. comm.), which facilitates 
gas-phase formation of heavy hydrocarbons. 

An alternative efficient route to accumulate complex hydrocarbons in disks is surface chemistry coupled to the 
destruction of hydrocarbon ices and other C-bearing species. In cold disk regions ($T\la 20$~K) it is mainly hydrogenation of
precursor species like C$_n$H$_m$  ($n=2-9$, $m\le2$). At low dust temperatures only hydrogen is mobile to scan the 
grain surface \citep[e.g.,][]{dHendecourtea85,HH93,Katz_ea99}. Ultimately, this process leads to formation of saturated 
ices, e.g. CH$_4$ and C$_2$H$_6$, which have $5\%$ chance to escape directly to the gas upon surface recombination 
(Sect.~\ref{chem_model}). Our model has a restricted 
set of hydrocarbon chemistry involving species with $\le 10$ carbon atoms and mostly having up to 4 hydrogen atoms only, and 
thus these processes do not appear in Table~\ref{tab:key_reac_C}. In the inner disk midplane ($r\la 20$~AU) warmed up by the 
accretion to the temperatures of $\sim40-50$~K, desorption of CH$_4$ 
dominates over its sticking to grains, forming an oasis of abundant gas-phase methane. At these elevated temperatures
heavier radicals like C, O, OH, etc. become partly mobile \citep[e.g.,][]{Garrod_Herbst06,Garrod_ea08b,Herbst_vanDishoeck09}. 
However, the surface growth of hydrocarbons by addition of atomic carbon is still not as efficient as the surface hydrogenation 
and the gas-phase chemistry, and accounts for only a few percent of the total surface production rate. 
The most favorable conditions for active surface hydrocarbon chemistry are met in the warm molecular layer, where 
atomic hydrogen is plentiful in the gas, and its accretion rate is fast even in comparison with rapid desorption. More
importantly,
complex ices are slowly destroyed by X-ray/CRP-driven photons as well as by partly absorbed stellar (inner disk) and interstellar 
UV (outer disk) photons, producing various reactive radicals. For example, C$_8$H is produced by photodestruction of
C$_8$H$_2$ ice, followed by evaporation (Table~\ref{tab:key_reac_C}). 
In the outer disk partly transparent to the IS UV
radiation these surface photoprocesses become important even in the midplane (see C$_2$H$_2$ in Fig.~\ref{fig:C}).

Note that our chemical network lacks the surface formation of benzene apart from its accretion from the gas.
The major production pathway for C$_6$H$_6$ is via ion-molecule reactions of 
C$_3$H$_4$ and C$_3$H$_4^+$ ($\alpha=7.5\,10^{-10}$~cm$^3$\,s$^{-1}$) as well as 
C$_2$H$_4$ and C$_6$H$_5^+$ ($\alpha=5.5\,10^{-11}$~cm$^3$\,s$^{-1}$), 
followed by the dissociative recombination, and freeze-out at $T\la 150$~K.
In turn, C$_3$H$_4^+$ and C$_6$H$_5^+$ are synthesized from neutral hydrocarbons by rapid charge transfer with 
ionized hydrogen and carbon or via protonation by H$_3^+$, H$_3$O$^+$, HCO$^+$, etc. The benzene building blocks, C$_3$H$_3$ 
and C$_3$H$_4$, are produced effectively via a number gas-phase and surface reactions as described above. 
The characteristic chemical timescale for benzene is thus mainly determined by the evolution of C$_3$H$_3$ and C$_3$H$_4$,
which involves slow surface routes.

As we already discussed in Sect.~\ref{timescales}, the surface chemistry timescales are the longest in the disk chemistry, 
typically exceeding several Myr (see also Tables~\ref{tab:tau_inner} and \ref{tab:tau_outer}). 
Photoprocessing of ices on dust grains has similar timescale in the molecular layer
and the disk midplane, $\tau_{\rm chem} \la 10^6$~years (Fig.~\ref{fig:timescales}). 
The overall pace of the surface chemistry is also dependent on the amount of involved
surface processes. Thus, the characteristic chemical timescales for hydrocarbons are beyond 1~Myr, particularly for the 
heaviest and most saturated ones in the network (e.g., C$_8$H$_4$, etc.). Not surprisingly, the 2D-turbulent mixing
strongly affects the hydrocarbon chemistry in disks. The evolution of heaviest carbon chains with more than 6 carbon atoms 
is so far from a steady-state that it becomes sensitive to transport processes even in the slow mixing model (see, e.g., the
C$_8$H$_2$ ice in Fig.~\ref{fig:C}). 
The vertical transport brings up dust grains from the midplane to 
the upper, more irradiated and warmer disk regions, allowing more efficient photoprocessing of ices, photodesorption of
frozen hydrocarbons, and their photodissociation. These processes increase relative abundances of gas-phase hydrocarbons
by up to several orders of magnitude. In turn, vertical transport downward allows to retain
newly formed hydrocarbon radicals in the icy mantles and facilitates their slow hydrogenation. The radial mixing does not
play a major role in the hydrocarbon chemistry. Consequently, abundances of the 
most saturated hydrocarbon ices in the network are increased by vertical mixing, in particular in the inner disk 
with the shortest dynamical timescales (e.g., benzene ice), 
whereas the intermediate products become less abundant (e.g., C$_8$H$_2$ ice).
Note also that the mixing allows CH$_2$ and C$_2$H to be more efficiently formed
in the outer disk atmosphere, where their second layers of high relative abundances are developed. 
This is caused by the turbulent transport upward of the molecular hydrogen from the molecular layer, 
which is otherwise slowly destroyed in the atmosphere by the X-rays and the cosmic ray particles.

\subsection{Oxygen-containing molecules}
\label{O-species}
\begin{figure*}
\includegraphics[width=0.48\textwidth]{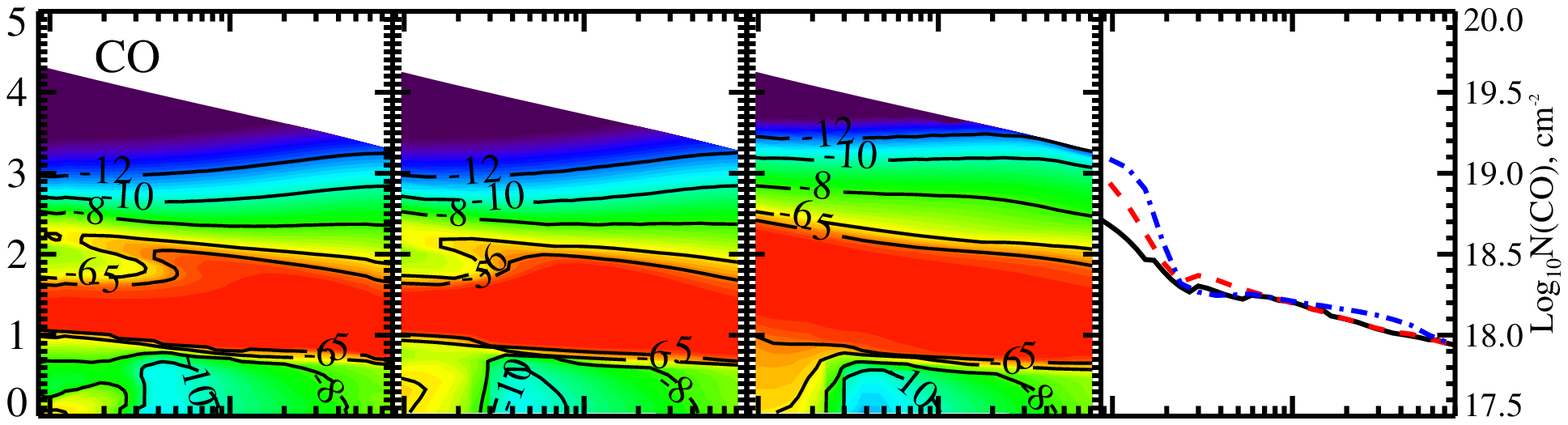}
\includegraphics[width=0.48\textwidth]{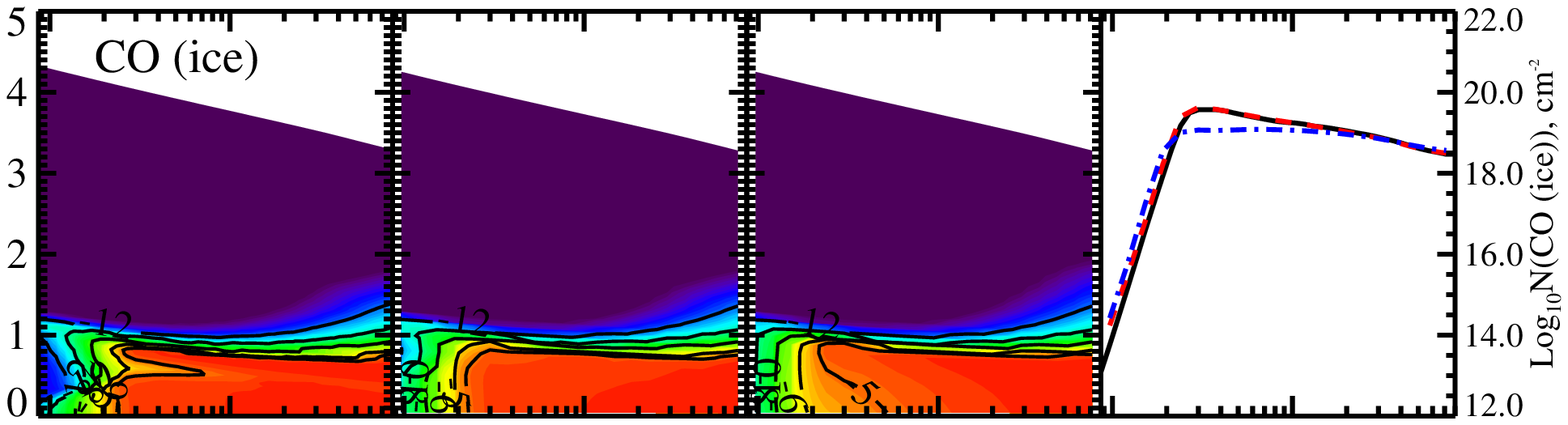}\\
\includegraphics[width=0.48\textwidth]{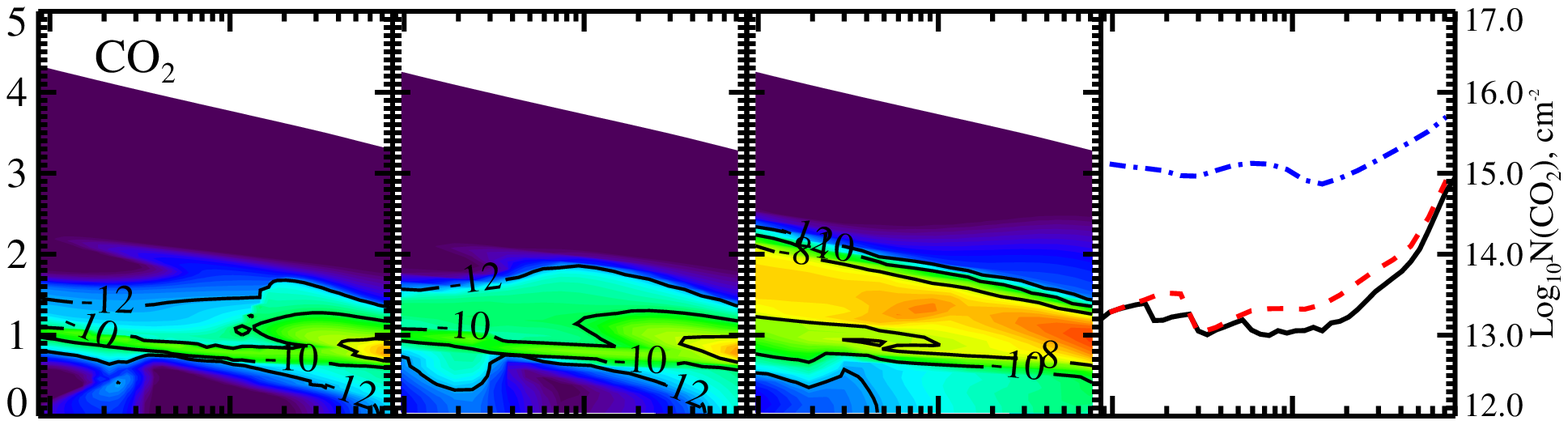}
\includegraphics[width=0.48\textwidth]{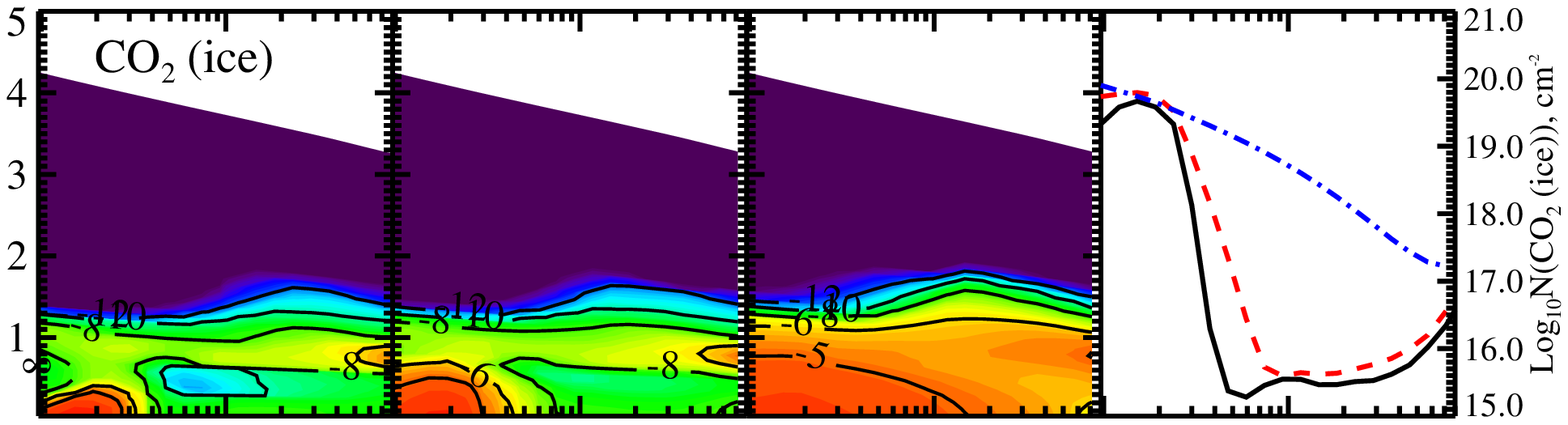}\\
\includegraphics[width=0.48\textwidth]{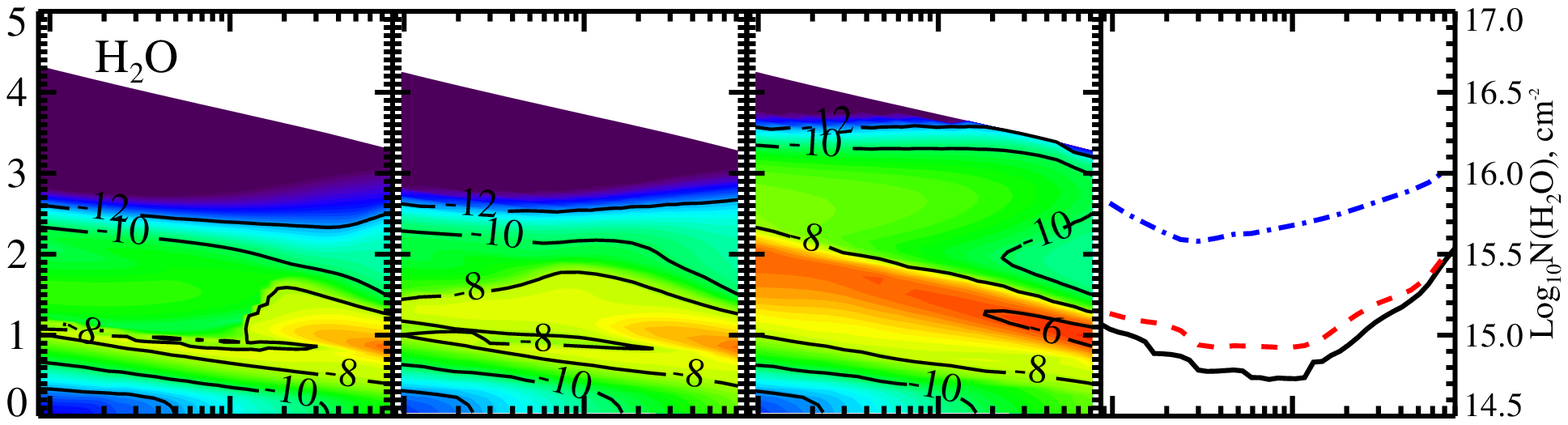}
\includegraphics[width=0.48\textwidth]{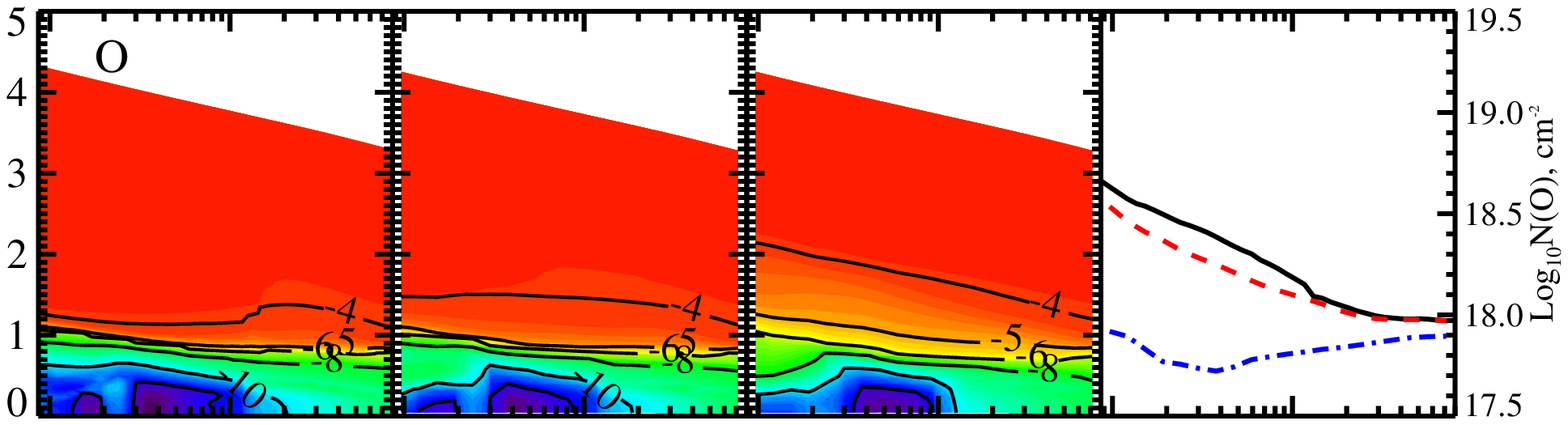}\\
\includegraphics[width=0.48\textwidth]{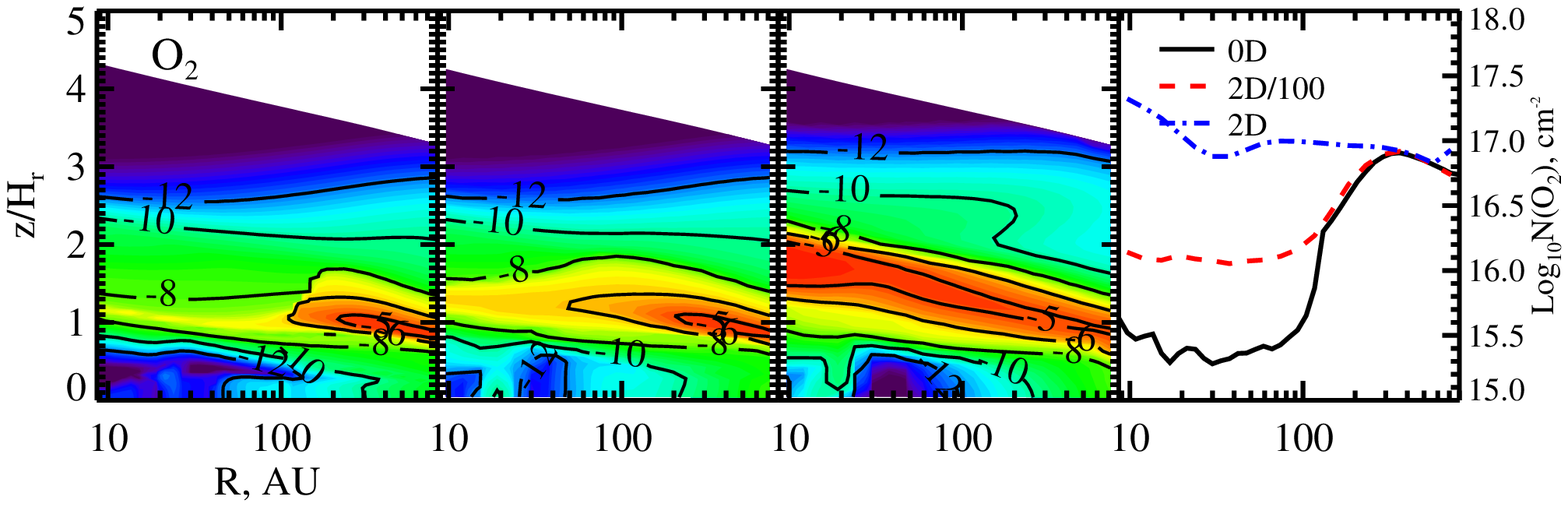}
\includegraphics[width=0.48\textwidth]{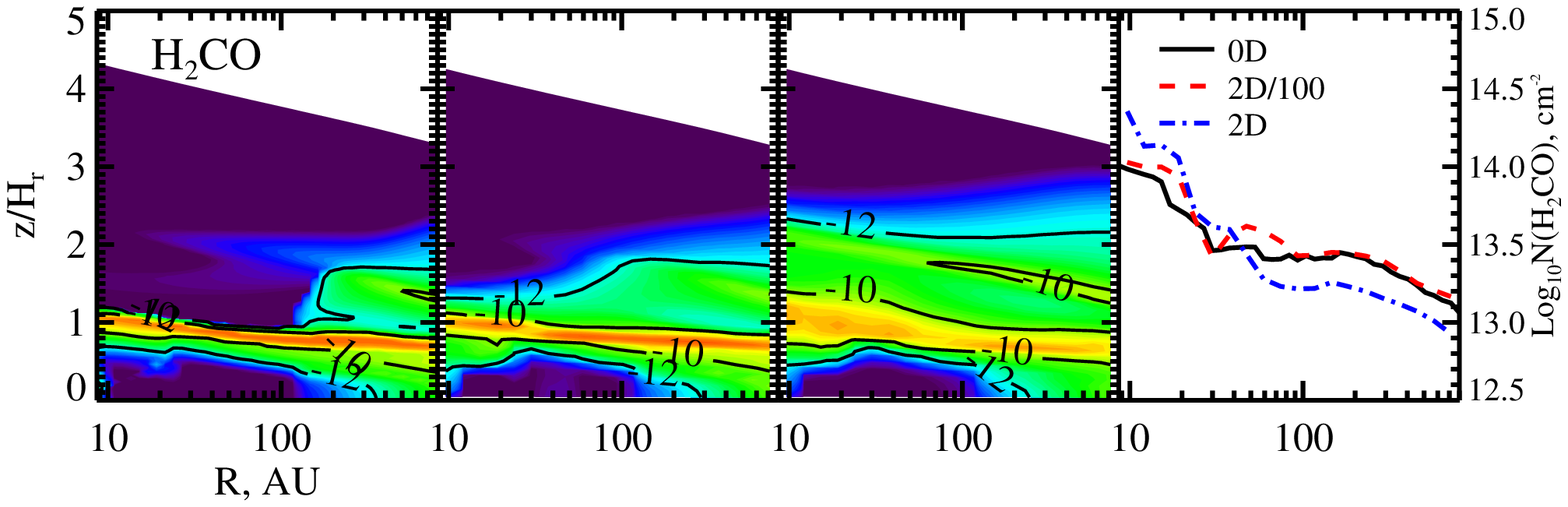}
\caption{The same as in Fig.~\ref{fig:ions} but for the O-containing
species. Results are shown for CO, CO ice, CO$_2$, CO$_2$ ice,
H$_2$O, O, O$_2$, and H$_2$CO.}
\label{fig:O}
\end{figure*}

\LongTables
\begin{deluxetable*}{llrrll}
\tabletypesize{\footnotesize}
\tablecaption{Key chemical processes: O-bearing species\label{tab:key_reac_O}}
\tablehead{
\colhead{Reaction} & \colhead{$\alpha$} & \colhead{$\beta$} & \colhead{$\gamma$}& \colhead{t$_{\rm min}$} & \colhead{t$_{\rm
max}$}\\
\colhead{} & \colhead{[(cm$^3$)\,s$^{-1}$]} & \colhead{} & \colhead{[K]} & \colhead{[yr]} & \colhead{[yr]}
}
\startdata
CO$_2$ ice   + h$\nu_{\rm CRP}$  $\rightarrow$  CO ice   + O ice   & $1.71\,(3)$   & $0$   & $0$   & $1.00$   & $5.00\,(6)$   \\
CO   + UV  $\rightarrow$  O   + C   & $0.20\,(-9)$   & $0$   & $3.53$   & $1.00$   & $5.00\,(6)$   \\
CO$_2$   + UV  $\rightarrow$  CO   + O   & $0.89\,(-9)$   & $0$   & $3.00$   & $1.00$   & $5.00\,(6)$   \\
H$_2$O   + UV  $\rightarrow$  OH   + H   & $0.80\,(-9)$   & $0$   & $2.20$   & $1.00$   & $5.00\,(6)$   \\
O   + grain  $\rightarrow$  O ice   & $1.00$   & $0$   & $0$   & $1.00$   & $5.00\,(6)$   \\
O$_2$   + grain  $\rightarrow$  O$_2$ ice   & $1.00$   & $0$   & $0$   & $1.00$   & $5.00\,(6)$   \\
CO   + grain  $\rightarrow$  CO ice   & $1.00$   & $0$   & $0$   & $1.00$   & $5.00\,(6)$   \\
CO$_2$   + grain  $\rightarrow$  CO$_2$ ice   & $1.00$   & $0$   & $0$   & $1.00$   & $5.00\,(6)$   \\
H$_2$CO   + grain  $\rightarrow$  H$_2$CO ice   & $1.00$   & $0$   & $0$   & $1.00$   & $5.00\,(6)$   \\
H$_2$O   + grain  $\rightarrow$  H$_2$O ice   & $1.00$   & $0$   & $0$   & $1.00$   & $5.00\,(6)$   \\
O ice  $\rightarrow$  O   & $1.00$   & $0$   & $8.00\,(2)$   & $1.00$   & $5.00\,(6)$   \\
O$_2$ ice  $\rightarrow$  O$_2$   & $1.00$   & $0$   & $1.00\,(3)$   & $1.00$   & $5.00\,(6)$   \\
CO ice  $\rightarrow$  CO   & $1.00$   & $0$   & $1.15\,(3)$   & $1.00$   & $5.00\,(6)$   \\
CO$_2$ ice  $\rightarrow$  CO$_2$   & $1.00$   & $0$   & $2.58\,(3)$   & $1.00$   & $5.00\,(6)$   \\
H$_2$CO ice  $\rightarrow$  H$_2$CO   & $1.00$   & $0$   & $2.05\,(3)$   & $1.00$   & $5.00\,(6)$   \\
H ice   + OH ice  $\rightarrow$  H$_2$O   & $1.00$   & $0$   & $0$   & $1.00$   & $5.00\,(6)$   \\
H ice   + HCO ice  $\rightarrow$  H$_2$CO   & $1.00$   & $0$   & $0$   & $1.00$   & $5.00\,(6)$   \\
OH ice   + CO ice  $\rightarrow$  CO$_2$ ice   + H ice   & $1.00$   & $0$   & $80.00$   & $1.00$   & $5.00\,(6)$   \\
OH ice   + CO ice  $\rightarrow$  CO$_2$   + H   & $1.00$   & $0$   & $80.00$   & $1.00$   & $5.00\,(6)$   \\
CH$_2$ ice   + O$_2$ ice  $\rightarrow$  H$_2$CO   + O   & $1.00$   & $0$   & $0$   & $1.96$   & $5.00\,(6)$   \\
CO$^+$   + H  $\rightarrow$  CO   + H$^+$   & $0.40\,(-9)$   & $0$   & $0$   & $1.00$   & $5.00\,(6)$   \\
CO$_2$   + H$^+$  $\rightarrow$  HCO$^+$   + O   & $0.30\,(-8)$   & $0$   & $0$   & $1.00$   & $5.00\,(6)$   \\
H$_2$O   + H$^+$  $\rightarrow$  H$_2$O$^+$   + H   & $0.73\,(-8)$   & $-0.50$   & $0$   & $1.00$   & $5.00\,(6)$   \\
O   + H$^+$  $\rightarrow$  O$^+$   + H   & $0.70\,(-9)$   & $0$   & $2.32\,(2)$   & $1.00$   & $5.00\,(6)$   \\
O$_2$   + H$^+$  $\rightarrow$  O$_2$$^+$   + H   & $0.12\,(-8)$   & $0$   & $0$   & $3.82$   & $5.00\,(6)$   \\
HCO$_2$$^+$   + CO  $\rightarrow$  CO$_2$   + HCO$^+$   & $0.25\,(-9)$   & $-0.50$   & $0$   & $1.00$   & $5.00\,(6)$   \\
O$^+$   + H  $\rightarrow$  O   + H$^+$   & $0.70\,(-9)$   & $0$   & $0$   & $1.00$   & $5.00\,(6)$   \\
O$^-$   + CO  $\rightarrow$  CO$_2$   + e$^-$   & $0.65\,(-9)$   & $0$   & $0$   & $7.48$   & $5.00\,(6)$   \\
CO   + He$^+$  $\rightarrow$  O   + C$^+$   + He   & $0.16\,(-8)$   & $0$   & $0$   & $1.00$   & $5.00\,(6)$   \\
H$_2$CO   + H$_3$$^+$  $\rightarrow$  H$_3$CO$^+$   + H$_2$   & $0.55\,(-8)$   & $-0.50$   & $0$   & $1.96$   & $5.00\,(6)$   \\
H   + OH  $\rightarrow$  H$_2$O   & $0.40\,(-17)$   & $-2.00$   & $0$   & $7.48$   & $5.00\,(6)$   \\
CO   + OH  $\rightarrow$  CO$_2$   + H   & $0.28\,(-12)$   & $0$   & $1.76\,(2)$   & $1.00$   & $5.00\,(6)$   \\
O   + C$_2$  $\rightarrow$  CO   + C   & $1.00\,(-10)$   & $0$   & $0$   & $1.00$   & $6.11\,(3)$   \\
O   + CH$_3$  $\rightarrow$  H$_2$CO   + H   & $0.14\,(-9)$   & $0$   & $0$   & $1.00$   & $5.00\,(6)$   \\
O   + HCO  $\rightarrow$  CO$_2$   + H   & $0.50\,(-10)$   & $0$   & $0$   & $1.00$   & $5.00\,(6)$   \\
O   + OH  $\rightarrow$  O$_2$   + H   & $0.75\,(-10)$   & $-0.25$   & $0$   & $1.00$   & $5.00\,(6)$   \\
O + H$_2$CO    $\rightarrow$  CO   + OH   + H   & $1.00\,(-10)$   & $0$   & $0$   & $1.00$   & $5.00\,(6)$   \\
O$_2$   + C$_3$  $\rightarrow$  CO$_2$   + C$_2$   & $1.00\,(-12)$   & $0$   & $0$   & $1.00$   & $5.00\,(6)$   \\
O$_2$   + C  $\rightarrow$  O   + CO   & $0.47\,(-10)$   & $-0.34$   & $0$   & $1.00$   & $5.00\,(6)$   \\
H$_3$O$^+$   + e$^-$  $\rightarrow$  H$_2$O   + H   & $0.11\,(-6)$   & $-0.50$   & $0$   & $1.00$   & $5.00\,(6)$   \\
HCO$_2$$^+$   + e$^-$  $\rightarrow$  CO$_2$   + H   & $0.60\,(-7)$   & $-0.64$   & $0$   & $1.00$   & $5.00\,(6)$   \\
H$_2$CO$^+$   + e$^-$  $\rightarrow$  H$_2$CO   & $0.11\,(-9)$   & $-0.70$   & $0$   & $14.60$   & $5.00\,(6)$   \\
H$_3$O$^+$   + e$^-$  $\rightarrow$  O   + H   + H$_2$   & $0.56\,(-8)$   & $-0.50$   & $0$   & $3.82$   & $5.00\,(6)$   \\
O$_2$$^+$   + e$^-$  $\rightarrow$  O   + O   & $0.19\,(-6)$   & $-0.70$   & $0$   & $1.96$   & $1.75\,(5)$   \\
H$_3$O$^+$   + grain(-)  $\rightarrow$  H$_2$O   + H   + grain(0)   & $0.25$   & $0$   & $0$   & $1.00$   & $5.00\,(6)$   \\
CH$_3$O$_2$$^+$   + grain(-)  $\rightarrow$  CO$_2$   + H$_2$   + H   + grain(0)   & $0.50$   & $0$   & $0$   & $28.60$   &
$3.42\,(5)$   \\
O$_2$$^+$   + grain(-)  $\rightarrow$  O   + O   + grain(0)   & $1.00$   & $0$   & $0$   & $1.96$   & $1.75\,(5)$   \\
\enddata
\end{deluxetable*}

The chemical evolution of O, O$_2$, CO, CO$_2$ and water is of
particular attention in astrophysics since the recent puzzling results obtained with {\it Spitzer}, {\it Herschel},
{\it SWAS}, and  {\it Odin} satellites \citep[e.g.,][]{Larsson_ea07,Carr_Najita08,Bergin_ea10a,Joergensen_vanDishoeck10},
and their potential relevance for astrobiology \citep[e.g.,][]{Selsis_ea02a,Kaltenegger_ea07a,Segura_ea07a}. 
In this subsection we analyze in detail chemical and mixing processes responsible for the evolution of oxygen-bearing
species in protoplanetary disks. We focus on simple neutral molecules that are composed of O, C, H only, and consider
complex organics in separate subsection.

There are 4 steadfast neutral O-bearing molecules (CO, OH, H$_2$CO, and the water ice), and 3 steadfast ions (H$_2$CO$^+$,
H$_3$CO$^+$, and H$_3$O$^+$)
(Table~\ref{tab:steadfast}). Their column densities in the laminar model and the fast mixing model differ by a factor of 3. 
Among the species sensitive to the turbulent transport (Table~\ref{tab:sens}) there are 5 molecules (C$_3$O, H$_2$O, HCO, O,
O$_2$),
3 ions (HCO$^+$, O$^+$, O$_2^+$),
and 4 solid species (CO, H$_2$CO, O, OH ices). Their column densities are changed by up to 2 orders of magnitude by the turbulent 
transport. The hypersensitive O-bearing species (Table~\ref{tab:hyps}) include 3 molecules (CO$2$, H$_2$O$_2$, O$_3$), 2 ions
(OH$^+$, H$_2$O$^+$) 
and 5 ices (CO$_2$, H$_2$O$_2$, O$_2$, O$_2$H, O$_3$). 
The turbulent diffusion alters their column densities by up to a factor of 4\,000 (CO$_2$ ice).
Apparently, neutral species with a large number of oxygen atoms (e.g., ozone) affected more strongly by the turbulent
mixing than the molecules containing a single O (e.g., CO). 
Similar to the hydrocarbon chemistry, abundances and column densities of O-bearing ices are stronger altered by the 
turbulent diffusion compared to the gas-phase molecules. Once again this is an indicator that the chemical evolution of 
the multi-oxygen molecules in protoplanetary disks is at least partly governed by slow chemical processes (surface 
recombination, photodissociation of ices, etc.).

In Fig.~\ref{fig:O} the distributions of the relative molecular abundances and column densities at 5~Myr of 
CO, CO ice, CO$_2$, CO$_2$ ice, H$_2$O, O, O$_2$, and H$_2$CO calculated with the laminar and the 2D-mixing models are 
presented. The abundance distribution of atomic oxygen shows a 2-layered structure, with a maximum of 
relative concentrations ($X({\rm O}) \approx 1.5\,10^{-4}$) in the disk atmosphere and an upper part of the molecular layer, 
and strong depletion in the midplane ($X({\rm O}) \la 10^{-11}$). On the other hand, CO and CO$_2$ ices are concentrated 
in the midplane and a lower part of the molecular layer (at $z\approx 1\,H_r$), with typical relative abundances of
$\la 10^{-5}$. The CO$_2$ ice abundance is particularly high in the inner warm midplane ($r\le 20$~AU). Its
distribution is also 2-layered. The rest of considered O-molecules show a 3-layered abundance structure similar to that of 
the polyatomic ions and hydrocarbons. The carbon monoxide spreads over a wide vertical heights in the disk
due to self-shielding \citep[e.g.,][]{vD88,Lyons_Young05,Visser_ea09b} and low desorption energy of $\sim 1\,000$~K
\citep[e.g.,][]{Bisschop_ea06}, $z\approx 1-2.5$ at $r=10$~AU and $z\approx 0.8-2$ at $r=800$~AU, with a typical abundance 
of $7.6\,10^{-5}$. The tip of low CO abundances in the inner low atmosphere at $2\,H_r$ is due to enhanced X-ray-driven 
dissociation in this region (see also Sect.~\ref{timescales}). 
Note that gas-phase CO in the laminar model is partially produced in the very inner disk midplane ($r<20$~AU).
The H$_2$O is also wide-spread throughout the disk, though due to photodissociation
and rapid freeze-out at $T\la120$~K its peak abundances of $10^{-8}-10^{-7}$ are confined to the bottom of the
molecular layer, with a maximum in the outer disk region at $r\ga200$~AU (Fig.~\ref{fig:O}). Gas-phase water comprises 
less than 0.01\% of the total water abundance that are locked in dust mantles in the midplane.
The O$_2$ abundance distribution has almost the same pattern as that of gas-phase water, with maximum values of 
$X({\rm O}_2) \approx 10^{-7}-3\,10^{-5}$. The peak CO$_2$ relative abundances are narrow and rather low, 
$\sim 10^{-10}-10^{-8}$, and restricted to the very outer disk ($r\ga400$~AU). In contrast, the H$_2$CO has a radially-uniform, 
narrow ($\approx 0.3\,H_r$) molecular layer, with maximum abundances of $\sim10^{-10-10^{-9}}$ at $z\approx 0.8-1\,H_r$.

Turbulent mixing does not considerably affect column densities of H$_2$CO and 
CO in gas and solid phases, although it populates the viscously-heated inner midplane with
carbon monoxide, and slightly widens their molecular layers upward (Fig.~\ref{fig:O}). Unlike atomic ions, 
the mixing softens the vertical gradient of the atomic oxygen 
abundances, and lowers its column density. Turbulent diffusion transports solid CO$_2$ from the inner midplane
radially  and vertically outward, enhancing its column densities and abundances by more than 3 orders of magnitude at 
$r\ga 30$~AU. Finally, column densities of gas-phase CO$_2$, H$_2$O, and O$_2$ are increased by the turbulence
by factors of 10-1\,000, with their molecular layers enriched and vertically expanded up to $\sim 1.2-2\,H_r$ and more
homogeneously distributed in the radial direction.

To better understand these results, we performed detailed analysis of the chemical evolution of 
CO, CO ice, CO$_2$, CO$_2$ ice, H$_2$O, O, O$_2$, and H$_2$CO 
in two disk vertical slices at $r=10$ and 250~AU (the laminar chemical model). The most important reactions responsible for
the evolution of their abundances in the midplane, the molecular layer, and the atmosphere are presented in 
Table~\ref{tab:key_reac_O}, both for the inner and outer disk regions. 
The final list contains only 
top 25 reactions per region (midplane, molecular layer, atmosphere) for the entire 5~Myr time span, with all repetitions 
removed. 

The chemical evolution of the selected O-bearing species is governed by a limited set of reactions. 
Atomic oxygen is present in the disk atmosphere and converted to CO, CO$_2$ and H$_2$O in the midplane and the molecular layer.
The rate of its convertion in the molecular layer is partly regulated by the evolution of H$_3^+$ and hydrocarbons, 
which is in turn determined by the slow X-ray irradiation of H$_2$ (Sect.~\ref{ions}) and the slow release of O from CO 
by the X-ray-ionized helium atoms in the inner disk (Sect.~\ref{C-species}), and the surface chemistry of O-bearing species. 
The resulting chemical timescale of $\ga 1$~Myr exceeds the transport timescale in the molecular layer 
(Tables~\ref{tab:tau_inner} and \ref{tab:tau_outer}).

The water ice forms on the dust surfaces via accretion of the gas-phase water in disk regions with $T\la120$~K, 
and via surface hydrogenation of frozen atomic oxygen and hydroxyle (minor route). 
The destruction pathways for water ice are thermal or 
photoevaporation (major route), and further hydrogenation to hydrogen peroxide ice (very minor route). 
In the gas, water formation begins by production of OH$^+$ from O and H$_3^+$,
followed by subsequent hydrogen abstraction reactions with H$_2$ till H$_3$O$^+$ is created. The protonated water
dissociatively recombines with electrons or negatively charged grains into H$_2$O (25\%) or OH (74\%) or O (1\%),
or de-protonates by ion-molecule reactions with other abundant neutral molecules (CH$_4$, CO, etc.). The gas-phase
formation of water is assisted by slow radiative recombination reaction between H and OH, and photoevaporation of water ice 
in the warm molecular layer. The gas-phase removal channels for H$_2$O include photodissociation, 
charge transfer reactions with H$^+$ followed by dissociative recombination, 
freeze-out, and reactions with He$^+$ (minor route). The solid water is a terminal species that serves as one of the sinks 
of the elemental oxygen in the disks. Thus, the key chemical processes leading to the evolution of water are fast,
$\tau_{\rm chem}\la1-10^2$~years, in the atmosphere and the midpane, whereas in the molecular layer it is regulated by
the late-time evolution of H$_3^+$ due to the X-ray ionization ($\tau_{\rm chem}\ga10^4$~years; see Sect.~\ref{ions}). 

The molecular oxygen is produced in the gas by neutral-neutral exothermic reactions of OH and O, and destroyed in 
ion-molecule and neutral-neutral combustion reactions with various radicals (mostly dehydrogenated hydrocarbons, e.g.,
C, C$_3$, etc.), by photodissociation, and via ion-molecule reactions of O$_2$ with H$_3^+$ and ionized C and H, 
followed by dissociative recombination of O$_2^+$ into atomic oxygen. On dust surfaces molecular oxygen 
is produced either directly from the recombination of oxygen atoms at conditions when surface O becomes 
mobile ($T\ga30$~K; the inner disk midplane and the warm molecular layer), or via surface oxidation reactions:
O (ice) + OH (ice) $\rightarrow$ O$_2$H (ice) and O (ice) + O$_2$H (ice) $\rightarrow$ O$_2$ (ice) + OH (ice). 
The surface O$_2$ can be further converted to O$_3$ ice. The characteristic chemical timescale for O$_2$
is set by the slow surface chemistry timescale ($\tau_{\rm chem}\ga10^6$~years).

CO molecules serve as a sink of almost all elemental carbon and about a half of elemental oxygen in disks.
Carbon monoxide is formed essentially in the gas-phase via reactions of atomic oxygen with CH, CH$_2$, and C$_2$, 
and other abundant hydrocarbons (see previous subsection and Table~\ref{tab:key_reac_C}). The removal pathway
of CO in the gas include freeze-out onto the grain surfaces (major channel), and the slow ion-molecule reaction with 
He$^+$ (minor channel). In the upper, heavily UV-irradiated and dilute disk atmosphere CO is also UV-photodissociated 
($z\ga 2.5\,H_r$). At $T\la 30$~K, in the outer midplane CO sticks to the grains and partly converted to H$_2$CO and CH$_3$OH 
ices via hydrogenation reactions, whereas in the warm midplane ($r\la30$~AU, T$\ga 30-40$~K) CO is transformed into CO$_2$ in 
a slightly endothermic reaction between the CO and OH ices (with a barrier of $80$~K). The chemistry of CO gas 
reaches a steady-state in the molecular layer and the cold outer midplane within less than $10^2-10^4$~years 
(see Fig.~\ref{fig:chem_ss}). The quasi-equilibrium CO chemistry in the molecular layer 
is restricted to protonation of CO molecules by H$_3^+$ into HCO$^+$, balanced by dissociative recombination. 
In the disk atmosphere CO chemistry is controlled by the photodissociation slowed down by self- and mutual-shielding
by H$_2$, and $\tau_{\rm chem}\la10^4-10^6$~years. Finally, in the inner warm midplane the surface conversion
of CO into CO$_2$ leads to a very long timescale of $\la 1$~Myr.

The carbon dioxide is a daughter molecule of CO, and is mainly produced in the gas phase via 
oxidation of HCO (O + HCO $\rightarrow$ CO$_2$ + H),
slow combustion of C$_3$ and C$_2$H, slow endothermic reaction of CO and OH with a barrier of 176~K 
(Table~\ref{tab:key_reac_O}), and desorption of CO$_2$ ice at $T\ga 60$~K 
or UV-photodesorption. The main removal gas-phase pathways include photodissociation, ion-molecule reactions with
C$^+$ and H$^+$ (forming CO$^+$ and HCO$^+$, respectively), and accretion to dust grains. The solid CO$_2$ is mainly
produced via endothermic reaction of surface CO and OH in the disk inner midplane and the low part of
the entire molecular layer, and through accretion of the gas-phase carbon dioxide. The CO$_2$ ice is destroyed by
the X-ray/CRP-induced UV photons in the inner midplane, and via thermal and UV-desorption. 
Thus, the CO$_2$ chemistry has particularly long timescale associated with surface reaction of CO and OH, and 
slow photoprocessing of the CO$_2$ ice ($\tau_{\rm chem}\ga1$~Myr). 

Finally, the H$_2$CO molecule is mostly produced in the gas-phase through reaction of CH$_3$ with oxygen atoms, and 
desorption of formaldehyde ice. The major gas-phase removal routes are sticking to dust in the disk regions with 
$T\la40-50$~K, photodissociation, reactions with ionized atomic C and H, oxidation by O (into CO, OH, and H), 
and protonation by H$_3^+$ followed by dissociative recombination back to H$_2$CO (33\%), or CO (33\%), or HCO (33\%). 
The surface evolution of H$_2$CO is governed by accretion and desorption processes, and 
a sequence of surface hydrogenation of CO into CH$_3$OH where formaldehyde ice is an intermediate product.
The timescale of key evolutionary processes for formaldehyde, namely, oxidation of CH$_3$ and H$_2$CO, 
as well as accretion and evaporation, is fast, $\la10^2-10^3$~years 
(Tables~\ref{tab:tau_inner}--\ref{tab:tau_outer}).

As a result, the column densities of formaldehyde are only slightly affected by the turbulent transport (see Fig.~\ref{fig:O}). 
The same holds true for very abundant gas-phase CO and the H$_2$O ice. Their global chemical
evolution is only slightly controlled by the surface chemistry, and the CRP/X-ray-irradiation becomes important only in 
the upper regions that do not contribute to the resulting vertical column densities.
The 2D-mixing enables more efficient production of formaldehyde in the atmosphere (at $\approx 1-2\,H_r$) thanks to
enhanced abundances of CH$_3$ (see Sect.~\ref{C-species}). The turbulent diffusion
lowers the column densities of atomic oxygen, and increases the column densities of gas-phase water, carbon dioxide in 
all phases, and molecular oxygen. The pace of the conversion of atomic oxygen into other O-bearing species is partly governed
by H$_3^+$, which is sensitive to transport (see discussion in Sect.~\ref{ions}). The gas-phase water
production involves this {slow process and slow desorption}, and thus gas-phase water becomes sensitive to the mixing. 
Finally, CO$_2$ and O$_2$ (and other multi-O species in the model) are the molecules which chemical evolution is at least 
partly determined by the  slow surface reactions. This makes them sensitive to the turbulent mixing. Note also that the 
chemical evolution of the CO$_2$ ice and, to a less degree, the CO ice, is influenced strongly by the radial mixing.

\subsection{Nitrogen-containing molecules}
\label{N-species}
\begin{figure*}
\includegraphics[width=0.48\textwidth]{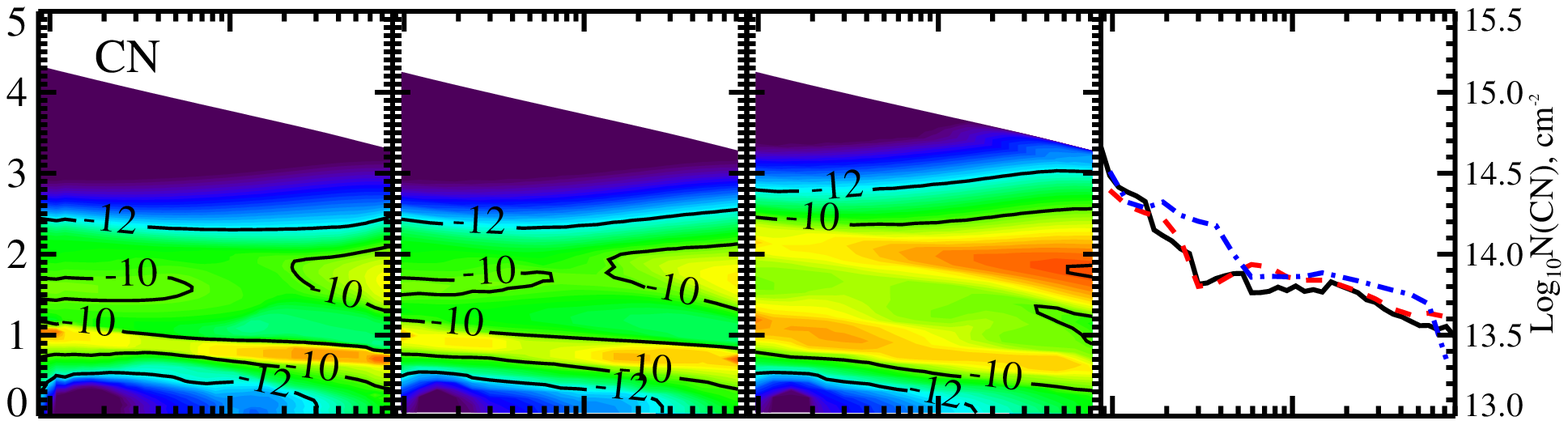}
\includegraphics[width=0.48\textwidth]{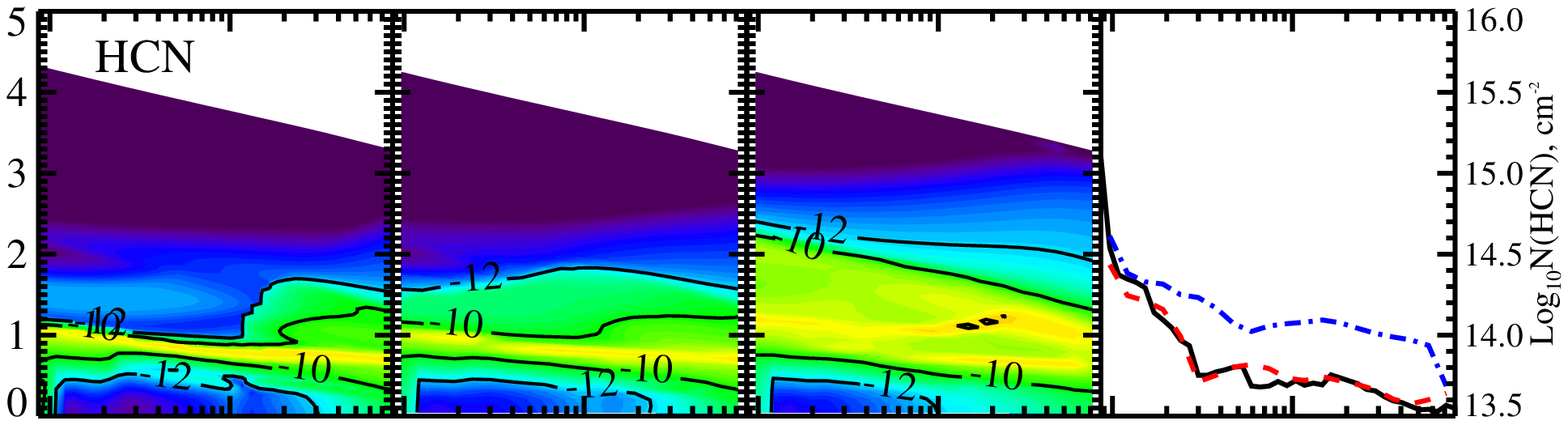}\\
\includegraphics[width=0.48\textwidth]{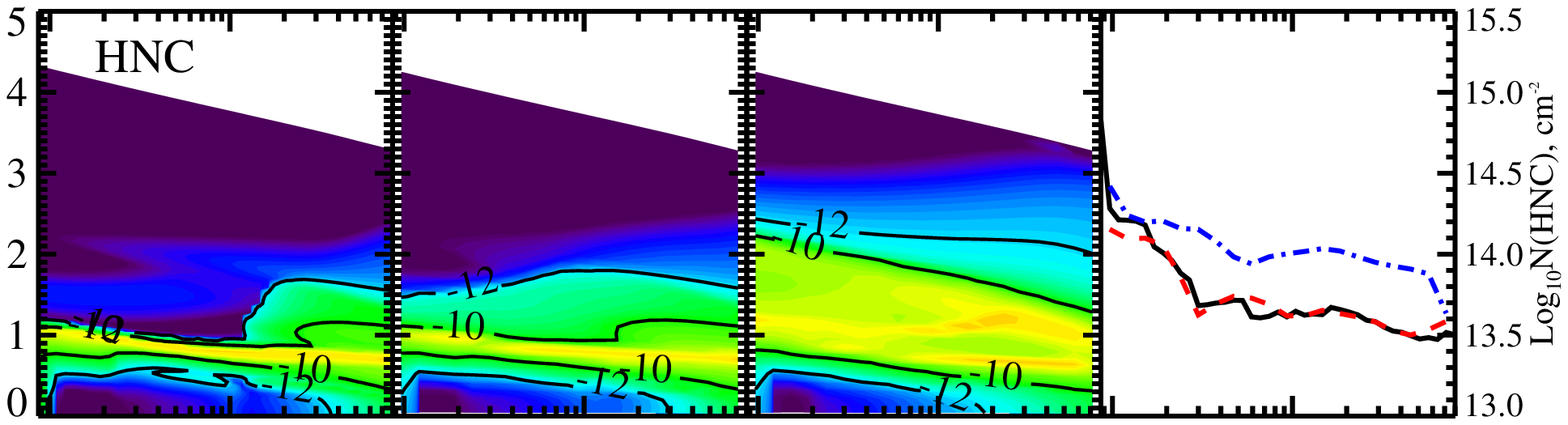}
\includegraphics[width=0.48\textwidth]{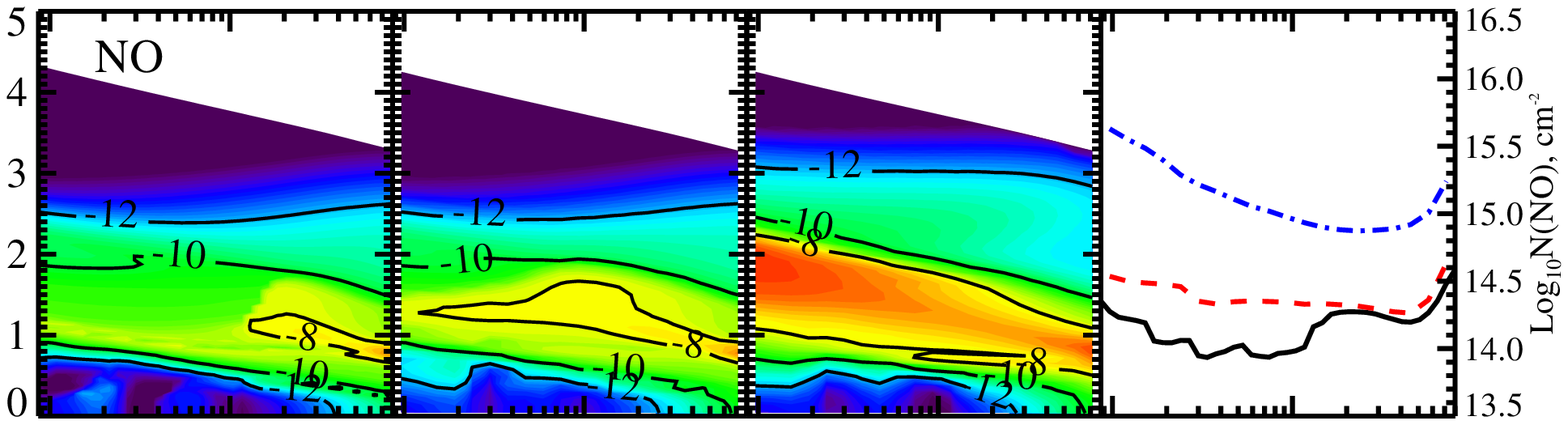}\\
\includegraphics[width=0.48\textwidth]{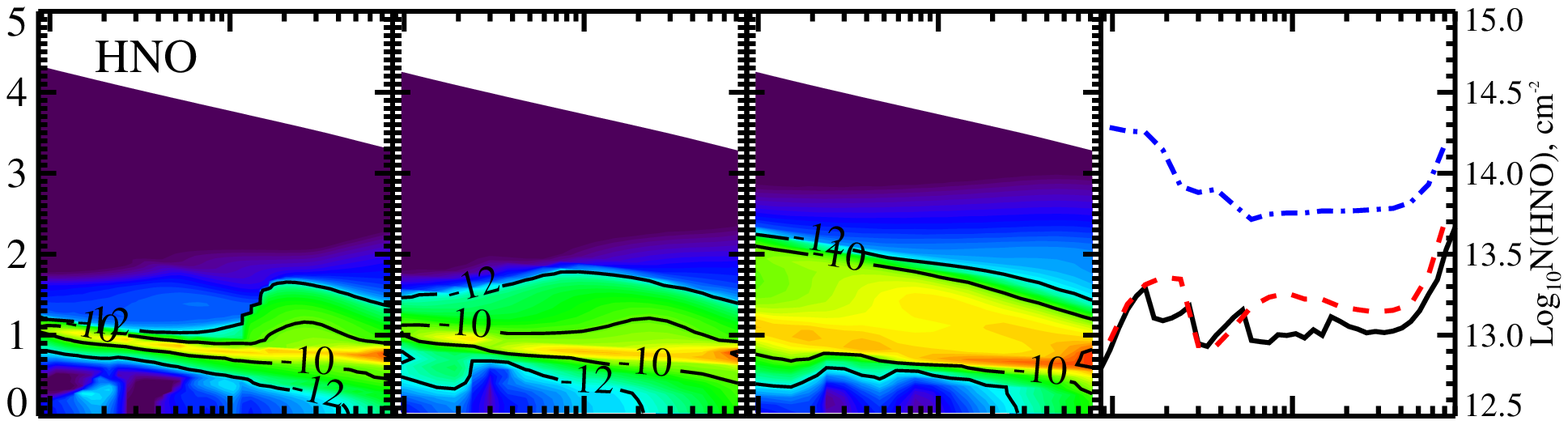}
\includegraphics[width=0.48\textwidth]{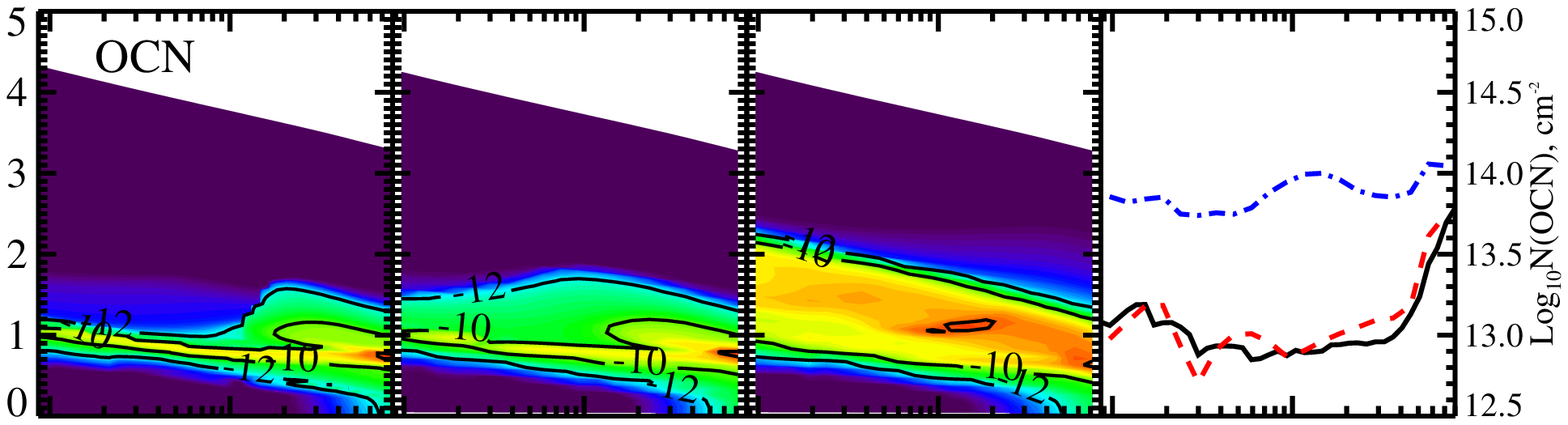}\\
\includegraphics[width=0.48\textwidth]{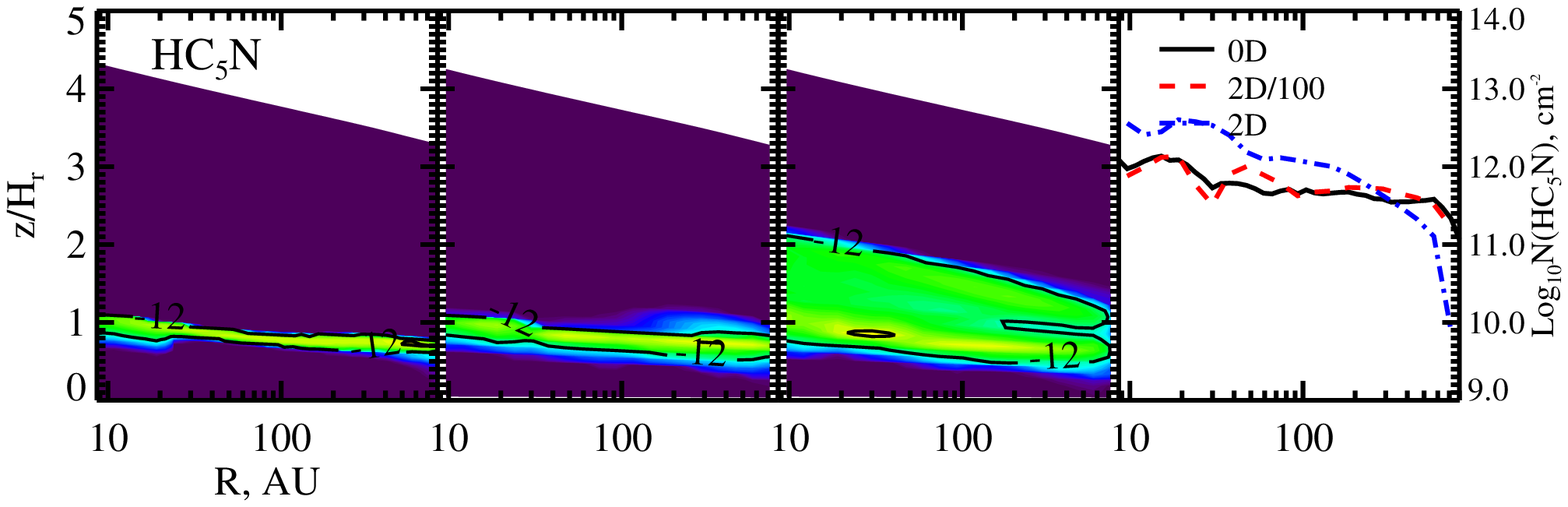}
\includegraphics[width=0.48\textwidth]{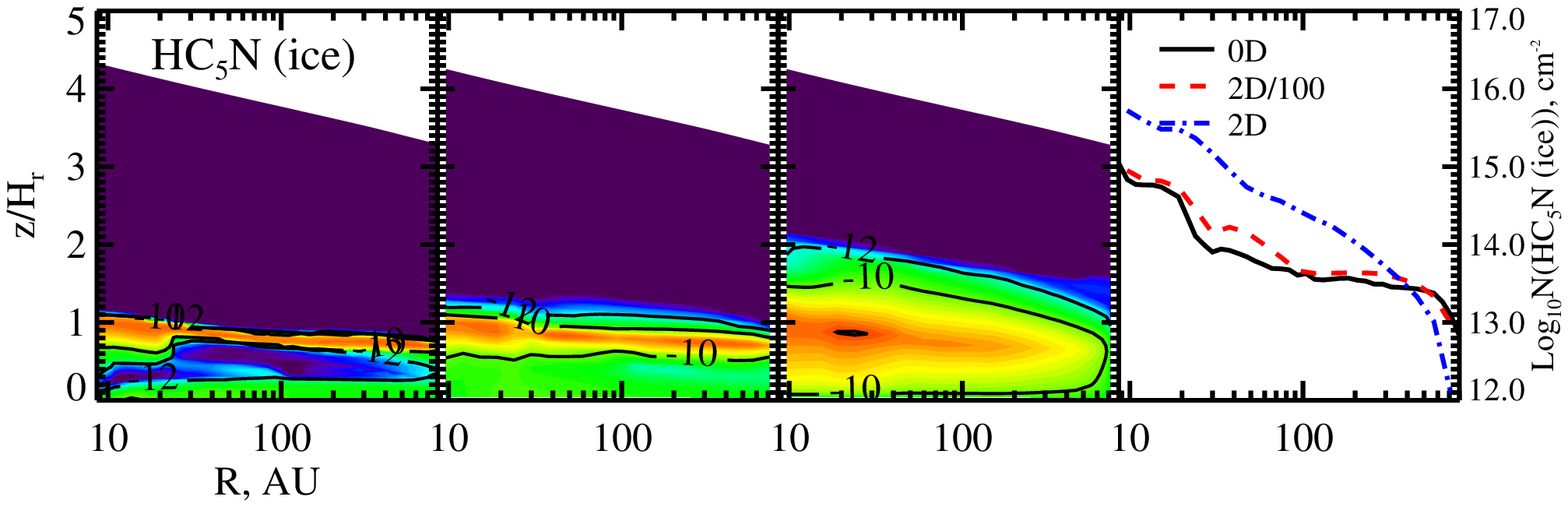}
\caption{The same as in Fig.~\ref{fig:ions} but for the N-containing
species. Results are shown for CN, HCN, HNC, NO, HNO,
OCN, HC$_5$N, and HC$_5$N ice.
} \label{fig:N}
\end{figure*}

\LongTables
\begin{deluxetable*}{llrrll}
\tabletypesize{\footnotesize}
\tablecaption{Key chemical processes: N-bearing species\label{tab:key_reac_N}}
\tablehead{
\colhead{Reaction} & \colhead{$\alpha$} & \colhead{$\beta$} & \colhead{$\gamma$}& \colhead{t$_{\rm min}$} & \colhead{t$_{\rm
max}$}\\
\colhead{} & \colhead{[(cm$^3$)\,s$^{-1}$]} & \colhead{} & \colhead{[K]} & \colhead{[yr]} & \colhead{[yr]}
}
\startdata
HC$_5$N ice   + h$\nu_{\rm CRP}$  $\rightarrow$  C$_4$H ice   + CN ice   & $1.75\,(3)$   & $0$   & $0$   & $1.00$   & $5.00\,(6)$
 \\
HCN   + UV  $\rightarrow$  CN   + H   & $0.16\,(-8)$   & $0$   & $2.69$   & $1.00$   & $5.00\,(6)$   \\
HNC   + UV  $\rightarrow$  CN   + H   & $0.55\,(-9)$   & $0$   & $2.00$   & $1.00$   & $5.00\,(6)$   \\
HNO   + UV  $\rightarrow$  NO   + H   & $0.17\,(-9)$   & $0$   & $0.53$   & $1.00$   & $5.00\,(6)$   \\
OCN   + UV  $\rightarrow$  O   + CN   & $1.00\,(-11)$   & $0$   & $2.00$   & $1.00$   & $5.00\,(6)$   \\
CN   + grain  $\rightarrow$  CN ice   & $1.00$   & $0$   & $0$   & $1.00$   & $5.00\,(6)$   \\
HCN   + grain  $\rightarrow$  HCN ice   & $1.00$   & $0$   & $0$   & $1.00$   & $5.00\,(6)$   \\
HNC   + grain  $\rightarrow$  HNC ice   & $1.00$   & $0$   & $0$   & $1.00$   & $5.00\,(6)$   \\
HNO   + grain  $\rightarrow$  HNO ice   & $1.00$   & $0$   & $0$   & $1.00$   & $5.00\,(6)$   \\
NO   + grain  $\rightarrow$  NO ice   & $1.00$   & $0$   & $0$   & $1.00$   & $5.00\,(6)$   \\
OCN   + grain  $\rightarrow$  OCN ice   & $1.00$   & $0$   & $0$   & $1.00$   & $5.00\,(6)$   \\
HC$_5$N   + grain  $\rightarrow$  HC$_5$N ice   & $1.00$   & $0$   & $0$   & $1.00$   & $5.00\,(6)$   \\
HCN ice  $\rightarrow$  HCN   & $1.00$   & $0$   & $2.05\,(3)$   & $1.00$   & $5.00\,(6)$   \\
HNC ice  $\rightarrow$  HNC   & $1.00$   & $0$   & $2.05\,(3)$   & $1.00$   & $5.00\,(6)$   \\
HC$_5$N ice  $\rightarrow$  HC$_5$N   & $1.00$   & $0$   & $6.18\,(3)$   & $1.00$   & $5.00\,(6)$   \\
H ice   + CN ice  $\rightarrow$  HCN   & $1.00$   & $0$   & $0$   & $1.00$   & $5.00\,(6)$   \\
H ice   + C$_5$N ice  $\rightarrow$  HC$_5$N ice   & $1.00$   & $0$   & $0$   & $1.00$   & $5.00\,(6)$   \\
H ice   + C$_5$N ice  $\rightarrow$  HC$_5$N   & $1.00$   & $0$   & $0$   & $28.60$   & $5.00\,(6)$   \\
H ice   + HC$_5$N ice $\rightarrow$  H$_2$C$_5$N ice   & $1.00$   & $0$   & $1.21\,(3)$   & $1.00$   & $5.00\,(6)$   \\
H ice   + NO ice  $\rightarrow$  HNO   & $1.00$   & $0$   & $0$   & $1.00$   & $5.00\,(6)$   \\
N ice   + C$_5$H ice  $\rightarrow$  HC$_5$N ice   & $1.00$   & $0$   & $0$   & $1.00$   & $5.00\,(6)$   \\
O ice   + CN ice  $\rightarrow$  OCN   & $1.00$   & $0$   & $0$   & $4.18\,(2)$   & $5.00\,(6)$   \\
O ice   + HNO ice  $\rightarrow$  NO   + OH   & $1.00$   & $0$   & $0$   & $1.00$   & $5.00\,(6)$   \\
CH$_3$C$_5$N   + H$^+$ $\rightarrow$  HC$_5$N   + CH$_3$$^+$   & $0.20\,(-7)$   & $-0.50$   & $0$   & $3.82$   & $5.00\,(6)$   \\
HC$_5$N   + H$^+$  $\rightarrow$  C$_5$HN$^+$   + H   & $0.40\,(-7)$   & $-0.50$   & $0$   & $1.00$   & $5.00\,(6)$   \\
HCN   + H$^+$  $\rightarrow$  HCN$^+$   + H   & $0.28\,(-7)$   & $-0.50$   & $0$   & $1.00$   & $5.00\,(6)$   \\
HNC   + H$^+$  $\rightarrow$  HCN   + H$^+$   & $0.25\,(-7)$   & $-0.50$   & $0$   & $1.00$   & $5.00\,(6)$   \\
HCN$^+$   + H  $\rightarrow$  HCN   + H$^+$   & $0.37\,(-10)$   & $0$   & $0$   & $1.00$   & $5.00\,(6)$   \\
NO   + C  $\rightarrow$  CN   + O   & $0.60\,(-10)$   & $-0.16$   & $0$   & $1.00$   & $5.00\,(6)$   \\
OCN   + C  $\rightarrow$  CN   + CO   & $1.00\,(-10)$   & $0$   & $0$   & $1.00$   & $5.00\,(6)$   \\
CN   + O$_2$  $\rightarrow$  OCN   + O   & $0.24\,(-10)$   & $-0.60$   & $0$   & $1.00$   & $5.00\,(6)$   \\
HNO   + O  $\rightarrow$  NO   + OH   & $0.38\,(-10)$   & $0$   & $0$   & $1.00$   & $5.00\,(6)$   \\
NH$_2$  + O$\rightarrow$  HNO   + H   & $0.80\,(-10)$   & $0$   & $0$   & $1.00$   & $5.00\,(6)$   \\
C$_7$N  + O $\rightarrow$  OCN   + C$_6$   & $0.40\,(-10)$   & $0$   & $0$   & $1.00$   & $5.00\,(6)$   \\
CN   + OH  $\rightarrow$  OCN   + H   & $0.70\,(-10)$   & $0$   & $0$   & $1.00$   & $5.00\,(6)$   \\
HNO   + H  $\rightarrow$  NO   + H$_2$   & $0.45\,(-10)$   & $0.72$   & $3.29\,(2)$   & $1.00$   & $5.00\,(6)$   \\
N   + CH  $\rightarrow$  CN   + H   & $0.17\,(-9)$   & $-0.09$   & $0$   & $1.00$   & $2.56\,(6)$   \\
N   + CH$_2$  $\rightarrow$  HNC   + H   & $0.40\,(-10)$   & $0.17$   & $0$   & $1.00$   & $5.00\,(6)$   \\
N   + HCO  $\rightarrow$  OCN   + H   & $1.00\,(-10)$   & $0$   & $0$   & $1.00$   & $5.00\,(6)$   \\
N + NO  $\rightarrow$  N$_2$   + O   & $0.30\,(-10)$   & $-0.60$   & $0$   & $1.00$   & $5.00\,(6)$   \\
N   + OH  $\rightarrow$  NO   + H   & $0.75\,(-10)$   & $-0.18$   & $0$   & $1.00$   & $5.00\,(6)$   \\
C$_3$H$_2$N$^+$   + e$^-$  $\rightarrow$  HNC   + C$_2$H   & $0.75\,(-7)$   & $-0.50$   & $0$   & $1.00$   & $5.00\,(6)$   \\
C$_3$HN$^+$   + e$^-$  $\rightarrow$  HCN   + C$_2$   & $0.30\,(-6)$   & $-0.50$   & $0$   & $1.00$   & $5.00\,(6)$   \\
C$_5$H$_2$N$^+$   + e$^-$  $\rightarrow$  HC$_5$N   + H   & $0.15\,(-6)$   & $-0.50$   & $0$   & $1.00$   & $5.00\,(6)$   \\
C$_5$H$_3$N$^+$   + e$^-$  $\rightarrow$  HC$_5$N   + H$_2$   & $1.00\,(-6)$   & $-0.30$   & $0$   & $7.48$   & $5.00\,(6)$   \\
C$_6$H$_4$N$^+$   + e$^-$  $\rightarrow$  HC$_5$N   + CH$_3$   & $1.00\,(-6)$   & $-0.30$   & $0$   & $1.00$   & $5.00\,(6)$   \\
CH$_2$CN$^+$   + e$^-$  $\rightarrow$  HCN   + CH   & $0.30\,(-6)$   & $-0.50$   & $0$   & $1.00$   & $5.00\,(6)$   \\
H$_2$CN$^+$   + e$^-$  $\rightarrow$  HCN   + H   & $0.19\,(-6)$   & $-0.65$   & $0$   & $1.00$   & $5.00\,(6)$   \\
H$_2$CN$^+$   + e$^-$  $\rightarrow$  HNC   + H   & $0.19\,(-6)$   & $-0.65$   & $0$   & $1.00$   & $5.00\,(6)$   \\
H$_2$NO$^+$   + e$^-$  $\rightarrow$  HNO   + H   & $0.15\,(-6)$   & $-0.50$   & $0$   & $1.00$   & $5.00\,(6)$   \\
H$_2$CN$^+$   + e$^-$  $\rightarrow$  CN   + H   + H   & $0.92\,(-7)$   & $-0.65$   & $0$   & $1.00$   & $5.00\,(6)$   \\
\enddata
\end{deluxetable*}

Similar to the oxygen-bearing molecules, the observations of the N-bearing species provide useful diagnostics on the 
physical conditions and chemical state of the cosmic objects. The ammonia doublet lines are used to 
measure kinetic temperatures \citep[e.g.,][]{Walmsley_Ungerechts83,Churchwell_ea90,Jijina_ea99}, while the CN and HCN relative 
line strengths are sensitive to the UV flux \citep[e.g.,][]{Bergin_ea04}. The linear HC$_3$N and heavier cyanopolyynes 
with large dipole moments serve as densitometers \citep[e.g.,][]{Pratap_ea97}, whereas OCN$^-$ has been proposed as a carrier 
of the $4.62\mu$m absorption feature in interstellar ices \citep[][]{Raunier_ea03,Bennett_ea10}.
In this Section we analyze in detail chemical and mixing processes responsible for the evolution of the neutral 
nitrogen-bearing species in protoplanetary disks. 

There are 4 steadfast neutral N-bearing molecules (CN, HCN, HNC, NH$_2$), 2 ions (H$_2$CN$^+$ and NH$_4^+$)
(Table~\ref{tab:steadfast}). Their column densities in the laminar and fast mixing models differ by a factor of 3. 
The sensitive nitrogen species include 19 molecules (e.g., C$_2$N, C$_3$N,..., C$_9$N, HC$_3$N),
6 ions (e.g., NO$^+$, N$_2$H$^+$, CNC$^+$), and 22 solid species (e.g, CH$_2$NH, HC$_3$N,.., HC$_9$N,
CN, NO; Table~\ref{tab:sens}). Their column densities are changed by up to 2 orders of magnitude by the turbulent 
transport. The N-bearing hypersensitive species are 4 molecules (C$_7$N, HC$_7$N, N$_2$O, and NO$_2$), N$^+$, 
and 8 ices (e.g., CH$_3$C$_3$N, HNO, N$_2$, NO$_2$, etc.), see Table~\ref{tab:hyps}. 
The turbulent diffusion alters their column densities by up to 7 orders of magnitude (H$_5$C$_3$N ice).
Similarly to hydrocarbons and oxygen-containing molecules, complex chains with multiple N, C, or H atoms  
are more strongly affected by the turbulent transport than chemically simpler species, though the trend is not
that clear (e.g., NO$_2$, HNCO, N$^+$, N$_2$O are the outliers). As \citet{Vasyunin_ea08} have found, nitrogen
chemistry in disks involves a larger number of key reactions, including many exothermic neutral-neutral reactions, 
surface processes, compared to the chemistries of the O- and C-containing species. 
We have selected several most interesting nitrogen species for detailed chemical analysis. 

In Fig.~\ref{fig:N} the distributions of the relative molecular abundances and column densities at 5~Myr of 
CN, HCN, HNC, NO, HNO, OCN, HC$_5$N, and HC$_5$N ice calculated with the laminar and the 2D-mixing models are 
presented. The relative abundance distributions of the all considered species show a 3-layered structure, 
with peak concentrations in the molecular layers at $\approx 0.8-1\,H_r$, with typical values of 
$10^{-10}-10^{-8}$ (with respect to the total amount of hydrogen nuclei). The molecular layers of N-bearing molecules 
are narrow, $\approx 0.2-0.5\,H_r$, similar to those of hydrocarbons, HCO$^+$ and H$_2$CO
(see Figs.~\ref{fig:C} and \ref{fig:O}). The photostable CN radical has a second molecular layer 
in the outer disk atmosphere, at $z\approx 1.8$ ($r\ga200$~AU), though it does not contribute much to the total
CN column density. Note that HC$_5$N ice is also concentrated in the molecular layer, at $\sim 1$ pressure scale height. 

Turbulent mixing does not affect column densities of steadfast CN, HCN, and HNC species (see 4th panels in
Fig.~\ref{fig:N}). Nonetheless, their molecular layers are broadened by diffusion, and the second, upper molecular layer of 
CN becomes more prominent (compare 1st and 3rd panels in the Figure). NO, HNO, OCN, HC$_5$N, and HC$_5$N ices are 
sensitive to the mixing, with their column densities increased by 2D-turbulent diffusion by up to a factor of 40 
(cyanodiacetylene; Table~\ref{tab:sens}). The corresponding abundance distributions are vertically extended up to 
the thicknesses of $\sim 0.5-1.5\,H_r$, and enhanced by the transport by up to several orders of magnitude.

To better understand these result, we investigate the evolution of CN, HCN, HNC, NO, HNO, OCN, HC$_5$N, and the surface HC$_5$N 
in the two disk vertical slices at $r=10$ and 250~AU  (the laminar chemical model). The most important reactions responsible for
the time-dependent evolution of their abundances in the midplane, the molecular layer, and the atmosphere are presented in 
Table~\ref{tab:key_reac_N}, both for the inner and outer disk regions. The final list contains only 
top 20 reactions per region (midplane, molecular layer, atmosphere) for the entire 5~Myr time span, with all repetitions 
removed. 

The evolution of cyanopolyynes is tightly connected with the evolution of carbon chains discussed in 
Section~\ref{C-species}. Their major production pathways in the gas include evaporation from icy mantles, neutral-neutral 
reaction of N with Renner-Teller hydrocarbons (C$_n$H), 
slow or slightly endothermic reaction of CN with C$_{n-1}$H$_2$ (e.g., CN + C$_4$H$_2$ $\rightarrow$ HC$_5$N), 
and at later times dissociative recombination of their protonated analogs formed by the ion-molecule reactions with H$_3^+$ 
(Table~\ref{tab:key_reac_N}). The main removal pathways are the freeze-out at temperatures $\la 70-170$~K, 
UV-photodissociation, reactive collisions with the ionized C, H, and He atoms. On dust surfaces, 
cyanopolyyne ices form either via addition of N to the frozen Renner-Teller hydrocarbons or through hydrogenation
of the C$_n$N ices. The major destruction routes for the cyanopolyyne ices are photoevaporation, photoprocessing
by the X-ray- or CRP-driven UV photons (e.g., HC$_5$N ice + h$\nu_{\rm CRP}$  $\rightarrow$  C$_4$H ice + CN ice),
and surface conversion to even more complex species (e.g., HC$_5$N ice + H ice  $\rightarrow$  H$_2$C$_5$N ice). 
In turn, dehydrogenated carbon chains with attached nitrogen atom are produced in the gas and on the dust surfaces by
rapid neutral-neutral reactions of N with the C$_n$H species, and dissociation of complex molecules with multiple
C and N atoms (e.g., CH$_3$C$_5$N ice + h$\nu_{\rm CRP}$  $\rightarrow$  C$_5$N ice + CH$_3$ ice).
Obviously, as in the case of complex carbon chains, characteristic timescales of the cyanopolyyne chemical evolution 
are regulated by the slow surface processes with $\tau_{\rm chem}$ exceeding million years.

In contrast, chemical histories of CN, HCN, and HNC are closely related and governed by a small set of reactions.
The production and destruction of CN proceeds entirely in the gas phase. The primal formation pathways are photodissociation
of hydrogen cyanide and isocyanide, rapid barrierless neutral-neutral reactions of N with CH, C$_2$, and C$_2$N, 
and NO with C (rate coefficients are $\sim10^{-11}-6\,10^{-10}$~cm$^3$\,s$^{-1}$), and dissociative recombination 
of HCNH$^+$ at later times ($t\ga 10^2-10^4$~years), see Table~\ref{tab:key_reac_N}. Another, less important formation pathway 
for CN is the neutral-neutral reaction between C and OCN leading to CN and CO. The destruction of CN is mostly caused by
photodissociation, fast neutral-neutral 
reactions with N, O, OH, and O$_2$, and freeze-out in the disk midplane at $T\la35-40$~K. In the atmosphere CN reacts
with H$^+$, forming CN$^+$, which is converted back to CN by the charge transfer with atomic hydrogen.
The fact that the CN chemistry involves only gas-phase routes implies its relatively short characteristic timescale, 
$\sim 10^3$~years (Tables~\ref{tab:tau_inner}--\ref{tab:tau_outer}). Only in the outer atmosphere 
$\tau_{\rm chem}({\rm CN})\ga10^5$~years, because the CN evolution there depends on the slow evolution of C$_2$H 
 (see Fig.~\ref{fig:C} and discussion in Sect.~\ref{C-species}).

The key production route for gas-phase HCN and HNC are the neutral-neutral reaction of nitrogen atoms with CH$_2$
and dissociative recombination of HCNH$^+$ (Table~\ref{tab:key_reac_N}). 
More energetically favorable isomer, HCN, is produced upon reactive collisions of ionized hydrogen with hydrogen isocyanide. The
minor formation channels in the disk midplane are direct surface recombination
of atomic hydrogen and cyanogen radical that leads to gas-phase HCN and H (with $5\%$ probability), 
and the charge transfer reaction of H and HCN$^+$. Due to their relatively large 
binding energies, thermal desorption of HCN and HNC does not occur until temperatures of $T\ga40$~K are reached.
The major destruction routes for the gas-phase HCN and HNC include accretion onto the dust grains, the charge transfer reaction
with 
H$^+$, the ion-molecule reaction with C$^+$ leading to CNC$^+$ or C$_2$N$^+$, and protonation reaction 
with H$_3^+$, HCO$^+$, and H$_3$O$^+$ (e.g., HCN  + HCO$^+$  $\rightarrow$  HCNH$^+$ + CO). 
Another important destruction channel for hydrogen (iso)cyanide in the upper disk layers at $\sim 1.5-2$ scale heights is
photodissociation. The characteristic timescales of the HCN and HNC evolution are $\la10^3-10^4$~years in the molecular layer. 
Their chemical timescales exceed $10^5-10^6$~years in the midplane, where their evolution is partly controlled by the slow 
surface formation, and in the inner upper molecular layer/low atmosphere subject to the slow X-ray-driven dissociation of
H$_2$ and release of oxygen from water and CO.

The evolution of NO, HNO, and OCN, similarly to that of CN, HCN, and HNC, is also governed by a set of 
rapid exothermic neutral-neutral reactions. The major formation pathways for nitrogen monoxide comprise
reactive collisions of atomic nitrogen with hydroxyle,
reactive collisions of atomic oxygen with nitrogen monohydride, 
gas-phase destruction of HNO by atomic oxygen, and, in the inner midplane, 
surface destruction of HNO by O, CH$_2$ and CH$_3$, with products (NO and OH, CH$_3$, or CH$_4$) 
directly injected into the gas. In the disk atmosphere NO is produced by the photodissociation of HNO.
In the inner disk midplane a source of the NO gas is thermal desorption of the NO ice.
The major removal processes for NO include photodissociation in the atmosphere, accretion onto the dust grains
in the disk regions with $T\la 30-35$~K, 
neutral-neutral reactions with atomic carbon and nitrogen (leading to CO and N$_2$), and charge transfer
reactions with H$^+$ and C$^+$. The characteristic timescale for NO is $\la10^3-10^4$~years 
in the molecular layer, and $\ga 10^5-10^6$~years in the midplane, where its evolution is partly controlled by the 
surface processes. In the upper molecular layer at $r\sim 100$~AU the NO chemical timescale is again long, $\sim 1$~Myr,
since it is related with the evolution of atomic oxygen that is slow at those disk heights.

The chemical evolution of nitric acid (HNO) is very similar to the evolution of NO. 
The major production terms for HNO are neutral-neutral reaction of atomic oxygen with nitrogen 
dihydride, and, 
in the warm molecular layer and the inner midplane, via surface hydrogenation of NO
directly to the gas-phase. The evaporation of the solid HNO plays only a minor role for production of gas-phase HNO
(in the laminar model).
The major destruction terms for NHO include photodissociation in the atmosphere, accretion onto the dust grains
in the disk regions with $T\la 40-50$~K, 
neutral-neutral reactions with atomic hydrogen and oxygen (leading to NO), and ion-molecule
reaction with H$^+$ (forming ionized NO and H$_2$). 
The distribution of the characteristic timescale for HNO over the disk is close to that of NO.

Finally, the chemical evolution of cyanate (OCN) is more diverse than that of NO and HNO.
The key formation routes comprise of neutral-neutral reactions of CN with OH 
and O$_2$, N and HNO,
and oxidation reactions with abundant cyanopolyynes, e.g. C$_7$N + O $\rightarrow$  OCN + C$_6$. 
It can also be produced by direct surface recombination 
in O and CN ices. Similar to HNO, desorption of the solid OCN is not a key formation process
for gas-phase cyanate (in the laminar model). The key removal routes include photodissociation in the atmosphere, 
sticking to the dust grains in the disk regions with $T\la 50-60$~K, neutral-neutral reactions with atomic oxygen and carbon
(leading to NO or CO), 
slow combustion reaction with O$_2$ (producing CO$_2$ and NO), and ion-molecule reaction with C$^+$ (forming 
ionized CO and CN). The characteristic timescale of the OCN evolution is $\la10^3-10^4$~years 
in the molecular layer, and $\ga 10^5-10^6$~years in the inner and outer midplane, where its evolution is partly 
controlled by the surface processes. In the atmosphere the OCN chemical timescale increases from $10^4$
till $10^6$~years, similar to the CO$_2$ timescale shown in Fig.~\ref{fig:chem_ss}.

Consequently, column densities of CN, HCN, and HNC are not much affected by diffusion
as their key evolutionary processes in their molecular layers proceed rapidly in the gas-phase
(Fig.~\ref{fig:N}). As for many molecules in the model, the turbulent transport expands their molecular layers in 
vertical direction. Since the chemistry of CN involves C$_2$H, which abundances 
are increased by the mixing in the atmosphere at $z\ga 2\,H_r$ (see Fig.~\ref{fig:C} and discussion in 
Sect.~\ref{C-species}), in the 2D-mixing case ($Sc=1$) the second molecular layer of CN is formed. Similarly, the 
hydrogen cyanide and isocyanide evolution is related to the evolution of CH$_2$, which is sensitive to the transport 
in the same upper disk region, though not as strong as ethynyl radical. 
The chemistry of the NO-containing species (e.g., NO, HNO, OCN) is sensitive to mixing as it depends on the evolution 
of O and O$_2$ that is in turn influenced by the transport, and since their production involves minor slow surface routes.
Moreover, thermal evaporation of these heavy species is inefficient, and thus photoevaporation of their ices becomes
an important production pathways for the gas-phase counterparts. Vertical diffusive mixing allows more proficient 
evaporation of NO, HNO, and OCN ices in the warm molecular layer, enhancing their gas-phase abundances. Finally,
due to extreme importance of surface chemistry for the chemical evolution of cyanopolyynes, their abundances and column 
densities are increased by the turbulent mixing, albeit not as strongly as for some heavy hydrocarbons.

\subsection{Sulfur-containing molecules}
\label{S-species}
\begin{figure*}
\includegraphics[width=0.48\textwidth]{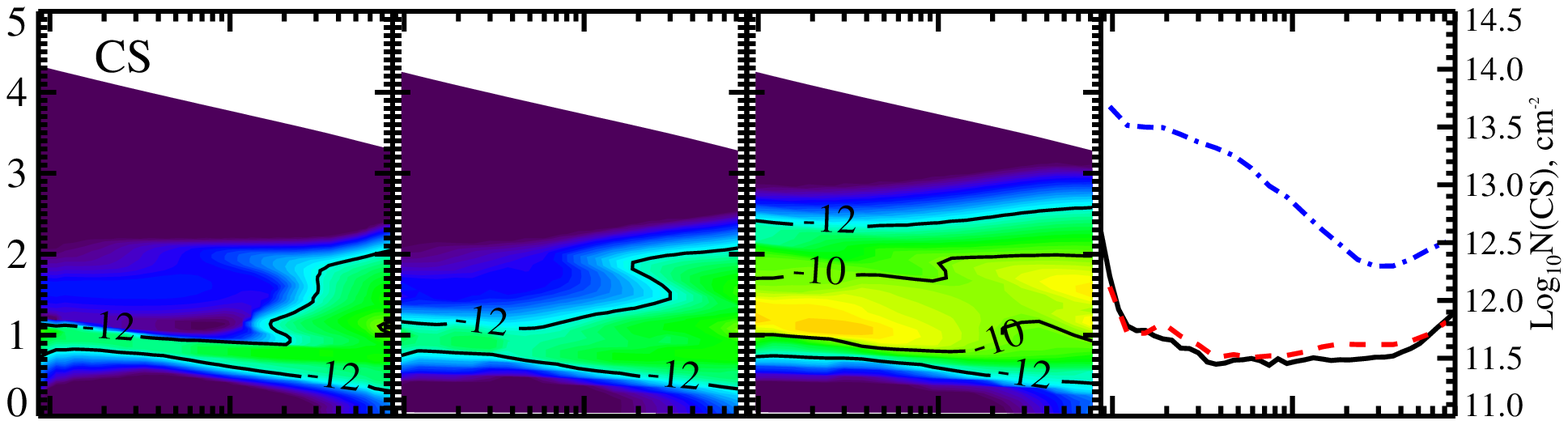}
\includegraphics[width=0.48\textwidth]{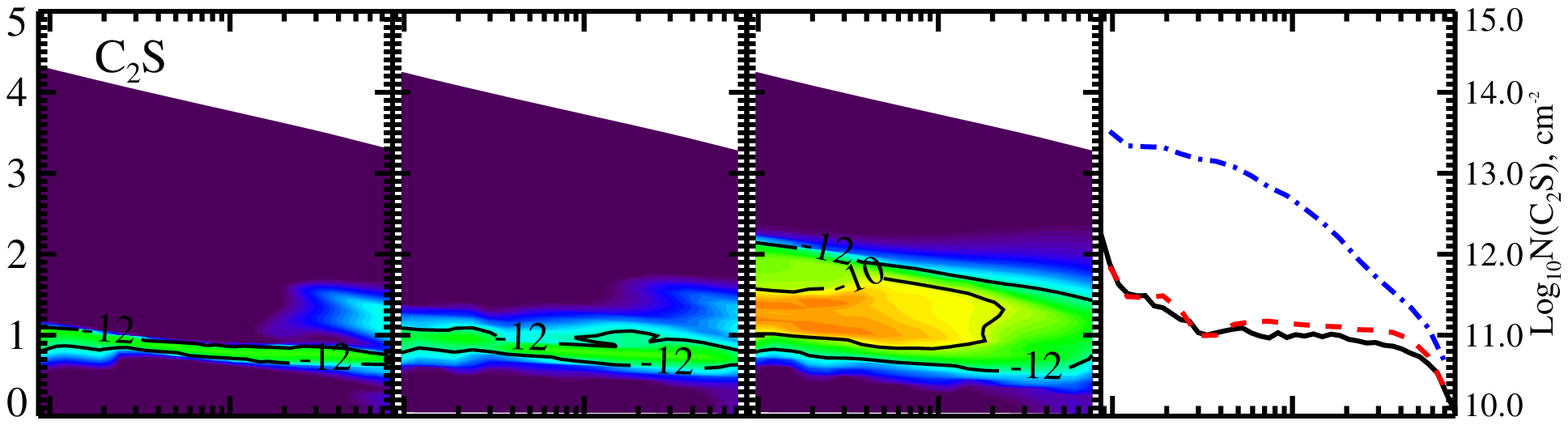}\\
\includegraphics[width=0.48\textwidth]{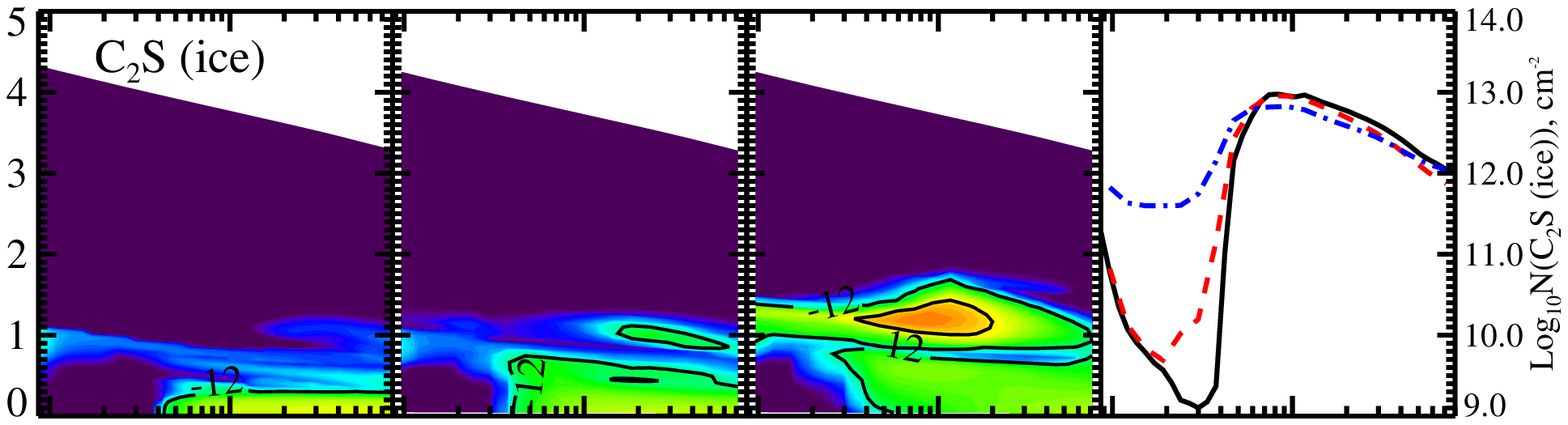}
\includegraphics[width=0.48\textwidth]{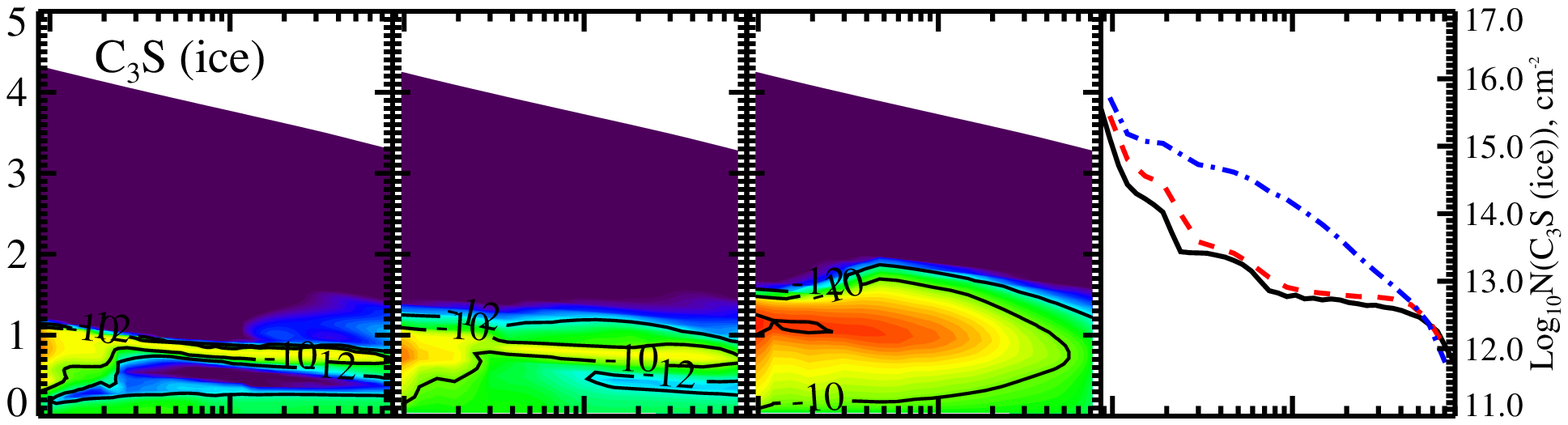}\\
\includegraphics[width=0.48\textwidth]{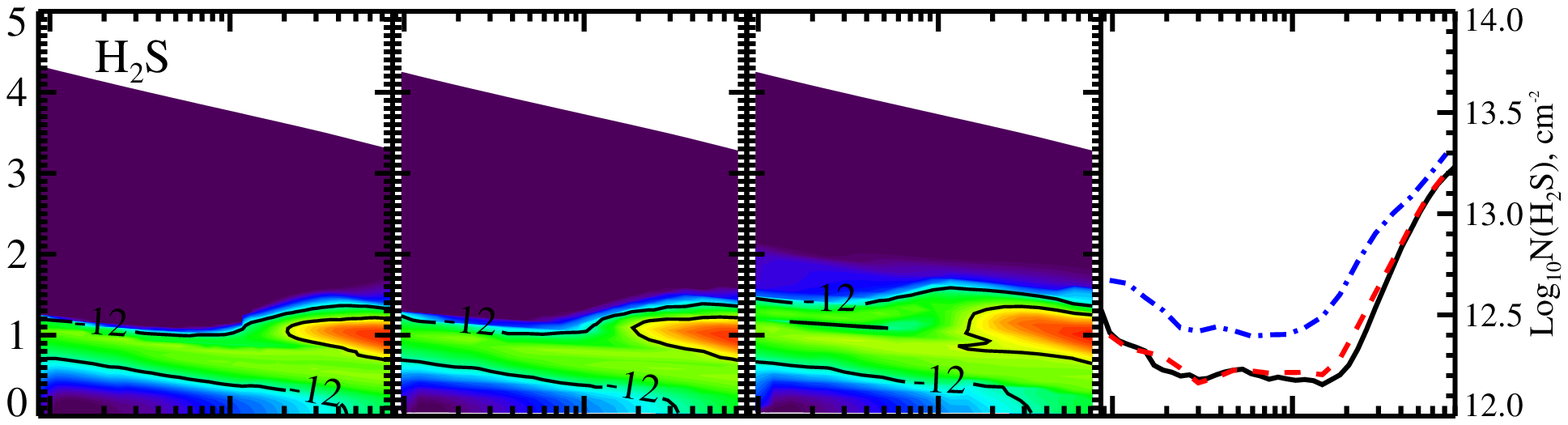}
\includegraphics[width=0.48\textwidth]{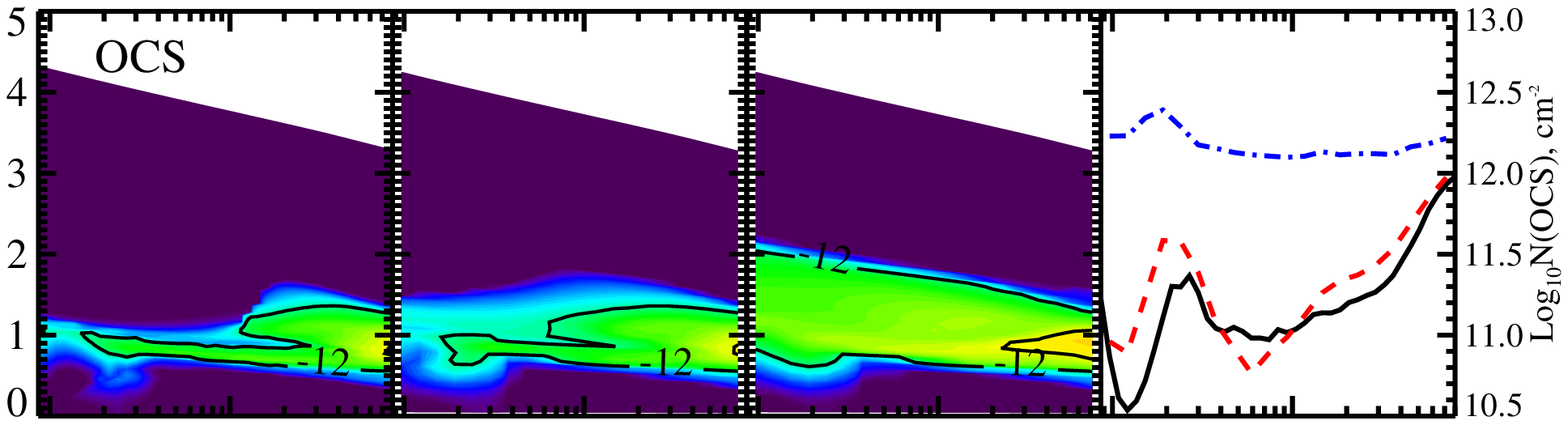}\\
\includegraphics[width=0.48\textwidth]{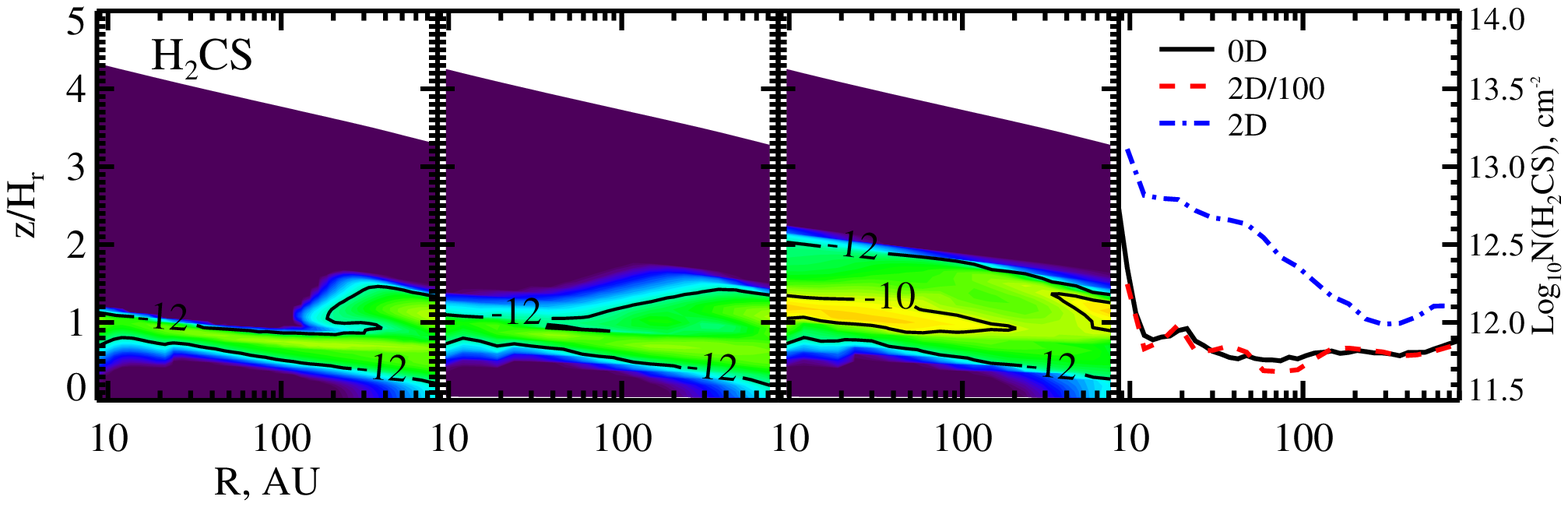}
\includegraphics[width=0.48\textwidth]{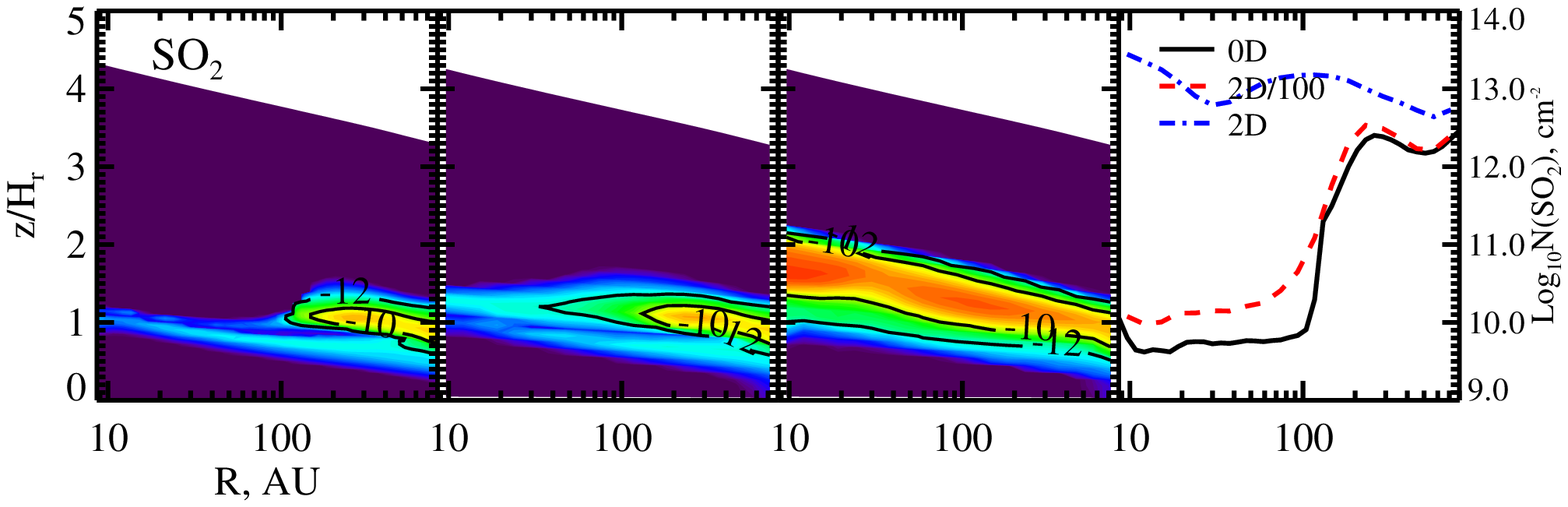}
\caption{The same as in Fig.~\ref{fig:ions} but for the S-containing
species. Results are shown for CS, C$_2$S, C$_2$S ice, C$_3$S ice,
H$_2$S, OCS, H$_2$CS, and SO$_2$.}
\label{fig:S}
\end{figure*}

\LongTables
\begin{deluxetable*}{llrrll}
\tabletypesize{\footnotesize}
\tablecaption{Key chemical processes: S-bearing species\label{tab:key_reac_S}}
\tablehead{
\colhead{Reaction} & \colhead{$\alpha$} & \colhead{$\beta$} & \colhead{$\gamma$}& \colhead{t$_{\rm min}$} & \colhead{t$_{\rm
max}$}\\
\colhead{} & \colhead{[(cm$^3$)\,s$^{-1}$]} & \colhead{} & \colhead{[K]} & \colhead{[yr]} & \colhead{[yr]}
}
\startdata
C$_3$S ice   + h$\nu_{\rm CRP}$  $\rightarrow$  C$_2$ ice   + CS ice   & $1.50\,(3)$   & $0$   & $0$   & $1.00$   & $5.00\,(6)$
\\
CS   + UV  $\rightarrow$  C   + S   & $0.98\,(-9)$   & $0$   & $2.43$   & $1.00$   & $5.00\,(6)$   \\
C$_2$S   + UV  $\rightarrow$  C$_2$   + S   & $1.00\,(-10)$   & $0$   & $2.00$   & $1.00$   & $5.00\,(6)$   \\
H$_2$CS   + UV  $\rightarrow$  CS   + H$_2$   & $1.00\,(-9)$   & $0$   & $1.70$   & $1.00$   & $5.00\,(6)$   \\
H$_2$S   + UV  $\rightarrow$  HS   + H   & $0.31\,(-8)$   & $0$   & $2.27$   & $1.00$   & $5.00\,(6)$   \\
OCS   + UV  $\rightarrow$  CO   + S   & $0.37\,(-8)$   & $0$   & $2.07$   & $1.00$   & $5.00\,(6)$   \\
SO$_2$   + UV  $\rightarrow$  SO   + O   & $0.19\,(-8)$   & $0$   & $2.38$   & $1.00$   & $5.00\,(6)$   \\
CS   + grain  $\rightarrow$  CS ice   & $1.00$   & $0$   & $0$   & $1.00$   & $5.00\,(6)$   \\
C$_2$S   + grain  $\rightarrow$  C$_2$S ice   & $1.00$   & $0$   & $0$   & $1.00$   & $5.00\,(6)$   \\
C$_3$S   + grain  $\rightarrow$  C$_3$S ice   & $1.00$   & $0$   & $0$   & $1.00$   & $5.00\,(6)$   \\
OCS   + grain  $\rightarrow$  OCS ice   & $1.00$   & $0$   & $0$   & $1.00$   & $5.00\,(6)$   \\
SO$_2$   + grain  $\rightarrow$  SO$_2$ ice   & $1.00$   & $0$   & $0$   & $1.00$   & $5.00\,(6)$   \\
H$_2$S   + grain  $\rightarrow$  H$_2$S ice   & $1.00$   & $0$   & $0$   & $1.00$   & $5.00\,(6)$   \\
H$_2$CS   + grain  $\rightarrow$  H$_2$CS ice   & $1.00$   & $0$   & $0$   & $1.00$   & $5.00\,(6)$   \\
CS ice  $\rightarrow$  CS   & $1.00$   & $0$   & $1.90\,(3)$   & $28.60$   & $5.00\,(6)$   \\
C$_2$S ice  $\rightarrow$  C$_2$S   & $1.00$   & $0$   & $2.70\,(3)$   & $1.00$   & $5.00\,(6)$   \\
C$_3$S ice  $\rightarrow$  C$_3$S   & $1.00$   & $0$   & $3.50\,(3)$   & $1.00$   & $5.00\,(6)$   \\
C$_2$S ice   + C ice  $\rightarrow$  C$_3$S ice   & $1.00$   & $0$   & $0$   & $1.00$   & $5.00\,(6)$   \\
H ice   + HS ice  $\rightarrow$  H$_2$S   & $1.00$   & $0$   & $0$   & $1.00$   & $5.00\,(6)$   \\
O ice   + CS ice  $\rightarrow$  OCS   & $1.00$   & $0$   & $0$   & $1.00$   & $5.00\,(6)$   \\
O ice   + SO ice  $\rightarrow$  SO$_2$   & $1.00$   & $0$   & $0$   & $1.00$   & $5.00\,(6)$   \\
S ice   + CO ice  $\rightarrow$  OCS   & $1.00$   & $0$   & $0$   & $1.00$   & $5.00\,(6)$   \\
CS   + H$^+$  $\rightarrow$  CS$^+$   + H   & $0.18\,(-7)$   & $-0.50$   & $0$   & $1.00$   & $5.00\,(6)$   \\
C$_2$S   + H$^+$  $\rightarrow$  C$_2$S$^+$   + H   & $0.11\,(-7)$   & $-0.50$   & $0$   & $1.00$   & $5.00\,(6)$   \\
H$_2$S   + H$^+$  $\rightarrow$  H$_2$S$^+$   + H   & $0.38\,(-8)$   & $-0.50$   & $0$   & $1.00$   & $5.00\,(6)$   \\
H$_2$CS   + H$^+$  $\rightarrow$  H$_2$CS$^+$   + H   & $0.64\,(-8)$   & $-0.50$   & $0$   & $1.00$   & $5.00\,(6)$   \\
OCS   + H$^+$  $\rightarrow$  HS$^+$   + CO   & $0.65\,(-8)$   & $-0.50$   & $0$   & $1.00$   & $5.00\,(6)$   \\
SO$_2$   + C$^+$  $\rightarrow$  SO$^+$   + CO   & $0.20\,(-8)$   & $-0.50$   & $0$   & $1.00$   & $5.00\,(6)$   \\
SO$_2$   + H$_3$$^+$  $\rightarrow$  HSO$_2$$^+$   + H$_2$   & $0.37\,(-8)$   & $-0.50$   & $0$   & $1.96$   & $5.00\,(6)$   \\
O   + SO  $\rightarrow$  SO$_2$   & $0.32\,(-15)$   & $-1.60$   & $0$   & $1.00$   & $5.00\,(6)$   \\
O   + HCS  $\rightarrow$  OCS   + H   & $0.50\,(-10)$   & $0$   & $0$   & $1.00$   & $5.00\,(6)$   \\
S   + CO  $\rightarrow$  OCS   & $0.16\,(-16)$   & $-1.50$   & $0$   & $1.00$   & $5.00\,(6)$   \\
S   + CH$_3$  $\rightarrow$  H$_2$CS   + H   & $0.14\,(-9)$   & $0$   & $0$   & $1.00$   & $5.00\,(6)$   \\
SO$_2$   + C  $\rightarrow$  CO   + SO   & $0.70\,(-10)$   & $0$   & $0$   & $1.00$   & $5.00\,(6)$   \\
OH   + SO  $\rightarrow$  SO$_2$   + H   & $0.86\,(-10)$   & $0$   & $0$   & $1.00$   & $5.00\,(6)$   \\
C$_3$S$^+$   + e$^-$  $\rightarrow$  C$_2$S   + C   & $1.00\,(-7)$   & $-0.50$   & $0$   & $1.00$   & $5.00\,(6)$   \\
C$_3$S$^+$   + e$^-$  $\rightarrow$  CS   + C$_2$   & $1.00\,(-7)$   & $-0.50$   & $0$   & $1.00$   & $1.20\,(4)$   \\
C$_4$S$^+$   + e$^-$  $\rightarrow$  C$_2$S   + C$_2$   & $1.00\,(-7)$   & $-0.50$   & $0$   & $1.00$   & $5.00\,(6)$   \\
H$_3$CS$^+$   + e$^-$  $\rightarrow$  H$_2$CS   + H   & $0.30\,(-6)$   & $-0.50$   & $0$   & $1.00$   & $5.00\,(6)$   \\
H$_3$S$^+$   + e$^-$  $\rightarrow$  H$_2$S   + H   & $0.30\,(-6)$   & $-0.50$   & $0$   & $1.00$   & $5.00\,(6)$   \\
HC$_2$S$^+$   + e$^-$  $\rightarrow$  C$_2$S   + H   & $0.15\,(-6)$   & $-0.50$   & $0$   & $1.00$   & $5.00\,(6)$   \\
HCS$^+$   + e$^-$  $\rightarrow$  CS   + H   & $0.18\,(-6)$   & $-0.57$   & $0$   & $1.00$   & $5.00\,(6)$   \\
HOCS$^+$   + e$^-$  $\rightarrow$  CS   + OH   & $0.20\,(-6)$   & $-0.50$   & $0$   & $1.00$   & $5.00\,(6)$   \\
OCS$^+$   + e$^-$  $\rightarrow$  CS   + O   & $0.48\,(-7)$   & $-0.62$   & $0$   & $1.00$   & $1.20\,(4)$   \\
H$_2$CS$^+$   + e$^-$  $\rightarrow$  H$_2$CS   & $0.11\,(-9)$   & $-0.70$   & $0$   & $14.60$   & $5.00\,(6)$   \\
H$_2$S$^+$   + e$^-$  $\rightarrow$  H$_2$S   & $0.11\,(-9)$   & $-0.70$   & $0$   & $1.00$   & $5.00\,(6)$   \\
H$_3$CS$^+$   + e$^-$  $\rightarrow$  CS   + H   + H$_2$   & $0.30\,(-6)$   & $-0.50$   & $0$   & $1.00$   & $5.00\,(6)$   \\
C$_3$S$^+$   + grain(-)  $\rightarrow$  C$_2$S   + C   + grain(0)   & $0.33$   & $0$   & $0$   & $3.82$   & $5.00\,(6)$   \\
HOCS$^+$   + grain(-)  $\rightarrow$  CS   + OH   + grain(0)   & $0.50$   & $0$   & $0$   & $3.82$   & $5.00\,(6)$   \\
H$_3$CS$^+$   + grain(-)  $\rightarrow$  H$_2$CS   + H   + grain(0)   & $0.50$   & $0$   & $0$   & $3.82$   & $5.00\,(6)$   \\
HC$_2$S$^+$   + grain(-)  $\rightarrow$  C$_2$S   + H   + grain(0)   & $0.50$   & $0$   & $0$   & $1.96$   & $5.00\,(6)$   \\
\enddata
\end{deluxetable*}

In this Section we analyze in detail chemical and mixing processes responsible for the evolution of 
sulfur-bearing species in protoplanetary disks. 
The chemistry of sulfur-bearing molecules is the least understood in astrochemistry. As other heavy elements, sulfur
is depleted from the gas, but the magnitude of the effect and in which form sulfur is bound is not yet
clearly determined \citep[e.g.,][]{Lodders_03,Flynn_ea06,Goicoechea_ea06}.
Moreover, many processes and reaction rate data for the S-bearing species are lacking accurate laboratory measurements or 
theoretical foundations, leading to considerable uncertainties in modeled abundances 
\citep[e.g.,][]{Vasyunin_ea08,Wakelam_ea10}. Thus, results in this subsection should be taken with special care. So far there are
only two sulfur-containing molecules detected in disks,
namely, CS \citep[e.g.,][]{DGG97} and, recently, SO \citep{Fuente_ea10a}. In the interstellar medium and the shells of
AGB stars other species have been discovered, e.g. C$_2$S, C$_3$S, and H$_2$CS 
\citep[e.g.,][]{Cernicharo_ea87,Yamamoto_ea87,Irvine_ea89}. So, here we restrict ourselves with discussion of neutral species
only.

There are 2 steadfast neutral S-bearing molecules (H$_2$S and HCS), 5 ices (HS, H$_2$S, H$_2$CS,
S, and S$_2$), and no ions (Table~\ref{tab:steadfast}). Their column densities in the laminar and the fast 2D-mixing model 
differ by a factor of $\le 3$. The sensitive sulfur-bearing species include 7 gas-phase molecules 
(e.g., CS, H$_2$CS, H$_2$S$_2$, HS, S), 2 ions (S$^+$ and SO$^+$), and 7 ices 
(e.g, CS, C$_3$S, C$_4$S, H$_2$s$_2$, SO; Table~\ref{tab:sens}). 
Their column densities are changed by up to 2 orders of magnitude by the turbulent transport. 
Among hypersensitive species (Table~\ref{tab:hyps}) are 4 gas-phase molecules (C$_2$S, C$_3$S, SO, and SO$_2$), 
no ions, and 3 ices (C$_2$S, NS, and SiS). 
The turbulent diffusion alters their column densities by up to a factor of 7\,000 (SO$_2$ ice).
Similarly to hydrocarbons and oxygen-containing molecules, complex S-chains with multiple carbon atoms or oxygen  
are more strongly affected by the turbulent transport than chemically simpler species. 
Unlike the case of C- and N-bearing species, there is no trend that the concentrations of the sulfur-bearing ices are
stronger altered by the transport than the abundances of their gas-phase analogs.

In Fig.~\ref{fig:S} the distributions of the relative molecular abundances and column densities at 5~Myr of 
CS, C$_2$S, C$_2$S ice, C$_3$S ice, H$_2$S, OCS, H$_2$CS, and SO$_2$ calculated with the laminar and the 2D-mixing 
models are shown. In the laminar model the relative abundance distributions of the gas-phase species show a
3-layered structure, with peak concentrations in the molecular layers at $\approx 0.8-1\,H_r$. In contrast, 
the C$_2$S and C$_3$S ices are more concentrated toward the disk midplane and the bottom of the molecular layer, 
with typical abundances of $10^{-12}-10^{-10}$.
The molecular layers of gas-phase sulfur-bearing molecules have various thicknesses, from $\approx 0.05\,H_r$
for C$_2$S and up to $\sim 2\,H_r$ for CS, and are not as pronounced as for carbon-, oxygen- and nitrogen-bearing species
(see Figs.~\ref{fig:C}--\ref{fig:N}). Their typical peak relative abundances are only $10^{-12}-10^{-10}$, though
in the outer molecular layer the H$_2$S abundances are as high as $\sim 10^{-8}$. 

Among the above molecules, only H$_2$S is steadfast (see 4th panels in Fig.~\ref{fig:S}).
The mixing enhances the H$_2$S concentration in the upper molecular layer at $z\ga 1.3\,H_r$ that does not contribute
to the total column density (compare 1st and 3rd panels in the Figure). The CS, C$_2$S, C$_2$S ice, C$_3$S ice, 
OCS, H$_2$CS, and SO$_2$ are sensitive and hypersensitive to the mixing.
The corresponding molecular layers are vertically broadened up to $\approx 2\,H_r$, and strongly molecularly enriched
by several orders of magnitude.

To better understand these results, we investigate the evolution of CS, C$_2$S, C$_2$S ice, C$_3$S ice, H$_2$S, OCS, 
H$_2$CS, and SO$_2$ in the two disk vertical slices at $r=10$ and 250~AU  (the laminar chemical model). The most important 
reactions responsible for the time-dependent evolution of their abundances in the midplane, the molecular layer, and 
the atmosphere are presented in Table~\ref{tab:key_reac_S}, both for the inner and outer disk regions. 
The final list contains top 20 reactions per region (midplane, molecular layer, atmosphere) for 
the entire 5~Myr time span, with all repetitions removed.

The chemical evolution of hydrogen sulfide in the atmosphere begins with radiative association
of ionized sulfur and H$_2$, leading to H$_2$S$^+$ that slowly radiatively
recombines into H$_2$S.
The major destruction reactions in the atmosphere is photodissociation and
rapid ion-molecule reaction with C$^+$ forming protonated CS and H (75\%) or H$_2$S$^+$ and C (25\%),
and charge transfer reaction with H$^+$. In the molecular layer and midplane hydrogen sulfide is initially
produced as in the atmosphere, but after 1-100~years the top formation pathway for gas-phase H$_2$S is a direct surface 
recombination of HS and H ices. This is caused by a relatively large binding energy of 
H$_2$S that prevents H$_2$S ice to return to the gas phase unless
temperatures exceed about 45-50~K. The HS radical is formed via neutral-neutral reaction of S with OH, and also via
direct surface recombination of H and S. The major destruction pathways for H$_2$S in the molecular layer and midplane
are accretion onto dust grain surfaces, ion-molecule reactions with HCO$^+$ and 
H$_3^+$ (leading to H$_3$S$^+$), and charge transfer reaction with atomic hydrogen. The characteristic timescale of 
the H$_2$S evolution is short in the midplane and molecular layer, 
$\la 10-10^4$~years since its initial production and freeze-out are 
relatively fast. Later, accretion is balanced out by the direct surface recombination, while desorption of the H$_2$S
is not significant. In the atmosphere the H$_2$S reaches a quasi steady-state within 
$10^4-10^6$~years, which is determined by the slow evolution of H$_2$, H$_2^+$, and H$_3^+$ (see discussion in Sect.~\ref{ions}).

The chemical evolution of CS proceeds entirely in the gas phase. The carbon monosulfide is 
synthesized by dissociative recombination of HCS$^+$, C$_2$S$^+$, C$_3$S$^+$, HOCS$^+$, and H$_3$CS$^+$ on electrons
or negatively charged grains in the inner midplane, and via desorption
of the CS ice at $T\ga 35-40$~K. The major destruction pathways
are photodissociation in the atmosphere, depletion in the molecular layer and the midplane, slow endothermic oxidation reaction
(CS + O $\rightarrow$ CO + S), charge transfer with ionized hydrogen
atoms, and ion-molecule reactions with primal ions, e.g. HCO$^+$, H$_3$O$^+$, and H$_3^+$, leading to HCS$^+$ 
(Table~\ref{tab:key_reac_S}). In turn, the protonated carbon monosulfide is initially synthesized through an ion-molecule
destruction of H$_2$ by C$^+$. The C$_2$S$^+$ is produced either by an ion-molecule reaction of ethynyl radical with ionized
atomic sulfur or via charge transfer reactions of C$_2$S with H$^+$ and C$^+$. The C$_3$S$^+$ chemistry is similar to that
of C$_2$S$^+$, with the production pathway that involves C$_3$H instead of ethynyl. The HOCS$^+$ is protonated
OCS, and is formed via ion-molecule reactions of OCS with dominant polyatomic ions (H$_3^+$, HCO$^+$, etc.).
Finally, H$_3$CS$^+$ is synthesized via protonation of H$_2$CS and via ion-molecule reactions of ionized sulfur or HS$^+$ with
methane. The steady-state for CS is attained at about $10^2-10^4$~years in the lower molecular layer/upper midplane, and later
in other disk parts. While the CS chemistry does not include surface processes directly, it is related with the evolution of
C$_2$S$^+$, C$_3$S$^+$, and HOCS$^+$, and, thus, their chemical ``parents'', namely C$_2$S, C$_3$S, and OCS. The latter 
molecules are partly synthesized via surface reactions, making their evolution slow in the cold midplane. 

The chemical evolution of all the C$_n$S species in the model ($n\le4$) is well coupled. 
For example, like in the case of CS, major formation channels for gas-phase C$_2$S include dissociative recombination
of C$_3$S$^+$, C$_4$S$^+$, HC$_2$S$^+$, and HC$_3$S$^+$ on negatively charged grains in the inner midplane and electrons.
The HC$_2$S$^+$ and HC$_3$S$^+$ ions are formed later in the disk by protonation of C$_2$S and C$_3$S.
The binding energy adopted for C$_2$S is 2\,700~K, and higher for more massive carbon chain sulfides, which precludes 
effective thermal desorption of their ices in the disk midplane and the molecular layer beyond 10~AU.
The major destruction channels for C$_2$S are photodissociation, accretion onto dust grains, charge transfer reactions 
with H$^+$ and C$^+$, and protonation by abundant polyatomic ions (leading to HC$_2$S$^+$).
The evolution of the C$_2$S ice is significantly simpler. It involves only 3 major processes: accretion of the gas-phase C$_2$S,
surface reaction of frozen atomic carbon with the C$_2$S ice producing the C$_3$S ice, and photodissociation of the C$_2$S
ice by the CRP/X-ray-induced UV photons (leading to C and CS ices). The evolution of the C$_3$S ice includes its 
surface synthesis from C$_2$ and C, accretion of gas-phase C$_3$S, and photodestruction to the C$_2$ and CS ices.
Consequently, the characteristic chemical timescales for gas-phase and, particularly, for surface C$_n$S
molecules are long, $\tau_{\rm chem} \ga 10^6$~years.

Thioformaldehyde, H$_2$CS, like CS and H$_2$CO, is synthesized in the gas. Its major production routes consist of 
neutral-neutral reaction between S and CH$_3$, and dissociative recombination of protonated H$_2$CS
at later times. The key destruction routes are photodissociation, accretion onto dust grains at $T\la45-50$~K, charge transfer
reaction with ionized atomic hydrogen, ion-molecule reaction with ionized atomic carbon, and proton addition in reactive
collisions with HCO$^+$, H$_3^+$, etc. The distribution of the H$_2$CS characteristic timescale over the disk closely
resembles $\tau_{\rm chem}({\rm CS})$.

The chemical evolution of carbonyl sulfide (OCS) begins with its formation upon oxidation of HCS and via radiative association 
of S and CO. At later times OCS is also produced by the direct surface recombination of
the S and CO ices, and the CS and O ices. The destruction of OCS involves photodissociation in the disk atmosphere, 
accretion onto dust in the molecular layer and the midplane, and ion-molecule reactions with ionized C and H 
(leading to either CS$^+$ and CO or HS$^+$ and CO). The protonation of OCS by the polyatomic ions leads to HOCS$^+$, 
which dissociatively recombines into OCS and H (50\%) or CS and H (50\%). The characteristic timescale for 
carbonyl sulfide is $\la 10^2-10^6$~years in the disk midplane, $\sim 10^3-10^4$~years in the lower molecular layer,
and $\ga 10^4-10^6$ elsewhere in the disk.

Finally, sulfur dioxide (SO$_2$) is synthesized by exothermic neutral-neutral reaction between SO and OH, slow radiative
association of SO and O, and the same reaction catalyzed by dust grains, with SO$_2$ directly
desorbed back to the gas (with 5\% probability). Its major removal channels are photodissociation, 
freeze-out in disk regions with $T\la 60-80$~K,  neutral-neutral
reaction with atomic carbon (leading to SO and CO), and ion-molecule reaction with C$^+$ (producing SO$^+$ and CO).
Later a quasi-equilibrium cycling of SO$_2$ prevails, which includes protonation of sulfur dioxide by abundant
polyatomic ions (H$_3^+$, HCO$^+$) followed by dissociation recombination back to SO$_2$ with the 66\% probability.
In turn, sulfur monoxide (SO) is produced by slow combustion of S, 
neutral-neutral reactions between S and OH and O with HS, and direct surface reaction of the HS and O ices. Consequently, the 
characteristic chemical timescale of SO$_2$ is governed by the slow evolution of molecular oxygen due to surface
processes (see Section~\ref{O-species}), and exceeds 1~Myr in the disk midplane, and $10^4-10^5$~years in the
molecular layer. 

Since chemical network for S-bearing species, especially complex chains, in our model is more limited than
that for the N- and C-containing molecules, the turbulent transport enhances abundances and column densities
even for transient S-species. All species considered above either formed via surface reactions or their synthesis
involves surface-produced radicals, and their chemical evolution proceed slower than the mixing (see Fig.~\ref{fig:S}). 
As for many molecules in the model, the turbulent transport expands their molecular layers in 
vertical direction. Complex S-ices are more readily produced in the mixing case as their formation is based on
heavy radicals, which become mobile at $T\ga 20-40$~K. Also, in the mixing model photodissociation of these ices is more
effective (see a peak in the relative abundances of C$_2$S ice at $r \sim 100$~AU, $z\sim 1.5\,H_r$). The diffusive transport
most strongly increases concentration of SO$_2$ molecules by producing more molecular oxygen needed for its 
synthesis. For SO$_2$, its radial distribution becomes sensitive to the transport. The least affected species is H$_2$S
(as well as HCS, and ices of S, HS, and H$_2$S) as its characteristic evolutionary timescale is shorter than the mixing
timescale in the disk region where it is most abundant (see Tables~\ref{tab:tau_inner}--\ref{tab:tau_outer}). 


\subsection{Complex organic molecules}
\label{complex}
\begin{figure*}
\includegraphics[width=0.48\textwidth]{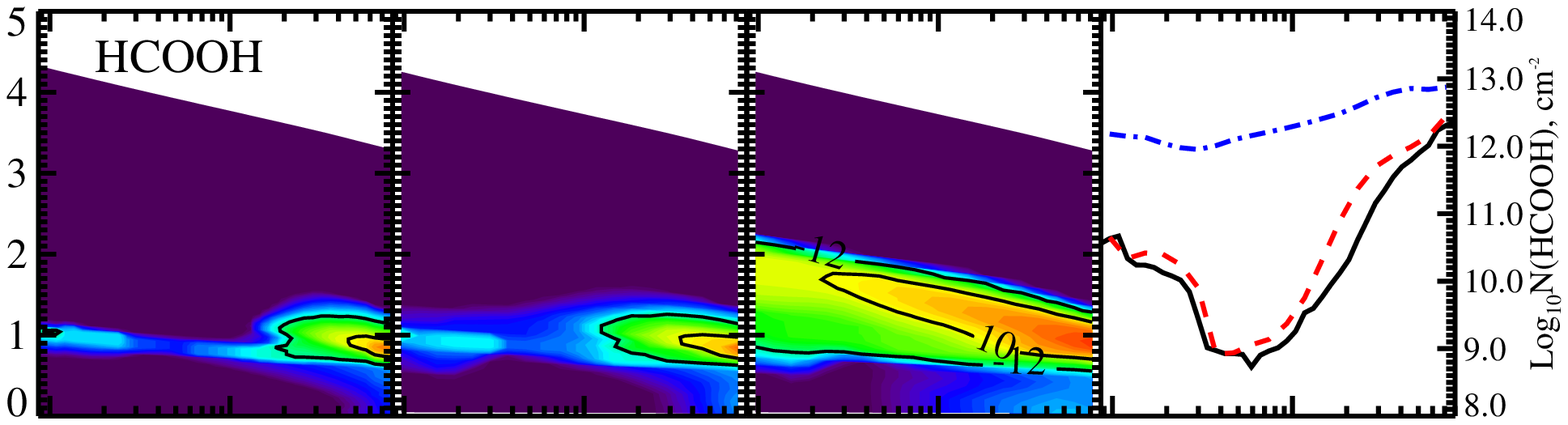}
\includegraphics[width=0.48\textwidth]{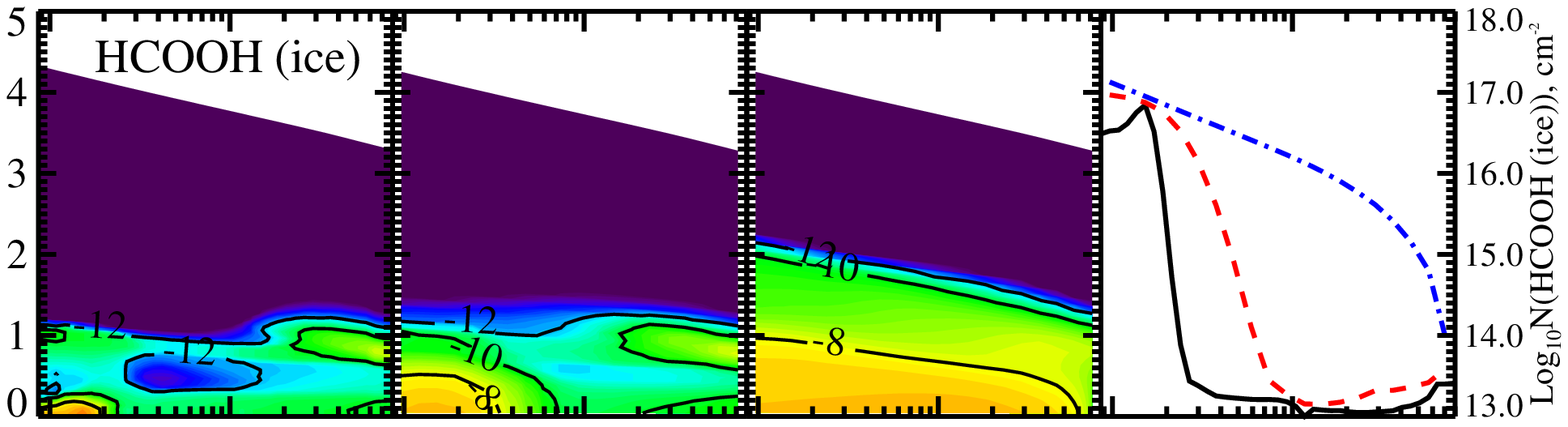}\\
\includegraphics[width=0.48\textwidth]{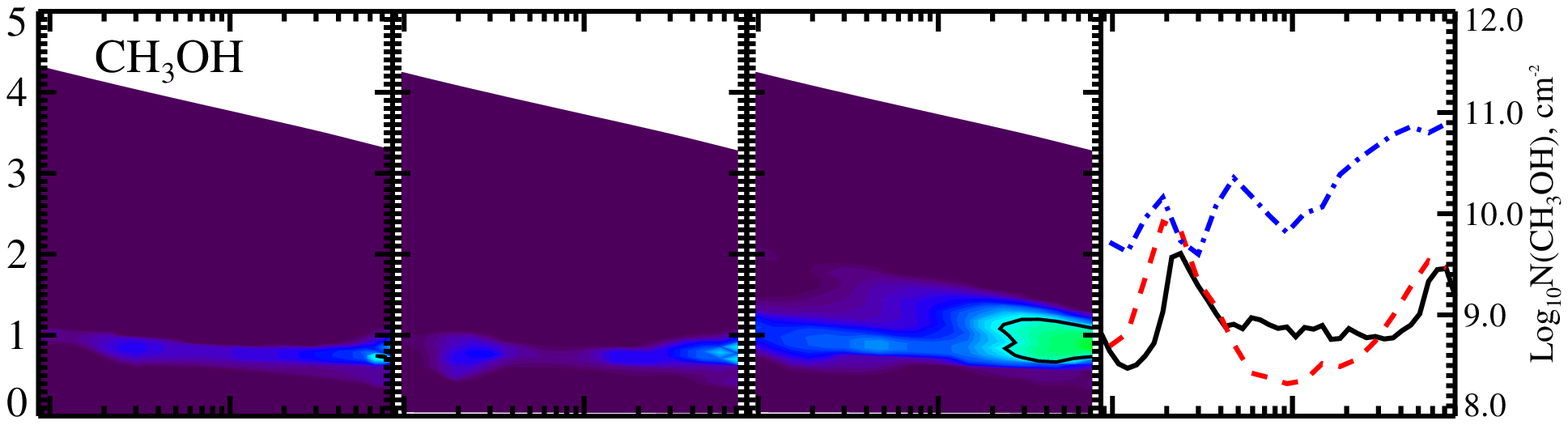}
\includegraphics[width=0.48\textwidth]{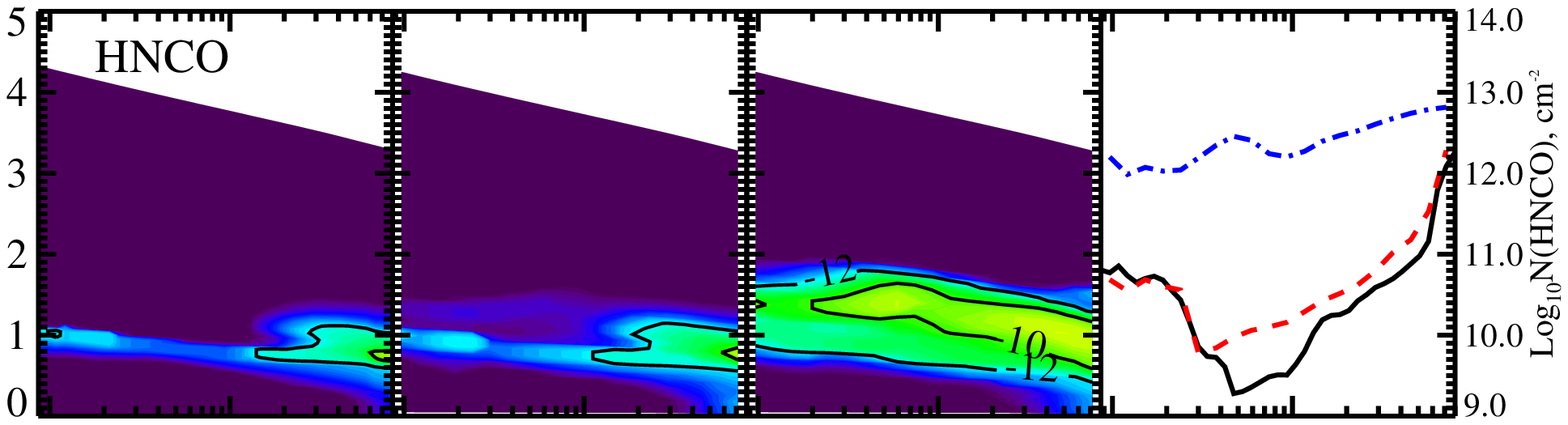}\\
\includegraphics[width=0.48\textwidth]{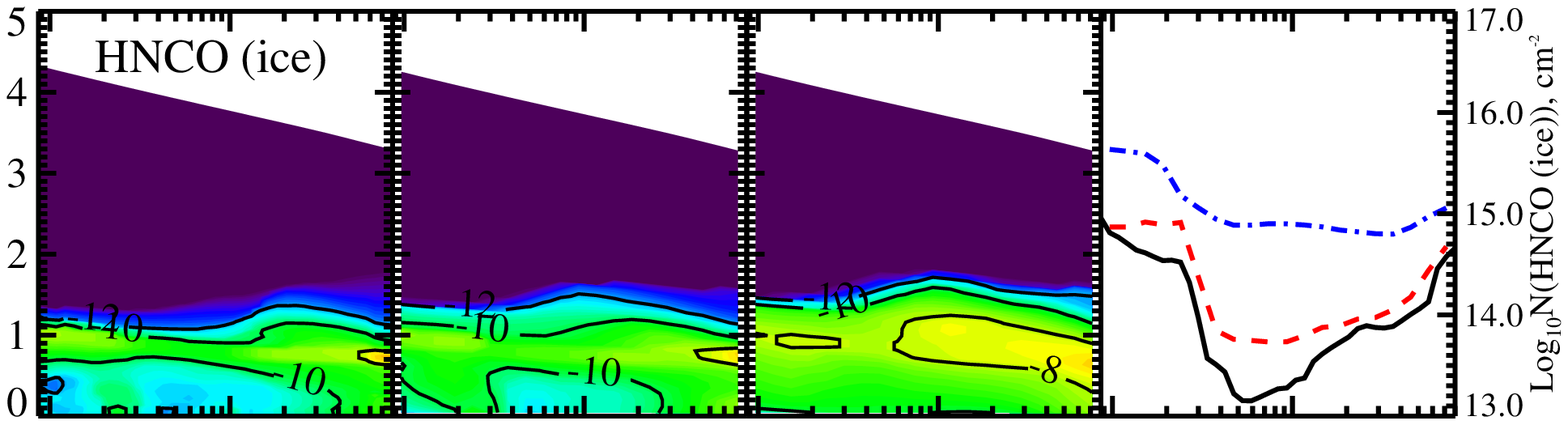}
\includegraphics[width=0.48\textwidth]{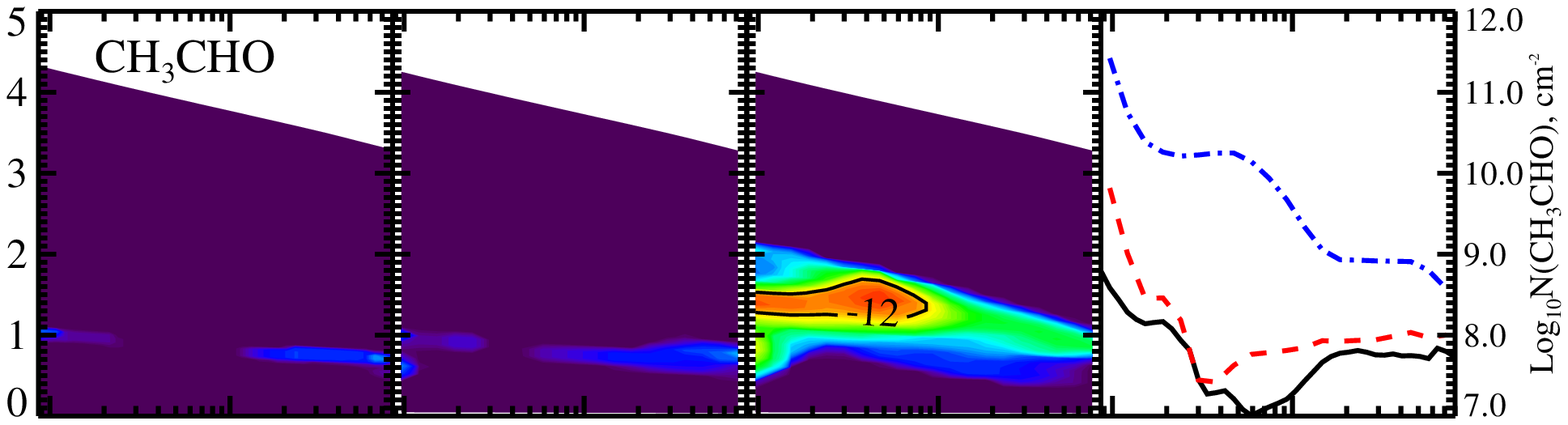}\\
\includegraphics[width=0.48\textwidth]{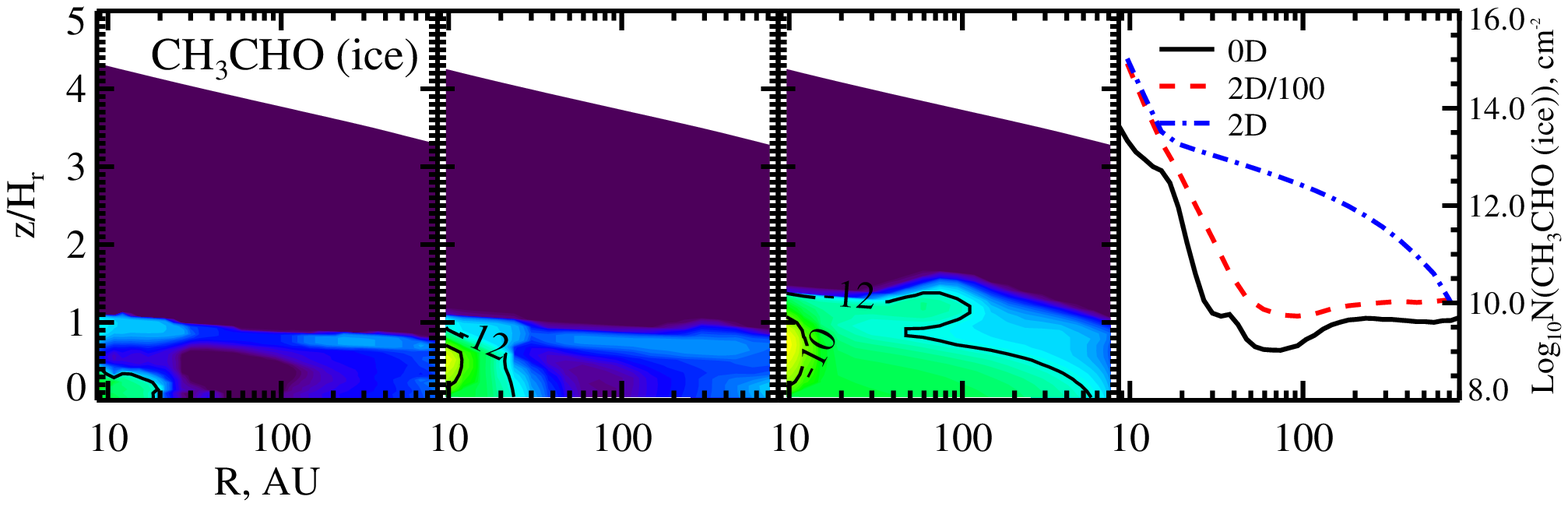}
\includegraphics[width=0.48\textwidth]{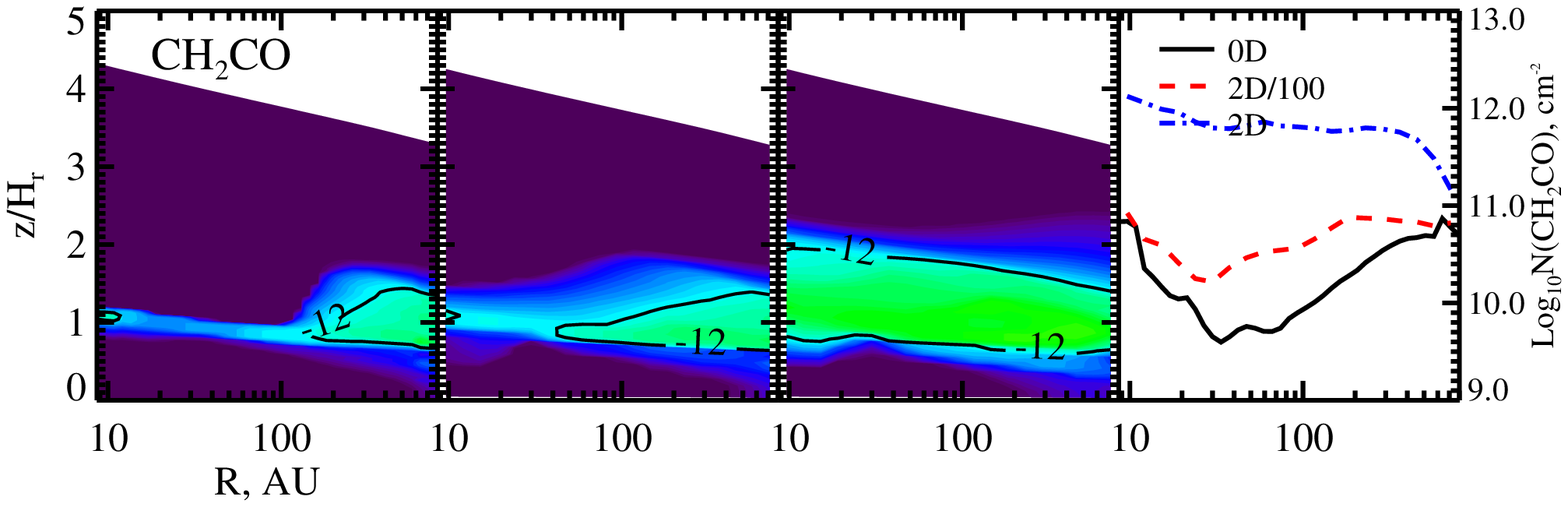}
\caption{The same as in Fig.~\ref{fig:ions} but for the complex (organic)
molecules. Results are shown for HCOOH, HCOOH ice, CH$_3$OH,
HNCO, HNCO ice, CH$_3$CHO, CH$_3$CHO ice, and CH$_2$CO.}
\label{fig:complex}
\end{figure*}

\LongTables
\begin{deluxetable*}{llrrll}
\tabletypesize{\footnotesize}
\tablecaption{Key chemical processes: organic species\label{tab:key_reac_organics}}
\tablehead{
\colhead{Reaction} & \colhead{$\alpha$} & \colhead{$\beta$} & \colhead{$\gamma$}& \colhead{t$_{\rm min}$} & \colhead{t$_{\rm
max}$}\\
\colhead{} & \colhead{[(cm$^3$)\,s$^{-1}$]} & \colhead{} & \colhead{[K]} & \colhead{[yr]} & \colhead{[yr]}
}
\startdata
CH$_2$CO ice   + h$\nu_{\rm CRP}$  $\rightarrow$  CH$_2$ ice   + CO ice   & $9.15\,(2)$   & $0$   & $0$   & $1.00$   & $5.00\,(6)$
  \\
CH$_2$CO ice   + h$\nu_{\rm CRP}$  $\rightarrow$  C$_2$ ice   + H$_2$O ice   & $4.07\,(2)$   & $0$   & $0$   & $1.96$   &
$5.00\,(6)$   \\
CH$_3$CHO ice   + h$\nu_{\rm CRP}$  $\rightarrow$  CH$_3$ ice   + HCO ice   & $5.25\,(2)$   & $0$   & $0$   & $1.96$   &
$5.00\,(6)$   \\
CH$_3$CHO ice   + h$\nu_{\rm CRP}$  $\rightarrow$  CH$_4$ ice   + CO ice   & $5.25\,(2)$   & $0$   & $0$   & $1.96$   &
$5.00\,(6)$   \\
CH$_3$OH ice   + h$\nu_{\rm CRP}$  $\rightarrow$  CH$_3$ ice   + OH ice   & $1.50\,(3)$   & $0$   & $0$   & $1.00$   & $5.00\,(6)$
  \\
CH$_3$OH ice   + h$\nu_{\rm CRP}$  $\rightarrow$  H$_2$CO ice   + H$_2$ ice   & $3.17\,(3)$   & $0$   & $0$   & $1.00$   &
$5.00\,(6)$   \\
C$_2$H$_5$OH ice   + h$\nu_{\rm CRP}$  $\rightarrow$  CH$_3$CHO ice   + H$_2$ ice   & $6.85\,(2)$   & $0$   & $0$   & $2.34\,(4)$
 & $5.00\,(6)$   \\
HCOOH ice   + h$\nu_{\rm CRP}$  $\rightarrow$  CO$_2$ ice   + H ice   + H ice   & $6.50\,(2)$   & $0$   & $0$   & $1.00$   &
$5.00\,(6)$   \\
HCOOH ice   + h$\nu_{\rm CRP}$  $\rightarrow$  HCO ice   + OH ice   & $2.49\,(2)$   & $0$   & $0$   & $3.82$   & $5.00\,(6)$   \\
HNCO ice   + h$\nu_{\rm CRP}$  $\rightarrow$  NH ice   + CO ice   & $6.00\,(3)$   & $0$   & $0$   & $1.00$   & $5.00\,(6)$   \\
CH$_3$OH ice   + UV  $\rightarrow$  H$_2$CO ice   + H$_2$ ice   & $0.72\,(-9)$   & $0$   & $1.72$   & $1.00$   & $5.00\,(6)$   \\
C$_2$H$_5$OH ice   + UV  $\rightarrow$  CH$_3$CHO ice   + H$_2$ ice   & $0.13\,(-9)$   & $0$   & $2.35$   & $4.18\,(2)$   &
$5.00\,(6)$   \\
HCOOH ice   + UV  $\rightarrow$  HCO ice   + OH ice   & $0.28\,(-9)$   & $0$   & $1.80$   & $1.00$   & $5.00\,(6)$   \\
HNCO   + h$\nu_{\rm CRP}$  $\rightarrow$  NH   + CO   & $6.00\,(3)$   & $0$   & $0$   & $1.00$   & $5.00\,(6)$   \\
CH$_3$CHO   + UV  $\rightarrow$  CH$_3$CHO$^+$   + e$^-$   & $0.26\,(-9)$   & $0$   & $2.28$   & $1.00$   & $5.00\,(6)$   \\
CH$_3$CHO   + UV  $\rightarrow$  CH$_4$   + CO   & $0.34\,(-9)$   & $0$   & $1.52$   & $1.00$   & $5.00\,(6)$   \\
CH$_3$CHO   + UV  $\rightarrow$  CH$_3$   + HCO   & $0.34\,(-9)$   & $0$   & $1.52$   & $1.00$   & $5.00\,(6)$   \\
HCOOH   + UV  $\rightarrow$  HCOOH$^+$   + e$^-$   & $0.17\,(-9)$   & $0$   & $2.59$   & $1.00$   & $5.00\,(6)$   \\
HCOOH   + UV  $\rightarrow$  HCO   + OH   & $0.28\,(-9)$   & $0$   & $1.80$   & $1.00$   & $5.00\,(6)$   \\
CH$_2$CO   + grain  $\rightarrow$  CH$_2$CO ice   & $1.00$   & $0$   & $0$   & $1.00$   & $5.00\,(6)$   \\
HCOOH   + grain  $\rightarrow$  HCOOH ice   & $1.00$   & $0$   & $0$   & $1.00$   & $5.00\,(6)$   \\
HNCO   + grain  $\rightarrow$  HNCO ice   & $1.00$   & $0$   & $0$   & $1.00$   & $5.00\,(6)$   \\
CH$_2$CO ice  $\rightarrow$  CH$_2$CO   & $1.00$   & $0$   & $2.20\,(3)$   & $1.00$   & $5.00\,(6)$   \\
CH$_3$CHO ice  $\rightarrow$  CH$_3$CHO   & $1.00$   & $0$   & $2.87\,(3)$   & $1.00$   & $5.00\,(6)$   \\
HNCO ice  $\rightarrow$  HNCO   & $1.00$   & $0$   & $2.85\,(3)$   & $1.00$   & $5.00\,(6)$   \\
H ice   + CH$_2$OH ice  $\rightarrow$  CH$_3$OH ice   & $1.00$   & $0$   & $0$   & $1.00$   & $5.00\,(6)$   \\
H ice   + HC$_2$O ice  $\rightarrow$  CH$_2$CO ice   & $1.00$   & $0$   & $0$   & $1.00$   & $5.00\,(6)$   \\
H ice   + OCN ice  $\rightarrow$  HNCO ice   & $1.00$   & $0$   & $0$   & $1.00$   & $5.00\,(6)$   \\
H ice   + OCN ice  $\rightarrow$  HNCO   & $1.00$   & $0$   & $0$   & $1.00$   & $5.00\,(6)$   \\
OH ice   + CH$_3$ ice  $\rightarrow$  CH$_3$OH ice   & $1.00$   & $0$   & $0$   & $1.00$   & $5.00\,(6)$   \\
OH ice   + HCO ice  $\rightarrow$  HCOOH ice   & $1.00$   & $0$   & $0$   & $1.09\,(2)$   & $5.00\,(6)$   \\
HCOOH   + H$^+$  $\rightarrow$  HCOOH$^+$   + H   & $0.28\,(-8)$   & $-0.50$   & $0$   & $1.00$   & $5.00\,(6)$   \\
HNCO   + H$^+$  $\rightarrow$  NH$_2$$^+$   + CO   & $0.15\,(-7)$   & $-0.50$   & $0$   & $1.00$   & $5.00\,(6)$   \\
CH$_3$CHO   + H$_3$$^+$  $\rightarrow$  CH$_3$CH$_2$O$^+$   + H$_2$   & $0.62\,(-8)$   & $-0.50$   & $0$   & $1.00$   &
$5.00\,(6)$   \\
CH$_3$CHO   + HCO$^+$  $\rightarrow$  CH$_3$CH$_2$O$^+$   + CO   & $0.25\,(-8)$   & $-0.50$   & $0$   & $1.00$   & $5.00\,(6)$
\\
HCOOH   + H$_3$$^+$  $\rightarrow$  HCO$^+$   + H$_2$O   + H$_2$   & $0.23\,(-8)$   & $-0.50$   & $0$   & $1.00$   & $5.00\,(6)$
\\
HCOOH   + HCO$^+$  $\rightarrow$  CH$_3$O$_2$$^+$   + CO   & $0.13\,(-8)$   & $-0.50$   & $0$   & $1.00$   & $5.00\,(6)$   \\
O   + C$_2$H$_5$  $\rightarrow$  CH$_3$CHO   + H   & $0.13\,(-9)$   & $0$   & $0$   & $1.00$   & $5.00\,(6)$   \\
OH   + H$_2$CO  $\rightarrow$  HCOOH   + H   & $0.20\,(-12)$   & $0$   & $0$   & $1.00$   & $5.00\,(6)$   \\
CH$_3$CH$_2$O$^+$   + e$^-$  $\rightarrow$  CH$_3$CHO   + H   & $0.15\,(-6)$   & $-0.50$   & $0$   & $1.00$   & $5.00\,(6)$   \\
CH$_3$O$_2$$^+$   + e$^-$  $\rightarrow$  HCOOH   + H   & $0.15\,(-6)$   & $-0.50$   & $0$   & $1.00$   & $5.00\,(6)$   \\
CH$_3$CH$_2$O$^+$   + grain(-)  $\rightarrow$  CH$_3$CHO   + H   + grain(0)   & $0.20$   & $0$   & $0$   & $55.90$   & $5.00\,(6)$
  \\
CH$_3$O$_2$$^+$   + grain(-)  $\rightarrow$  HCOOH   + H   + grain(0)   & $0.50$   & $0$   & $0$   & $1.00$   & $5.00\,(6)$   \\
CH$_3$CHO   + grain  $\rightarrow$  CH$_3$CHO ice   & $1.00$   & $0$   & $0$   & $1.00$   & $5.00\,(6)$   \\
CH$_3$OH   + grain  $\rightarrow$  CH$_3$OH ice   & $1.00$   & $0$   & $0$   & $1.00$   & $5.00\,(6)$   \\
\enddata
\end{deluxetable*}

The presence of amino acids and other  complex organics in the early Solar system is a well-established
through detailed mass-spectrometry of carbonaceous meteorites \citep[e.g.,][]{Glavin_ea10}, via analysis of 
the cometary dust sampled by the {\it Giotto} spacecraft in the Halley comet
\citep[e.g.,][]{JCK88}, and the recent identification of glycine in the {\it Stardust} cometary dust samples
\citep[][]{Elsila_ea09}. The ground-based search for simple gas-phase organic species, such as methanol and formic acid,  in
nearby protoplanetary disks 
with radiotelescopes have so far been fruitless, though
formaldehyde has been detected in DM Tau \citep[e.g.,][]{DGH07} and \object{LkCa 15} \citep[][]{Aikawa_ea03}. 
The simple organic ices, e.g. HCOOH ice, has been identified in the {\it Spitzer} spectra of several low-mass 
Class~I/II objects \citep[e.g.,][]{Zasowski_ea08}. In contrast, rich organic gas-phase molecules have been detected
in hot massive cores and corinos, where conditions are appropriate for steady sublimation and production of heavy 
complex ices \citep[e.g.,][]{Belloche_ea08,Garrod_ea08b}. Nowadays synthesis of complex organics in cosmic objects
is thought to proceed solely via surface processes, such as photoproduction of reactive radicals by high-energy irradiation
of complex precursors and their surface recombination at elevated temperatures, \citep[e.g.,][]{Herbst_vanDishoeck09}.
In this Section we analyze in detail chemical processes responsible for the formation and destruction of the complex 
organic species in protoplanetary disks. These are meant to be polyatomic neutral molecules consisting of
at least several H, C, and O atoms, and heavier elements.

There is a single organic species insensitive to the turbulent diffusion, namely, formaldehyde (see Sect.~\ref{O-species}).
The transport-sensitive organic species include 2 gas-phase molecules (CH$_2$CO and HNCO) and 5 ices 
(CH$_2$CO, HNCO, CH$_3$OH, H$_2$CO, NH$_2$CHO; Table~\ref{tab:sens}). Their column densities are changed by up to 2 orders 
of magnitude. One gas-phase molecule 
(HCOOH) and 3 ices (HCOOH, H$_2$C$_3$O, and CH$_3$CHO) are among hypersensitive species in Table~\ref{tab:hyps}. The turbulent
diffusion alters their column densities by up to 
a factor of 3\,000 (solid formic acid). The chemical evolution of all the considered organic species is influenced
by the transport as their major production and removal routes require surface processes. 

In Fig.~\ref{fig:complex} the distributions of the relative molecular abundances and column densities at 5~Myr of 
HCOOH, HCOOH ice, CH$_3$OH, HNCO, HNCO ice, CH$_3$CHO, CH$_3$CHO ice, and CH$_2$CO calculated with the laminar and 
the 2D-mixing models are presented. In the laminar model the relative abundance distributions of the gas-phase species show a
3-layered structure, with very narrow molecular layers located at $\approx 0.8-1\,H_r$. Note that their abundances and
column densities are substantially lower than those of simpler C- or O-bearing species considered above, so that
some heavy organic species are not included in Tables~\ref{tab:steadfast}--\ref{tab:hyps}.
The HCOOH, HNCO, and CH$_2$CO
molecular layers are wider beyond $\sim 200$~AU, and have higher abundances. Remarkably, abundance distributions of
complex ices are maximum either at the bottom of the molecular layer (HNCO ice, HCOOH ice) or in the inner warm midplane
(CH$_3$CHO ice, HCOOH ice). The overall tendency can be easily explained as heavy ices are hard to evaporate thermally,
so photoevaporation is necessary, while gas-phase organic molecules are rather photofragile. For efficient production 
of heavy ices surface mobility of radicals (O, C, CH, etc.) necessitates warm temperatures ($T\ga30$~K) and/or
photoprocessing of their precursor molecules by the X-ray/CRP-induced UV photons.

To better understand these results, we investigate the evolution of HCOOH, HCOOH ice, CH$_3$OH, HNCO, HNCO ice, CH$_3$CHO, 
CH$_3$CHO ice, and CH$_2$CO in two disk vertical slices at $r=10$ and 250~AU  (the laminar chemical model). 
The most important reactions responsible for the time-dependent evolution of their abundances in the midplane, the molecular 
layer, and the atmosphere are presented in Table~\ref{tab:key_reac_organics}, both for the inner and outer disk regions. 
The final list contains only top 25 reactions per region (midplane, molecular layer, atmosphere) for 
the entire 5~Myr time span, with all repetitions removed.

The chemical evolution of formic acid (HCOOH) starts with a single production channel via dissociative recombination of 
CH$_3$O$_2^+$ either on electrons or negatively charged grains (Table~\ref{tab:key_reac_organics}). In turn, protonated 
formic acid is produced by radiative association of HCO$^+$ and H$_2$O, 
and by ion-molecule reaction of methane with ionized molecular oxygen.
The primal removal channels for HCOOH are photodissociation and photoionization, depletion onto dust grains at 
$T\la 100-120$~K, charge transfer with ionized atomic hydrogen, and 
protonation by abundant polyatomic ions (HCO$^+$, H$_3^+$, H$_3$O$^+$). The HCOOH ice is essentially synthesized by
accretion of gas-phase formic acid, and destroyed by secondary UV photons in the disk midplane and molecular layer,
and photoevaporation. The surface as well as gas-phase formation of HCOOH via neutral-neutral reaction of OH and H$_2$CO
is only a minor channel in the laminar model. Consequently, the distribution of evolutionary timescales for gas-phase 
and surface HCOOH is similar and dominated by $\tau_{\rm chem}\ga 10^5$~years.

Gas-phase methanol (CH$_3$OH) is produced by direct surface recombination of frozen H and CH$_2$OH
as well as frozen OH and CH$_3$, evaporation of methanol ice, dissociative recombination of protonated methanol, and 
dissociative recombination of H$_5$C$_2$O$_2^+$ (minor route). Its key removal pathways are surface accretion
in the disk regions with $T\la 100-120$~K, photodissociation and photoionization, and 
protonation by the dominant polyatomic ions. On dust surfaces methanol is produced via a sequence of hydrogenation
reactions, starting from the CO ice, and is photodissociatied by the UV radiation. Similarly to HCOOH, the characteristic
chemical timescale for methanol is fully controlled by the slow surface chemistry ($\tau_{\rm chem}\ga 10^5$~years).

The chemical evolution of the isocyanic acid (HNCO) is also dominated by surface processes. In the gas-phase it is produced either
by evaporation of HNCO ice at $T\la 50-60$~K or by direct surface recombination of
surface H and OCN. The major gas-phase destruction pathways are accretion to grains in the midplane and the molecular
layer, and ion-molecule reaction with H$^+$ (leading to NH$_2^+$ and CO; Table~\ref{tab:key_reac_organics}). The
HNCO ice is produced by the surface reaction involving H and OCN, re-accretion of the HNCO gas, and destroyed by
the UV. Their characteristic timescales alike those of HCOOH and methanol.

The chemistry of acetaldehyde (CH$_3$CHO) involves gas-phase production by reactive collisions between O and C$_2$H$_5$,
accretion to dust grains at $T\la 50-60$~K, and desorption at $T\la 60$~K, and removal 
via charge transfer reactions with C$^+$ and ion-molecule reactions with C$^+$, H$_3^+$, and H$_3$O$^+$. At later times
acetaldehyde is re-produced from its protonated analog by dissociative recombination (albeit with low probability).
The frozen acetaldehyde is synthesized by accretion of gas-phase CH$_3$CHO, via surface recombination of CH$_3$ and HCO
ices, and destroyed by the UV-dissociation. As for other complex organics, the chemical timescale for acetaldehyde is 
typically longer than 1~Myr in the disk.

Finally, the gas-phase ethenone (CH$_2$CO) is produced via oxidation of C$_2$H$_3$,
direct surface recombination of the H and HC$_2$O ices, and dissociative recombination of protonated ethenone. 
Thermal evaporation of ethenone ice is effective when dust temperatures exceed $\sim100$~K. 
Key removal channels include photodissociation and photoionization, freeze-out in the midplane and the molecular layer,
charge transfer reaction with C$^+$ and H$^+$, and, at later times, protonation by polyatomic ions. The 
chemical timescale for CH$_2$CO exceeds $10^5-10^6$~years.

Not surprisingly that turbulent mixing enhances abundances and column density of these organic species, 
given their long evolutionary timescales governed by surface reactions and photodissociation of ices by the CRP- and
X-ray-induced secondary UV radiation field. The HCOOH and the HCOOH ice are more abundant since water ice abundances 
and HCO$^+$ abundances are increased by the mixing, leading to more efficient production of the parental ion, CH$_3$O$^+$.
Both for gas-phase and solid formic acid radial transport is important. The effect is less pronounced for HNCO as
its precursor species, OCN, is enhanced by transport only by less than 1 order of magnitude. The HNCO ice, produced in the
lower part of the molecular layer, is transported by diffusion to the cold midplane where it cannot be synthesized otherwise.
Relative abundances of gas-phase acetaldehyde are greatly enriched by the disk mixing in the middle of the molecular layer
($z\approx 1.5\,H_r$), within the inner 100~AU. As its synthesis proceeds via surface recombination of CH$_3$ and HCO,
in the laminar model appropriate conditions are only met in the very inner midplane/molecular layer. In the fast mixing
model larger quantities of solid acetaldehyde can be accumulated as more icy grains reach the warm disk regions. The vertical
mixing brings the CH$_3$CHO ice in the inner midplane to the molecular layer, where it photoevaporates, and then radially 
transported to larger distances. The ethenone abundances are vertically broaden by the mixing, and moderately
enriched as the ethenone ice production is enhanced by the turbulent mixing.

\section{Discussion}
\label{diss}

\subsection{Comparison with previous studies and future developments}
\label{sec:diss:studies}
In this Section we discuss the results and the drawbacks of our model in the context of other studies
of the chemo-dynamical evolution of protoplanetary systems. 

{\em (Models of the inner Solar nebula: the role of advective transport)}
Historically, the interest to this topic has been 
initiated by the cosmochemical community studying the initial stages of the formation of the Solar system.
The presence of high-temperature condensates, like CAIs and crystalline silicates, in pristine chondritic meteorites 
(formed within a few AU from the Sun) and comets (formed within 10--20~AU) as well as almost perfect isotopic homogeneity
of the inner Solar nebula ($r\la10-20$~AU) at a bulk level both require an efficient mixing mechanism with a transport speed 
of $\la 1$~AU per $5\,10^4$~yr or 10~cm\,s$^{-1}$ \citep[e.g.,][]{Wooden_ea07,Brownlee_ea08,Ciesla_09}. 
According to the current knowledge, the Solar nebula and other protoplanetary
disks represent an outcome of the viscous evolutionary stage that follows the initial collapse of a molecular cloud
core \citep[e.g.,][]{CassenMoosman81,cameron1995,Larson2003}. 

The ability of turbulent or advective transport 	
to cause radial mixing of materials in the inner early Solar system has been proposed and investigated by Morfill 
\citep[e.g.,][]{Morfill_83,Morfill_Voelk84}, with a simple 1D analytical disk model and passive tracers. In
a similar manner, 2D radial mixing of gaseous and solid water in the inner nebula has been studied by \citet{Cyr_ea98}.
In a long series of papers the group of Prof.~H.-P. Gail (Heidelberg University) has investigated various aspects of the
chemo-dynamical 
evolution of the inner protosolar nebula ($1-10$~AU), progressing from stationary accretion disk models and 
crude gas-phase chemistry toward a self-consistent 2D radiative-hydrodynamical model with the coupled 
C-, H-, O- gas-phase chemistry \citep[e.g.,][]{bauer,FG97,Gail98,G01,G02,Wehrstedt_Gail02,Gail_04,Keller_Gail04,TG07}.
In \citet{bauer} the gas-phase C-, H-, N-, O-chemistry driven by dust destruction and evaporation of ices in the presence of 
slow radial transport has been modeled, utilizing the semi-analytical one-zone disk model of \citet{Duschl_ea96}. The authors
have found that silicates tend to evaporate under equilibrium conditions, whereas carbon dust is slowly destroyed by
combustion by OH, and that at $r\la 1$~AU the CO concentration is higher than that of water. In \citet{FG97} and 
\citet{G01} the one-zone stationary disk model has been considered, whereas in \citet{G02} 
and \citet{Wehrstedt_Gail02} the one-zone time-dependent $\alpha$-disk model with a larger chemical network of 106 species, 
annealing of amorphous silicates, and 
more detailed description of carbon dust oxidation by O and OH has been used. The major finding is that the radial
transport enriches the outer, $r>10$~AU nebular regions with methane and acetylene produced by oxidation of carbon dust
at $r\la1$~AU, as observed in comets of Hyakutake and Hale-Bopp. The advection also transports crystalline silicates
to the $10-20$~AU region. In the later papers \citep[][]{Keller_Gail04,TG07}, a more accurate model has
been developed, with a self-consistent 2D prescription of advective transport and turbulent diffusion, grey radiative transfer, 
proper dust and gas opacities, and disk hydrodynamical evolution with parametric viscosity. The model was coupled to a limited 
gas-phase C-, H-, O-chemistry involving nearly $90$ neutral-neutral reactions. They have found that quasi-stationary accretion
flows develop a 2D structure such that in the disk midplane gas moves outward, carrying out the angular momentum, whereas
inward mass accretion goes through the surface layers. The radial advection dominates the diffusive mixing in outer disk
regions at $r>5$~AU. Thus, chemical species produced by the ``warm'' chemistry in the inner nebula can reach its outer
region, where they are intermixed by turbulence with surrounding matter and freeze-out. They have considered neither X-ray-
nor UV-driven processes, no ion-molecule and surface chemistry, and assumed sublimation-evaporation equilibrium for ices. 
Their key result is that the slow water-shift reaction transforming  CO and water into CO$_2$ and H$_2$ at high pressure and 
temperature is important for the oxygen chemistry. The H$_2$ and CO evolution is insensitive to transport, while abundance
distributions of e.g. O, O$_2$, CO$_2$ are altered by the disk dynamics within 300~years of the evolution.

Contrary to these studies, our modeling is related to outer nebular regions beyond 10~AU and does not include 
advective transport.
However, in the presence of steep chemical gradient for a molecule, radial diffusive mixing is able to transport it
from 100~AU to 1~AU within about $1-2\,10^6$~years, and from 1~AU to 10~AU within about $\la 3\,10^5$~years, which is still 
less than a typical disk lifetime of several Myrs (see Eq.~(\ref{eq:tau_phys}) and Fig.~\ref{fig:disk_struc}).
Our chemical network incorporates much more species and reactions, including surface 
chemistry, and the high-energy processing of the gas and icy 
mantles. Nonetheless, the calculated H$_2$ and CO column densities are not sensitive to the diffusive mixing in our model, 
similar to what was found by \citet{TG07} (Fig.~\ref{fig:O} and Table~\ref{tab:steadfast}). Also, in agreement with Gail et al.
results, 
we show that the synthesis of hydrocarbons, 
including CH$_4$ and C$_2$H$_2$, is intensified by the mixing, particularly in the zone of comet formation around 10--20~AU
(Fig.~\ref{fig:C}). Furthermore, CO$_2$ ice, 
which is produced via surface reactions in the warm inner midplane and can be considered as an example of a ``warm'' species, 
is transported outward by the 2D-turbulent mixing. In addition, we also found that oxidation chemistry is important for the 
evolution of O-bearing species, though we do not account for the carbon dust destruction, unlike \citet{TG07}.

{\em (Models of protoplanetary disks: the role of advective transport)}
Next, we focus on disk models that has been used to simulate chemical evolution of protoplanetary disks in the astrophysical
context. \citet{Aea99} have utilized an isothermal $\alpha$-disk model of \citet{Lynden-BellPringle74} with 
$\dot{M}= 10^{-8}\,M_\odot$\,yr$^{-1}$, and calculated its density structure assuming hydrostatic equilibrium in vertical 
direction. 
They have used the early  version of the UMIST\,95 chemical ratefile \citep{umist95}, considered sticking and 
thermal desorption of molecules and surface formation of molecular hydrogen, and grain re-charging (about 250 species
and 2\,300 reactions). No X-ray-driven and surface chemistry, other than H$_2$ formation,  has been included. 

These authors have modeled the chemical evolution in the presence of steady inward accretion, using both atomic and molecular
initial abundances, 
and found similar results. We also find that the difference in the calculated abundances between the adopted ``low metals''
atomic 
initial abundances and those from a molecular cloud is negligible in our model. 
\citet{Aea99} have found that it takes about 3~Myr to transport disk material from 400~AU to 10~AU (similar to the timescale 
of radial mixing in our model). The inward advection leads to higher concentrations of heavy hydrocarbons at $\la20$~AU, 
whereas methane is a dominant hydrocarbon in the outer disk region. In our laminar model methane prevails over heavier 
hydrocarbons in the entire disk, while in the presence of radial and vertical mixing a substantial fraction of CH$_4$ is 
converted into heavy carbon chains (especially at $10-50$~AU; see Fig.~\ref{fig:C}).
\citet{Aea99} have concluded that the radial transport  leads to simultaneous existence of the reduced (e.g. CH$_4$) and 
oxidized (e.g. CO$_2$) ices, as observed in comets, and as found by our chemo-dynamical modeling.
Finally, they have studied in detail chemistries of O-, N-, and C-bearing species and found that H$_3^+$ and He$^+$ produce 
e.g. CO$_2$, CH$_4$, and NH$_3$ from CO and N$_2$, and that grain properties and ionization rate are crucial factors for 
molecular evolution. This is consistent with our findings, but with an additional notion that X-rays further strengthen He$^+$ and
H$_3^+$ influence on the
chemical evolution of turbulent protoplanetary disks.

Later, \citet{Woods_Willacy07} have investigated the genesis of benzene in protoplanetary disks at $r\la 35$~AU.
They have used a D'Alessio-like flaring $\alpha$-disk model with accurate UV radiative transfer and detailed calculations of the
heating and cooling balance of gas. The X-ray ionization and dissociation have not been taken into consideration.
The  chemical network has been adopted from a sub-set of the UMIST\,99 \citep{umist99}
ratefile augmented with gas-grain and surface reactions (in total 200 species and 2\,400 reactions). The passage of a gas
parcel from $>35$~AU onto the central star has been calculated. \citet{Woods_Willacy07} have found that the radial transport 
results in efficient production of benzene at $\la 3$~AU, mostly due to ion-molecule reactions between C$_3$H$_3$ and 
C$_3$H$_4^+$, followed by grain dissociative recombination. These results are very sensitive to the adopted value of 
C$_6$H$_6$ binding energy. Similarly, we find that benzene is produced via the same ion-molecule reaction because
the surface network is limited for heavy hydrocarbons and does not include surface routes to C$_6$H$_6$, whereas 
the gas-phase production is efficient. Nevertheless, as in the paper of Woods \& Willacy, the C$_6$H$_6$ abundances and column 
densities in the inner disk at 10--30~AU are increased by vertical turbulent diffusion in our model by up to 2 orders of
magnitude,
see Fig.~\ref{fig:C}.

\citet{Nomura_ea09}, using a disk chemical model with radial advection, studied the evolution of the inner
region ($\la 12$~AU). The $\alpha$-model of a disk around a $6\,000$~K star with the mass accretion rate of 
$10^{-8}\,M_\odot$\,yr$^{-1}$ has been considered, with gas and dust temperatures computed 
independently. They have modeled penetration of the UV and X-ray radiation into the disk, and utilized the RATE\,06 
\citep{Woodall_ea07} with gas-grain interactions. The chemical evolution of gas parcels following various inward 
trajectories has been calculated. The authors have found that the abundances are sensitive to the transport speed, and that the 
fast transport allows gaseous molecules to reach disk regions where they would otherwise be depleted. They have concluded
that dynamical disk model facilities synthesis of methanol, ammonia, hydrogen sulfide, acetylene, etc., and predicted the 
observability of the $J=3_0-2_0$ methanol line at 145~GHz with ALMA. As we have shown above,
our 2D-mixing model follows the same tendency, and enriches molecular content of the protoplanetary disk with complex species.

\citet{Visser_ea09} has developed a 2D semi-analytical evolutionary model that follows collapse of a dense core into a disk 
(a transition from the Class~0 to Class~I phase) and tracks by trajectories infall motions of gas. They have used time-dependent 
RT to compute dust temperature consistently, and considered only freeze-out and evaporation of CO and water ices. 
Their major results are that the CO ice evaporates during collapse and later re-accretes to dust, while the water ice remains
almost intact (unless heated by a protostar at $r<10$~AU). Material inside 5--30~AU in the early disk
spends $\la 10^4$~years at temperatures of $\sim 20-40$~K, which is sufficient to produce complex organics. With our disk 
physical model that is at a steady-state and incorporates a detailed gas-surface chemistry we indeed see a rise in abundances 
of many complex molecules, as their surface production at $T\la 30$~K and evaporation are enhanced by the diffusive mixing.

{\em (Models of protoplanetary disks: the role of turbulent transport)}
Now we compare our results to more closely related studies of disk chemical evolution with turbulent mixing. In the first 
paper on this topic, \citet{IHMM04} have investigated the influence of the 1D vertical mixing and the radial advection on
chemistry of a steady 1+1D-$\alpha$-disk model ($\alpha=0.01$, $\dot{M}= 10^{-8}\,M_\odot$\,yr$^{-1}$). The modeling of a 
reactive flow system has been discussed in detail. They have adopted  
\citet{Xie_ea95} description for the  diffusion and Lagrangian description for advective transport (trajectories from 10 to 
1~AU). \citet{IHMM04} have used a subset of the UMIST\,95 database with gas-grain interactions and no surface chemistry 
($\approx 240$ species, 2\,500 reactions). Ilgner et al. have considered 3 steady-state solutions for the 
disk physical structure, and investigated the disk evolution by adding advective and turbulent transport processes. 
Their key findings are that
the chemistry is sensitive to the disk thermal profile, that the vertical mixing removes vertical abundance gradients, and 
that local changes in species concentrations due to mixing can be radially transported by advection. \citet{IHMM04} have 
concluded that diffusion has a limited effect on the disk regions dominated by gas-grain kinetics, though it enhances 
abundances of atomic oxygen and thus alters the evolution of related species (SO, SO$_2$, CS, etc.). The advective 
transport without mixing results in destruction of oxygen and OH at $r<5$~AU. Our model, despite being limited to outer disk 
regions (beyond 10~AU), confirms their results that oxygen abundances are increased by diffusion (Fig~\ref{fig:O}), 
further propagating into the chemistry of many related species, particularly sulfur-bearing species like SO and SO$_2$. We find 
that turbulent mixing may steepen or 
soften chemical gradients (e.g., for atomic ions and most of molecules, respectively), and  it definitely affects more
strongly the evolution of species produced via gas-surface kinetics. This fundamental feature of the presented chemical
model is due adopted extended set of surface photo- and recombination processes.
This further emphasizes the need in the extended surface reaction network. 
Some conclusions, especially those related to complex organics do depend on
the complexity of the adopted surface chemistry model.

Next, \citet{Willacy_ea06} have utilized a steady-state $\alpha$-disk model similar to that of Ilgner et al. and used the Xie 
approach to account for the disk viscosity. A plane-parallel model of the UV irradiation, no stellar X-ray radiation, and a 
subset of the UMIST\,95 database with gas-grain interactions and surface kinetics from \citet{HHL92} have been adopted. Our 
chemical network is about 3 times larger, and uses recently updated reaction rates. They have 
studied the impact of the 1D-vertical mixing on the chemical evolution of the outer disk ($r>100$~AU). Overall, they have 
found that the vertical transport can increase column densities by up to 2 orders of magnitude. Another result is 
that the 3-layered disk structure is preserved in the mixing model, albeit depths of many molecular layers are increased.
They have observed that the higher the diffusion coefficient, the higher the impact of the turbulent mixing.
Our results confirm these general findings, with the chemical network, that is about 3 times larger and based on recently updated
reaction rates. In contrast to the results of Willacy et al., we obtain that
the evolution of the ionization degree, ammonia and N$_2$H$^+$ are sensitive to the mixing, whereas column densities
of CO, H$_2$CO, CN, C$_2$H are steadfast (Tables~\ref{tab:steadfast}--\ref{tab:hyps}). 
This is because in our model the X-ray irradiation of the disk,
production of reactive radicals by photodissociation of ices, and UV-photodesorption are taken into account. In the
absence of photoevaporation, e.g., the CO production in the model of Willacy et al. becomes responsive to transport of ices
from midplane and atomic C and O from atmosphere. In our model this is true only for heavy ices with large binding
energies ($\la 3\,000-5\,000$~K), like hydrocarbons. Also, our disk study focuses on the large distances from 800 till 10~AU
that has not been considered by Willacy et al. If we restrict our model to the distances beyond 100~AU, the ionization degree
and column densities of many other species become steadfast. Furthermore, our study supports 
the result that the column densities of complex molecules like methanol are greatly enhanced by the disk dynamics. 

In 2006 Ilgner \& Nelson have investigated in detail ionization chemistry in disks and its sensitivity to various
physical and chemical effects at $r<10$~AU, like the X-ray flares from the young T~Tauri star \citep{Ilgner_Nelson06b}, 
vertical mixing \citep{Ilgner_Nelson06a}, amount of gas-phase metals, and validity of various chemical networks 
\citep{Ilgner_Nelson06}. Using an $\alpha$-disk model with stellar X-ray-irradiation, they have demonstrated that
the simple \citet{OD74} network made of 5 species tend to overestimate the ionization degree since 
metals exchange charges with polyatomic species, and that magnetically decoupled ``dead'' region may exist in disks unless 
small grains and metals are removed from gas. Next, the influence of vertical diffusion on the ionization fraction has been 
studied. They have found that the mixing has no effect on $X({\rm e}^-)$ if metals are absent in the gas since recombination 
timescales
are fast, whereas at $X({\rm Me})\la10^{-10}-10^{-8}$ $\tau_{\rm chem}>\tau_{\rm mix}$ and diffusion drastically reduces the 
size of the ``dead'' zone.
Finally, it has been shown that in the disk model with sporadic X-ray flares (by up to a factor of 100 in the luminosity) 
the outer ``dead'' zone disappear, whereas the inner ``dead'' zone evolves along with variations of the X-ray flux. 
Indeed, our extended disk chemical model  
shows that ionization degree and abundances of charged atoms and molecules are sensitive to the transport in the disk regions
with $r\la 100-200$~AU, and that polyatomic ions are important charge carriers in the disks \citep[see also][]{Red2}.
Also, the role of the X-ray photons is crucial for the disk ionization fraction and its molecular composition
\citep[see also][]{Schreyer_ea08,Henning_ea10}.

The first attempts to model self-consistently disk chemical, physical, and turbulent structures in full 3D have been
performed by \citet{Turner_ea06} and \citet{Ilgner_ea08}. Both studies have employed a shearing-box approximation to 
calculate a patch of a 3D MHD disk at radii of $\sim 1-5$~AU, treated the development of the MRI-driven turbulence, and 
focused on the multi-fluid evolution of the disk ionization state. In the study of Turner et al. ohmic resistivity, vertical 
stratification, the CRP-ionization, and the Oppenheimer-Dalgarno time-dependent ionization network have been considered. 
They have found that turbulent mixing transports free charges into the dark disk midplane faster than these recombine, 
coupling the midplane matter to magnetic fields. As a result, accretion stresses in the disk ``dead'' 
zone are only several times lower than in surface layers, with a typical timescale of about 1--5~Myr.
\citet{Ilgner_ea08} have adopted a similar approach but considered the X-ray ionization with $L_X=10^{31}$~erg\,s$^{-1}$.
They have re-confirmed their earlier findings \citep{Ilgner_Nelson06a,Ilgner_Nelson06} that turbulent mixing has no effect on
the disk ionization structure in the absence of the gas-phase metals. The presence of metals, however, prolongs the
recombination timescale, and the mixing is thus able to enliven the ``dead'' zone at $r\ge5$~AU (with the resulting 
$\alpha=1-5\,10^{-3}$). Both these studies are based on a pure gas-phase chemical network, whereas in our extended 
gas-grain model metals are fully depleted in the disk regions with $T\la150-200$~K. Consequently, our modeling corresponds 
to the ``no metals'' case of \citet{Ilgner_ea08}, so turbulent transport does not much affect the ionization fraction
in the disk midplane (see Fig.~\ref{fig:ions}, 1st panel).
 
Finally, in the recent paper by \citet{Heinzeller_ea11} the chemical evolution of the protoplanetary disk along with
radial viscous accretion, vertical mixing, and vertical wind transport has been investigated. The steady-state disk 
model with $\alpha=0.01$, $\dot{M}= 10^{-8}\,M_\odot$\,yr$^{-1}$, and $L_X=10^{30}$~erg\,s$^{-1}$ has been adopted.
The heating and cooling processes have been included to calculate consistently the gas temperature in the disk surface region.
In the inner disk region the gas and dust temperatures start to diverge when the $z/r$ ratio reaches about 0.2, which for our 
disk model translates to about $1.5\,H_r$. Thus our approximation of equal dust and gas temperatures is accurate in the
disk midplane and the lower part of the molecular layer, and becomes unrealistic in the disk atmosphere. Yet it
barely affects column densities of many molecules as these are dominated by the molecular concentrations in the dense regions
where $T_{\rm dust}\approx T_{\rm gas}$.

\citet{Heinzeller_ea11} have adopted the gas-grain RATE\,06 \citep{Woodall_ea07} network, no surface reactions apart from 
the H$_2$
formation, and the X-ray and UV-photochemistry (375 species and about 4350 reactions). Heinzeller et al. have concluded 
that water and  hydroxyle abundances in the disk surface regions may increase substantially via neutral-neutral reactions
with molecular hydrogen produced via chemisorbed-assisted surface recombination. It has been found that the disk wind has
a negligible effect on disk chemistry as its upward transport speed is too low compared to the longest chemical timescale
attributed to the adsorption. Note that in our model the adsorption timescale is also long, but the surface chemistry timescales
are even longer ($\ga 1$~Myr). They have pointed out that the radial accretion flow alters the molecular abundances
in the cold midplane, whereas diffusive turbulent mixing affects the disk chemistry in the warm molecular layer 
($r=1.3$~AU, $T\sim 200$~K).
We find that the radial turbulent transport can affect abundances in the midplane at $r\ga 10$~AU 
(see, e.g., CO$_2$ ice in Fig.~\ref{fig:O}), while effective vertical mixing operates from midplane well through the molecular 
layer and lower disk atmosphere. They have concluded that diffusive mixing smoothens the chemical gradients and that the
abundances of NH$_3$, CH$_3$OH, C$_2$H$_2$ and sulfur-containing species are the most enhanced. This has been related to
increased ammonia abundances in the transport model, in which NH$_3$ is effectively produced via gas-phase reactions with more 
abundant oxygen.
In our simulations temperatures are usually well below 200~K, but we also see that sulfur-bearing molecules are among the 
most sensitive species to the disk dynamics, along with complex organics (like methanol), hydrocarbons (like acetylene), and
other 
species (like ammonia). This is due to their dependence on slow surface processes and evaporation (for heavy gaseous species).

{\em (Future directions of development for disk chemical models)}
Overall, there is a considerable progress over the last decade in constructing more feasible chemo-dynamical models of the 
early Solar nebula and other protoplanetary disks. It will gain further momentum in light of the forthcoming ground-breaking 
Atacama Large Millimeter Array (ALMA) currently under construction in Chile. In what follows, we surmise possible future
directions of the development of the nebular and protoplanetary disks chemical models. 
The full 3D chemo-MHD models of the entire disk evolution within 1-5~Myr are beyond possible even with modern 
computational resources available for astronomers. Only local 3D models with radically reduced chemistry and a short
evolutionary time span of $1\,000-10\,000$~years or semi-analytical models of the entire disk with extended chemistry 
and $t\sim 1$~Myr will be feasible in the foreseeable future. The inner, planet-forming disk regions subject to disk-planet(s)
interactions
(gaps, spiral waves, shocks) may well be far from axial symmetry and require 3D modeling approach 
\citep[e.g.,][]{Wolf_ea02,Fea04,Pietu_ea05}. The accurate opacities of dust and gas covering a wide range of temperatures, 
densities, gas composition, and dust topological and mineralogical properties will have to be developed 
\citep[e.g.,][]{RP_opacities,Helling_Lucas09}.

Likely, the steady-state disk physical model adopted in the present study needs 
to be re-adjusted into an evolutionary model with consistent calculations of grain evolution 
\citep[e.g., ][]{Birnstiel_ea10a,Fogel_ea11,Vasyunin_ea11}, photoevaporation \citep[e.g.,][]{Gorti_ea09}, and 
accurate gas temperature \citep[e.g.,][]{Gorti_Hollenbach04,Woitke_ea09}.
The grain coagulation, fragmentation, and sedimentation lead to re-distribution of the total dust surface area, 
higher UV ionization rates, and thus to shift of molecular layers toward the midplane, with a significant increase of 
molecular column densities in the intermediate layer and decrease of their depletion zones \citep{Fogel_ea11,Vasyunin_ea11}. 
In the evolutionary disk models the total disk mass and size, and thus its density, thermal, and ionization structures 
change with time, altering conditions at which chemical processes proceed both in the gas phase and onto the dust 
surfaces \citep{Gorti_ea09}. In evolutionary models covering formation and build-up disk phases, a large amount of 
gas and dust materials experience events of heating, cooling, re-condensation, annealing, and varying irradiation 
intensities \citep[e.g.,][]{G02,Gail_04,Visser_ea09}. 

A more realistic 2D or full 3D prescription of the X-ray and UV radiation transfer modeling with scattering 
has to be included. The accurately calculated UV spectrum with the L$_\alpha$ line is vital to compute realistic
photodissociation and photoionization rates, and shielding factors for CO and H$_2$  
\citep[see, e.g.,][]{vZea03,Bethell_ea07,Visser_ea09b}. The realistic (may be, variable) stellar X-ray spectrum is required to 
calculate gas temperature in the disk atmosphere, and for modeling disk ionization structure and ion-molecule chemistry
\citep[e.g.,][]{Glassgold_ea05,Meijerink_ea08,Aresu_ea10a}. At densities of $\la 10^{4}-10^{6}$~cm$^{-3}$ gas temperature 
decouples from that of dust, 
which will affect the chemical and dynamical evolution in the upper disk layers
\citep[e.g.,][]{Kamp_Dullemond94,Owen_ea11a}. The dissipation of Alfvenic waves generated by the MHD processes inside the disk can
also heat gas in the disk atmospheres \citep[e.g.,][]{Hirose_Turner11a}.

In comparison to the status of the disk physical structure where many key issues have been realized and partly solved,  
the advances in modeling disk chemistry are less straightforward. First of all, chemical studies employ various 
networks fully or partly based on astrochemical ratefiles that have been developed to model
distinct astrophysical environments. Hopefully, this will be relieved with the advent of the public Kinetic Database for 
Astrochemistry (KIDA)\footnote{http://kida.obs.u-bordeaux1.fr} that includes state-of-the-art rate data \citep{Wakelam_ea10}.
Unfortunately, the intrinsic uncertainties in gas-phase reaction rates hamper the accuracy of chemical predictions both on
abundances 
and column densities \citep[see, e.g.,][]{Wakelam_ea06a,Vasyunin_ea08,Wakelam_ea10a}. The pace at which these
quantities are measured in laboratories or calculated by quantum chemical models precludes the rapid progress for thousands 
of astrochemically-relevant processes \citep{Savin_ea11a}. 

In the inner ($\la 5$~AU) disk regions with $T\ga100-300$~K and
densities exceeding $10^{12}$~cm$^{-3}$ three-body processes, many reverse reactions, and neutral-neutral 
reactions with large barriers are activated, and have to be fully taken into account \citep[e.g.,][]{Aea02,TG07,Harada_ea10a}.
The X-ray-driven ionization and dissociation of molecules other than H$_2$ are poorly understood, but important
for disk chemistry \citep[e.g.,][]{Glassgold_ea09}.
In the upper disk regions, upon gas-phase or surface recombination or due to ionization/dissociation 
an excess of energy may translate into (ro-)vibrational excitation of a product molecule, which may react differently with 
other species \citep[e.g.,][]{Pierce_AHearn10a}. This aspect has so far been almost completely neglected in astrochemical models.

In cool outer disk regions gas-grain interactions and surface reactions are essential, yet the latter are often disregarded
in disk models. The same is true for nuclear-spin-dependent chemical reactions involving ortho- and para-states of 
key species, such as H$_3^+$, H$_2$, H$_2$O, etc. \citep[e.g.,][]{Pagani_ea92,Pagani_ea09a,Crabtree_ea11a}.
Usually weak physisorption of molecules is considered in chemical models, while molecules can also form chemical bonds 
with dust surfaces, leading to heterogeneous surface chemistry active both in cold and hot regions, particularly on PAHs 
and carbonaceous grains \citep[e.g.,][]{Fraser_ea05,Cazaux_ea05,Cuppen_Hornekaer08a}.
Despite the recent laboratory efforts to measure binding energies  of key molecules like CO, N$_2$, water, etc. 
to various astrophysical ices and their photodesorption yields, many of these values are still lacking accurate estimates
\citep[][]{Bisschop_ea06,Oeberg_ea09a,Oeberg_ea09b}. The dynamics, reactivity, photodissociation, and desorption of ices 
embedded into dust mantles are hard to measure or model, and even harder to interpret 
\citep[see, e.g., results for the water ice][]{Andersson_ea06a,Andersson_vD08,Bouwman2_ea11a}. 	
For example, sub-surface diffusion may increase desorption efficiency in the case of well-mixed ices within the H$_2$O 
ice matrix. Along with thermal, UV-, CRP- and X-ray-triggered desorption \citep[][]{leger,Najita_ea01,Walsh_ea10}, 
typically considered in disk chemical studies, other non-thermal mechanisms,
like grain-grain destruction and explosive desorption \citep[][]{Shalabiea_Greenberg94} can be operative.  
The surface recombination on porous grain surfaces or within heterogeneous ices is also non-trivial to model accurately
\citep[e.g.,][]{Cuppen_ea09a,Fuchs_ea09a,Ioppolo_ea11b}. The surface chemistry can be restricted to several uppermost monolayers
of the grain, while often in chemical models the corresponding rates are calculated assuming it is active everywhere
in the ice mantle \citep[e.g.,][]{HH93}. In addition, the presence of large, photostable PAHs in the gas 
assists synthesis of polyatomic molecules by providing surface area for surface recombination, ability to re-radiate energy
releazed upon non-destructive ion-PAH recombination, and due to the combustion chemistry in the inner disk region
\citep[e.g.,][]{Wakelam_Herbst08,Kress_ea10a,PerezBecker_Chiang11a}. Finally, calculated molecular concentrations depend on
the adopted set of elemental abundances that may vary from region to region, and underlying assumptions about depletion of
heavy elements from the gas phase \citep[e.g.,][]{Wakelam_ea10a}.

\subsection{Cold molecules in DM Tau}
\label{sec:diss:cold_gases}
\begin{figure*}
 \begin{center}
\includegraphics[height=8cm]{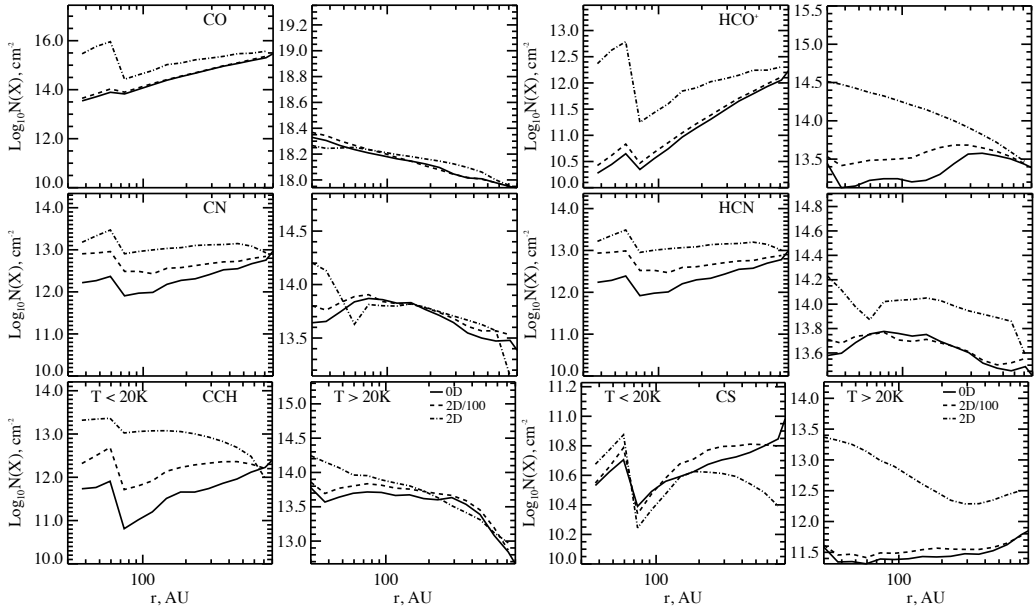}
\caption{(Top to bottom) The vertical column densities of CO, HCO$^+$, CN,
HCN, CCH, and CS at $t=5$~Myr in the cold ($T<20$~K; left) and
warm ($T>20$~K; right) disk regions. Results are shown for the 3 disk
models: (1) the laminar chemistry (solid line), (2)  the 2D-mixing model with
$Sc=100$ (dashed line), and (3) the 2D-mixing model with
$Sc=1$ (dash-dotted line).}
\label{fig:cold_gasses}
 \end{center}
\end{figure*}

In the following two subsections we elaborate on feasibility of the present study to explain observations of diagnostic molecules
and ices in protoplanetary disks and the Solar system.

The kinetic temperature distributions have been measured in nearby disks of \object{DM Tau}, \object{LkCa 15}, \object{MWC 480},
and 
\object{AB Aur} by \citet{DDG03,Pietu_ea05,Pietu_ea07} and \citet{Henning_ea10}, using the Plateau de Bure Interferometer and 
the (1-0) and (2-1) emission lines of CO isotopologues,
HCO$^+$, and C$_2$H. These emission lines have different opacities and thus sample gas temperature at various heights above the
disk equatorial plane \citep[see][]{DDG03}. CO lines are particularly suitable for such studies because they are easily excited
at 
low densities of $10^4$~cm$^{-3}$, while the CO isotopologue abundances vary by orders of magnitude. 
Consistently with predictions from a passively heated, flaring model \citep[e.g.,][]{CG97,DAea99,DD04}, vertical temperature
gradients
have been found. In disks around hot stars midplane temperatures are quite high. At $\sim 100$~AU, the disk around the 10\,000~K
Herbig~A0 \object{AB Aur} has gas temperatures ranging 
from $20\pm3$~K (as measured by C$^{18}$O $J=1-0$) to $35-40$~K ($^{13}$CO $1-0$, $2-1$) and to $68$~K ($^{12}$CO $2-1$), 
see \citet{Pietu_ea05}. A similar trend has been obtained for a disk surrounding the cooler, $8\,500$~K Herbig~A4 \object{MWC
480} 
star, with CO temperatures growing from $21\pm4$~K in the midplane to $48\pm1$~K in the molecular layer 
at 100~AU from the star \citep{Pietu_ea07}. 


In the bright, large 800~AU disk around cold, $3\,750$~K (M1) \object{DM Tau} star very cold CO temperatures of $\approx 15$~K 
at 100~AU have been reported by \citet{DDG03}, based on analysis of the PdBI $^{13}$CO (1-0) spectral map. 
In the later study by \citet{Pietu_ea07} even lower temperatures of 8 and 15~K probed by the (1-0) and (2-1) transitions of
$^{13}$CO, and $T_{\rm kin}=14\pm2$~K probed by the HCO$^+$ (1-0) line have been derived for the outer \object{DM Tau}
disk. In contrast, in the same disk optically thick $^{12}$CO lines give kinetic temperature of about 25~K.
\citet{Henning_ea10} found very low kinetic temperature of $\approx 7$~K for C$_2$H in the \object{DM Tau} system,
using the CCH (1-0) and (2-1) PdBI data. Finally, in Chapillon et al. (2011, submitted) similarly low values for
the CN (1-0) temperatures have been inferred. While derived low temperatures are characteristic for midplane regions of
protoplanetary disks surrounding cool T~Tauri stars, the presence of gaseous molecules at such extreme conditions is puzzling.
The evaporation temperatures for CO, C$_2$H, and CN are about 20, 30, and 40~K, respectively (see Table~\ref{tab:des_E}),
and in dense, dark midplane ($T\sim 10-20$~K) they are rapidly depleted, $\tau_{\rm acc}\la 10^3-10^4$~years
(Tables~\ref{tab:tau_inner}--\ref{tab:tau_outer}). 

In Paper~I \citep{Semenov_ea06} the problem of the cold CO gas reservoir in the \object{DM Tau} disk has been tackled
using the 1+1D \citet{DAea99} $\alpha=0.01$ flaring disk model with the 2D radial-vertical turbulent diffusion.
We have found that the CO column density of about $10^{16}$~cm$^{-2}$ is required in the disk midplane
to explain the puzzling observations of the cold CO in \object{DM Tau}. Such CO column densities in the disk region
with $T\le25$~K have been obtained in the model with vertical and radial mixing at all radii, while in the laminar 
model these values only appeared at $r\ga 500$~AU. The CO molecules are transported from the warm molecular layer
to the cold disk midplane, maintaining concentration of CO even at late times, $t\ga10^5-10^6$~years. 

\citet{Aikawa_07} have used a steady-state $\alpha$-disk structure and a 1D-vertical mixing model to show that 
the warm CO gas from the intermediate layer can be transported down to the cold midplane at a rate that can be competitive 
with the CO adsorption rate. The efficiency of this process has been found particularly pronounced for the disk model
when moderate grain growth is allowed, with grain sizes $\la 1\mu$m. A substantial grain growth leads to a decrease of the
averaged dust surface area and thus less active freeze-out \citep[see also results of][]{Fogel_ea11,Vasyunin_ea11}.

Later, \citet{Hersant_ea09} have utilized a power-law,
$\alpha$-disk model with an isothermal midplane, strong UV scattering, and 1D-vertical mixing. They have found that the
UV-desorption of CO and other ices prevails over their upward transport by the vertical mixing to warm regions, 
leading to a large amount of gaseous CO, HCO$^+$, HCN in the midplane even in the laminar model. 
This is due to the adopted high UV-desorption yield of $\sim 0.1\%$ \citep[e.g.,][]{Oeberg_ea09a} 
and assumed high UV penetration efficiency caused by the UV scattering on small dust grains in the disk atmosphere. 
  
Compared to Paper~I, we now use the full chemical network based on the OSU model, higher photodesorption yields of 
$10^{-3}$, and the slow surface hydrogenation rates based on the results of \citet{Katz_ea99}, with $5\%$ of products 
to be released back to the gas upon surface recombination \citep{Garrod_ea07}. In Figure~\ref{fig:cold_gasses} we show
the column densities of CO, HCO$^+$, CN, HCN, C$_2$H, and CS at 5~Myr in the disk regions with $T<20$~K (odd rows) and 
$T>20$~K (even rows) calculated with the laminar, the slow 2D-, and the fast 2D-mixing models. Note that in the inner region 
at $r\la30$~AU kinetic temperatures are higher than 20~K even in the midplane (Fig.~\ref{fig:disk_struc}, 1st panel). 

As can be clearly seen, turbulent transport does enhance molecular abundances and column densities in the \object{DM Tau} disk 
midplane for the all species apart from CS. Compared to the results of Paper~I, the CO column densities in the midplane 
region ($T<20$~K)
are increased by the diffusive mixing by 2 orders of magnitude, to values between $10^{15}$ and $10^{16}$~cm$^{-2}$,
while in the laminar model these are still relatively high, $\ga 10^{14}$~cm$^{-2}$. As in the study of \citet{Hersant_ea09}
powerful CRP-induced and UV-desorption keeps some molecules in the gas phase even in the midplane. Transport from the
disk regions with $T\ga 20$~K, like the molecular layer and the inner, accretion-heated midplane at $r\la30$~AU, where
CO can be produced on grains, replenishes gaseous CO in the cold midplane. The column densities of ``warm'' CO are 
much higher, $\approx 10^{18}$~cm$^{-2}$ and are not affected by the disk dynamics. This is because chemical evolution of CO is
relatively fast compared to the transport speeds, while CO is also widespread through the disk, canceling out the chemical 
gradient needed for its diffusion, see Section~\ref{O-species}. The amount of cold CO in the mixing model is hardly sufficient
to explain the puzzling $^{13}$CO PdBI observations of \object{DM Tau}. On the other hand, there is enough warm CO to explain the 
presence of the $T\approx 26$~K CO gas derived from the optically thick $^{12}$CO data tracing the disk upper region. 

HCO$^+$ shows an increase of midplane column densities by up to a factor of 100 in the mixing case,
while in the molecular layer its column densities are enhanced by a factor of $\la 30$. This is due to the sensitivity
of H$_3^+$ ion, required for HCO$^+$ production, to the turbulent transport (see discussion in Section~\ref{ions}). 
In the mixing model, H$_3^+$ is particularly abundant at elevated disk heights, $z\ga 1-2\,H_r$, and so is HCO$^+$. 
The ratio of column densities of warm and cold HCO$^+$ is large, and nearly the same for both the laminar 
and 2D-mixing models, a factor of $\approx 30$ 
and $10-100$, respectively. With a typical column density of $10^{13}-10^{14}$~cm$^{-2}$ at $r\ga100$~AU, 
the HCO$^+$ (1-0) line remains optically thin in the outer disk, sampling both the disk midplane and the warm molecular 
layer. 

On the other hand, the critical density required to excite a molecular line from a linear molecule can be estimated as:
\begin{equation}
\label{eq:ncrit}
 n_{\rm cr} = \frac{A_{ul}}{\sigma v},
\end{equation}
where $A_{ul}$ is the Einstein coefficient for spontaneous emission from the level $u$ to $l$, and $\sigma$ and $v$ are the 
cross-section and velocity of a collisional partner, respectively. For the H$_2$-dominated disk midplane with $T=10-20$~K 
$\sigma v \approx 3\,10^{-11}$~cm$^3$\,s$^{-1}$ \citep[e.g.,][]{Wilson_ea09}. We adopt $A_{10}=4.25\,10^{-5}$~s$^{-1}$ 
from the LAMDA database \citep{lamda}, and find that the critical density for the HCO$^+$ $J=1-0$ excitation is 
$\approx 1.5\,10^6$~cm$^{-3}$. The region where densities exceed $10^6$~cm$^{-3}$ extends vertically up to 
$\la 3$ and $1\, H_r$ in the inner and outer disk, respectively, and covers the midplane and the molecular layer 
(see Fig.~\ref{fig:disk_struc}). Thus, HCO$^+$ (1-0) emission would originate from the $T>20$~K disk region, and the resulting 
temperature derived from this transition would likely exceed 15~K obtained from interferometric observations by
\citet{Pietu_ea07}.

The behavior of C$_2$H, CN, and HCN is similar as their chemical evolution is tightly linked (Section~\ref{N-species}). 
Their vertical column densities are not much altered by the turbulent mixing, 
having a typical value of $\sim 3\,10^{13}-10^{14}$~cm$^{-2}$ at 100~AU. Since absolute concentrations of C$_2$H, CN, and 
HCN are highest in the molecular layer where chemical timescales are relatively short, 
their $T>20$~K column densities are not strongly increased by diffusion, less than by a factor of 3. Contrary, the column
densities 
in the \object{DM Tau} midplane ($T<20$~K) are raised by the transport by an order of magnitude for CN and HCN, and by 2 orders of
magnitude for C$_2$H. This is caused by the long evolutionary timescales in the midplane associated with gas-surface kinetics
(e.g., gaseous HCN is produced in the midplane upon recombination of the H and CN ices).
The resulting column densities through the midplane are $\ga 10^{13}$~cm$^{-2}$. Thus, in the 
fast 2D-mixing model column densities of warm and cold C$_2$H, CN, and HCN differ 
by only a factor of several, which further decreases outward. Using expression~(\ref{eq:ncrit}) and the LAMDA
database, we estimate that the optically thin 1-0 rotational lines of C$_2$H, CN, and HCN are excited at densities 
of about $4-8\,10^5$~cm$^{-3}$. Consequently, in the presence of
strong turbulent mixing low-lying transitions of C$_2$H, CN, and HCN 
likely trace the cold midplane of \object{DM Tau}, supporting the observational evidence.

Finally, turbulent transport decreases CS column densities in the midplane region by a factor of 
$\la$ 4, simultaneously increasing it in the $T>20$~K zone by up to 2 orders of magnitude. The chemistry of sulfur-bearing
molecules is among the most altered by the mixing due to slow surface processes associated with these heavy
species (see Section~\ref{S-species}). However, even in the laminar model column densities of 
cold ($T<20$K) CS are lower than that of warm ($T>20$K) CS by an order of magnitude. The overall CS column densities
in the outer disk are $\approx 3\,10^{11}-3\,10^{12}$~cm$^{-2}$, and the low-$J$ CS lines are optically thin.
The critical densities of the excitation for the (2-1) and (3-2) CS transitions at 98 and 147~GHz are $\sim 5\,10^6$ 
and $2\,10^6$~cm$^{-3}$, respectively. Therefore, if our model is correct, the kinetic temperatures derived from the 
low-lying CS emission lines should be above 20~K in the \object{DM Tau} disk.

Clearly, our modeling shows a potential of turbulent transport as a cause for the presence of molecules in the 
cold midplanes in T~Tauri disks. The non-thermal broadening of the CO lines by $\la 150$~m\,s$^{-1}$ 
in the \object{DM Tau} system has been reported in \citet{Pietu_ea07}, which corresponds to the turbulent velocities of 
$\la 10\%$ of the sound speed. 
However, the increase in concentrations of the cold gases in our 2D-mixing model is not strong enough to
rule out other explanations. First of all, more accurate modeling of the UV scattering toward the midplane
would result in faster photoevaporation of ices in the lower molecular layer and upper midplane \citep[][]{vZea03,Hersant_ea09}.
Second, as mentioned in \citet{Aikawa_07}, moderate grain growth beyond $1~\mu$m  in disk central regions
lengthens the depletion time for molecules such that the gas-phase CO abundances may remain high even after $1$~Myr of the
evolution. Indeed, \citet{Vasyunin_ea11} and \citet{Fogel_ea11} have modeled chemical evolution in disks taking grain evolution
into account and found that the grain growth substantially decreases the depletion zones of molecules. Moreover,
from analysis of the SED slopes in millimeter and centimeter wavelengths large grain sizes of $1$~mm 
have been inferred for many young systems in various star-forming regions, including Taurus-Auriga association
\citep[e.g.,][]{Rodmann_ea06,Lommen_ea10a,Ricci_ea10a}. The \object{DM Tau} disk has an 
inner hole of $\sim4-20$~AU with grains as large as 1~mm, and hence is in a transitional phase
\citep[e.g.,][]{Calvet_ea05,Sargent_ea09,Andrews_ea11a}. From advanced theoretical models of grain coagulation, 
fragmentation, sedimentation, and turbulent stirring the grain evolution should proceed everywhere in the disk, 
though the mean grain sizes remain smaller in the outer disk compared to the inner disk region
\citep[e.g.,][]{Brauer_ea08a,Birnstiel_ea10a}. Recently, \citet{Guilloteau_ea11a} have used high-resolution, multi-frequency
interferometric PdBI observations to discern more accurately the dust emissivity slopes at millimeter wavelengths in a sample
of young stars. Their analysis has indeed shown that the outer disk of \object{DM Tau} contains large grains with sizes 
$\ga1~\mu$m. We advocate for a combined action of turbulent transport and grain growth as a mechanism to maintain 
a sizable reservoir of cold gases in the \object{DM Tau} system.
$^{12}$CO(3-2)

\subsection{Comparison with observations}
\label{sec:diss:obs}
\begin{deluxetable*}{llcccc}
\tablewidth{0pt}
\tablecaption{Observed and modeled column densities in DM Tau at 250~AU
\label{tab:best_fit}}
\tablehead{
Species & Observed & Refs & \multicolumn{3}{c}{Modeled} \\
& & & Laminar & 2D-mixing ($Sc=100$) & 2D-mixing ($Sc=1$)}
\startdata
    $^{12}$CO$^*$  &   3.0 (17)       & (1) & 3.0 (17) & 3.0 (17) & 3.0 (17)\\
\hline
    HCO$^+$        &   1.7 (13)       & (1,2) & 8.0 (12) & 1.2 (13) & 2.5 (13)\\
    H$_2$CO        &   1.0-2.0 (13)   & (3) & 6.2 (12) & 6.7 (12) & 4.0 (12)\\
    N$_2$H$^+$     &   4.0 (11)       & (2) & 3.4 (11) & 4.7 (11) & 1.2 (12)\\
      CS           &   4.0 (12)       & (3) & 8.1 (10) & 1.1 (11) & 5.6 (11)\\
      CN           &   4.0 (13)       & (4) & 1.4 (13) & 1.5 (13) & 1.8 (13)\\
     HCN           &   8.0 (12)       & (4,5) & 1.2 (13) & 1.2 (13) & 2.7 (13)\\
     HNC           &   3.0 (12)       & (4,5) & 1.0 (13) & 1.0 (13) & 2.4 (13)\\
     C$_2$H        &   3.0 (13)       & (6) & 1.1 (13) & 1.3 (13) & 1.4 (13)\\
\hline
Agreement$^{**}$	&                &        & 7/8     & 7/8     & 6/8  \\
\enddata
\tablenotemark{*}\tablenotetext{*}{The calculated column densities are scaled down to match the observed values for $^{12}$CO.
The renormalization factor is applied to the column densities of other species.}
\tablenotemark{**}\tablenotetext{**}{Agreement is achieved when the observed
and calculated column densities for a molecule do not differ by more than a factor of 4).}
\tablerefs{(1) Pietu et al.~(2007); (2)  Dutrey et al.~(2007); (3) Dutrey et al. (2011, submitted);
(4) Chapillon et al. (2011, submitted); (5) Schreyer et al.~(2008); (6) Henning et al.~(2010)}
\end{deluxetable*}

\begin{deluxetable}{lclccc}
\tablewidth{0pt}
\tablecaption{Observed and modeled column densities of ices in DM Tau at $r=10-30$~AU
\label{tab:comets}}
\tablehead{Species & Relative  & Ref & 10~AU & 20~AU & 30~AU\\
                   & abundance &     &          &          &      }
\startdata
H$_{2}$O           & 100       &     & 100      & 100      & 100\\
CO	           & $5-30$    & (1,2,3,4) &  $<0.01$ &  0.7       & 16 \\
CO$_{2}$           & $3-20$     & (1,4) & 10       & 20       & 0.6\\
H$_{2}$CO          & $0.04-1$  & (1,2,3,4) & 0.02      & 0.015     & $<0.01$\\
CH$_3$OH           & $2$       & (2,3,4) & 0 & $<0.01$ & $<0.01$\\
HCOOH              & $0.06-0.09$    & (3,4) & 0.02 & 0 & 0\\
HCOOCH$_3$         & $0.06-0.08$    & (3,4) & 0 & 0 & 0\\
CH$_{4}$           & $0.6-5$     & (1,3,4) & 0  & 40       & 30 \\
C$_2$H$_2$         & $0.1-0.5$     & (3,4) & 0  & 0       & 0 \\
C$_2$H$_6$         & $0.3-0.4$     & (3,4) & 0.07  & 0       & 0 \\
NH$_{3}$           & $0.1-2$ & (1,3,4) & 33       & 20       & 15 \\
HCN                & $\le0.25$ & (1,2,3,4) & 12        & 3.6        & 0.5 \\
HNC                & $0.01-0.04$ & (4,4) & 0.5        & 0.02        & 0 \\
HNCO               & $0.06-0.1$ & (3,4) & 0        & 0       & 0 \\
N$_{2}$	           & $\sim 0.02$ & (1)& 0       & 0        & 0 \\
CH$_3$CN           & $0.01-0.02$ & (3,4) & 0        & 0       & 0 \\
H$_3$CN            & $0.02$ & (3,4) & 0.003        & 0       & 0 \\
NH$_2$CHO          & $0.01-0.02$ & (3,4) & 0        & 0       & 0 \\
H$_2$S	           & $0.75-1.5$ & (1,2,3,4)& 0.15     & 0.09     & 0.06 \\
CS	           & $\sim 0.1$ & (2)& 0       & 0        & 0 \\
SO	           & $0.2-0.8$ & (1,3,4)& 0       & 0        & 0 \\
SO$_{2}$           & $0.1-0.2$ & (1,3,4)& 0       & 0        & 0 \\
OCS	           & $0.1-0.4$ & (1,3,4)& 0       & 0        & 0 \\
H$_2$CS	           & $0.02$ & (3,4)& 0.006     & 0     & 0 \\
S$_2$              & $0.005$ & (4)& 0     & 0     & 0 \\
\enddata
\tablerefs{(1) Aikawa et al.~(1999); (2) Biver et al.~(1999); (3) Bockel{\'e}e-Morvan et al.~(2000);
(4) Crovisier \& Bockel{\'e}e-Morvan~(1999)}
\end{deluxetable}

{\em (Molecules in protoplanetary disks)}
Some previous studies of the laminar disk chemistry have reported reasonable quantitative agreement 
with observationally-inferred column densities \citep[e.g.,][]{Aea02,Semenov_ea05,Dutrey_ea07,Schreyer_ea08}.
On the other hand, the radial profiles of the column densities derived from fitting high-resolution interferometric data
have not been fully reproduced by conventional disk models. In the 1D-vertical mixing study of \citet{Willacy_ea06}
the modeling results have been compared with single-dish and interferometric observations of several molecules in the
disks around \object{DM Tau}, \object{LkCa 15}, and \object{TW Hya}. The good agreement between theoretical and observed column
densities for
the fast mixing model has been inferred from the single-dish data, whereas the interferometric data have not been reproduced.
Calculating infrared emission lines for the disk model with radial advective and vertical mixing transport,
\citet{Heinzeller_ea11} have shown that it improves agreement with the {\it Spitzer} observations of the inner disks around 
\object{AA Tau}, \object{DR Tau}, and \object{AS 205}, compared to predictions of the laminar chemical model. 

We compile a table with column densities at 250~AU derived from the analysis of high-quality PdBI interferometric
observations and compare them with the results of our laminar, slow, and fast 2D-mixing models (see Table~\ref{tab:best_fit}).
Calculated column densities have intrinsic uncertainties of a factor $\sim3-5$, caused by reaction rate uncertainties
\citep{Vasyunin_ea08}, whereas observational data suffer from calibration inaccuracies ($\sim 10-20\%$), 
distance uncertainties ($10-20\%$), etc. Therefore, we assume that the agreement is good when observed and modeled 
values differ by a factor of $\la 4$ and bad otherwise. The disk total surface density cannot be accurately derived
from the continuum data, as these suffer from poorly known dust opacities at (sub-)millimeter wavelengths that can vary 
by factors of several \citep{RP_opacities}. Instead, the observed CO column density distribution can be used as a proxy 
to the disk gas density structure. The problem here is that the $^{12}$CO rotational lines are optically thick and thus probe
localized disk region, whereas from optically thin CO isotopologue lines the total CO column densities can only be recovered 
if $^{12}$C/$^{13}$C or $^{16}$O/$^{18}$O isotope ratios are known. We use $^{13}$CO and $^{12}$CO column densities 
obtained for the outer \object{DM Tau} disk from the interferometric observations by \citet{Pietu_ea07} and constrain the 
total CO column density at 250~AU to $\approx 3\,10^{17}$~cm$^{-2}$. The corresponding theoretical value is almost the same in 
the laminar and mixing models, $N(\rm {CO})\approx 10^{18}$~cm$^{-2}$ (Fig.\ref{fig:O}). 
To match this value, modeled column densities in Table~\ref{tab:best_fit} were renormalized
accordingly.

Unlike \citet{Willacy_ea06}, all the considered models match quite well the observed column densities, apart from the 
CS data. The stellar and disk parameters are well known for the nearby \object{DM Tau} system, so its physical structure can be
reliably reconstructed. The detected species are simple and their chemical evolution is governed by a limited 
amount of mostly gas-phase processes, with many accurately acquired rate constants. However, CS column densities are 
underpredicted by factors of $\sim 7-50$, though situation considerably improves for the fast 2D-mixing
model. Dutrey et al. (2011, submitted) have observed SO, H$_2$S, and CS in \object{DM Tau}, \object{LkCa 15}, and \object{MWC 480}
with 
the IRAM 30-m and PdBI interferometer. They have derived upper limits for H$_2$S and SO in the \object{DM Tau} disk at 300~AU, 
namely, $N({\rm H}_2{\rm S})\la 2\,10^{11}$~cm$^{-2}$ and $N({\rm SO})\la 8\,10^{11}$~cm$^{-2}$, whereas CS has been firmly 
detected with the column density of $N({\rm CS})\la 3.5\,10^{12}$~cm$^{-2}$. Dutrey et al. have found that the sulfur content 
of the three disks cannot be explained by modern chemical models as these result in higher abundances of H$_2$S and SO 
compared to the CS abundances. This is also the case in our modeling. While having too less CS, our disk models overproduce 
column densities of sulfur monoxide and hydrogen sulfide. The reason for such a disagreement, as we discussed in 
Section~\ref{S-species}, is that the sulfur chemistry is hampered by poorly known reaction data and may lack key reactions, 
and the depletion of elemental sulfur from the gas is also crudely constrained.

Generally, the fast 2D-mixing model produces more molecules per CO compared to the laminar disk chemistry, and overproduces 
HNC with respect to the observations. The utilized chemical network leads to the synthesis of hydrogen cyanide and isocyanide
with almost equal probability in the all disk models. 
On the other hand, the observed column density of HCN is higher than that of HNC by a factor of $\sim 2-3$. 
The HCN/HNC ratios of $\la1$ are typical for dark, dense cores \citep[][]{Tennekes_ea06a,HilyBlant_ea10a}, while in 
irradiated environments like PDR regions this ratio exceeds 1 \citep[e.g.,][]{Loenen_ea08a}. Indeed, in the study of 
\citet{Fogel_ea11} with accurate description of the UV RT including the L$_\alpha$ radiation and the moderate grain growth 
the resulting ratio of HCN/HNC is about 1.5. The HCN/HNC ratio of 3 has been obtained by \citet{Vasyunin_ea11} for the 
disk model with consistently calculated grain evolution and the UV penetration. An increase in the HCN/HNC ratio to 3-5 
due to lowering the disk mass or including dust sedimentation has been reported by \citet{ah1999}.
Furthermore, \citet{Sarrasin_ea10a} have 
accurately calculated the collisional data for HCN and HNC with He and found that the use of the HCN rates to interpret 
HNC observations in dark clouds may lead to the underestimation of the HCN/HNC ratio. Apparently, the use of ISM-like 
dust grains or the crude approach to calculate UV penetration utilized in our disk model make the UV opacities too high
in the molecular layer to result in HCN/HNC$>1$.

Column densities or upper limits have also been reported for other species.
Using the IRAM 30-m antenna, \citet{DGG97} have not been able to detect molecular lines of SiO, SiS, HC$_3$N, C$_3$H$_2$,
CH$_3$OH, CO$^+$, SO$_2$, HNCS, HCOOCH$_3$ in \object{DM Tau}. Among these species, only SiO, C$_3$H$_2$, and SO$_2$ have 
large modeled column densities between $10^{12}$ and $10^{13}$~cm$^{-2}$ at 250~AU. The SO$_2$ lacks permanent dipole moment 
and has no pure rotational spectra, unless oxygen atoms are isotopically different. Its ro-vibrational lines are excited
at $T\ga 50-100$~K, so that the emission arises only from an inner disk region ($r\la 5-10$~AU). 
The C$_3$H$_2$ molecule can be in ortho- or para-state 
\citep[e.g.,][]{Madden_ea89a,Morisawa_ea06a}, 
though in our model these details are not taken into account. The linear isomer H$_2$CCC (propadienylidene) has a large 
dipole moment of about 4~D and rotational spectrum starting at cm wavelengths. However, the IRAM 30-m observations of TMC1 and 
IRC +10216 by \citet{Cernicharo_ea91a} have revealed that the abundance of the linear isomer is only $\sim 1\%$ of that of 
the cyclopropenylidene. The cyclic isomer has a rich ro-vibrational spectrum, with emission lines at (sub-)millimeter
wavelengths that are excited at $n_{\rm cr}\sim 10^6$~cm$^{-3}$ \citep[LAMDA database;][]{Green_ea87a,lamda}.
Due to energy partition the individual lines are not as strong as in the case of HCO$^+$, making them hard to detect.
Finally, the outflow tracer, SiO, has a dipole moment of 3.1~D and strong rotational lines at (sub-)mm,
excited at $\sim 10^5$~cm$^{-3}$ \citep[][]{lamda}. The calculated column density of SiO at 250~AU 
is $4\,10^{12}$~cm$^{-2}$ in the laminar model and $10^{13}$~cm$^{-2}$ in the fast 2D-mixing model. These values are 
sensitive to the abundance of the elemental silicon remained in the gas and may not be representative of \object{DM Tau} in 
the ``low metals'' set of initial abundances of \citet{Lea98} adopted in our modeling.

The recent non-detection/tentative detection of cold water vapor in the DM Tau disk by the {\it Herschel}/HIFI
have been reported by \citet{Bergin_ea10a}. They have interpreted the observational data with an advanced chemical disk and 
the line radiative transfer model, and inferred the disk-averaged water column densities within 
$\sim 5\,10^{12}-3\,10^{13}$\,cm$^{-2}$. Our calculated disk-averaged H$_2$O column densities are $5\,10^{14}$\,cm$^{-2}$
and $2\,10^{15}$\,cm$^{-2}$ in the laminar and fast mixing models, respectively. Therefore, all our models overestimate
the abundance of gaseous water in \object{DM Tau} by a factor of at least 15. Similarly large values have been obtained in the
studies 
of disk chemistry with grain evolution by \citet{Vasyunin_ea11} and \citet{Fogel_ea11}. Bergin et al. have concluded that most 
water ice is trapped in large dust grain aggregates that sedimented toward the \object{DM Tau} midplane and thus cannot be easily 
photodesorbed. However, if this is the case, other molecules like CO, CN, etc. have to be depleted from the 
gas phase much more severely, and should not be present in the cold midplane. The presence of large reservoir of cold CO, CN, HCN,
C$_2$H observed 
in the \object{DM Tau} midplane contradicts with such hypothesis, unless other molecules begin accreting onto dust grains later
than 
water, when grains are already large, $\ga 1\,\mu$m.

{\em (Molecules in comets of the Solar system)}
We calculate column densities of ices at 10, 20, and 30~AU and compare them with molecules
observed in comets as done in \citet{Aea99}, see Table~\ref{tab:comets}. Observed values are taken from 
\citet[and references therein]{Aea99,Biver_ea99a,BM_ea00a,Crovisier_BM99a}.
The difference in calculated ice column densities between the laminar 
and mixing models is smaller than their radial variations, so in Table~\ref{tab:comets} only results of the
laminar model are presented. The comets have formed in the early Solar nebula 
at distances of $\ga 10-20$~AU and later have been expelled outwards \citep{Lissauer_87a}. Comets are 
reservoirs of pristine compounds, and their chemical composition is indicative of evolutionary history of the outer 
Solar system during the first several million years \citep[e.g.,][]{Ehrenfreund_Charnley00}. However, molecules detected 
in their comae can be partly photodissociated or are products of immediate coma chemistry and very long irradiation of bulk 
cometary ices by cosmic ray particles, so direct comparison with model predictions is not straightforward. 
The presence of both reduced and oxidized ices in comets have been explained by \citet{Aea99} as due to CRP-ionization
of the disk, whereas in cosmochemical models the Fischer-Tropsch catalysis or cloud-nebula evolutionary 
scenario have been invoked \citep[e.g.,][]{Greenberg_82,Prinn_Fegley89}. We find that our disk model incorporating the 
high-energy radiation is able to reproduce relatively high abundances of water, CO, and CO$_2$ ices, as well as
abundances of CH$_4$ and C$_2$H$_6$ ices. Also, abundances of H$_2$CO, HCOOH, HCN, HNC, H$_2$S,
and H$_2$CS are reproduced. However, the model overproduces NH$_3$ ice by a factor of $\ga 7$, 
and severely underpredicts abundances of most of the sulfur-bearing ices (CS, SO, SO$_2$, OCS, S$_2$) and complex 
organics (CH$_3$OH, HCOOCH$_3$, HNCO, NH$_2$CHO). The chemical model with fast turbulent transport gives better results
for heavy sulfur-bearing and organic molecules, but is still far below the observed values.
As we discussed above, the sulfur chemistry is rather dubious in modern
astrochemical databases. The adopted chemical network has a limited number of surface reactions leading to complex organic 
molecules and not many endothermic neutral-neutral reactions. Calculated ice abundances are also sensitive to the adopted 
binding energies, many of which are not accurately derived.

Apart from that, we conclude, that our disk chemical models with and without transport processes are equally and reasonably well
agree to the 
high-quality interferometric observations of \object{DM Tau} and the chemical composition of comets in the Solar system.

\subsection{Observable molecular tracers of dynamical processes}
\label{sec:diss:tracers}
\begin{deluxetable}{ll}
\tablecaption{Detectable tracers of turbulent mixing\label{tab:tracers}}
\tablehead{
\colhead{Steadfast} & \colhead{Hypersensitive}
}
\startdata
CO          &  Heavy hydrocarbons (e.g.,C$_6$H$_6$) \\
H$_2$O ice   &  C$_2$S \\
       &  C$_3$S \\
       &  CO$_2$ \\
       &  O$_2$ \\
      &  SO \\
      &  SO$_2$\\
      &  OCN \\
      &  Complex organics (e.g., HCOOH)\\
\enddata
\end{deluxetable}

In Table~\ref{tab:tracers} we show most promising tracers of transport processes in protoplanetary disks as found with our 
modeling. The quantity least biased by observational and modeling uncertainties is the ratio of the observed column 
densities of an abundant steadfast species to that of a hypersensitive molecule. Among the insensitive species 
(Table~\ref{tab:steadfast}) the most promising are CO and the water ice as these terminal species incorporate substantial 
fractions of the elemental carbon and oxygen in disks, and are easy to observe. Cold molecular hydrogen is not observable, 
whereas concentrations of C$^+$, light hydrocarbons, CN, HCN, and HNC are more sensitive not to mixing, but to the stellar X-ray and UV radiation 
\citep[e.g.,][]{Fogel_ea11,Aresu_ea10a,Kamp_ea11a}. Other molecules and ices unresponsive to mixing have low abundances
or also model-dependent (e.g., S-bearing ices). Both pure rotational CO lines at (sub-)mm
and ro-vibrational CO lines at IR wavelengths have been detected in inner and outer disk regions 
\citep[e.g.,][]{Pietu_ea07,Oberg_11a,Salyk_ea11a}. Water ice absorption feature at $3\,\mu$m has also been 
detected in protoplanetary disks \citep[e.g.,][]{Terada_ea07a} and envelopes around young protostars 
\citep[e.g.,][]{Boogert_ea08a}.

Among species sensitive and hypersensitive to the turbulent transport (Tables~\ref{tab:sens}--\ref{tab:hyps}) most 
promising are gaseous and solid heavy hydrocarbons, C$_2$S, C$_3$S, SO, SO$_2$, CO$_2$, O$_2$, and complex organic 
molecules (e.g., HCOOH). Basically, any non-terminal, abundant molecule produced mostly via grain-surface kinetics can be 
used as a tracer of dynamical transport in  protoplanetary disks. The OH$^+$ and H$_2$O$^+$ ions are sensitive to the 
disk ionization structure and thus cannot be reliable tracers of turbulent mixing. Due to atmospheric opacity, molecular oxygen 
can only be observed from space \citep[e.g.,][]{Larsson_ea07}. Gas-phase molecules actually detected in disks are 
warm CO$_2$ at IR \citep[e.g.,][]{Salyk_ea11a} and SO at millimeter wavelengths \citep{Fuente_ea10a}. Rotational lines of 
C$_2$S, C$_3$S and various hydrocarbons and hydrocarbon anions have been detected in cold dense cores 
\citep[e.g.,][]{Dickens_ea01a,Kalenskii_ea04,Sakai_ea10a}. The emission lines of SO, SO$_2$, and several organic
molecules have been identified in submillimeter spectra of young stellar objects \citep[e.g.,][]{Joergensen_ea05a}.
CO$_2$ ice feature at $15.2\,\mu$m has been detected in molecular clouds \citep[e.g.,][]{Kim_ea11a} and toward embedded 
young low-mass stars \citep[e.g.,][]{Pontoppidan_ea08b}. In addition, plethora of other ices like CH$_4$, SO$_2$, HCOOH, 
H$_2$CO, and CH$_3$OH have been observed in Class~I/II objects \citep[e.g.,][]{Zasowski_ea08}. 

The column density ratios of these sensitive molecules to the steadfast CO and H$_2$O ice vary between our laminar and fast 
2D-mixing models by 2--4 orders of magnitude (see CDR values in Table~\ref{tab:hyps}). In the not so far distant future 
forthcoming observational facilities like ALMA, extended VLA, and James Webb Space Telescope will allow us to 
observe many of the mixing tracers along with CO and water ice in nearby protoplanetary disks of various ages, masses, and 
sizes, and testify our predictions. 

%

\section{Summary and Conclusions}
\label{concl}
We study the influence of dynamical processes on the chemical evolution of protoplanetary disks.
Our analysis is based on the 2D flared $\alpha$-model of a $\sim5$~Myr \object{DM Tau}
disk coupled to the large-scale gas-grain chemical code. To account for
production of complex molecules, the chemical network is supplied with
a set of surface reactions (up to HCOOH, CH$_3$OH, CH$_3$OCH$_3$, etc.)
and photoprocessing of ices. For the first time our disk model covers a wide
range of radii, 10--800~AU, and includes warm planet-forming zone and cold outer
region. Turbulent transport of gases and ices is modeled using the mixing-length
approximation in full 2D (based on the $\alpha$-prescription for
viscosity). Since turbulent transport efficiency of
molecules in disks is not well known, we consider two dynamical models with the
Schmidt number of 1 and 100. We come up with a simple analysis for laminar
chemical models that allows to highlight the potential sensitivity of a molecule to
turbulent transport. It is shown that the higher the ratio
of the characteristic chemical timescale to the turbulent transport timescale
for a given molecule, the higher the probability that its column density
will be affected by dynamical processes.
With our chemo-dynamical models, we find that the turbulent transport
influences abundances and column densities of many gas-phase species and
especially ices. The results of the chemical model with reduced turbulent
diffusion are much closer to those from the laminar model, but not completely.
Mixing is important in disks since a chemical steady-state is not
reached for many species due to long timescales
associated with surface chemical processes and slow evaporation of heavy molecules ($t\ga10^5$years).
When a grain with an icy mantle is transported from a cold disk
midplane into a warm intermediate/inner region, the warm-up makes heavy radicals
mobile on dust surface, enriching the mantle with complex ices, which
can be released into the gas in appropriate temperature regions.
In contrast, simple radicals and molecular ions, which chemical
evolution proceed solely in the gas phase, are not much affected by dynamics.
We divide all
molecules into 3 distinct groups with respect to the sensitivity of their column
densities to the diffusive mixing. The molecules that are
unresponsive to dynamical transport include such observed and potentially
detectable molecules as C$_2$H, C$^+$, CH$_4$, CN, CO, HCN, HNC, H$_2$CO, OH,
as well as water and ammonia ices. Their column densities computed with the laminar and
fast 2D-mixing model do not differ by more than the factor of $2-5$ (``steadfast'' species). 
The molecules which vertical column densities in the
laminar and dynamical models differ by no more than 2 order of magnitude include,
e.g., C$_2$H$_2$, some carbon chains, CS, H$_2$CS,
H$_2$O, HCO$^+$, HCOOH, HNCO, N$_2$H$^+$, NH$_3$, CO ice, H$_2$CO ice, CH$_3$OH
ice, and electrons (``sensitive'' species). Molecules which column densities are modified by 
diffusion by more than 2 orders of magnitude include, e.g., C$_2$S,
C$_3$S, C$_6$H$_6$, CO$_2$, O$_2$, SiO, SO, SO$_2$, long carbon chain ices,
CH$_3$CHO ice, HCOOH ice, O$_2$ ice, and OCN ice (``hypersensitive'' species).
The sulfur-bearing molecules, along with polyatomic (organic) molecules frozen
onto the dust grains, are among the most sensitive species to the turbulent
mixing. The chemical evolution of assorted molecules in the laminar and turbulent models is 
thouroughly analyzed and compared with previous studies.
We find that the observed column densities in the DM Tau disk are well
reproduced by both the laminar and the mixing disk models. 
The observed abundances of reduced and oxidized cometary ices are also successfully reproduced by the both models.
A combination of efficient UV photodesorption, grain growth, and turbulent mixing leads to 
non-negligible amount of molecular gases in the cold disk midplane.
We propose several observable or potentially detectable tracers of
dynamical processes in protoplanetary disks, e.g. ratios of the CO$_2$,
O$_2$, SO, SO$_2$, C$_2$S, C$_3$S column densities to that of CO and the water ice.
Some of these tracers have been observed in disks by
the current radiointerferometers and infrared facilities (e.g. PdBI, SMA,
{\it Spitzer}, Keck, VLT) and some will be targeted by the {\it Herschel} telescope.
The detection of complex species (e.g., dimethyl ether, formic acid, methyl
formate, etc.) in protoplanetary disks with ALMA and JWST will be a strong indication
that chemical evolution of these objects is influenced by transport processes.

\acknowledgments
This research made use of NASA's Astrophysics Data System.
DS acknowledges support by the {\it Deutsche Forschungsgemeinschaft} through
SPP~1385: ``The first ten million years of the solar system - a
planetary materials approach'' (SE 1962/1-1).
DW acknowledges support from the the Federal Targeted Program ``Scientific and 
Educational Human Resources of Innovation-Driven Russia'' for 2009-2013.

\bibliographystyle{apj}
 \bibliography{references}

\end{document}